\documentclass[twoside,12pt]{PhDthesisPSnPDF}

\usepackage{lineno}
\usepackage{amsbsy}
\usepackage{amsmath}
\usepackage{lscape}
\usepackage{deluxetable}
\usepackage{xspace}
\usepackage{sidecap}
\usepackage{wtmmPkg}
\usepackage{natbib}
\usepackage{multirow}
\usepackage{paralist}
\usepackage{fancyhdr} 
\usepackage{hyperref}
\usepackage{graphicx}
\usepackage{epstopdf}
\usepackage{bibunits}
\usepackage[margin=10pt,font=small,labelfont=bf,labelsep=colon]{caption} 
\hypersetup{colorlinks, citecolor=black, filecolor=black, linkcolor=black, urlcolor=black}

\usepackage[intoc]{nomencl} 
\renewcommand{\nomname}{Glossary} 
\makenomenclature 

\makeatletter

\newcommand{\Rmnum}[1]{\expandafter\@slowromancap\romannumeral #1@}
\makeatother
\newcommand{\ion}[2]{#1~\Rmnum{#2}}

\newcommand{\rsun}{R$_{\odot}$}

\renewcommand{\vec}[1]{ {\mathbf #1} }

\newcommand{\curl}{ {\bf \nabla} \times}
\newcommand{\mdiv}{ {\bf \nabla} \cdot}

\newcommand{\corr}[1]{{{\textcolor{black}{#1}}}}

\newcommand{\aap}{    {Astronomy \& Astrophysics}}

\newcommand{\apj}{    {The Astrophysical Journal}}

\newcommand{\apjl}{   {The Astrophysical Journal Letters}}

\newcommand{\ssr}{    {Space Science Reviews}}

\ifpdf  
    \pdfcatalog { /PageMode (/UseOutlines)
                  /OpenAction (fitbh)  }
\fi

\title{Kinematic Properties of Globally-Propagating Waves in the Solar Corona}

\ifpdf
  \author{\href{mailto:longda@tcd.ie}{David M. Long}, B.A.~(Mod.)}
  \collegeordept{\href{http://www.physics.tcd.ie}{School of Physics}}
  \university{\href{http://www.tcd.ie}{Trinity College Dublin, Ireland}}

  \crest{
  \vspace{1cm}
  \includegraphics[width=0.3\textwidth]{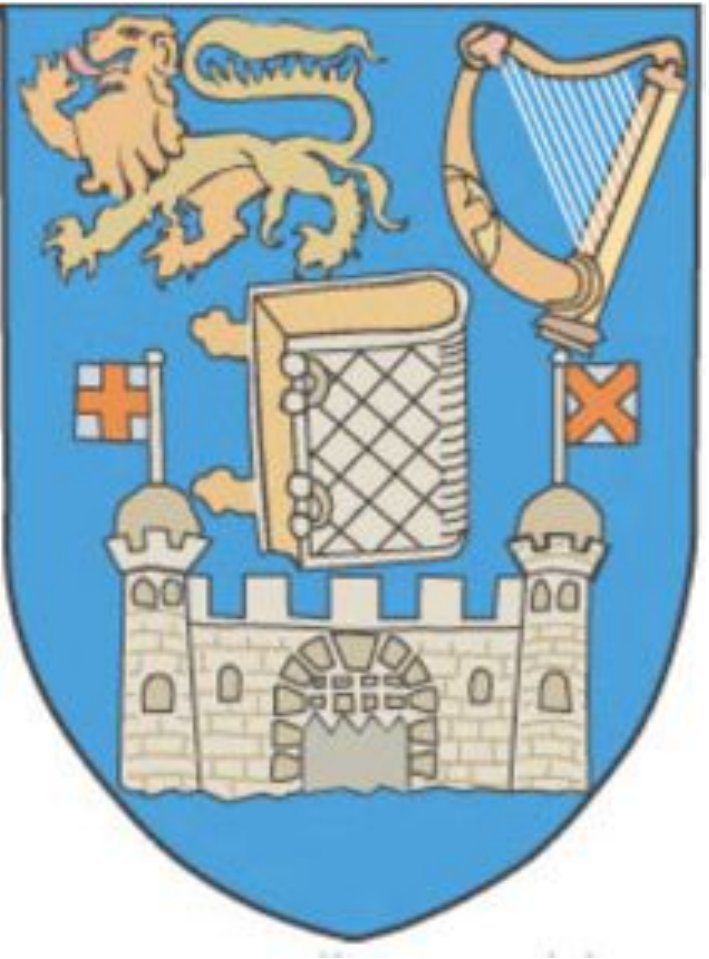}
   \vspace{1cm}
    }
  
\else
  \author{\href{mailto:longda@tcd.ie}{David M. Long}}
  \collegeordept{School of Physics}
  \university{University of Dublin, Trinity College}
  \crest{
  \vspace{1cm}
  \includegraphics[width=0.3\textwidth]{TCD_Crest}
   \vspace{1cm}
    }
\fi
\degree{Philosophi\ae\ Doctor (PhD)}
\degreedate{October 2011}

\usepackage{setspace}
\begin{document}


\renewcommand\baselinestretch{1.2}
\baselineskip=18pt plus1pt


\maketitle  

\newpage
\noindent
Please note that this is the online version of this thesis. As a result, some figures may have lower resolution than the printed version to reduce the file size.

\pagestyle{plain}
%









\frontmatter
\cleardoublepage



\begin{declaration}        

I, David M. Long, hereby certify that I am the sole author of this thesis and that all the work presented in it, unless otherwise referenced, is entirely my own. I also declare that this work has not been submitted, in whole or in part, to any other university or college for any degree or other qualification. 

\vspace{5mm}

The thesis work was conducted from October 2007 to October 2011 under the supervision of Dr. Peter T. Gallagher at Trinity College, University of Dublin.

\vspace{5mm}
In submitting this thesis to the University of Dublin I agree to deposit this thesis in the \corr{University's} open access institutional repository or allow the library to do so on my behalf, subject to Irish Copyright Legislation and Trinity College Library conditions of use and acknowledgement.

\textbf{Name:} David M. Long	
 \begin{center}
 \textbf{Signature:}  ........................................		\textbf{Date:}  ..............
\end{center}

\end{declaration}


\setcounter{page}{4}
\cleardoublepage
\begin{abstracts}        

Globally--propagating bright fronts in the solar corona are a poorly understood phenomena despite almost 15~years of research. Here, the kinematics and morphology of these disturbances are studied using Extreme UltraViolet (EUV) observations from the \emph{Solar TErrestrial RElations Observatory} (\emph{STEREO}) and \emph{Solar Dynamic Observatory} (\emph{SDO}) spacecraft. 

The first observations of a coronal bright front (CBF) made using \emph{STEREO} are presented, with the pulse observed in all four EUV passbands, namely 171, 195, 284, and for the first time, 304~\AA. The pulse exhibited similar kinematics in all passbands studied, with a velocity peak of 238$\pm$20~km~s$^{-1}$ within $\sim$28 minutes of launch and acceleration varying from 76~m~s$^{-2}$ to -102~m~s$^{-2}$ observed in the 304~\AA\ passband. Similar kinematics are found in the 195~\AA\ passband, while lower values were found in the lower cadence 284~\AA\ passband. The higher cadence 171~\AA\ passband shows much stronger variation, with the velocity peaking at 475$\pm$47~km~s$^{-1}$ within $\sim$20 minutes of launch, and the acceleration varying from 816~m~s$^{-2}$ to -413~m~s$^{-2}$. The derived pulse velocity and acceleration were found to be strongly influenced by the observing cadence, implying that previous kinematics may have been underestimated. 

Several different image processing techniques for identifying CBFs were tested to determine the optimal method for CBF pulse analysis. It was found that the techniques traditionally used to identify CBFs are fundamentally flawed and prone to undefined user--dependent errors. The kinematics of a simulated data--set were derived using several different numerical differencing techniques. These techniques were found to be inadequate for dealing with the typical data--set sizes encountered when studying CBFs. A semi--automated identification and tracking algorithm is proposed to provide a statistically rigorous analysis of CBFs.

This algorithm was applied to \emph{STEREO} observations of four CBF events and used to determine their kinematics. Following launch at initial velocities of $\sim$240--450~km~s$^{-1}$ each of the four events studied showed significant negative acceleration ranging from $\sim -$290 to $-$60~m~s$^{-2}$. The spatial and temporal widths of the CBF pulses were also found to increase from $\sim$50~Mm to $\sim$200~Mm and $\sim$100~s to $\sim$1500~s respectively, indicating that the pulses are dispersive in nature. The variation in position-angle averaged pulse-integrated intensity with propagation was examined to identify evidence of dissipation, but exhibited no clear trend across the four events studied.

A CBF pulse was also observed using contemporaneous data from both \emph{STEREO} and \emph{SDO}. The CBF exhibited a lower initial velocity with weaker deceleration in \emph{STEREO} observations ($\sim$340~km~s$^{-1}$ and $-72$~m~s$^{-2}$) compared to \emph{SDO} ($\sim$410~km~s$^{-1}$ and $-279$~m~s$^{-2}$). The CBF kinematics from \emph{SDO} were found to be highly passband-dependent, with an initial velocity ranging from $379\pm12$~km~s$^{-1}$ to $460\pm28$~km~s$^{-1}$ and acceleration ranging from $-128\pm28$~m~s$^{-2}$ to $-431\pm86$~m~s$^{-2}$ in the 335~\AA\ and 304~\AA\ passbands respectively. These kinematics were used to estimate a quiet coronal magnetic field strength range of $\sim$1--2~G. Significant pulse broadening was also observed, with expansion rates of $\sim$130~km~s$^{-1}$ (\emph{STEREO}) and $\sim$200~km~s$^{-1}$ (\emph{SDO}). By treating the CBF as a linear superposition of sinusoidal waves within a Gaussian envelope, the resulting dispersion rate of the pulse was found to be $\sim$8--13~Mm$^2$~s$^{-1}$. 

All of these results indicate that coronal bright fronts are best interpreted as fast--mode magnetoacoustic waves propagating in an inhomogeneous medium. 

\end{abstracts}


\begin{dedication} 


For Mum and Dad\\
Thank you.

\end{dedication}

\begin{acknowledgements}      




First and foremost, I wish to thank my supervisor Dr. Peter Gallagher for all of the help and encouragement he has given me over the last four years. I would not be where I am now without him and for that I am truly grateful. A huge thank you must also go to Dr. Edward DeLuca for making me feel like part of the family while I was at the CfA. I am forever in your debt. 

To James and Shaun, thank you for everything. Thank you for letting me throw around ideas and for your patience and assistance when those ideas should never have been thrown. You are the most infuriatingly rigorous people to work with and the example I strive to emulate. That said, I will not forget those hours spent arguing over the positioning of punctuation... I would also like to thank Jack, Alex, Mike and Ryan for all of their support and guidance over the years. I had a great time at NASA and the credit must go to you all.

It would be remiss of me not to thank everyone who did their time in the office. Thank you to David, Paul, Claire, Jason, Larisza, Shane, Higgo, Sophie, Dan, Eamon, Eoin, Pietro, Aidan, Paddy and Peter. There are too many memories from the fantastic to the ridiculous, but I will miss morning coffee, procrastination, shenanigans and of course, the blue box.

To everyone at the CfA who helped to make my time there so amazing, thank you. Thank you to everyone in the SSXG, especially (in no particular order) Trae, Alisdair, Alec, Kathy, Steve, Kelly, Mark, Jonathan, Suli, Yingna, Leon, Kamen, Tony, Arnaud, Paolo, Paola and Nick. I hope to return some day soon and see you all again. Thank you as well to all of the pre-docs, post-docs and grad students, especially Katha, Roberto, Alexa, Daniel, Petri, Robin, Tana, Ellie and Katherine. I will miss both Predoc coffee and the subsequent deferral to Grendels..

A special thank you to Karen, Lauren, Aleida, Kandi, Amanda, Natalie, Sam, Vinnie and Sam. You all made starting out in a foreign country so much easier than it should have been.

I would also like to thank Alan, Karen, Joe and Hugh for giving a hand when I looked for help and wanting a break when I needed one. A very special thank you must go to Lina. You have always been there for me in good times and bad. I don't have the words to describe how important you are to me and none of this would have been possible without you.

Finally, thank you to Mum, Dad, Joan and Neil for everything. Thank you for all of the support that you have given me and for making sure I kept my feet on the ground. I am who I am because of you.

\end{acknowledgements}


\setcounter{secnumdepth}{3} 
\setcounter{tocdepth}{3}    
\tableofcontents            


\chapter{Publications}
\label{chapter:publications}


\begin{enumerate}

\item Gallagher, P.~T., Bloomfield, D.~S., Carley, E., Higgins, P.~A., \textbf{Long, D.~M.}, Maloney, S.~A., Murray, S.~A., O'Flannagain, A., P\'erez-Su\'arez, D., Ryan, D., Zucca, P. \\
\emph{A Comprehensive Overview of the 2011~June~7 Solar Storm}, \\
\corr{2012,\aap, {\it in preparation}}

\item \textbf{Long, D.~M.}, DeLuca, E., Gallagher, P.~T.,\\
\emph{The Wave Properties of Coronal Bright Fronts Observed Using SDO/AIA}, \\
2011,\apjl,  \corr{\textbf{741}, L21}

\item Ma, S., Raymond, J.~C., Golub, L., Lin, J., Chen, H., Grigis, P., Testa, P., \textbf{Long, D.~M.},\\
\emph{Observations and Interpretation of a Low Coronal Shock Wave Observed in the EUV by the SDO/AIA}, \\
2011,\apj, \textbf{736}(2), 160
	
\item \textbf{Long, D.~M.}, Gallagher, P.~T., McAteer, R.~T.~J., Bloomfield, D.~S.,\\
\emph{Deceleration and Dispersion of Large-scale Bright Fronts in the Solar Corona}, \\
2011,\aap, \textbf{531}, A42 

\item Gallagher, P.~T. \& \textbf{Long, D.~M.},\\
\emph{Large-scale Bright Fronts in the Solar Corona: A Review of ``EIT waves"}, \\
2011,\ssr, \textbf{158}, 365--396

\item \textbf{Long, D.~M.}, Gallagher, P.~T., McAteer, R.~T.~J., Bloomfield, D.~S.,\\
\emph{The Kinematics of a Globally Propagating Disturbance in the Solar Corona}, \\
2008,\apjl, \textbf{680}(1), 81-84

\end{enumerate}

\listoffigures  

\listoftables  



\nomenclature{\emph{SOHO}}{\emph{SOlar and Heliospheric Observatory}}
\nomenclature{\emph{STEREO}}{\emph{Solar TErrestrial RElations Observatory}}
\nomenclature{\emph{SDO}}{\emph{Solar Dynamics Observatory}}
\nomenclature{\emph{TRACE}}{\emph{Transition Region And Coronal Explorer}}
\nomenclature{EIT}{Extreme ultraviolet Imaging Telescope}
\nomenclature{EUVI}{Extreme UltraViolet Imager}
\nomenclature{AIA}{Atmospheric Imaging Assembly}
\nomenclature{SECCHI}{Sun--Earth Connection Coronal and Heliospheric Investigation}
\nomenclature{FASR}{Frequency Agile Solar Radiotelescope}
\nomenclature{LASCO}{LArge Solar COronagraph}
\nomenclature{GOES}{Geostationary Operations Environmental Satellite}
\nomenclature{XRT}{X--Ray Telescope}
\nomenclature{LOFAR}{LOw Frequency Active Radiotelescope}
\nomenclature{CME}{Coronal Mass Ejection}
\nomenclature{CBF}{Coronal Bright Front}
\nomenclature{AEC}{Automatic Exposure Control}
\nomenclature{RD}{Running Difference}
\nomenclature{BD}{Base Difference}
\nomenclature{PBD}{Percentage Base Difference}
\nomenclature{EUV}{Extreme UltraViolet}
\nomenclature{SXR}{Soft X--Ray}
\nomenclature{FITS}{Flexible Image Transport System}
\nomenclature{JSOC}{Joint \emph{SDO} Operations Center}
\nomenclature{NOAA}{National Oceanic and Atmospheric Administration}
\nomenclature{PA}{Position Angle}
\nomenclature{CCD}{Charged Coupled Device}
\nomenclature{EIS}{Extreme ultraviolet Imaging Spectrometer}

\clearpage
\markboth{\nomname}{\nomname}
\printnomenclature


\mainmatter

\allowdisplaybreaks

\onehalfspacing
\pagestyle{fancy}


\chapter{Introduction}
\label{chap:intro}

\ifpdf
    \graphicspath{{1_introduction/figures/PNG/}{1_introduction/figures/PDF/}{1_introduction/figures/}}
\else
    \graphicspath{{1_introduction/figures/EPS/}{1_introduction/figures/}}
\fi


\noindent 
\\ {\it 
Here comes the Sun \\ And I say it's alright
\begin{flushright}
George Harrison \\
\end{flushright}
 }

\vspace{15mm}
No physical object (aside from the Earth) has had more influence on human civilization than the Sun, with the heat and light that it provides allowing life to flourish on Earth. Every individual culture has worshipped the Sun in its own way, telling epic tales of the Sun Gods (such as Helios, Ra and Surya; the Greek, Egyptian and Hindu Sun Gods respectively), building temples that align with the rising or setting Sun (e.g.,\ Newgrange) and by treating its heroes, kings and saints as human incarnations of the divine Sun (cf. religious halos and ``Le Roi Soleil'').

While science and technology have removed much of the mystique that has traditionally surrounded the Sun, humanity has not stopped its worship of our closest star. Every year millions of people travel to warmer parts of the globe to bask in the healing powers of the Sun, while the suntan has become symbolic of health and vigour. Although the ability of the Sun to burn is a cause of some caution and alarm, humanity still finds it \corr{impossible} to live without the heat and light that only the Sun can provide.

The Sun may have been present throughout humanity's stay on the Earth, but it is only recently that it has become a source of endless investigation and countless unanswered questions. The first inkling that the Sun may have an effect on Earth that extends beyond the provision of light and heat came from observations of a flash on the Sun made by \citet{Carrington:1859ab}. Carrington observed a flash of white light in a particularly large sunspot region when making his daily observation of the Sun. It was also noted that the magnetic field experiment at Kew Observatory in London exhibited a strong reaction to some unknown driver at about the same time. Several days later, electrical and magnetic equipment all around the world began behaving erratically with no apparent cause, with a telegraph operator on the transatlantic telegraph from Valentia to Newfoundland thrown across the room by a large electric shock coming from the wires while aurorae were observed as far south as Cuba. Although the contemporary nature of these events was interesting, it was not suggested that they were linked; ``one swallow does not make a summer''.

Since then the Sun-Earth connection has been the source of detailed investigation, with the Sun shown to have a major influence on the Earth \corr{(beyond just heat and light)}. The advent of larger and more powerful ground-- and space--based telescopes has shown that the Sun, once thought of as a serene globe around which the world revolves (or which revolves around the Earth), is instead a turbulent roiling ball of plasma and relatively powerful magnetic fields. The \corr{outer} solar atmosphere, once only visible during the pseudo-night of a solar eclipse, is now observed regularly using artificial means to block the face of the Sun, revealing magnetic structures and vast ejections of plasma that are fired into the heliosphere. 

\begin{figure}[!t]
\centering
   \includegraphics[width=0.6\textwidth,clip=,trim=10mm 15mm 10mm 15mm,angle=270]{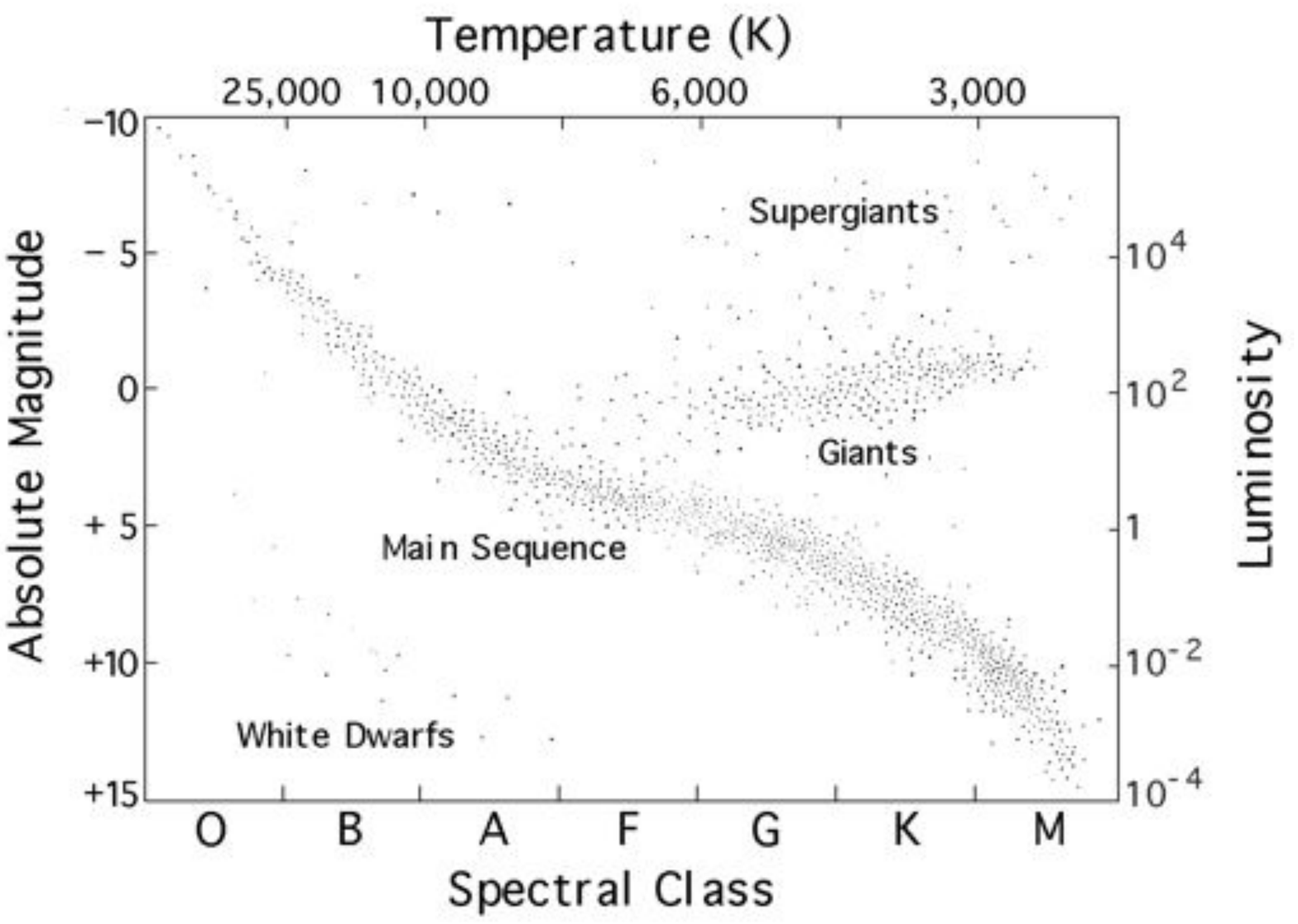}
\caption[The Hertzsprung--Russell diagram]{The Hertzsprung--Russell diagram. The Sun is a G2 star located on the main sequence, with a surface temperature of $\sim$6000~K. Note that stellar luminosity values are measured relative to the solar luminosity ($L=L_{star}/L_\odot$; picture courtesy \emph{heasarc.gsfc.nasa.gov}).}
\label{fig:hertzsprung_russell}
\end{figure}

Scientifically, the Sun is classified as a G2\corr{\Rmnum{5}} yellow dwarf star found on the main sequence of the Hertzsprung--Russell diagram (a graphical representation of stars obtained by plotting the absolute magnitude against the spectral class; see Figure~\ref{fig:hertzsprung_russell}) with a mass of \corr{$\sim$2$\times10^{30}$~kg} and a radius of $\sim$6.96$\times10^{8}$~m. The Sun has a chemical composition consisting primarily of hydrogen ($\sim$75~\%) and helium ($\sim$25~\%) with some trace amounts of heavier elements. The high concentration of hydrogen, combined with its large size, provides the conditions necessary for the production of energy via nuclear fusion deep within the heart of the Sun.

%
\section{The Solar Interior}
\label{sect:sun_interior}

\begin{figure}[!t]
\centering
   \includegraphics[width=0.85\textwidth,clip=,trim=0mm 30mm 0mm 35mm]{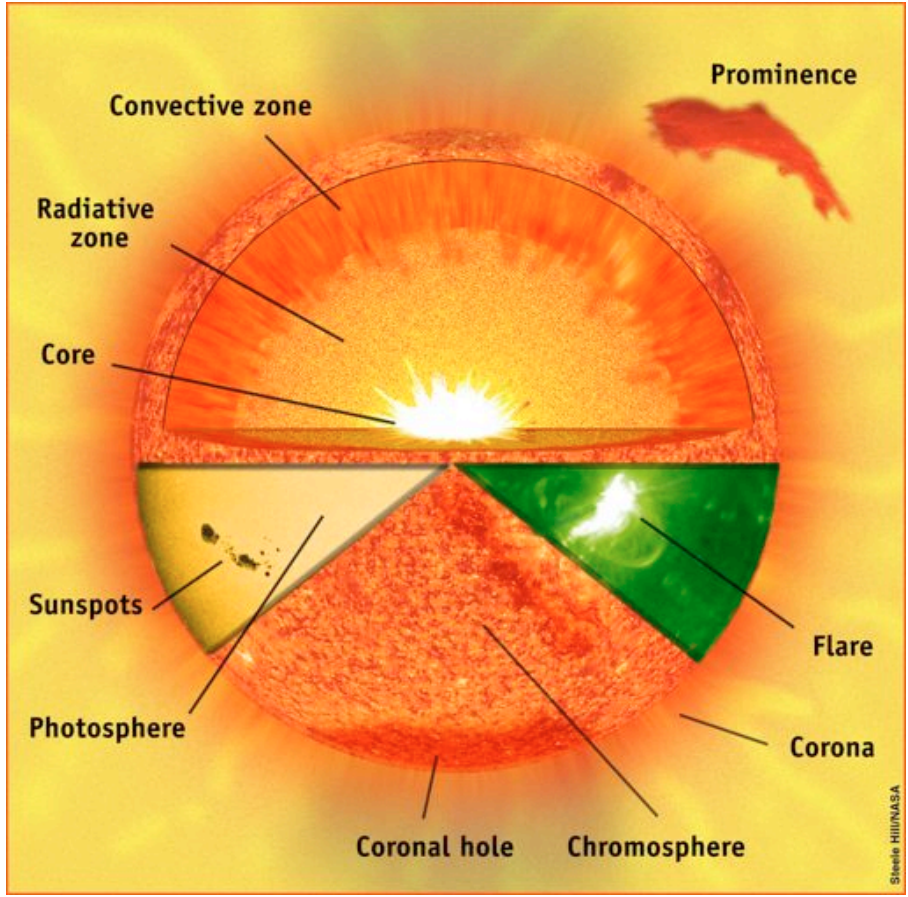}
\caption[Cartoon showing the different regions of the Sun]{Cartoon showing the different regions of the Sun along with some of the features observable in the solar atmosphere (picture courtesy \emph{sohowww.nascom.nasa.gov}).}
\label{fig:sun_interior}
\end{figure}

The solar interior consists of three main regions; the core, the radiative zone and the convective zone (see Figure~\ref{fig:sun_interior}). Each \corr{region is} determined using the variation in physical values such as density and temperature, with the result that their absolute limits are not well defined. 

In the core, the temperatures and densities are sufficiently high ($\sim$10--15~MK and $\sim$10--150~g~cm$^{-3}$ respectively) that proton--proton chain nuclear fusion can occur. This \corr{typically occurs through the pp~\Rmnum{1} chain}\footnote{\corr{It is also possible to get fusion through the pp~\Rmnum{2} chain using \ion{Li}{7} as an intermediate. However this process typically requires higher temperatures than those available and as a result makes a minimal contribution to the energy produced in the core.}} according to the steps,
\begin{eqnarray}
^{1}H + \ ^{1}H &\rightarrow& ^{2}H + e^{+} + \nu \\
^{2}H + \ ^{1}H &\rightarrow& ^{3}He + \gamma \\
^{3}He + \ ^{3}He &\rightarrow& ^{4}He + 2\ ^{1}H + \gamma
\end{eqnarray}
where $^{1}H$ is a hydrogen nucleus, $^{2}H$ is the deuteron isotope of hydrogen, $^{3}He$ and $^{4}He$ are the 1-- and 2--neutron isotopes of helium respectively, $e^{+}$ is a positron, $\nu$ is a neutrino and $\gamma$ is a gamma-ray photon. The overall reaction therefore looks like,
\begin{equation}
4\ ^{1}H \rightarrow\ ^{4}He + 2e^{+} + 2\nu + \gamma.
\end{equation}
\corr{The positrons ($e^{+}$) are then annihilated through interaction with electrons after a very short period of time, producing 1.02~MeV.} One set of these interactions produces \corr{a total} energy of $\sim$26.7~MeV, emitted primarily as high-energy $\gamma$-ray photons. Although a temperature of $\sim$10$^{10}$~K is required to overcome the Coulomb repulsion between the protons through purely thermal interactions, this reaction can occur at lower temperatures comparable to those in the core by including quantum mechanical effects \citep[cf. the Gamow factor;][]{Gamow:1928fk}.

This process occurs within $\sim$0.25~\rsun\ of Sun centre where the temperature and density are conducive to fusion. At $\sim$0.25~\rsun, the temperature and density drop sufficiently low that fusion can no longer occur. This decrease in density makes the plasma optically thin, allowing the transfer of energy outwards from the core through radiative diffusion. However, the temperature is still too high to allow neutral atoms to form, with the result that the energetic photons produced by fusion in the core are scattered by the free electrons, neutrons and protons in the radiative zone.  This scattering results in the photons performing a random walk through the radiative zone that can take \corr{$\sim$10$^{4}$}~years to complete. 

At \corr{$\sim$0.7~\rsun \citep{Christensen-Dalsgaard:1991fk}}, the temperature drops sufficiently to allow the formation of atoms, with the plasma becoming opaque. The constant influx of energetic photons from the radiative zone produces a temperature gradient, causing the plasma to become convectively unstable. This results in the formation of giant convective cells (akin to those found in a boiling kettle) that carry energy towards the solar surface. The density continues to drop with distance from Sun centre until the plasma is no longer opaque. The point at which this occurs for visible light defines the surface of the Sun, where the energy carried by convective cells from deep within the Sun can be radiated away as visible photons.

As a massive body containing constantly moving charged ions, the Sun is magnetically active with a magnetic field that can be treated as a dipole similar to that of the Earth. However, unlike the Earth's magnetic field, it does vary from a dipolar to quadrupolar configuration on an 11~year timescale. This is a result of the variation of the solar rotation rate from that of a solid body in the core and radiative zone to a differential rotation in the convective zone. The interface between these two rotation types is called the tachocline, and it is at this point that the solar magnetic field is wound up by the differential solar rotation. The magnetic field lines then become buoyant, and are carried to the surface by convective motion before emerging into the solar atmosphere \citep[the $\alpha$--$\Omega$ effect;][]{Babcock:1961yq}.

%
\section{The Solar Atmosphere}
\label{sect:sun_atmosphere}

The solar atmosphere, similar to the Earth's atmosphere, is defined as the region starting above the \corr{visible} solar surface and extending out into the heliosphere \corr{(defined as the region in space directly influenced by the solar wind)}, technically to the heliopause at the edge of the solar system\corr{, where the solar wind pressure is equal to the pressure of the ambient interstellar medium}. It consists of a tenuous, highly structured, ionised plasma that is threaded and modified by the solar magnetic field. Although difficult to see due to the dominant photospheric emission, the structure of the outer atmosphere can be discerned during a solar eclipse.

\begin{figure}[!t]
\centering
   \includegraphics[width=1\textwidth,clip=,trim=20mm 140mm 20mm 25mm]{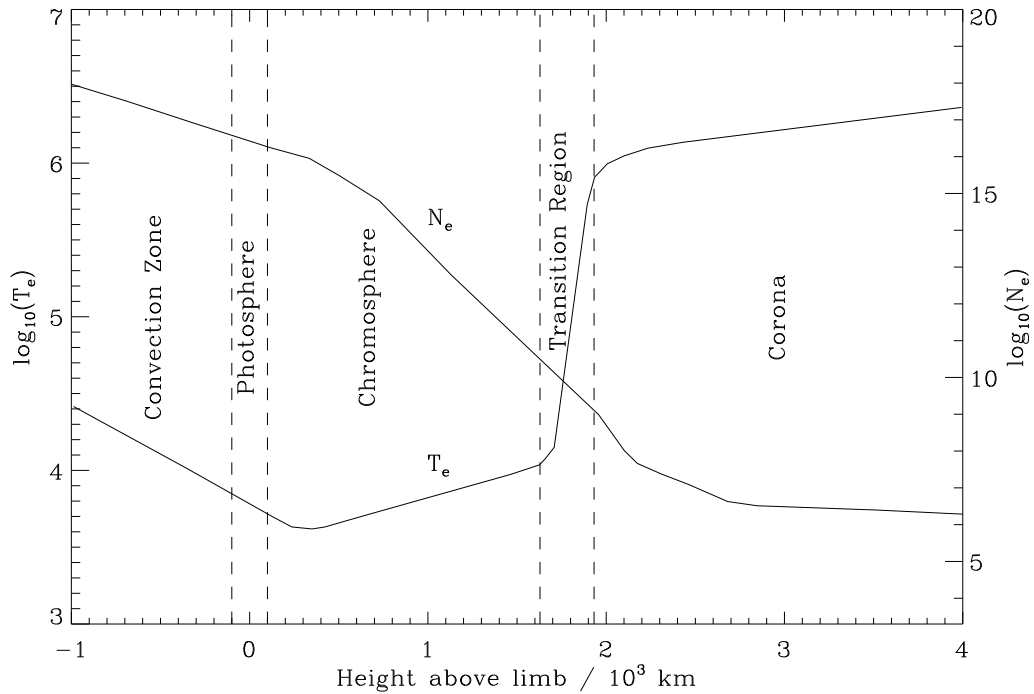}
\caption[Variation in the temperature and density of the solar atmosphere with height]{Plane parallel model showing the variation in solar temperature and density with height. The different regions of the solar atmosphere are marked, with both temperature (T$_{e}$; K) and density (N$_{e}$; cm$^{-3}$) variation shown using logarithmic scales. Figure from \citet{Gallagher:2000fk} after \citet{Gabriel:1982fk}.}
\label{fig:solar_model}
\end{figure}

The different regions of the solar atmosphere are shown in Figure~\ref{fig:sun_interior}, while the variation in the density and temperature with height is shown in Figure~\ref{fig:solar_model}. Each of the regions of the solar atmosphere are defined according to variations in density and temperature and these are noted in Figure~\ref{fig:solar_model}. However, the dynamic nature of the solar atmosphere means that these are not firm definitions with clearly defined layers but general trends defined by different plasma properties.

\subsection{The Photosphere}
\label{subsect:photosphere}

\begin{figure}[!t]
\centering
   \includegraphics[width=0.99\textwidth,clip=,trim=10mm 100mm 10mm 50mm]{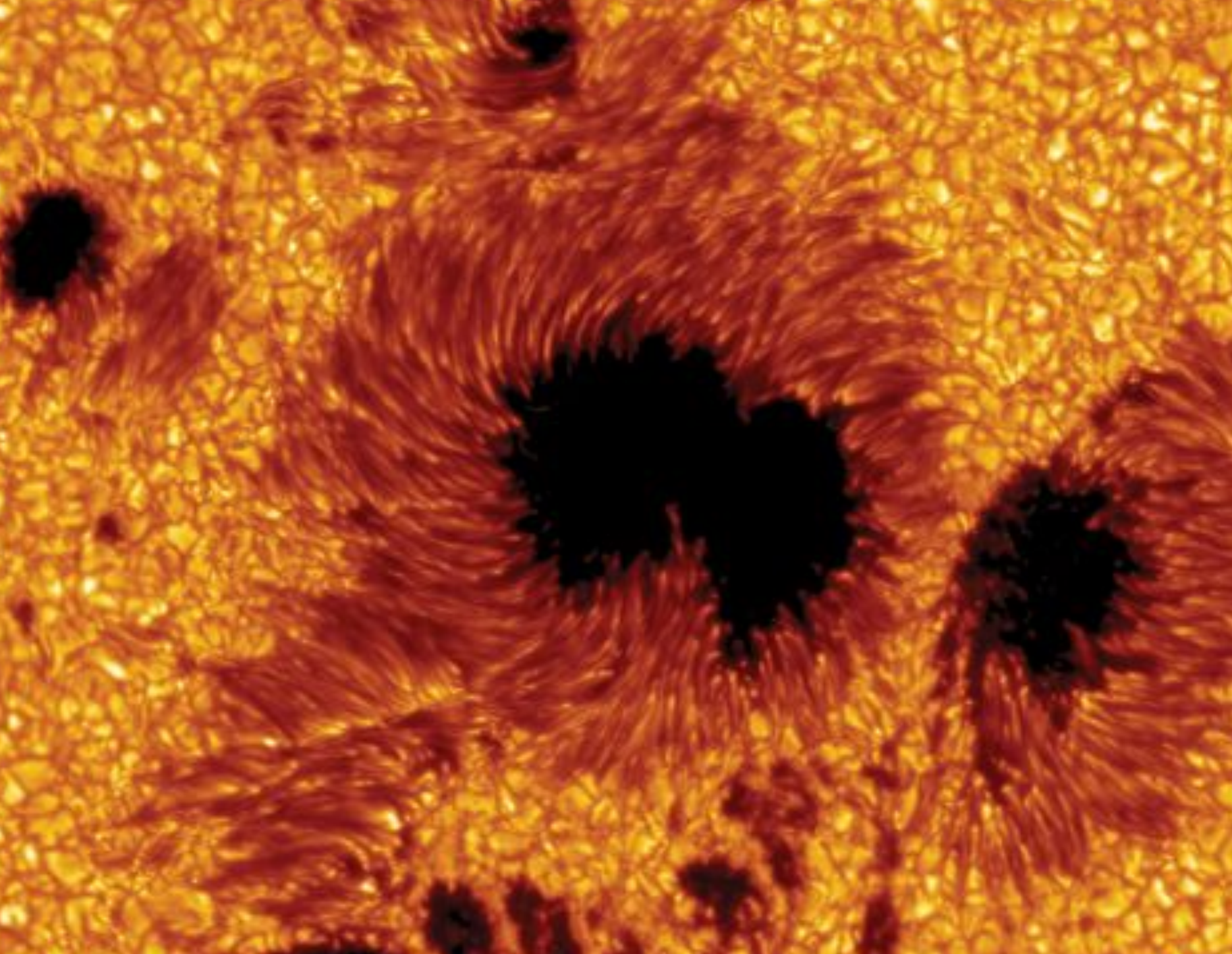}
\caption[Close--up of a sunspot observed by the Swedish Solar Telescope.]{A close--up of a sunspot in the photosphere observed by the Swedish Solar Telescope in the Canary Islands (figure courtesy The Royal Swedish Academy of Sciences/The Institute for Solar Physics).}
\label{fig:photosphere}
\end{figure}

The photosphere is the visible surface of the Sun, with a black-body temperature of $\sim$6000~K. As it is situated at the top of the convective zone, the photosphere is home to small-scale granulation arising from small convective flows resulting from temperature gradients below the surface. These features are visible in Figure~\ref{fig:photosphere}, which shows a close--up image of a sunspot taken using the Swedish Solar Telescope (SST) in the Canary Islands. The dark sunspot is seen in the centre of the image, with smaller dark pores around it while the mottled ``bubbles'' show the granulation resulting from the small--scale convective flows. The solar magnetic field also begins to play a significant role here, with the buoyant magnetic field lines from deep within the Sun emerging through the photosphere. Upon breaching the photosphere, these regions of concentrated magnetic field (which are cooler than the surrounding photospheric regions) appear as dark sunspots. 

The photosphere comprises a very thin portion of the solar atmosphere at a temperature of $\sim$6000~K. This region of the Sun is best observed using visible light such as that observed by the Solar Optical Telescope (SOT) onboard the \emph{Hinode} spacecraft or by ground-based telescopes such as the SST.

\subsection{The Chromosphere}
\label{subsect:chromosphere}

Above the photosphere, the temperature and density both continue to drop with height. Although the density continues to drop into the outer atmosphere (see Figure~\ref{fig:solar_model}), the temperature reaches a minimum at a height of $\sim$200~km above the photosphere before beginning to increase once more. The region of the solar atmosphere where this occurs is called the chromosphere, taking its name from the coloured appearance it has during a solar eclipse (from the Greek word \emph{khr\={o}ma}, meaning colour).

\corr{Despite the gradual increase in temperature, the steady drop in density with height in the chromosphere results in} a drop in the gas pressure of the chromosphere (as $p_{gas} = n_{e}k_{B}T_{e}$, where $n_{e}$ is the density, $k_{B}$ is the Boltzmann constant and $T_{e}$ is the temperature). This drop in the plasma pressure produces a situation where the magnetic pressure (\corr{in cgs units, $p_{B} = B^{2}/8\pi$, where $B$ is the magnetic field strength in gauss, or in SI units, $p_{B} = B^{2}/2\mu$, where $B$ is the magnetic field strength in tesla and $\mu$ is the permeability of free space}) approaches and begins to dominate the gas pressure, with the result that the magnetic field begins to dominate the structure of the chromosphere. This is illustrated by the variation of the plasma--$\beta$ value from the chromosphere into the corona, where the plasma--$\beta$ is defined as,
\begin{equation}
\beta = \frac{p_{gas}}{p_{B}} = \frac{n_{e}k_{B}T_{e}}{B^{2}/8\pi}
\end{equation}

\begin{figure}[!p]
\centering
   \includegraphics[width=0.75\textwidth,clip=,trim= 5mm 5mm 5mm 5mm]{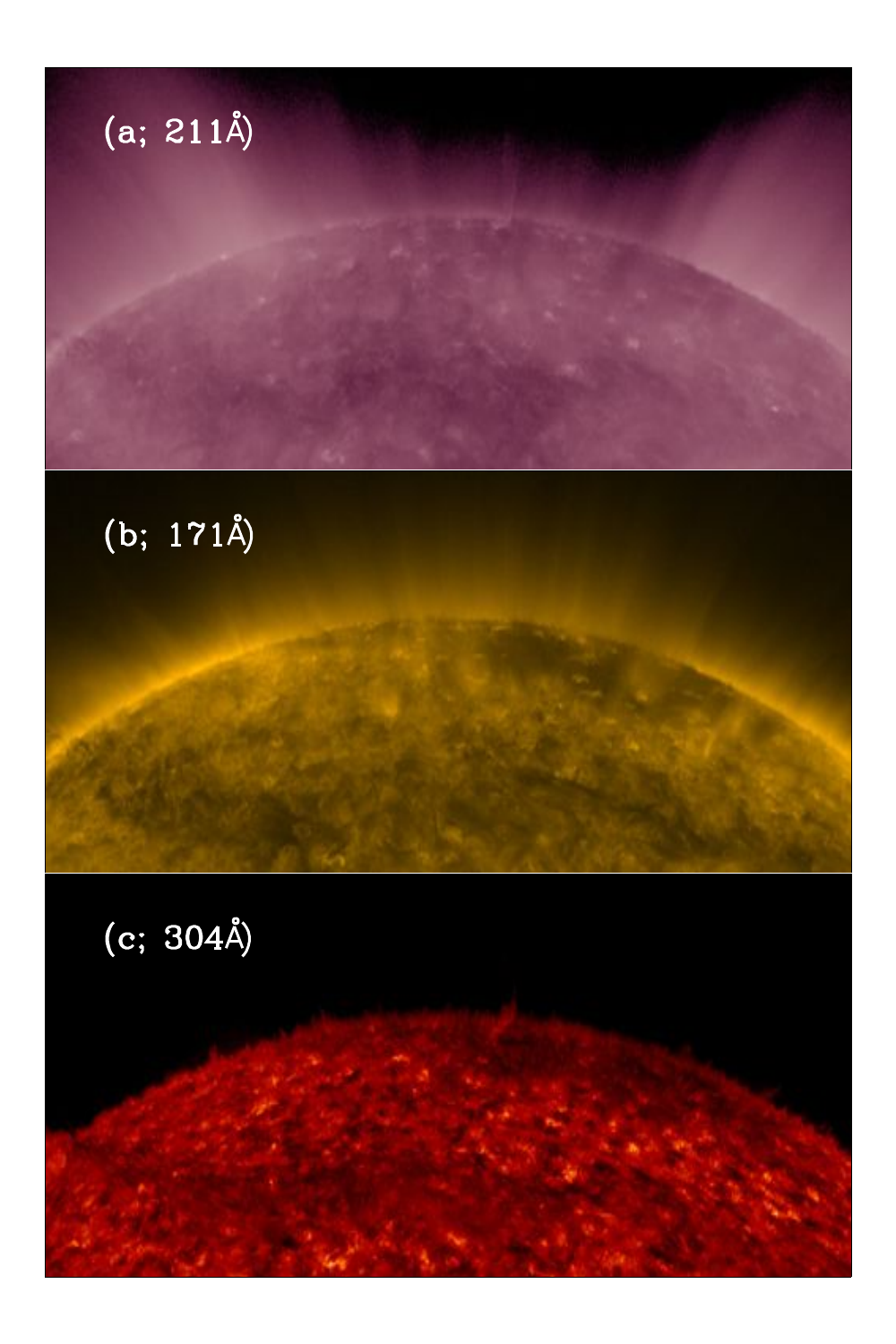}
\caption[The solar atmosphere observed in the Extreme UltraViolet (EUV) \corr{by \emph{SDO}/AIA}.]{Close-up images showing the dynamic nature of the solar atmosphere observed in the Extreme UltraViolet (EUV) \corr{by \emph{SDO}/AIA}. Panel (a) shows the corona (211~\AA; T $\sim$2~MK), panel (b) shows the upper transition region and low corona (171~\AA; T $\sim$0.7~MK), while panel (c) shows the upper chromosphere and lower transition region (304~\AA; T $\sim$0.06~MK). Each panel has been coloured using the relevant colour tables for clarity.}
\label{fig:atmosphere}
\end{figure}

The gradual domination of the solar magnetic field can be seen in observations of the solar chromosphere, which exhibits a very dynamic structure in all observable passbands including H$\alpha$ and EUV emission (such as 304~\AA; panel c of Figure~\ref{fig:atmosphere}). The relatively cool nature of the chromosphere means that it has been possible to observe it from ground-based telescopes using H$\alpha$ filters for many years, with many observatories taking daily synoptic H$\alpha$ observations of the solar chromosphere (a practice that still continues today, cf. Kanzelh\"{o}he Observatory in Austria).

It was using H$\alpha$ observations of a solar flare that the first globally-propagating disturbance in the solar atmosphere was observed by \citet{Moreton:1960p44}. Called the ``Moreton wave'', this disturbance was observed to have a velocity of $\sim$1000~km~s$^{-1}$ and be strongly associated with solar flare eruptions and Type~\Rmnum{2}\footnote{\corr{There are five types of radio bursts typically observed in solar physics. Type~\Rmnum{1} are associated with solar flares and active regions, Type~\Rmnum{2} indicate moving shocks in the corona, Type~\Rmnum{3} indicate particle emission, Type~\Rmnum{4} correspond to stationary shocks and Type~\Rmnum{5} indicate short--lived continuum emission}} radio bursts in the solar corona, indicating a moving shock. There were some issues with the observations, particularly regarding the high velocity of the pulse, which greatly exceeds the estimated Alfv\'{e}n speed for the solar chromosphere. To overcome this problem, \citet{Uchida:1968qv} proposed that the Moreton wave was the chromospheric signature of a fast-mode magnetoacoustic wave propagating in the solar corona; this is discussed in more detail in Section~\ref{sect:cbf_obs}. 

\begin{figure}[!t]
\centering
   \includegraphics[width=1\textwidth,clip=,trim=0mm 0mm 0mm 0mm]{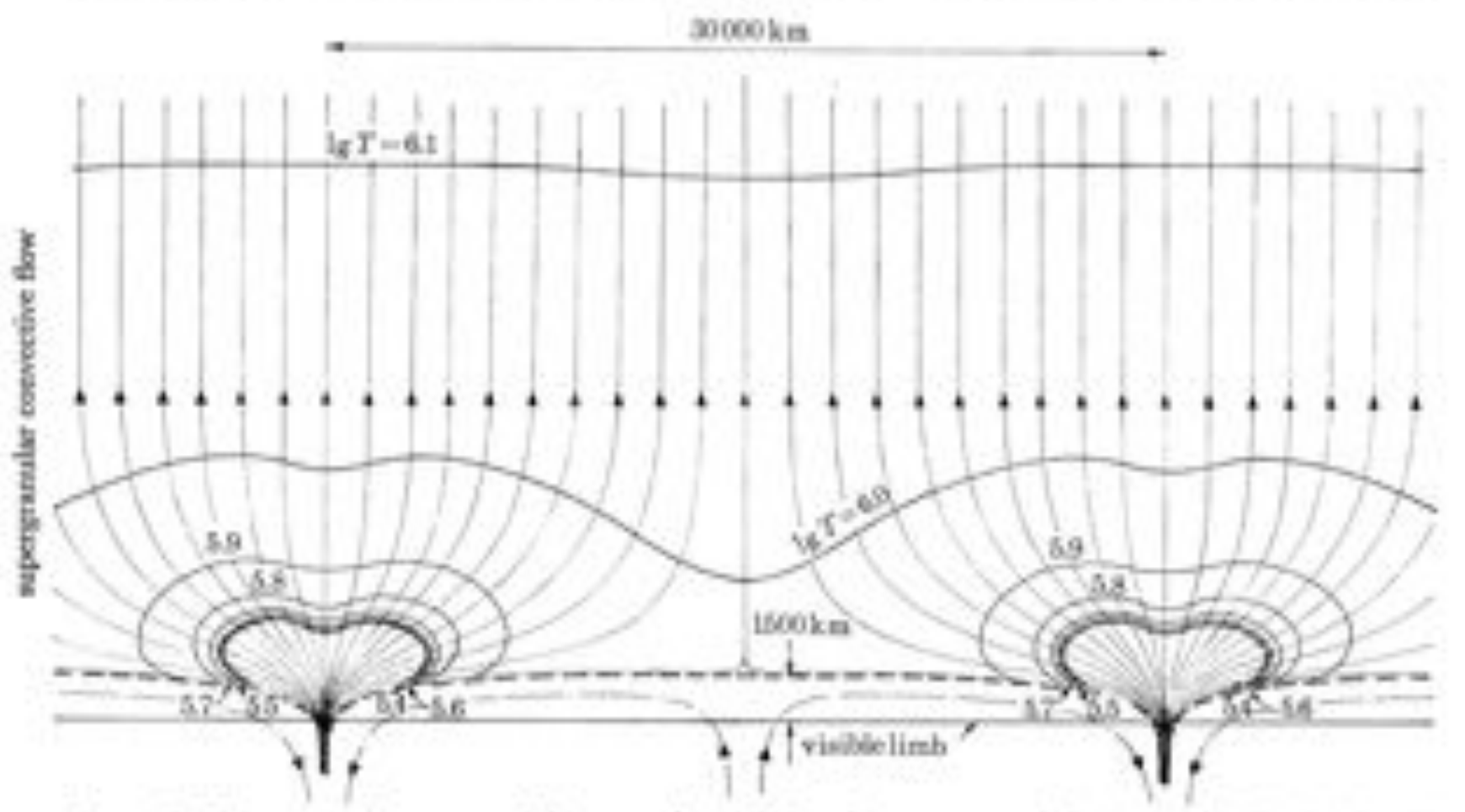}
\caption[The structure of the solar atmosphere based on an energy balance model proposed by \citet{Gabriel:1976uq}]{The structure of the solar atmosphere based on an energy balance model proposed by \citet{Gabriel:1976uq}. The magnetic field lines emerging through the photosphere fan out in the chromosphere and expand radially outwards from the Sun through the corona.}
\label{fig:corona_model}
\end{figure}

The region between the chromosphere and the corona is known as the transition region (see Figure~\ref{fig:solar_model}), with the temperature showing a dramatic increase over a height range of $\sim$100~km. The reason for this sudden temperature increase (the coronal heating problem) remains uncertain, and is one of the major topics of investigation in modern astrophysics. \corr{However, there are two main branches of theories that have been proposed to resolve this issue. The first deals with the dissipation of magnetic stresses caused by the gradual displacement of coronal loop foot--points by photospheric motions, producing direct heating (the direct current or DC models). The other branch of theories proposes that the displacement of coronal loop foot--points occurs on a rapid timescale compared to the end--to--end Alfv\'{e}n travel time, generating waves which dissipate to heat the local coronal plasma (the alternating current or AC models). For more information, see the review of coronal heating processes by \citet{Klimchuk:2006uq}.}


The anomalous nature of the upper chromosphere and transition region has important implications for a thorough analysis of the low corona, with the diffuse structures found in the corona having their foundation in this region. The concentrated magnetic field lines emerging from the photosphere fan out through the chromosphere into the corona as a consequence of the decreasing plasma--$\beta$ value. This is shown in Figure~\ref{fig:corona_model}, which provides a graphical outline of the structure of the solar atmosphere based upon an energy balance model \citep{Gabriel:1976uq}. Although this has since been superseded by a more dynamic model, it remains a useful guide to the basic structure of the solar atmosphere and the relationships between the various atmospheric regions.

\subsection{The Corona}
\label{subsect:corona}

\begin{figure}[!t]
\centering
   \includegraphics[width=1\textwidth,clip=,trim=0mm 0mm 0mm 0mm]{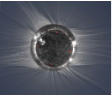}
\caption[A combined EUV and white light image of a total solar eclipse from 2010~July~11 reproduced from \citet{Pasachoff:2011uq}]{A combined EUV and white light image of a total solar eclipse from 2010~July~11 reproduced from \citet{Pasachoff:2011uq}. The EUV image of the Sun has been processed using a radial gradient filter, while the white-light corona has been processed using the technique proposed by \citet{Druckmuller:2006kx} to highlight features.}
\label{fig:corona}
\end{figure}

The solar corona comprises the outer part of the solar atmosphere, extending from $\sim$2000~km above the solar photosphere out into the heliosphere. The intensity of the corona is several orders of magnitude less than that of the Sun itself, with the result that it is best observed by blocking out emission from the solar disk. The best example of this occurs during a total solar eclipse, when the \corr{near--perfect alignment of the Moon and the Sun leads to the solar disk being blocked, thus allowing} the corona to be observed (see Figure~\ref{fig:corona}).

\begin{figure}[!t]
\centering
   \includegraphics[width=0.9\textwidth,clip=,trim=20mm 20mm 20mm 20mm,angle=90]{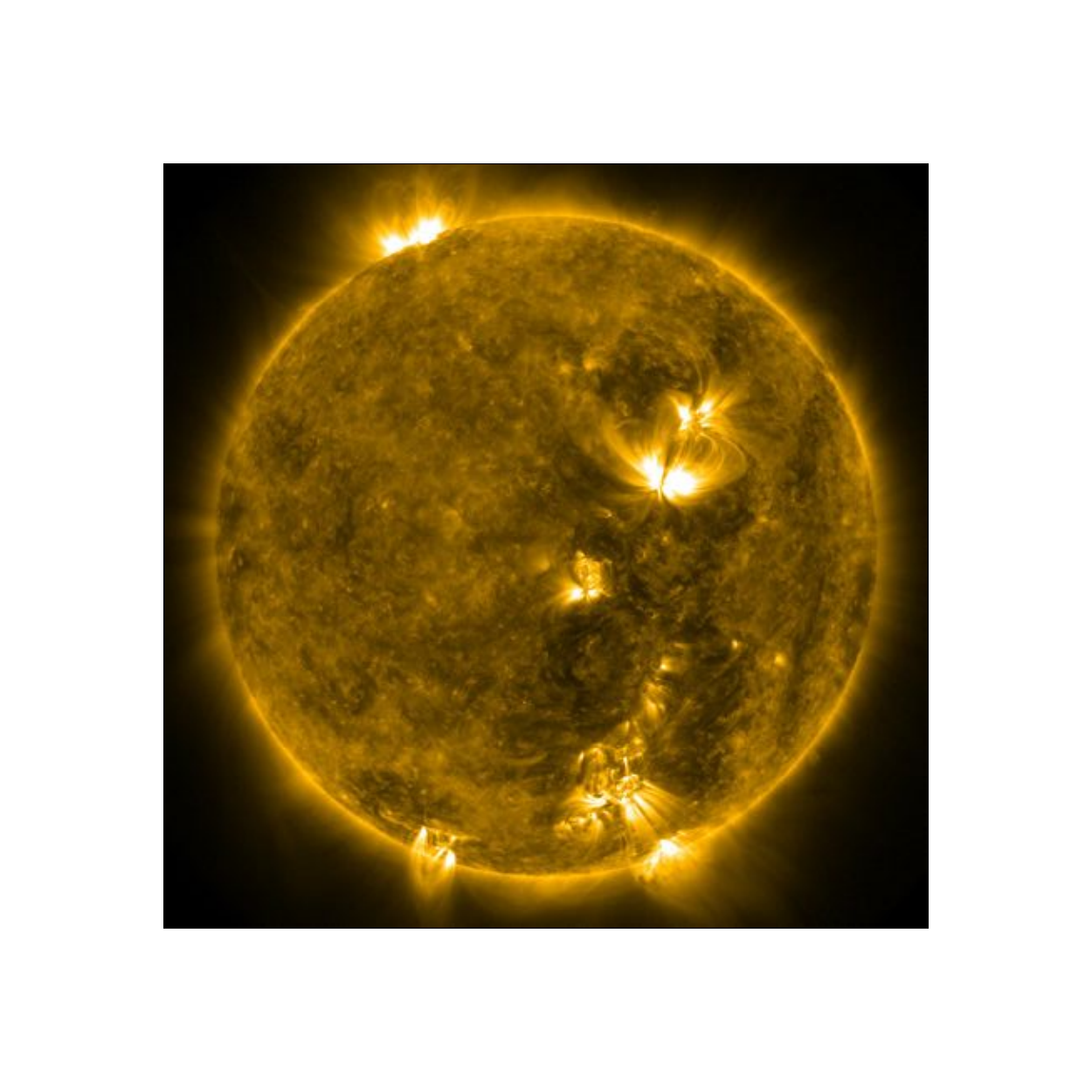}
\caption[EUV image from the 171~\AA\ passband \corr{taken by \emph{SDO}/AIA}.]{An EUV image from the 171~\AA\ passband showing the solar corona \corr{as observed by \emph{SDO}/AIA}. The image has been displayed using the appropriate colour table for clarity.}
\label{fig:euv_corona}
\end{figure}

As previously shown in Figure~\ref{fig:solar_model}, the temperature of the corona is $\sim$10$^{6}$~K, much hotter than the photosphere and chromosphere below it, while the density is $\sim$10$^{6}$--10$^{9}$~cm$^{-3}$. As a result, much of the emission from the low corona comes from highly ionised heavy elements such as iron. This emission is primarily in the extreme ultraviolet range of the electromagnetic spectrum, and consequently must be observed using space-based instruments. This is apparent in Figure~\ref{fig:euv_corona}, which shows the EUV corona as observed in the 171~\AA\ passband \corr{which is dominated by emission from the \ion{Fe}{9} line}. 

The solar corona is a low--$\beta$ plasma, and as a result is dominated by the solar magnetic field, with the plasma ``stuck'' to the magnetic field lines. This is apparent in Figure~\ref{fig:corona}, where the magnetic structure of the inner and outer corona is quite visible, particularly above the bright active regions and at the north and south poles. The fine structure in Figure~\ref{fig:euv_corona} is also a result of the low--$\beta$ nature of the corona. The dominance of the magnetic field is such that it defines the different regions within the corona; active regions, coronal holes and the quiet corona.

Coronal active regions (seen as the brighter regions in Figures~\ref{fig:corona} and \ref{fig:euv_corona}) are regions of intense magnetic field that are associated with photospheric sunspot regions. Here, magnetic field strengths can reach up to several hundreds of Gauss, while the interaction of different field lines can result in magnetic reconnection, producing solar flares and coronal mass ejections (CMEs). Coronal holes lie at the opposite end of the spectrum, with open magnetic field lines that radiate out into the heliosphere (this is most apparent at the poles of the Sun in Figure~\ref{fig:corona}). This allows plasma to propagate away from the Sun along the open field lines as the \corr{fast} solar wind \citep[cf.][]{Parker:1958kx}, with this continuous loss of plasma a reason why coronal holes are darker than the surrounding corona in EUV passbands.

The quiet corona is the intermediate region between active regions and coronal holes. Here, the topology consists of a combination of open and closed magnetic field lines and small scale structures that are changing constantly on a variety of timescales. This ``magnetic carpet'' can produce small solar flares and CME eruptions, with a density and temperature that is higher than a coronal hole, but lower than an active region.

The concentrated magnetic fields and increased densities in active regions compared to the quiet Sun produce the increased emission apparent in Figures~\ref{fig:atmosphere}, \ref{fig:corona} and \ref{fig:euv_corona}. Similarly, the low densities and open magnetic field lines result in coronal holes being much darker than the surrounding quiet Sun regions when observed using EUV and X-ray images. The different passbands typically observed by space-based observatories allow these different regions of the solar atmosphere to be studied by examining different emission lines. This is most apparent in Figure~\ref{fig:atmosphere}, where the decrease in temperature from panel (a) to (c) \corr{highlights} different structures and regions of the atmosphere. 

The heights of the different regions of the solar atmosphere observed by each passband can be estimated given the temperature of the plasma observed. This is done following the example of \citet{Chapman:1957p1254} and assuming that the solar corona is in hydrostatic equilibrium. In this case, 
\begin{equation}
\frac{dp\corr{_{gas}}}{dh} = -\rho g \label{eqn:hydrostatic}
\end{equation}
which indicates that the change in \corr{gas} pressure $p\corr{_{gas}}$ with height $h$ is dependent on the mass density $\rho$ and solar gravity $g$. The solar gravity can be calculated using Newton's Law as,
\begin{equation}
g = \frac{GM_{\odot}}{R_{\odot}^2},
\end{equation}
where $G$ is the gravitational constant ($G=6.67\times10^{-11}$~m$^{3}$~kg$^{-1}$~s$^{-2}$), $M_{\odot}$ is the solar mass ($M_{\odot} = 2\times10^{30}$~kg) and \rsun\ is the solar radius (\rsun$=6.96\times10^{8}$~m). This produces an acceleration due to gravity of $\sim$270~m~s$^{-2}$ at the solar surface. The mass density $\rho$ can be determined using the ideal gas law $p\corr{_{gas}} = nk_{B}T$, where $n$ is the particle number density, $k_{B}$ is the Boltzmann constant ($k_{B} = 1.38\times10^{-23}$~m$^{2}$~kg~s$^{-2}$~K$^{-1}$) and $T$ is the temperature. The ideal gas law can then be rewritten as,
\begin{equation}
p\corr{_{gas}} = nk_{B}T = \frac{\rho k_{B}T}{m} \rightarrow \rho = \frac{m}{k_{B}T}p\corr{_{gas}},
\end{equation}
where $m$ is the proton mass, $1.67\times10^{-27}$~kg. This can then be substituted into Equation~\ref{eqn:hydrostatic} to give,
\begin{equation}
\frac{dp\corr{_{gas}}}{dh} = -\frac{mg}{k_{B}T}p\corr{_{gas}}
\end{equation}
\corr{If isothermal equilibrium is assumed, t}his can be solved by rearranging the equation and integrating as,
\begin{equation}
\int_{p_0}^{p_{\corr{gas}}} \frac{dp\corr{_{gas}}}{p\corr{_{gas}}} = -\frac{mg}{k_{B}T} \int_0^h dh,
\end{equation}
where $p_0$ is the surface \corr{gas} pressure and $p\corr{_{gas}}$ is the \corr{gas} pressure at the height \corr{$h$} above the surface. Solving this equation gives the estimate of \corr{gas} pressure $p\corr{_{gas}}$ at a height $h$ above the surface as,
\begin{equation}
p\corr{_{gas}}(h) = p_0 e^{-h/H}
\end{equation}
where $H$ is the scale height and is defined as,
\begin{equation}
H = \frac{k_{B}T}{mg}.
\end{equation}
Using the values defined above for a typical coronal temperature of $\sim$1--2~MK produces an estimated scale height of \corr{$\sim$30--60~Mm}, while typical chromospheric temperatures of \corr{$\sim$10$^{4}$--10$^{5}$~K produce scale heights of 0.3--3~Mm}. These values correspond quite well with those estimated in Figure~\ref{fig:solar_model}, indicating that the passbands shown in Figure~\ref{fig:atmosphere} can be used to study the upper chromosphere and low corona.

Although the temperature and density of the corona can be estimated using emission lines, the magnetic field in the corona is more difficult to study. The effects of the magnetic field are quite apparent in the solar structures shown in Figure~\ref{fig:euv_corona}, but the strength of the magnetic field must be inferred using photospheric magnetograms or by calculating the speed of coronal shocks using Type~\Rmnum{2} radio bursts. These techniques produce an estimated magnetic field strength of $\sim$10~G in the quiet corona, with values of the order $\sim$100~G in active regions. 

While the magnetic field strength, temperature and density variation in the corona can be estimated using various techniques, these estimates do have large associated errors. A technique that could potentially be used to probe the physical structure of the corona directly is coronal seismology, which uses the oscillation of coronal structures such as loops to estimate the magnetic field strength and density \citep[e.g.,][]{Ballai:2004p1998}. However, this technique requires high temporal cadence observations of the solar corona with very high spatial resolution.

An understanding of the magnetic field strength variation is vital for understanding the structure of the solar corona as a direct consequence of the fact that the corona is a low--$\beta$ plasma. Although the structure of the corona is undeniably complex (as shown in Figures~\ref{fig:atmosphere}, \ref{fig:corona} and \ref{fig:euv_corona}), it is possible to derive a mathematical framework that can be used to study it. This is \corr{done} by combining the electromagnetic equations of \citet{Maxwell:1865ab} with the fluid equations, producing the magnetohydrodynamic equations.

\section{Magnetohydrodynamics}
\label{sect:mhd_theory}

The concept of examining the electromagnetic--hydrodynamic waves resulting from the motion of a conducting liquid in a constant magnetic field was first proposed by \citet{Alfven:1942fk}. This involves the use of magnetohydrodynamics (MHD), which combines the theories of electromagnetism and fluid dynamics, allowing the evolution of magnetic plasma to be studied. At its most basic level, MHD theory quantifies the interplay of forces involved in the motion of a conducting plasma in the presence of a magnetic field. As a result, MHD is of vital importance for a thorough analysis of the solar corona, which is a highly ionised plasma threaded by the solar magnetic field and in a constant state of flux. In this section, the MHD equations will be defined and used to \corr{examine} the types of waves that can result in this environment. \corr{Note that a number of different books were used as reference sources for this section; these may be found in the \nameref{chap:bibliography}.}

\corr{Maxwell's} equations constitute the bulk of the electromagnetic contribution to the set of MHD equations and allow the relationship between the electric currents and magnetic fields to be quantified. They are typically given as,
\begin{eqnarray}
\curl \vec{E}  &=& -\frac{\partial \vec{B}}{\partial t} \label{eqn:max_1} \\
\mdiv \vec{E} &=& \frac{1}{\epsilon_0}\rho \label{eqn:max_2} \\
\curl \vec{B} &=& \mu_0 \vec{j} \label{eqn:max_3} \\
\mdiv \vec{B} &=& 0 \label{eqn:max_4}
\end{eqnarray}
where $\vec{E}$ is the electric field vector, $\vec{B}$ is the magnetic field vector, $\epsilon_0$ is the permittivity of free space, $\rho$ is the charge density, $\mu_0$ is the magnetic \corr{permeability} and $\vec{j}$ is the current density. Faraday's law (Eqn.~\ref{eqn:max_1}) indicates that a spatially varying electric field can induce a magnetic field, while \corr{Amp\`{e}re's} law (Eqn.~\ref{eqn:max_3}; simplified here by assuming that the plasma velocity in an MHD system is much smaller than the velocity of light) shows that gradients in a magnetic field induce electric currents. Gauss' law (Eqn.~\ref{eqn:max_2}) covers the conservation of electric charge, while the solenoidal constraint (Eqn.~\ref{eqn:max_4}) indicates that magnetic monopoles cannot exist. The generalised Ohm's law is also included in addition to the above equations,
\begin{equation}
\vec{j} = \sigma(\vec{E} + \vec{v}\times \vec{B}) \label{eqn:max_5}
\end{equation}
where $\sigma$ is the electrical conductivity and $\vec{v}$ is the velocity of the plasma. This allows the fluid and electromagnetic equations to be coupled through the shared plasma velocity parameter.

When considering MHD in solar physics, it is often useful to replace the electric field $\vec{E}$ with the magnetic field $\vec{B}$. This can be done using \corr{Ohm's} Law (Equation~\ref{eqn:max_5}) and \corr{Amp\`{e}re's} law (Equation~\ref{eqn:max_3}) as,
\begin{equation}
\vec{E} = \frac{1}{\sigma}\vec{j} - \vec{v}\times\vec{B} = \frac{1}{\mu_{0}\sigma}\curl\vec{B} - \vec{v}\times\vec{B}.
\end{equation}
This can then be substituted into \corr{Faraday's} Law (Equation~\ref{eqn:max_1}) to give,
\begin{equation}
\frac{\partial\vec{B}}{\partial t} = \curl(\vec{v}\times\vec{B}) - \curl(\eta\curl\vec{B}),
\end{equation}
where the magnetic diffusivity $\eta=1/\mu_{0}\sigma$. This is assumed to be constant, and vector identities are then used to reduce this equation to, 
\begin{equation}
\frac{\partial\vec{B}}{\partial t} = \curl(\vec{v}\times\vec{B}) + \eta \nabla^2 \vec{B}.
\end{equation}
This is the induction equation, and is an important equation in MHD as it allows the magnetic field $\vec{B}$ to be calculated if the velocity $\vec{v}$ is known. In addition, it can be used to determine the behaviour of the magnetic field through the Magnetic Reynolds number $R_{m}$. This is defined as,
\begin{equation}
R_{m} = \frac{\curl(\vec{v}\times\vec{B})}{\eta \nabla^2 \vec{B}},
\end{equation}
and indicates the ratio of the advection and diffusion terms of the induction equation. If $R_{m}$ is \corr{larger than unity}, then the magnetic field is frozen into the plasma and the motion of the magnetic field is defined by plasma motion, whereas if $R_{m}$ is \corr{smaller than unity}, then the magnetic field tends to diffuse through the plasma. $R_{m}$ \corr{is analogous to the Reynolds number used in fluid dynamics, and} can be estimated \corr{using},
\begin{equation}
R_{m} \sim \frac{l_0 v_0}{\eta},
\end{equation}
where $\eta=10^9 T^{-3/2}$~m$^{2}$~s$^{-1}$. Using typical coronal values of T$\sim$10$^6$~K, $l_0\sim$10$^5$~m and $v_0\sim$10$^3$~m~s$^{-1}$ gives a value of $R_{m}\sim$10$^8$. The magnetic field in the solar corona is therefore dominated by plasma motion. 

As well as the electromagnetic equations, the fluid equations are also used in MHD to study the mass motion of the plasma. The conservation of mass of the plasma is defined as,
\begin{equation}
\frac{\partial \rho}{\partial t} + \mdiv(\rho \vec{v}) = 0 \label{eqn:mass_1}
\end{equation}
where $\rho$ is the plasma density. This indicates that the density at a point increases if mass flows into and decreases if mass flows out of the surrounding region. Equation~\ref{eqn:mass_1} can also be rewritten using vector identities to give,
\begin{equation}
\frac{\partial \rho}{\partial t} + \vec{v}\cdot\nabla\rho + \rho\mdiv\vec{v} = 0. \label{eqn:mass_2}
\end{equation}
However, for an incompressible fluid along a path with velocity $\vec{v}$ the convective time derivative $D/Dt = \partial/\partial t + \vec{v}\cdot\nabla = 0$. As a result, Eqn.~\ref{eqn:mass_2} reduces to, 
\begin{equation}
\mdiv \vec{v} = 0,
\end{equation}
indicating that there are no sources or sinks in an incompressible fluid.

Next, the equation of motion of the fluid must be considered,
\begin{equation}
\rho \frac{D\vec{v}}{Dt} = -\nabla p + \vec{F} \label{eqn:mot_1}
\end{equation}
where $p$ is the pressure and $\vec{F}$ is the external force acting on the fluid. In this case, the external force consists of the Lorentz force $\vec{j}\times\vec{B}$ and the gravitational force $\rho \vec{g}$. Equations~\ref{eqn:mot_1} can therefore be written as,
\begin{equation}
\rho \frac{\partial\vec{v}}{\partial t} + \rho(\vec{v}\cdot\nabla)\vec{v} = -\nabla p + \vec{j}\times\vec{B} + \rho \vec{g} \label{eqn:mot_2}
\end{equation}
where the electromagnetic and fluid equations have been combined through the Lorentz force and $D/Dt$ has been rewritten as shown above.

As well as the equations outlined above, the MHD equations also include the ideal gas law,
\begin{equation}
p = n k_B T \label{eqn:gas_1}
\end{equation}
where $n$ is the density, $k_B$ is the Boltzmann constant ($1.38\times10^{-23}$~m$^2$~kg~s$^{-2}$~K$^{-1}$) and $T$ is the temperature. This allows the plasma to be treated as an ideal gas, simplifying analysis. The energy equation for the plasma is also included, 
\begin{equation}
\frac{D}{Dt}\left(\frac{p}{\rho^{\gamma}}\right) = 0 \label{eqn:energy_1}
\end{equation}
where $\gamma$ is the ratio of specific heats ($\gamma = c_p/c_v$\footnote{$c_p$ is the specific heat at constant pressure while $c_v$ is the specific heat at constant volume}; usually taken as 5/3) and the plasma has been assumed to be adiabatic in order to simplify calculations. This can then be rewritten as,
\begin{equation}
\frac{\partial p}{\partial t} + (\vec{v}\cdot \vec{\nabla})p = -\gamma p\mdiv \vec{v} \label{eqn:energy_2}
\end{equation}
using vector identities and Equation~\ref{eqn:mass_1}.

With the MHD equations defined, they can now be used as a basis from which to examine the evolution of a magnetised plasma and identify the different waves that result from this evolution.

\subsection{Acoustic Waves}
\label{subsect:acoustic_waves}

To determine the different MHD wave types possible in a magnetised plasma, the above equations are linearised by considering small perturbations. This is achieved by assuming that the wave amplitude is small and that each parameter consists of an equilibrium value (with subscript zero in the following equations; $\vec{B}_0$) and a perturbed value with a spatial and temporal component (with subscript one; $\vec{B}_1$). The different parameters can therefore be described as, 
\begin{eqnarray}
\vec{B} &=& \vec{B_0} + \vec{B_{1}}(\vec{r}, t)  \label{eqn:lin_1} \\
\vec{v} &=& \vec{v_0} + \vec{v_{1}}(\vec{r}, t) = \vec{v_{1}}(\vec{r}, t) \textrm{  (as $\vec{v}_{0} = 0$)} \label{eqn:lin_2} \\
\rho &=& \rho_{0} + \rho_{1}(\vec{r}, t) \label{eqn:lin_3} \\
p &=& p_0 + p_{1}(\vec{r}, t)  \label{eqn:lin_4}
\end{eqnarray}
\corr{With the aid of} these assumptions, the MHD equations presented above can be linearised to give,
\begin{align}
\frac{\partial \vec{B_{1}}}{\partial t} &= \curl(\vec{v_{1}} \times \vec{B_{0}}) \label{eqn:mhd_lin_1} \\
\mdiv \vec{B_{1}} &= 0 \label{eqn:mhd_lin_2} \\
\frac{\partial \rho_{1}}{\partial t} + \mdiv (\rho_{0} \vec{v_{1}}) &= 0 \label{eqn:mhd_lin_3} \\
\rho_{0} \frac{\partial \vec{v_{1}}}{\partial t} &= -\vec{\nabla} p_{1} + \frac{1}{\mu}(\curl \vec{B_{1}}) \times \vec{B_{0}} \nonumber\\
	& \qquad{} + \frac{1}{\mu}(\curl \vec{B_{0}}) \times \vec{B_{1}} + \rho_{1} \vec{g} \label{eqn:mhd_lin_4} \\
\frac{\partial p_{1}}{\partial t} + \vec{v_{1}} \cdot \vec{\nabla} p_{0} &= -\gamma p_{0} \mdiv \vec{v_{1}} \label{eqn:mhd_lin_6}
\end{align}

The first step in determining the different wave types present in a magnetised plasma is an identification of the properties of acoustic waves. To do this, it is assumed that the plasma is not magnetised ($B_{0} = 0$) and that there are no gravitational effects ($g = 0$). The equilibrium pressure gradient is also assumed to be zero, giving a constant equilibrium pressure ($p_{0}$ = constant and $\rho_{0}$ = constant). The linearised MHD equations then become,
\begin{eqnarray}
\frac{\partial \rho_{1}}{\partial t} + \rho_{0} \vec{\nabla} \cdot \vec{v_{1}} &=& 0 \label{eqn:fin_mhd_lin_1} \\
\rho_{0}\frac{\partial \vec{v_{1}}}{\partial t} &=& -\vec{\nabla} p_{1} \label{eqn:fin_mhd_lin_2} \\
\frac{\partial p_{1}}{\partial t} &=& -\gamma p_{0} \vec{\nabla} \cdot \vec{v_{1}} \label{eqn:fin_mhd_lin_3}
\end{eqnarray}
Fourier analysis is then used to search for plane wave solutions to these equations by assuming a solution of the form $Ce^{i(kx - \omega t)}$, where $k$ is the wavenumber\corr{, $\omega$ is the angular frequency and $C$ is a constant}. This form of analysis is used as the mathematical operators may be simplified as,
\begin{align}
\frac{\partial}{\partial t} = -i\omega \hspace{10mm} \frac{\partial^2}{\partial t^2} = -\omega^2 \notag \\
\frac{\partial}{\partial x} = \text{\corr{ik}} \hspace{10mm} \frac{\partial^2}{\partial x^2} = -k^2 \label{eqn:operators} \\
\vec{\nabla} \cdot = i\vec{k} \cdot \hspace{5mm} \vec{\nabla} \times = i\vec{k} \times \hspace{5mm} \vec{\nabla} = i\vec{k} \notag
\end{align}

Equations~\ref{eqn:fin_mhd_lin_1} to \ref{eqn:fin_mhd_lin_3} can then be solved using this approach\corr{, giving the equations,
\begin{eqnarray}
\vec{v}_1 &=& \frac{p_{1}}{\omega \rho_{0}}\vec{k} \label{eqn:ac_1} \\
\vec{k}\cdot\vec{v}_1 &=& \frac{\omega \rho_1}{\rho_0} \label{eqn:ac_2} \\
\vec{k}\cdot\vec{v}_1 &=& \frac{\omega p_{1}}{\gamma p_{0}}. \label{eqn:ac_3}
\end{eqnarray}
These equations indicate that the} result is a longitudinal wave, with the velocity vector $\vec{v}$ parallel to the wave vector $\vec{k}$. The resulting dispersion relation for this plane wave solution \corr{can then be found by rearranging equations~\ref{eqn:ac_1} to \ref{eqn:ac_3} to give,}
\begin{equation}
\omega^2 = k^2 c_s^2
\end{equation}
where $c_{s}$ is the sound speed of the plasma and has a value given by,
\begin{equation}
c_{s}^{2} = \frac{\gamma p_{0}}{\rho_{0}}. \label{eqn:sound_speed}
\end{equation}
It is possible to estimate the sound speed in the solar corona using typical coronal values for this equation. From the ideal gas law in Equation~\ref{eqn:gas_1} above, \corr{Equation~\ref{eqn:sound_speed} can be rewritten} as, 
\begin{equation}
c_s = \sqrt{\frac{\gamma k_B T}{m_p}}.
\end{equation}
For T$\sim$10$^6$~K, $\gamma\sim$1, k$_{B}=1.38\times10^{-23}$~m$^2$~kg~s$^{-2}$~K$^{-1}$ and m$_{p}\sim$10$^{-27}$~kg, this produces an estimated sound speed of c$_{s}\sim$100~km~s$^{-1}$ for the solar corona.

This indicates that there is a wave solution to the MHD equations that depends entirely on the pressure and density of the plasma, i.e.,\ an acoustic wave. Analogously, it is possible to ignore the effects of pressure and density and determine if there is a purely magnetic wave solution to the MHD equations.

\subsection{Magnetic Waves}
\label{subsect:mag_waves}

To identify pure magnetic waves, density and pressure variations must first be ignored, setting $p_{0} = 0$\corr{, $\rho_{0} \neq 0$, $p_{1} =$ const, $\rho_{1} =$ const and $g = 0$}. For simplicity of calculations, the magnetic field is also assumed to be uniform and directed in the $\vec{z}$ direction. The linearised MHD equations (equations~\ref{eqn:mhd_lin_1} to \ref{eqn:mhd_lin_6} above) then become,
\begin{eqnarray}
\mdiv \vec{v_{1}} &=& 0 \\
\rho_{0} \frac{\partial \vec{v_{1}}}{\partial t} &=& \frac{1}{\mu} (\curl \vec{B_{1}}) \times \vec{B_{0}} \\
\frac{\partial \vec{B_{1}}}{\partial t} &=& \curl (\vec{v_{1}} \times \vec{B_{0}}) \\
\mdiv \vec{B_{1}} &=& 0
\end{eqnarray}

Once again, plane wave solutions to these equations may be found using Fourier analysis\corr{, giving,
\begin{eqnarray}
i \vec{k} \cdot \vec{v}_1 &=& 0 \\
-i \omega \rho_0 \vec{v}_1 &=& \frac{1}{\mu}(i\vec{k} \times \vec{B}_1)\times\vec{B}_0 \\
-i \omega \vec{B}_1 &=& i\vec{k} \times (\vec{v}_1 \times \vec{B}_0) \\
i\vec{k} \cdot \vec{B}_1 &=& 0.
\end{eqnarray}
Here, the plane wave solution has a velocity vector $\vec{v}_1$ that is perpendicular to the wave vector $\vec{k}$, and also to the direction of the magnetic field vector $\vec{B_{1}}$. These equations may then be manipulated using vector identities to produce,
\begin{equation}
\omega^2 = \frac{(\vec{k} \cdot \vec{B}_0)^2}{\mu \rho_0},
\end{equation}
which can then be used to produce the dispersion relation for a purely magnetic wave,}
\begin{equation}
\omega = \pm v_A k_z
\end{equation}
where $k_z$ is the wave-vector in the z-direction. The wave velocity in this case ($v_A$) is called the Alfv\'{e}n speed and is derived as
\begin{equation}
v_{A} = \frac{B_{0}}{\sqrt{\mu \rho_{0}}}, \label{eqn:alfven_speed_0}
\end{equation}
in S.I. units, where $\mu$ is the magnetic permeability. It is typically assumed that the plasma is fully ionised, with solar physics also using cgs nomenclature by convention, allowing Equation~\ref{eqn:alfven_speed_0} to be rewritten as,
\begin{equation}
v_{A} = \frac{B_{0}}{\sqrt{4\pi n_{i} m_{p}}}, \label{eqn:alfven_speed_1}
\end{equation}
where $B_{0}$ is in units of \corr{gauss}, $n_{i}$ is the ion \corr{number} density (cm$^{-3}$), and $m_{p}$ is the proton mass (g). Once again, it is possible to estimate a value for the Alfv\'{e}n speed in the solar corona using typical coronal values. For $n_{i}\sim$10$^8$~cm$^{-3}$, $m_{p}=1.67\times10^{-24}$~g and $B\sim$10~G, this produces an Alfv\'{e}n speed of $v_A \sim$2000~km~s$^{-1}$ for the corona.

These waves are called Alfv\'{e}n waves and are pure magnetic waves that result from the oscillation of magnetic field lines. They are anisotropic, incompressible waves that propagate along magnetic field lines.

\subsection{Magneto-acoustic Waves}
\label{subsect:mag_acous_waves}

As the solar corona contains variations in magnetic field, density, and pressure, any waves present will contain a contribution from both acoustic and magnetic wave modes, giving magneto-acoustic waves. Identifying these waves requires the combination of magnetic and gas pressure, although gravitational effects can be ignored as they are negligible in comparison. The linearised MHD equations (particularly Equations~\ref{eqn:mhd_lin_1}, \ref{eqn:mhd_lin_4} and \ref{eqn:mhd_lin_6}) are then used to determine the dispersion relation for a magnetoacoustic wave.

Both the magnetic and gas pressure are assumed to be uniform and of the form $\vec{B_{0}} = B_{0}\vec{z}$ for the magnetic pressure and $p_{0} = C$ (where $C$ is a constant) for the gas pressure. By combining Equations~\ref{eqn:mhd_lin_1}, \ref{eqn:mhd_lin_4} and \ref{eqn:mhd_lin_6} with the mathematical operators shown in \ref{eqn:operators}, it is possible to derive the simplified equation,
\begin{equation}
\omega^{4} - \omega^{2}k^{2}(c_{s}^{2} + v_{A}^{2}) + c_{s}^{2}v_{A}^{2}k^{4}\cos^{2}\theta_{B} = 0, \label{eqn:disp_rel}
\end{equation}
where $\theta_{B}$ is the angle between the wave-vector $\vec{k}$ and the direction of the magnetic field $\vec{B}$. This is a quadratic equation in terms of $\omega^{2}/k^{2}$ and may be solved using the quadratic formula to produce the dispersion relation for magnetoacoustic waves,
\begin{equation}
\frac{\omega}{k} = \left(\frac{1}{2}(c_{s}^{2} + v_{A}^{2}) \pm \frac{1}{2} \sqrt{c_{s}^{4} + v_{A}^{4} - 2c_{s}^{2}v_{A}^{2}\cos 2\theta_{B}}\right)^{\frac{1}{2}}.
\end{equation}

There are two distinct solutions to this equation. Solving for the negative sign produces the slow-mode MHD wave, while solving for the positive sign produces the fast-mode MHD wave. The equation also reduces to the dispersion relations for an acoustic and magnetic wave when magnetic and pressure effects respectively are ignored. 

To examine the differences between the fast-mode and slow mode MHD waves, first consider the case where $\theta_{B} = 0$ (i.e.,\ $\vec{k} \parallel \vec{B}$). This gives the solutions,\corr{
\begin{equation}
\frac{\omega}{k} = v_{fast-mode} = v_{A} \hspace{3mm} \textrm{and} \hspace{3mm} \frac{\omega}{k} = v_{slow-mode} = c_{s}
\end{equation}
}
for the fast--mode (left) and slow--mode (right) MHD waves. This means that the fast--mode wave travels at the \corr{Alfv\'{e}n} speed while the slow--mode wave travels at the \corr{local sound} speed when propagating along the magnetic field. For the case where $\theta_{B} = \pi/2$ (i.e.:\ $\vec{k} \perp \vec{B}$) the solutions are,
\begin{equation}
\left(\frac{\omega}{k}\right)^{2} = v_{fast-mode}^{2} = c_{s}^{2} + v_{A}^{2} \hspace{3mm} \textrm{and} \hspace{3mm} \frac{\omega}{k} = v_{slow-mode} = 0
\end{equation}
for the fast--mode (left) and slow--mode (right) respectively. This means that the fast--mode wave velocity when propagating across the magnetic field is dependent on both the sound speed and the Alfv\'{e}n speed, while the slow--mode wave cannot propagate across the magnetic field.

\begin{figure}[!t]
\centerline{\includegraphics[width=0.9\textwidth,clip=,trim= 2mm 4mm 0mm 2mm]{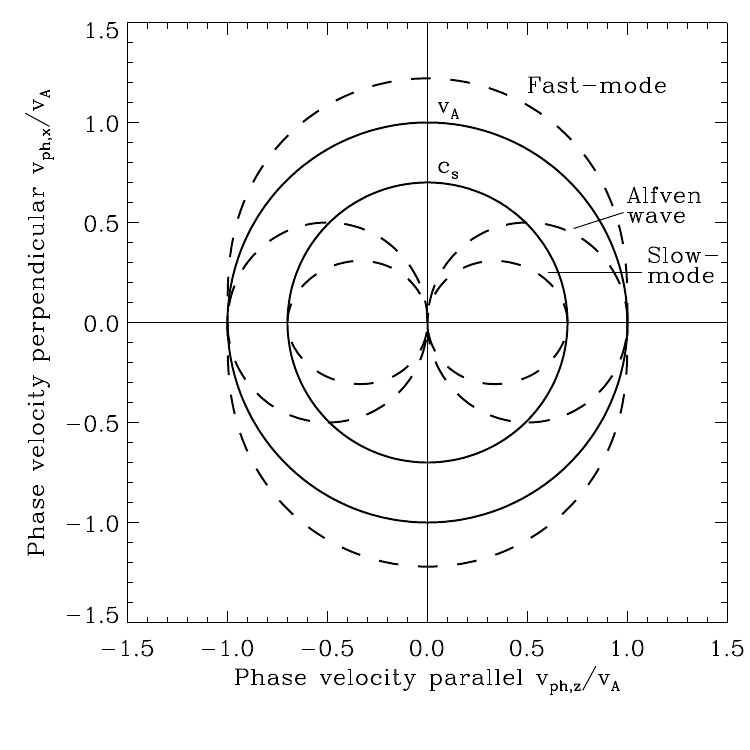}}
\caption[The phase velocities of magnetoacoustic waves]{Polar plot showing the phase velocities of magnetoacoustic waves for a ratio $c_{s}/v_{A} = 0.7$. The sound \corr{and Alfv\'{e}n speeds are shown as solid lines, with the slow--mode, Alfv\'{e}n and fast--mode wave speeds shown as dashed lines \citep[cf.][]{Aschwanden:2004fk}}.}
\label{fig:polarplot}
\end{figure}

The solutions to equation~\ref{eqn:disp_rel} for all angles $\theta$ are shown in Figure~\ref{fig:polarplot} for the ratio $c_{s}/v_{A} = 0.7$ \corr{with the Alfv\'{e}n and sound speeds} also shown for comparison. It is \corr{clear} that the Alfv\'{e}n and sound velocity (v$_{A}$ and c$_{s}$) do not show any variation as the orientation of the magnetic field changes. In contrast, the fast--mode, slow--mode and Alfv\'{e}n wave velocities are highly dependent on the \corr{magnetic field} orientation, a result that has major implications for the study of the solar corona.

It should be noted that in ideal MHD as discussed here the different wave velocities are constant, with any variation between events dependent on the plasma parameters. These waves are also non--dispersive and non--dissipative, with the result that the MHD equations must be extended to include resistive terms when considering phenomena such as flares or magnetic reconnection in the solar atmosphere. However, recent work by \citet{Murawski:2001ab} and \citet{Nakariakov:2001p677} suggests that the randomly structured nature of the solar corona may result in the deceleration, dispersion and dissipation of a propagating MHD wave. This behaviour was observed using simulations of MHD waves and may be applicable to observations of the solar corona. \corr{The variation of the morphology of larger--amplitude waves in a plasma has been studied by \citet{Vrsnak:2000p15,Vrsnak:2000p10}. In these examples, the pulse exhibits non--linear behaviour, with a propagation velocity that is dependent on the amplitude, while deceleration, dispersion and dissipation of the pulse follow as a natural consequence.}

\section{Thesis Outline}
\label{sect:outline}

The work outlined in this thesis deals with the physical properties of large--scale propagating disturbances observed in the low solar corona, with particular attention paid to their kinematics and temporal variation in morphology. The kinematics of these disturbances remain a source of much debate as a consequence of conflicting observations and the relatively low observing cadence of the instruments used to study them. It is shown here that this low observing cadence may have produced an underestimation of the kinematics previously estimated for these pulses. The different techniques traditionally used to derive the pulse kinematics are also examined and compared to a bootstrapping technique, with the bootstrapping technique shown to produce a better estimate of the true pulse kinematics. A semi--automated technique for identifying, tracking and analysing this phenomenon is presented and used to show that the disturbances exhibit clear deceleration and pulse dispersion with propagation. Finally, the dispersion relation of a pulse is calculated using high cadence observations of a CBF pulse.

An overview of previous observations of globally--propagating disturbances in the corona is presented in Chapter~\ref{chap:cbfs}, with the different theories proposed to explain the phenomenon also discussed. The different instruments used in this work are outlined in Chapter~\ref{chap:inst}, along with a discussion of the software used to process the data. The first observations of a globally--propagating disturbance made by the \emph{STEREO} spacecraft are presented in Chapter~\ref{chap:first_obs}, with some of the issues resulting from the analysis of this event discussed in Chapter~\ref{chap:methods}. A semi-automated technique for studying these disturbances outlined in Chapter~\ref{chap:methods} is used in Chapter~\ref{chap:wave_properties} to provide an unbiased analysis of four separate events observed by the \emph{STEREO} spacecraft. This technique is also applied to an event observed by the \emph{SDO} spacecraft in Chapter~\ref{chap:sdo_cbf} and used to examine the physical properties of the pulse and the corona itself. Finally, the major conclusions of this thesis and some possible directions for future work are provided in Chapter~\ref{chap:concs}.	 

\chapter{Observations and Theory of Coronal Bright Fronts}
\label{chap:cbfs}
\ifpdf
    \graphicspath{{2/figures/PNG/}{2/figures/PDF/}{2/figures/}}
\else
    \graphicspath{{2/figures/EPS/}{2/figures/}}
\fi

\noindent 
\\ {\it 
I am not accustomed to saying anything with certainty after only one or two observations.
\begin{flushright}
Andreas Vesalius \\
\end{flushright}
 }

\vspace{15mm}
In this chapter, a brief history of the observations of globally--propagating disturbances in the solar atmosphere is presented along with an overview of the myriad theories proposed to explain them. The first observation of a globally propagating disturbance in the EUV solar corona was made by \citet{Moses:1997vn} using the Extreme ultraviolet Imaging Telescope \citep[EIT;][]{Delaboudiniere:1995p51} onboard the \emph{SOlar and Heliospheric Observatory} \citep[\emph{SOHO};][]{Domingo:1995fk} spacecraft. They noted the appearance of ``a localized increase in emission propagating across the solar disk'' associated with a Coronal Mass Ejection (CME) eruption from 1996~December~23. The observed pulse was labelled an ``EIT wave'', and became a source of much interest as a result of its association with other solar phenomena such as flares and CMEs. 

``EIT waves'' have been studied in detail for almost 15~years and while advances have been made with regard to their physical interpretation, many unanswered questions still remain. The various observations of these pulses are presented in Section~\ref{sect:cbf_obs}, with the two main branches of theories proposed to explain them outlined in Section~\ref{sect:cbf_theories}. This section is complemented by recent reviews of ``EIT waves'' by \citet{Warmuth:2007p1991,Warmuth:2010p2822}, \citet{Wills-Davey:2010ab}, \citet{Vrsnak:2008p622} and \citet{Zhukov:2011rt}, with much of the material used here taken from the review paper published in \emph{Space Science Reviews} by \citet{Gallagher:2011fk}.

It should be noted at this point that ``EIT waves'' are so-called for their wave-like appearance and their first observation using \emph{SOHO}/EIT. Since their initial discovery, they have been observed using myriad instruments across different passbands, while their true physical nature remains ambiguous. Although this name is still widely used for historical reasons, in this work we shall refer to ``EIT waves'' as ``coronal bright fronts'', or CBFs, following the lead of \citet{Gallagher:2011fk}.

\section{Observational Properties}
\label{sect:cbf_obs}

\begin{figure}[!t]
\centering
\includegraphics[angle=90,clip=,trim=5mm 5mm 96mm 5mm,width=0.99\textwidth]{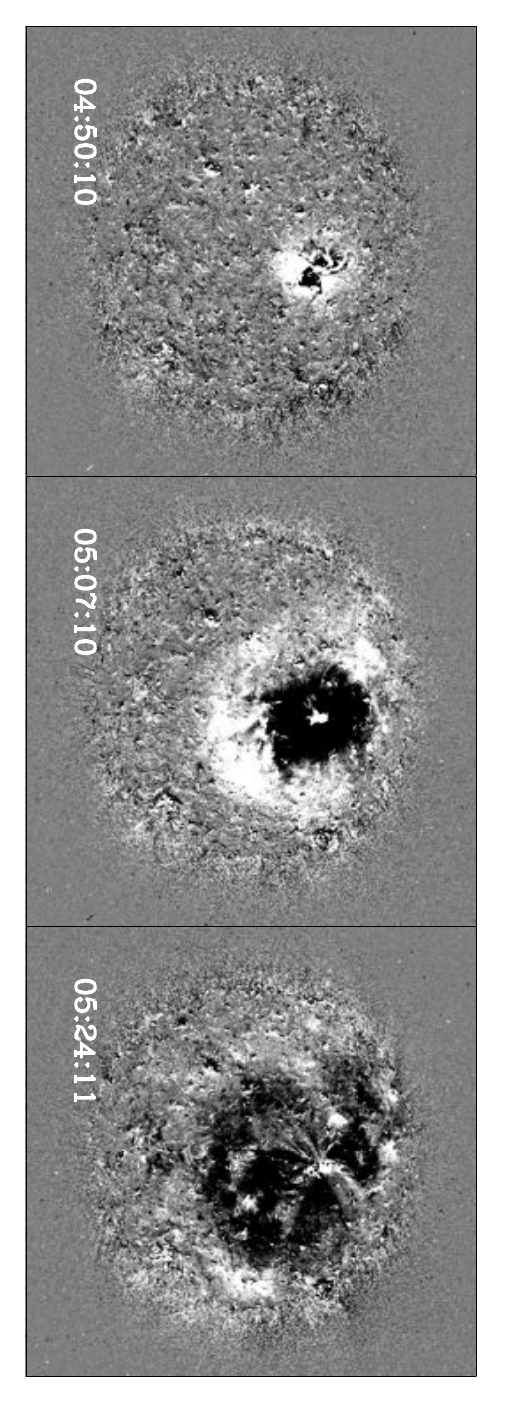}
\caption[Running difference images showing the 1997~May~12 ``EIT wave'' event]{Running difference images showing the 1997~May~12 ``EIT wave'' event studied by \citet{Thompson:1998sf}. The time of the leading image is given in the bottom left of each panel.}
\label{fig:may1997}
\end{figure}

After the initial identification by \citet{Moses:1997vn}, an event was observed on 1997~May~12 (see Figure~\ref{fig:may1997}), and studied in much greater detail by \citet{Thompson:1998sf}. This is traditionally treated as the original CBF identification paper, as it classifies the different coronal signatures associated with the event, including the formation of dimming regions close to the event origin, a post--eruption arcade and the bright wavefront propagating quasi--radially away from the source region.

The work of \citet{Thompson:1998sf} was subsequently followed by a detailed analysis of a CBF observed using the \emph{Transition Region And Coronal Explorer} \citep[\emph{TRACE};][]{Handy:1999p185} spacecraft. Here, \citet{Wills-Davey:1999ve} studied the higher cadence observations from both the 171 and 195~\AA\ passbands available using \emph{TRACE} and found that the pulse propagated non--uniformly, with variations in both trajectory and velocity along the front of the pulse. The pulse was also \corr{noted} to have a Gaussian form, indicating a single compressional front.

\corr{With the assistance of} the continuous data stream available from \emph{SOHO}/EIT, the number of observed CBF events steadily increased, with numerous events observed and catalogued \citep[see][for a catalogue of events from 1997--1998, but not published until 2009]{Thompson:2009yq}. These observations were used for statistical studies to try and determine the relationship of CBFs to other solar phenomena with either CMEs or flares thought to be the initial driver. 

This uncertainty in origin was overcome with the work of \citet{Biesecker:2002lq} and \citet{Cliver:2005bd} who used statistical studies of CBFs to study the relationship between CMEs, flares and CBFs. After correcting for statistical bias and observational effects, \citet{Biesecker:2002lq} found that CBFs were always associated with a CME, whereas the converse was not necessarily true. This was followed by \citet{Cliver:2005bd}, who found that approximately half of the CBF sample studied were associated with small-scale flares of GOES intensities below C-class. This suggested that flares did not have the requisite energy to drive a CBF pulse, indicating that CMEs were the most likely driver. 

Although most apparent in EUV observations as a result of the data available from \emph{SOHO}/EIT, globally-propagating disturbances have been observed in other passbands including soft X-ray \citep[SXR;][]{Hudson:2003gd,Warmuth:2005vf, Khan:2002qq}, radio \citep{White:2005ss}, \ion{He}{1} 10830~\AA\ \citep{Gilbert:2004hl}, and H$\alpha$ \citep[][as mentioned in Section~\ref{subsect:chromosphere}]{Moreton:1960p44}.

\begin{figure}[!t]
\begin{center}
\includegraphics[keepaspectratio, width=0.5\textwidth,clip=,trim=35mm 10mm 35mm 10mm,angle=270]{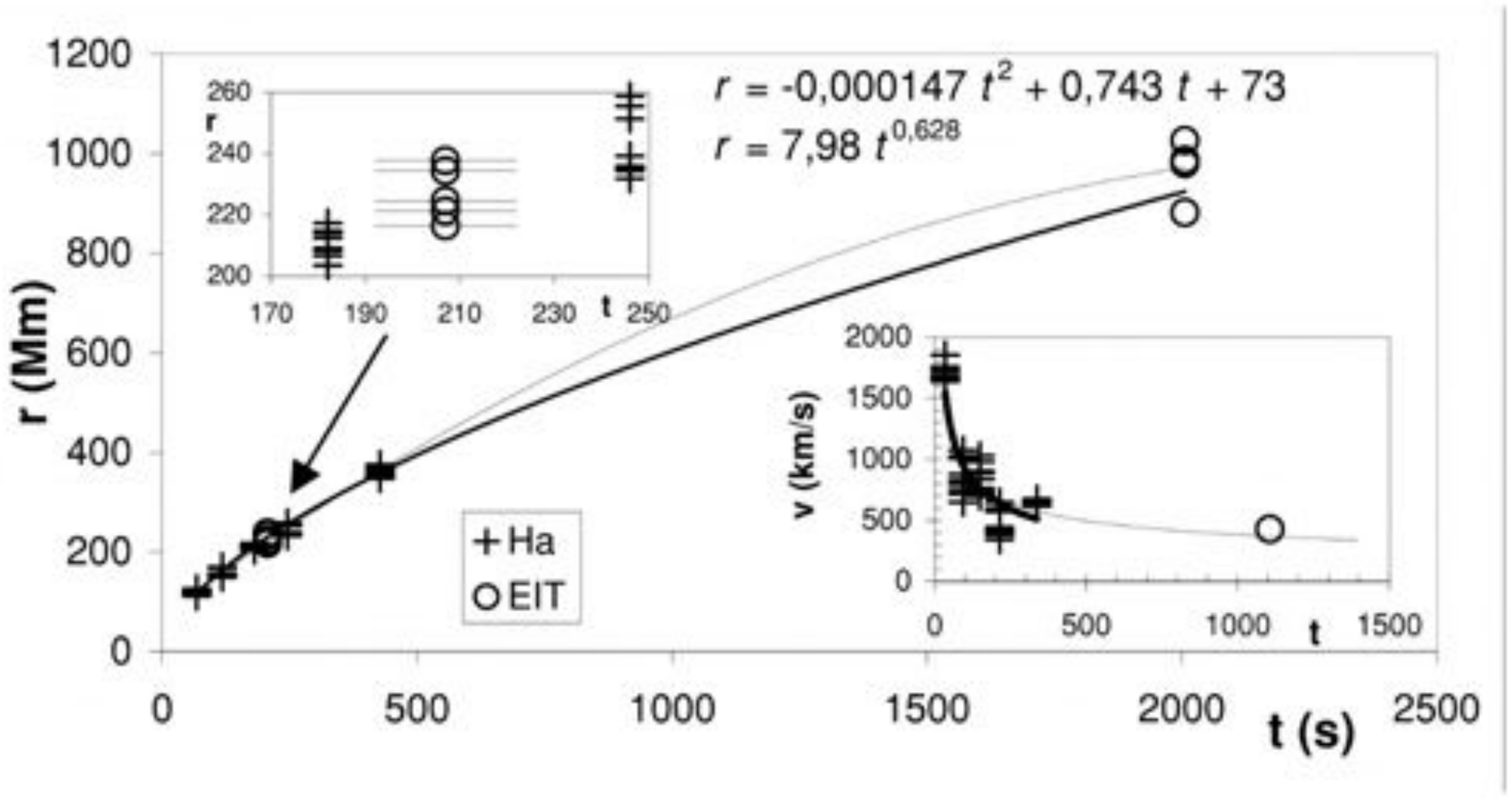}
\includegraphics[keepaspectratio, width=0.8\textwidth,clip=,trim=10mm 80mm 10mm 80mm]{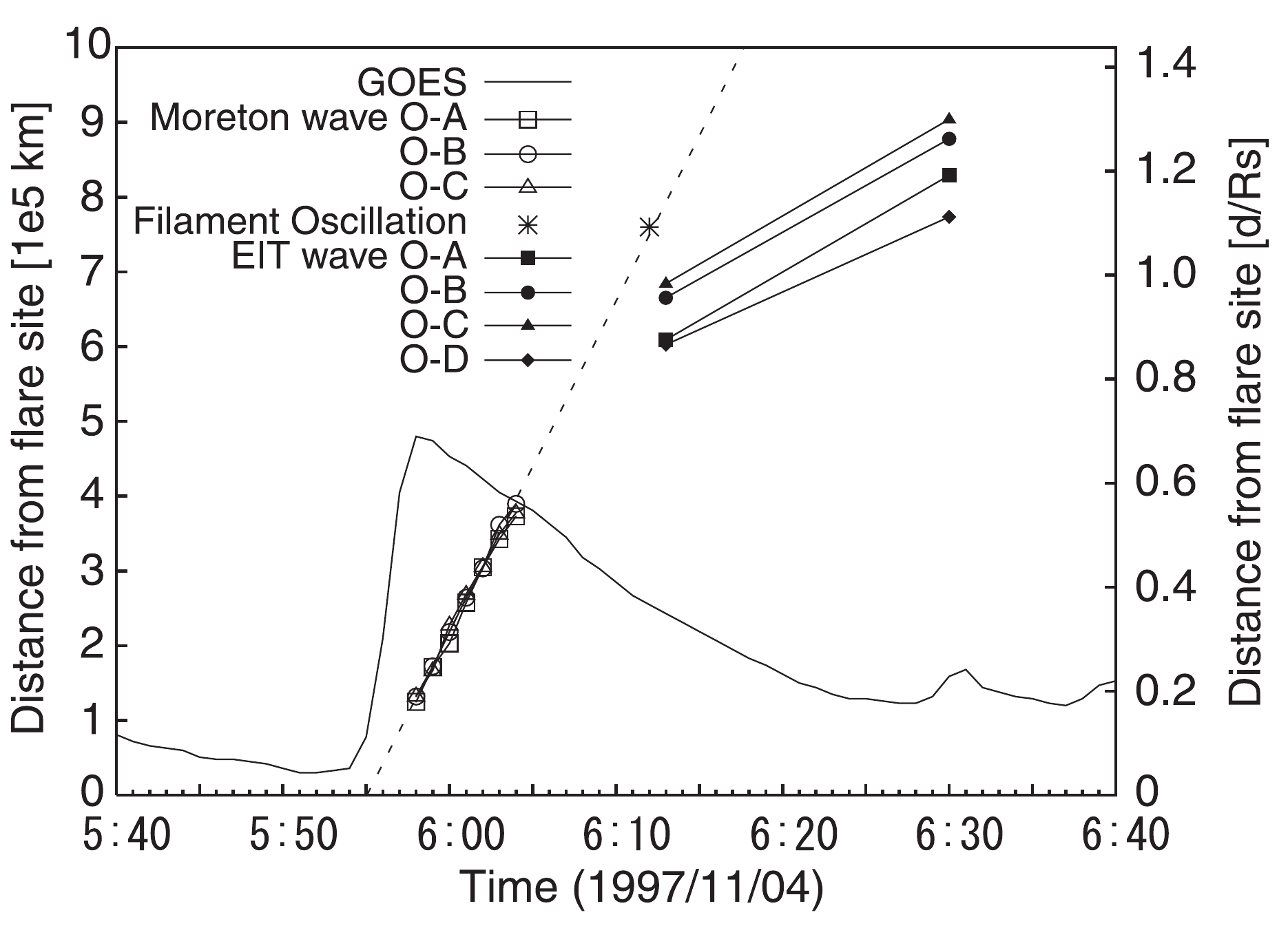}
\caption[Distance-time plots from \citet{Warmuth:2001p77} (top) and \citet{Eto:2002p120} (bottom)]{\emph{Top}: Distance-time plot from \citet{Warmuth:2001p77} showing strong correspondence between Moreton wave and CBF kinematics for an event from 1997~November~3. \emph{Bottom}: Distance-time plot from \citet{Eto:2002p120} showing the different kinematics of a Moreton wave and a CBF for an event from 1997~November~4.}
\label{fig:cbf_moreton}
\end{center}
\end{figure}

The relationship between CBFs and Moreton waves has been a particular focus of investigation, with many authors trying to identify the physical connection (if any) between the two phenomena. The first work to investigate H$\alpha$ and EUV observations of the same eruptive events \citep{Warmuth:2001p77} found compelling evidence that the Moreton waves and CBFs travelled co-spatially, suggesting that they could be different forms of the same disturbance (see the top panel of Figure~\ref{fig:cbf_moreton}). However, this work was quickly repudiated by \citet{Eto:2002p120}, who found no overlap between the visibility times of a Moreton wave and a CBF for an event from 1997~November~4 (bottom panel of Figure~\ref{fig:cbf_moreton}), while a morphological and kinematical investigation of the same event showed no similarities between the different disturbances. Using a statistical approach, \citet{Okamoto:2004cq} found that the majority of CBFs were not associated with a Moreton wave, \corr{while \citet{Warmuth:2004rm,Warmuth:2004ab}, found a 100~\% correlation with CBF events using a list of 12 observed Moreton waves as a starting point. This suggests that while CBF events may be relatively common, an additional constraint must be met before a Moreton wave can occur.}

\begin{figure}[!t]
\begin{center}
\includegraphics[keepaspectratio, width=0.38\textwidth,clip=,trim=60mm 5mm 40mm 5mm,angle=270]{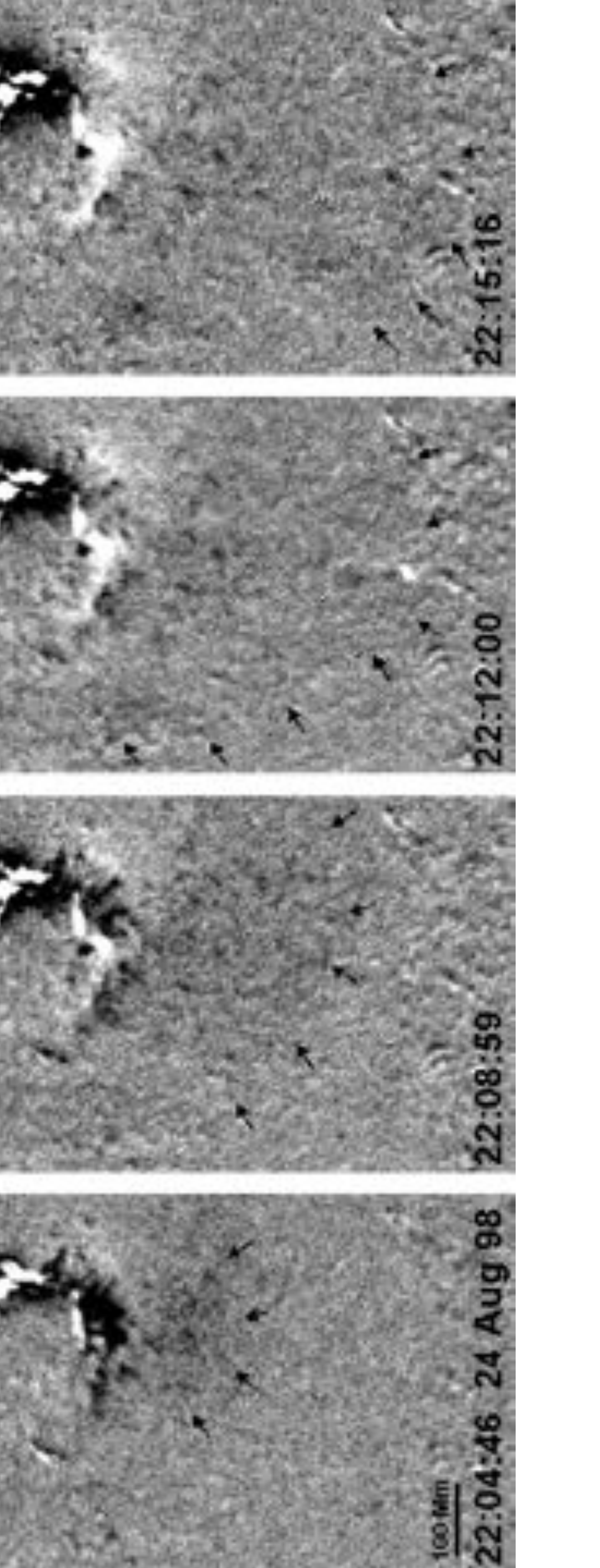}
\includegraphics[keepaspectratio, width=0.5\textwidth,clip=,trim=30mm 0mm 30mm 0mm,angle=270]{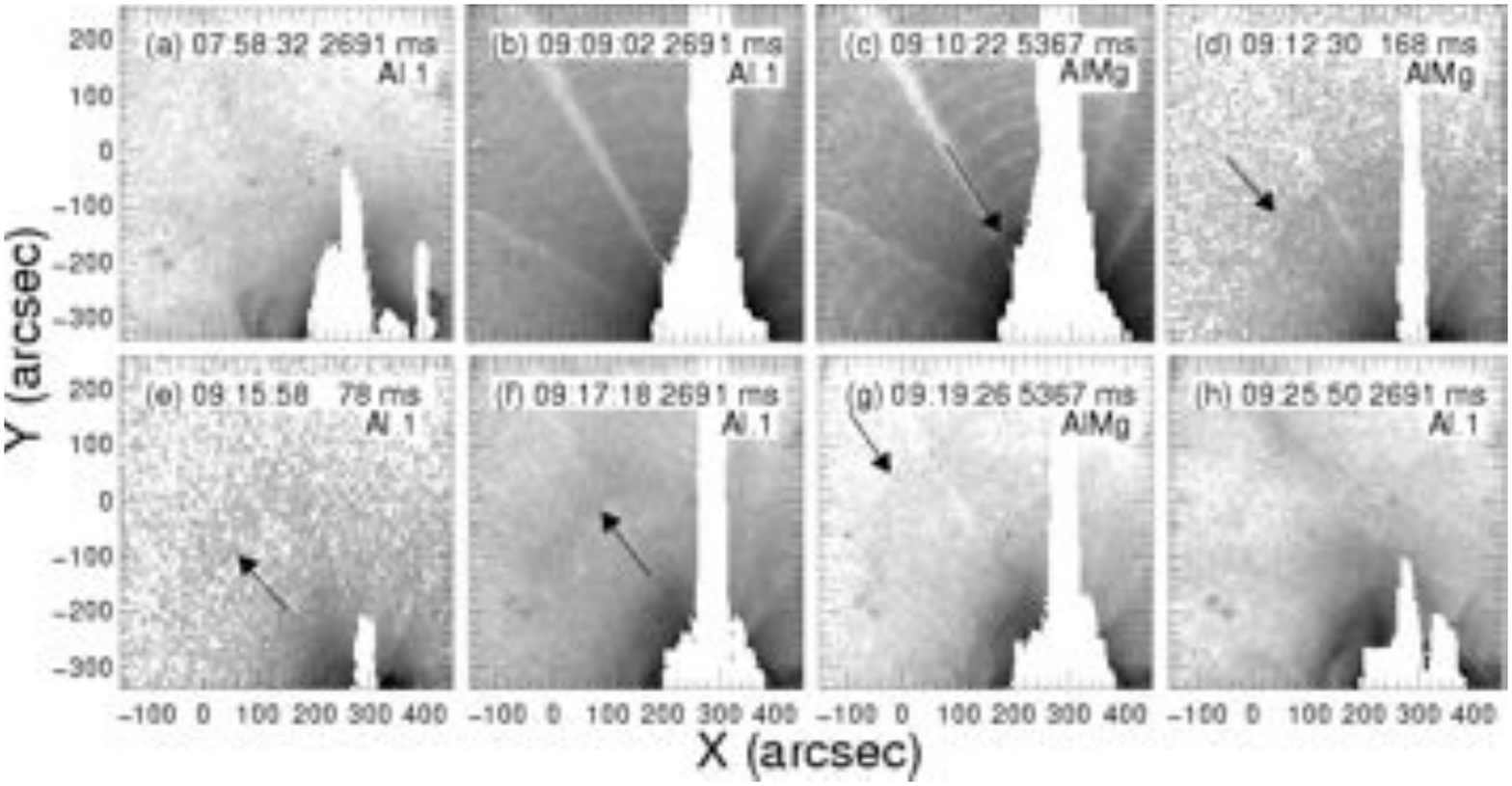}
\caption[Image panels showing a \ion{He}{1} pulse from \citet{Vrsnak:2002p151} (top) and a soft X-ray pulse from \citet{Khan:2002qq} (bottom)]{\emph{Top}: Difference images from \citet{Vrsnak:2002p151} showing a pulse observed using the \ion{He}{1} emission line from 1998~August~24. \emph{Bottom}: Plain images from \citet{Khan:2002qq} showing a pulse observed on 1997~November~3 using soft X-ray emission from \emph{Yohkoh}.}
\label{fig:he1_xray}
\end{center}
\end{figure}

A possible link between the coronal EUV observations of a CBF and the chromospheric Moreton wave was observed by \citet{Vrsnak:2002p151} using the chromospheric \ion{He}{1}~10830~\AA\ emission line (see the top panel of Figure~\ref{fig:he1_xray}). This emission line provides a way to probe the upper chromosphere, potentially allowing it to be used to combine H$\alpha$ and EUV observations, as initially proposed by \citet{Vrsnak:2002p151}. The importance of this emission line was emphasised by \citet{Gilbert:2004ab}, who observed 2 CBF pulses traveling co-spatially with \ion{He}{1} pulses. These observations implied that the \ion{He}{1} pulse was an imprint of the CBF pulse in the chromosphere as it propagates in the corona, rather than being a distinct wave formed by the same process as the CBF. Although it is tempting to view \ion{He}{1} observations as the ``missing link'' between the Moreton wave and the CBF, the complex formation mechanism of the \ion{He}{1} line, combined with the low observation rate in this passband makes this difficult \citep{Gilbert:2004hl}. However, these observations must be explained as part of any theory attempting to explain the CBF phenomenon.

While globally propagating disturbances have been observed \corr{directly} using radio data \citep[cf.][]{White:2005ss}, the most apparent radio signature of a coronal disturbance is the Type~\Rmnum{2} burst. A strong correlation was noted between Type~\Rmnum{2} bursts and chromospheric Moreton waves \citep{Kai:1970p1036}, while a statistical survey of observed Type~\Rmnum{2} bursts and Moreton waves found that $\sim$70\% of Moreton waves were associated with a type~\Rmnum{2} burst \citep{Pinter:1977p369}. \corr{\citet{Klassen:2000p7} compared a list of recorded Type~\Rmnum{2} radio bursts with data from \emph{SOHO}/EIT, finding that $\sim$90~\% of the observed radio events had an associated CBF, although the converse is not necessarily true \citep{Biesecker:2002lq}.} Microwave emission at 17 and 34~GHz from the Nobeyama Radioheliograph was used by \citet{White:2005ss} to derive a velocity of $\sim$830~km~s$^{-1}$ for a CBF event, leading the authors to suggest that the low cadence of \emph{SOHO}/EIT may be producing an underestimation of the pulse kinematics. However, the observations in this case were poorly constrained with a low signal-to-noise ratio. This poor signal-to-noise ratio is a major contributing factor to the sparse observations of radio pulses, although it may be rectified using modern radio observatories such as the LOw Frequency ARray (LOFAR) or the Frequency Agile Solar Radiotelescope \citep[FASR;][]{Bastian:2004ab} systems.

Globally propagating disturbances have also been observed in soft X-ray (SXR) emission although they have been difficult to observe due to the relatively low signal-to-noise of the pulse. The first observation was made using data from the \emph{Yohkoh} spacecraft \citep[][bottom panel of Figure~\ref{fig:he1_xray}]{Khan:2002qq}, with the studied event also showing a Moreton wave and a CBF. The X-ray wave was found to have a derived velocity of $\sim$550~km~s$^{-1}$, while the observation times of all three pulses were found to be comparable. The extrapolated leading edge of the X-ray wave was found to match the Moreton wave and CBF very closely, suggesting that all three disturbances were morphologically related. X-ray waves have since been observed by both \citet{Hudson:2003gd} and \citet{Warmuth:2005vf}, with both authors suggesting that the emission is consistent with a propagating coronal shock wave.

This interpretation has been questioned by \citet{Attrill:2009ab}, who combined X-ray observations from the X-Ray Telescope \citep[XRT;][]{Golub:2007ab} onboard the \emph{Hinode} spacecraft \citep{Kosugi:2007uq} with EUV observations from the Extreme UltraViolet Imager \citep[EUVI;][]{Wuelser:2004bs} onboard the \emph{Solar TErrestrial RElations Observatory} \citep[\emph{STEREO};][]{Kaiser:2008p16} spacecraft. The X-ray emission corresponding to \corr{an X--ray} pulse was shown to match emission from an EUV pulse, with both matching the edge of the erupting CME. These observations lead \citet{Attrill:2009ab} to suggest that the  X-ray and EUV emission was a result of the CME edge shocking the coronal plasma as it erupts rather than being due to a freely propagating wave. The rare nature of X-ray waves means that this remains an open question.

\section{Theories of Coronal Bright Fronts}
\label{sect:cbf_theories}

\corr{The multitude of observations across many diverse passbands outlined in the previous section underlines the difficulties associated with elucidating ``EIT waves'', and has resulted in the evolution of two sets of theories that attempt to explain them.} On one hand, CBFs are interpreted \corr{as waves} using \corr{either} magnetohydrodynamic wave theory \corr{or shock wave theory. Alternatively, they are visualised as} brightenings resulting from the reorganisation of the coronal magnetic field during the eruption of the associated CME. In this case, the ``EIT wave'' is not a true wave but instead \corr{a brightening due to small--scale magnetic reconnection or Joule heating}. Both of these sets of theories are discussed below.

\subsection{Wave Interpretations}
\label{subsect:cbfs_waves}

A globally-propagating wavefront in the corona was initially proposed by \citet{Uchida:1968qv} to explain the Moreton wave phenomenon first observed by \citet{Moreton:1960fj}. This was a propagating front moving in the chromosphere at velocities much greater than the predicted local Alfv\'{e}n velocity. \citet{Uchida:1968qv} proposed that the observed signatures could be explained by the propagation of a fast-mode MHD wave in the solar corona, expounding on this theory in later work \citep{Uchida:1970p2,Uchida:1973p139,Uchida:1974p5}. As a fast-mode MHD wave propagates in the corona, the variation of the Alfv\'{e}n speed at the corona/chromosphere intersection should result in a circular intersection line that is observed as a Moreton wave. This matches the ``up-down'' swing in the H$\alpha$ line observed by \citet{Dodson:1968fk} corresponding to the passage of a Moreton wave; indicative of a pulse compressing the chromosphere as it passes.

With the discovery of the ``EIT wave'' in \emph{SOHO}/EIT data \citep{Moses:1997vn,Dere:1997p159} it was thought that this could be the coronal counterpart to the chromospheric Moreton wave originally predicted by \citet{Uchida:1968qv}. The first thorough analyses of the coronal bright front were made by \citet{Thompson:1998sf} and \citet{Thompson:1999cd}, who studied events from 1997~May~12 and 1997~April~7 respectively. These papers were the first to document in detail the different observational features associated with the CBF event, including the formation of dimming regions close to the event origin, a post-eruption arcade and the bright wavefront which propagated quasi-radially away from the source region. The pulses are noted to be quite diffuse, anisotropic, and inconsistent with the propagation of a shock, while the estimated velocities are quite low ($\sim$250~km~s$^{-1}$). 

These observations of the so-called ``EIT wave'' did not match the expected kinematics of a fast-mode wave that could explain the Moreton wave. In particular, the low velocities (typical CBF velocities are $\sim$2--3 times lower than that of a Moreton wave) and morphology (CBFs are observed to be broad and diffuse with Moreton waves thin and pronounced) are quite different. Despite these disagreements, Moreton waves and CBFs do show some similarities, with \citet{Veronig:2006fy} outlining evidence of refraction at a coronal hole boundary for a Moreton wave and an associated CBF.

A detailed examination of a CBF pulse was made by \citet{Wills-Davey:1999ve} using observations of an event made by \emph{TRACE}, which allowed a much higher cadence, higher resolution observation to be made using a smaller field-of-view in both the 171 and 195~\AA\ passbands. The pulse was observed to propagate non-uniformly, with variations in both trajectory and velocity along the front of the pulse, indicating Alfv\'{e}nic behaviour with a strong compressional component. An intensity slice across the pulse revealed a Gaussian form, indicating a single compressional front, while the intensity ratio between the 171 and 195~\AA\ passbands was used to show the presence of plasma heating and constrain the temperature through which the pulse propagated to $\sim$1--1.4~MK.

The wave interpretation of CBFs was supported by an analysis of multiple events performed by \citet{Warmuth:2004rm,Warmuth:2004ab}, who found that the kinematics of CBFs lay on a similar kinematical curve to those of a Moreton wave. This result strongly implied that the observed CBF could be directly related to the Moreton wave and is supported by observations of CBFs and Moreton waves traveling co-spatially \citep{Thompson:2000rm,Okamoto:2004cq}. 

\begin{figure}[!t]
\centering{
\includegraphics[keepaspectratio, width=0.9\textwidth,clip=,trim=0mm 0mm 0mm 0mm]{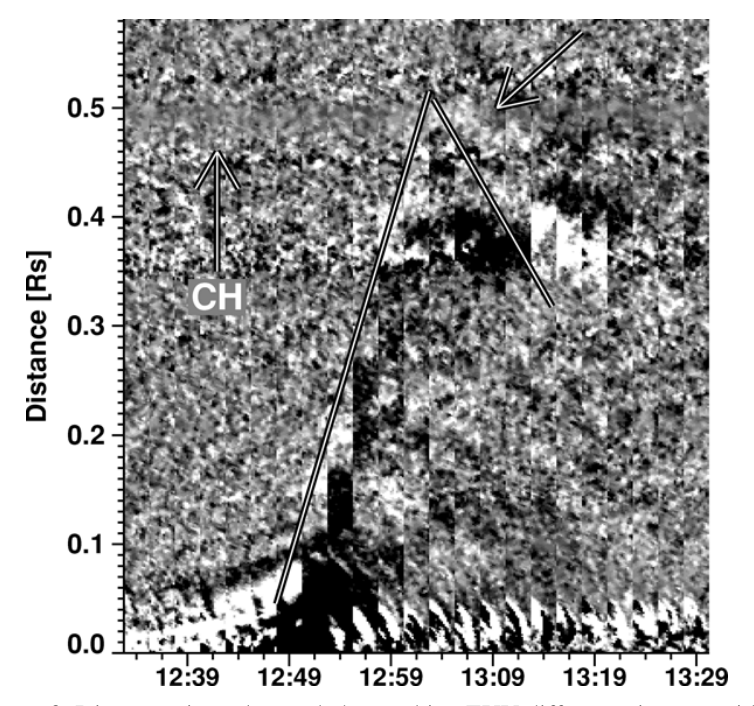}
}
\caption[Distance-time plot showing a reflected CBF pulse from \citet{Gopalswamy:2009p1527}]{Distance-time stack plot showing the variation in intensity along a defined line from the erupting active region (N01W08; bottom) towards a nearby coronal hole (S10W45; top). The propagating CBF pulse is indicated, along with the reflected pulse at $\sim$13:00~UT \citep{Gopalswamy:2009p1527}.}
\label{fig:gopal_reflection}
\end{figure}

The wave interpretation of CBFs has also been supported by observations of reflection and refraction \citep[][see Figure~\ref{fig:gopal_reflection}]{Long:2008eu,Veronig:2008p216,Gopalswamy:2009p1527} at active region and coronal hole boundaries. The kinematics of CBFs have also been determined to a high degree of accuracy using the higher cadence available from the \emph{STEREO} spacecraft \citep[cf.,][]{Long:2008eu,Veronig:2008p216,Kienreich:2011ab,Long:2011ab}, with the derived kinematics showing a decelerating pulse that also appears to exhibit dispersion with propagation. These observations show good agreement with simulations of fast-mode wave propagation \citep{Wang:2000tg,Ofman:2002pi} and are also consistent with the simulated propagation of a fast-mode wave through a randomly structured magnetic field \citep{Murawski:2001ab}.

A slight modification to the fast-mode interpretation has been proposed by \citet{Grechnev:2008p629}, who modeled a CBF event using self-similar dimensional analysis \citep[cf.][]{Taylor:1950p164,Taylor:1950p173,Sedov:1959fk}. They considered a strong point-like explosion in a region of constant density and a region with a radial density drop-off from the source (i.e.,\ $\rho \propto r^{-\alpha}$). For a region of constant density, the energy of a strong shock can be determined using dimensional analysis to be
\begin{equation}
E \text{ \corr{$\propto$} } \rho_0 r^3 v^2 ,
\end{equation}
where $r$ is the radius of the shock, $v$ is the shock velocity and $\rho_0$ is the density. This produces a shock velocity,
\begin{equation}
v \text{ \corr{$\propto$}} \left(\frac{E}{\rho_0 r^3}\right)^{\corr{1/2}} \propto r^{-3/2},
\end{equation}
with $r \propto t^{2/5}$. For a region with a radial density fall-off $\rho = br^{-\alpha}$, the velocity $v$ is given by,
\begin{equation}
v \propto r^{[-(3-\alpha)/2]},
\end{equation}
where $r \propto t^{[2/(5-\alpha)]}$. As a result, a strong spherical shock should show deceleration for $\alpha \le 3$ and acceleration for $\alpha \ge 3$.

\begin{figure}[!t]
\includegraphics[keepaspectratio, width=1\textwidth]{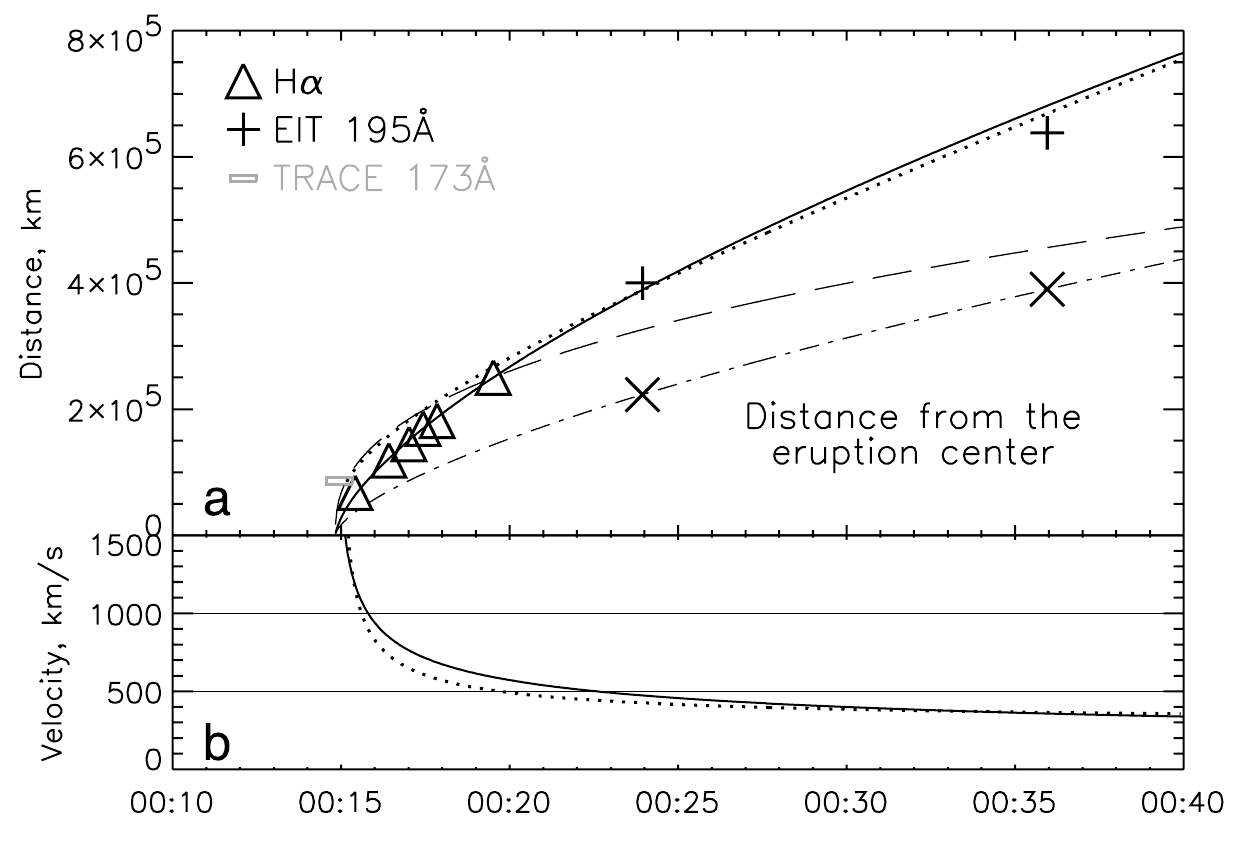}
\caption[Distance--time plot with derived kinematics for a Moreton wave and CBF event from \citet{Grechnev:2008p629}]{Distance--time plot with derived kinematics for a Moreton wave and CBF event from \citet{Grechnev:2008p629}. Panel~\emph{a} shows positions of Moreton (triangles) and ``EIT wave'' (crosses) measured in the same direction from the eruption source. Panel~\emph{b} shows the derived kinematics for constant $\rho$ (dotted line) and $\rho \propto r^{-2}$ (solid line).}
\label{fig:blast_wave}
\end{figure}

\citet{Grechnev:2008p629} derived kinematics for an event from 2004~July~13 (see Figure~\ref{fig:blast_wave}), arguing that the CBF could be best explained as a coronal blast wave \corr{(i.e., a shock wave with a low Alfv\'{e}nic Mach number resulting from a sudden release of energy in the corona)}. This event was quite distinct, with the erupting CME breaking apart and some material falling back onto the Sun. A Moreton wave and a CBF were both observed for this event, with the authors suggesting that their derived kinematics are consistent with an initially strong or moderate intensity shock that later damps to a moderate wave pulse. However, \citet{Grechnev:2008p629} do note that this mechanism may not be universally consistent across all observations of CBFs.

The fast-mode wave solution has been decried as insufficient by multiple authors. \citet{Wills-Davey:2007oa} proposed that the physical properties of CBFs are more consistent with an MHD soliton than a fast-mode MHD wave. In particular, they claim that the single pulse nature, stable morphology and varying kinematics can be best explained by a soliton approximation \citep[cf.,][]{Russell:1844fk}. As the velocity of a soliton is dependent on the amplitude of the pulse, this would account for the large range of derived CBF velocities \citep{Thompson:2009yq}, while the inherent physical cohesiveness of a soliton would allow it to propagate across large regions of the solar atmosphere relatively coherently.

This interpretation has failed to gain much traction in the literature, as more recent results suggest that CBF pulses may not be as stable as previously thought, with clear evidence of dispersion noted by multiple authors \citep{Veronig:2010ab,Kienreich:2011ab,Long:2011ab,Muhr:2011fk}. Widespread pulse deceleration has also been noted \citep[e.g.,][]{Warmuth:2004rm,Warmuth:2004ab,Long:2011ab} that may help to account for the wide range of derived pulse velocities.

\corr{A number of simulations have also been performed indicating that CBFs can be best described as fast--mode MHD waves. In particular, the simulations of \citet{Wang:2000tg} and \citet{Wu:2001dz} produce pulses that propagate across the quiet Sun, avoiding coronal holes and active regions. In both cases, a photospheric magnetogram was used to extrapolate a realistic coronal magnetic field through which the pulse was allowed to propagate. However, both sets of simulations have been criticised for using plasma--$\beta$ values that were unrealistic \citep{Chen:2002rw}. A more recent simulation by \citet{Schmidt:2010ab} used the BATS--R--US simulation software combined with photospheric magnetograms to simulate an event from 2007~May~19 which was observed by the \emph{STEREO} spacecraft. In this case, the simulation is in good agreement with the observed eruption, producing a CME and a CBF pulse, which exhibited reflection and refraction from coronal hole and active region boundaries similar to the actual event. The advancement of simulation software combined with the higher spatial and temporal resolution data available from the \emph{SDO} spacecraft will allow greatly improved simulations of CBF events.} 

\subsection{Pseudo-wave Theories}
\label{subsect:cbf_pseudowave_theory}

Several authors have raised issues with the interpretation of CBFs as waves, instead choosing to explain them in terms of the magnetic field restructuring during the eruption of a coronal mass ejection. This is supported by the results of \citet{Biesecker:2002lq}, who found that every CBF is associated with a CME, although the converse is not necessarily true. A number of observations inconsistent with the behaviour of a magnetoacoustic wave have also been reported, including the presence of stationary bright fronts \citep{Delannee:1999mz} and a dependence of the rotation of the CBF on the helicity of the erupting active region \citep{Attrill:2007vn}. Numerical simulations of CME eruptions have also produced evidence of co-existing CBFs and Moreton waves, where the propagation of increased emission due to successive opening magnetic field lines is observed as the CBF \citep{Chen:2002rw}.

The first non-wave theory for the interpretation of CBFs was promoted by \citet{Delannee:1999mz}. This resulted from the analysis of an event observed by \emph{SOHO}/EIT to contain both moving and stationary bright fronts. Here, the CBF was explained as plasma compression near the footpoints of opening field lines adjacent to a separatrix during the eruption of a CME. As the CME erupts and field lines open farther from the source, the CBF would be expected to propagate, but crucially, would stop and remain stationary when faced with regions containing vertical field lines (such as loop footpoints and coronal holes).

\begin{figure}[!t]
\includegraphics[keepaspectratio, width=1\textwidth]{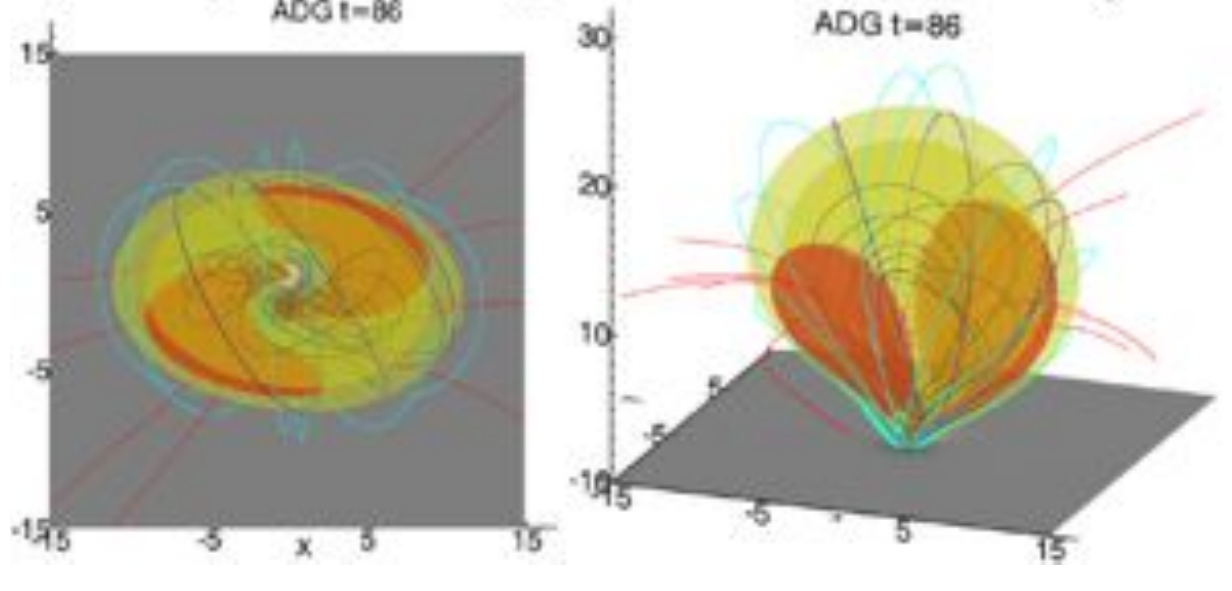}
\caption[Simulation illustrating the current shell model proposed by \citet{Delannee:2008uq}]{Simulation illustrating the current shell CBF model proposed by \citet{Delannee:2008uq}. Left-hand image is a top-down view, right hand image is a side-on view. The current shell is visible as the yellow transparent surface.}
\label{fig:current_shell_model}
\end{figure}

The theory was readdressed and extended in further analyses \citep{Delannee:2007kx,Delannee:2008uq}, with the bright CBF ascribed to Joule heating at the edges of a current shell containing the erupting CME flux rope (see Figure~\ref{fig:current_shell_model}). A 3--dimensional MHD code was used to simulate the CME eruption corresponding to two well-observed CBF events (from 1997~April~7 and 1997~May~12), with a twisted flux rope in a bipolar magnetic field configuration used as the initial condition. This flux rope was then allowed to erupt, compressing the surrounding potential field and producing return electric currents that evolve into a large-scale current shell. An integration along this current shell produces a structure similar in appearance to the observed CBF pulse. 

There are some issues with this interpretation for CBFs \corr{\citep[cf.][]{Patsourakos:2009bc}}. The current shell model implies that the bright front observed as the CBF should form at the edges of the erupting CME. This would produce successively higher emission as the CME propagates out into the heliosphere. However, quadrature observations using the \emph{STEREO} spacecraft have indicated that the CBF \corr{emission} is confined to $\sim$90~Mm above the photosphere \citep{Kienreich:2009ab,Patsourakos:2009ab}, a region consistent with the scale height of observed EUV emission. In spite of this criticism, the current shell model has been invoked to explain multiple CBF events. Most recently, \citet{Schrijver:2011vn} explained a CBF/CME/flare event from 2011~February~15 using the current shell model. Here, the CBF is formed by plasma compression and Joule heating at the interface between the erupting CME and an overarching helmet streamer structure. The observed emission signatures and derived temperatures are also noted to show good correspondence with the predicted behaviour of the current shell model.

An alternative to the current shell model was proposed by \citet{Chen:2002rw} based on the results of numerical simulations of a CME eruption. \citet{Chen:2002rw} found evidence of two disturbances that propagated laterally away from the source of the eruption as the CME flux rope expanded out into the heliosphere. As the CME flux rope rises, it induces a piston-driven shock forward of the flux rope. The legs of this shock extend to and sweep through the low corona and chromosphere, prompting \citet{Chen:2002rw} to interpret this as the manifestation of a Moreton wave. Behind these shock legs, a region of plasma compression caused by successive opening of field lines covering the flux rope is observed propagating laterally away from the source at a velocity three times lower than the shock front. It is suggested that this moving compression front is the observed ``EIT wave'', with \citet{Chen:2002rw} pointing to observed discrepancies in derived kinematics for Moreton waves and CBFs for confirmation \citep{Smith:1964p1077,Klassen:2000p7}.

\begin{figure*}[!t]
\begin{center}
\includegraphics[width=1\textwidth]{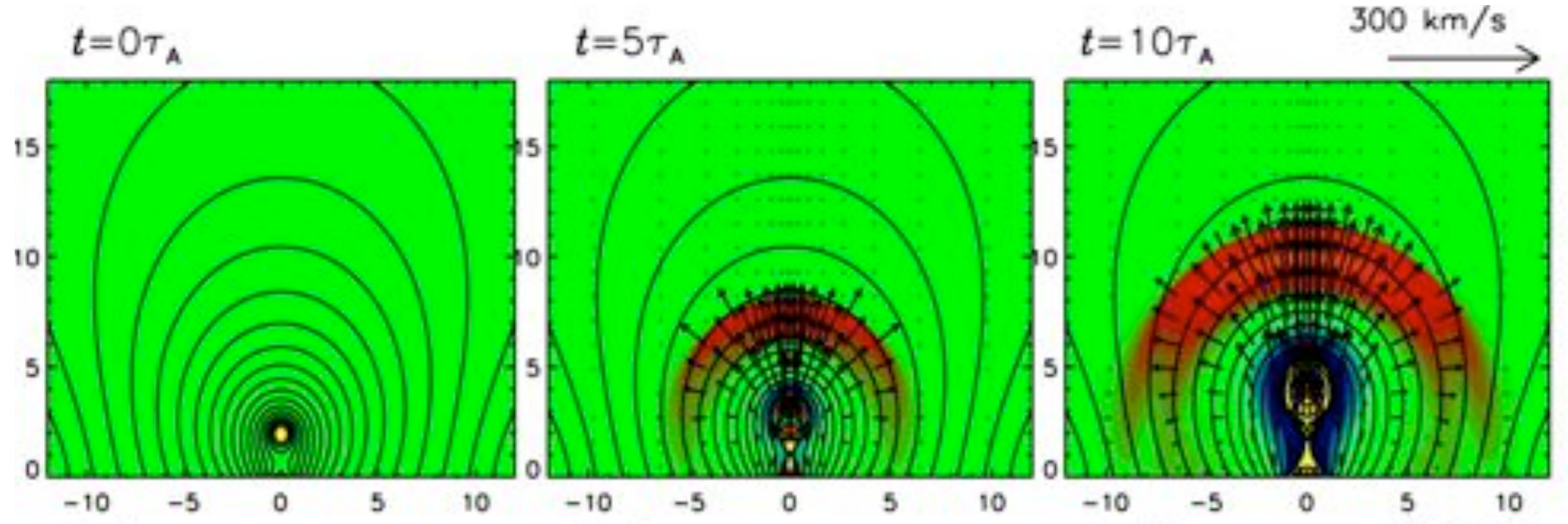}
\includegraphics[width=0.8\textwidth]{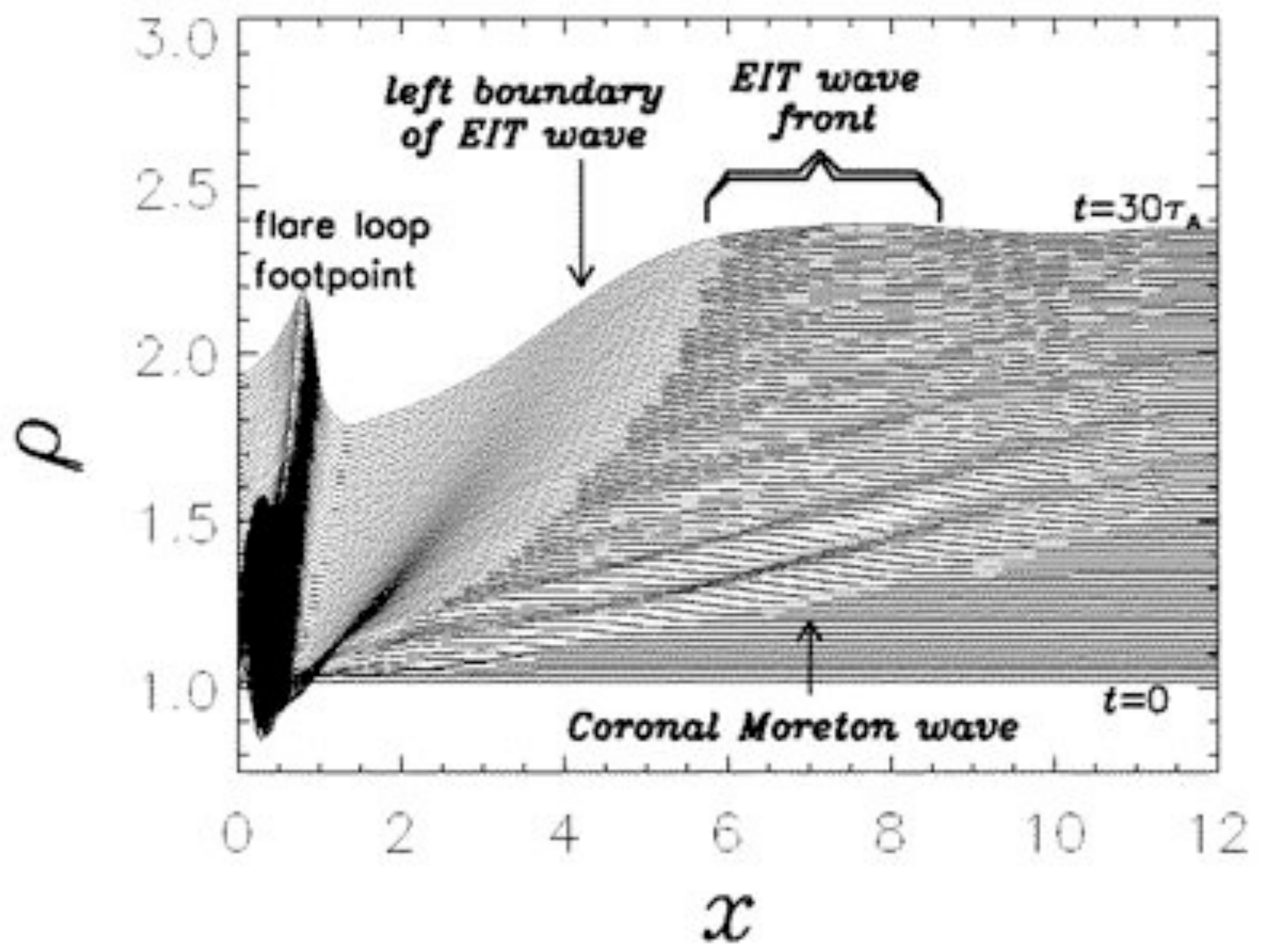}
\caption[The model for CBFs proposed by \citet{Chen:2002rw}]{{\it Top:} The Chen model for CBFs. As the flux tube rises a shock front is induced in front of it (red in the figure). The legs of this shock front sweep the chromosphere and are observed as Moreton waves. Simultaneously, the CBF is observed as a brightening as the field lines covering the flux rope open. {\it Bottom:} Stack plot showing  the evolution of density ($\rho$) with propagation distance \citep{Chen:2002rw}.}
\label{fig:chen_model}
\end{center}
\end{figure*}

This model is illustrated in Figure~\ref{fig:chen_model}, with the top part showing several steps of the simulation and the bottom part showing the variation in density ($\rho$) with distance ($x$) for successive times along a cut of constant height. The piston-driven shock is visible in the upper part of Figure~\ref{fig:chen_model}, while the CBF is apparent in the lower panel as an increase in density following behind the Moreton wave. The theory was extended in work by \citet{Chen:2005xe} and \citet{Chen:2005p157} who used MHD simulations to account for the observational features of CBFs, including stationary bright fronts and higher velocities in the quiet Sun compared to active regions \citep{Foley:2003p69}. The piston-driven shock model outlined by \citet{Chen:2005xe} predicts that two pulses should be identifiable in coronal data; the CBF and a coronal pulse corresponding to the Moreton wave. Stationary bright fronts are explained as a consequence of the opening magnetic field lines associated with the erupting flux rope coming into contact with a neighbouring active region. In this case, the magnetic field lines connect to the active region, producing a long-lived bright region observed as a stationary bright front.

While the piston driven shock model does account for the low observed CBF velocities and observations of stationary bright fronts, it has been difficult to definitively prove or disprove as a result of the relatively low cadence of available data. The model predicts two pulses associated with each CBF event, with a coronal Moreton wave moving at a velocity that is $\sim$3 times faster than the CBF pulse. For CBF velocities of $\sim$200--400~km~s$^{-1}$ this would require a coronal Moreton wave with a velocity $\sim$600--1200~km~s$^{-1}$, too fast to be observed by \emph{SOHO}/EIT. However, only single pulses have been observed by \emph{STEREO}/EUVI and the Atmospheric Imaging Assembly \citep[AIA;][]{Lemen:2011uq} onboard the \emph{Solar Dynamics Observatory} \citep[\emph{SDO;}][]{Pesnell:2011ab} spacecraft. This is despite the fact that \emph{SDO}/AIA would be expected to observe very fast pulses as a result of its high cadence \citep{Kienreich:2011ab,Long:2011ab}. A double pulse event was apparently observed by \citet{Chen:2011p3855}, although the evidence remains unconvincing.

An alternative proposal for the reconfiguration of the magnetic field during a CME eruption, resulting in a CBF was put forward by \citet{Attrill:2006vn}. As the CME erupts, the magnetic field lines undergo successive reconnection with adjacent quiet Sun loops, producing bright regions through small-scale chromospheric evaporation that are observed as the CBF. The concept was extended in later work \citep{Attrill:2007vn,Attrill:2009ab}, allowing stationary bright regions to be explained as the result of continuous reconnection between the overlying magnetic field lines and the the open field lines of a coronal hole.

\begin{figure}[!t]
\includegraphics[width=1\textwidth]{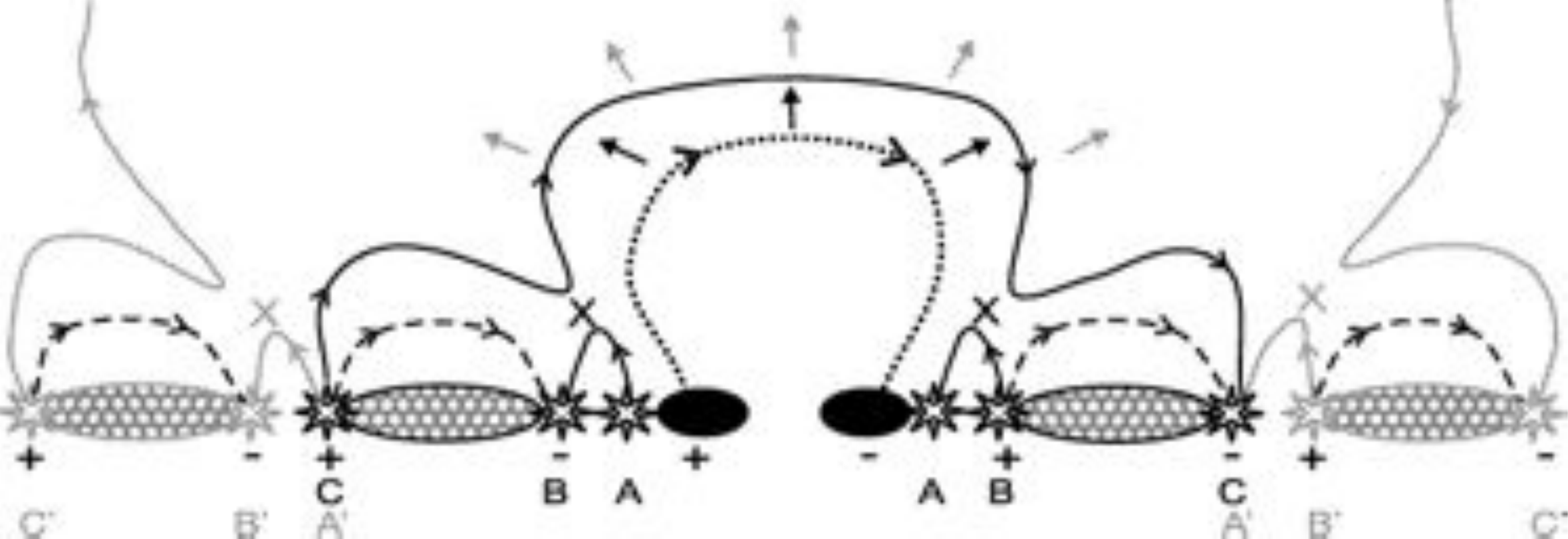}
\caption[Cartoon illustrating the Attrill CME model for the 1997 April 07 event]{Cartoon illustrating the Attrill CME model for the 1997 April 07 event.}
\label{fig:attrill_model}
\end{figure}

A cartoon illustrating the Attrill concept is shown in Figure~\ref{fig:attrill_model}. The erupting flux rope (black dotted line) pushes the overlying magnetic field line (black solid line), resulting in magnetic reconnection with adjacent quiet Sun loops and chromospheric brightening at points A, B and C either side of the erupting flux rope. This process is then repeated, with the overlying magnetic field line  (black solid line) reconnecting with the grey solid line, producing brightening at points A', B' and C' and so on. 

The Attrill cartoon is not without its detractors, with \citet{Delannee:2009p1113} in particular questioning the validity of the successive reconnection mechanism. They examined the 1997~May~12 event previously studied by \citet{Attrill:2007vn} using an extrapolated magnetic field topology to examine the interconnectivity of the coronal magnetic field. Whereas \citet{Attrill:2007vn} claimed that the erupting flux rope was magnetically linked to adjacent quiet Sun loops, \citet{Delannee:2009p1113} found that this was not the case, with the erupting active region linked to magnetic field lines $\sim$300~Mm away. There have also been some conflicts between the predicted and observed pulse morphology for this model, with quadrature observations of CBFs indicating that the emission is produced too high in the solar atmosphere to be a result of small-scale chromospheric brightening \citep{Patsourakos:2009ab,Kienreich:2009ab}.

Several authors have proposed that CBFs may be a combination of both wave and pseudo--wave modes rather than a single interpretation. This was first suggested by \citet{Zhukov:2004kh}, who studied several CBF events using the 171, 195, and 284~\AA\ passbands available from \emph{SOHO}/EIT (in the process making the first CBF observation in the 284~\AA\ passband). The events studied were found to display both wave and non--wave characteristics, with the isotropic nature of the pulses implying a wave nature, while the extent of the coronal dimming region behind the pulse and the low pulse velocities indicate that this interpretation is insufficient.

Later work also concluded that this scenario may be the optimum solution to the CBF problem. \citet{Cohen:2009p1864} used the BATS--R--US MHD code to simulate the eruption of a coronal mass ejection from 2009~February~13, comparing the simulation with observations of the eruption made by \emph{STEREO}/EUVI. It was found that the CBF front was best represented by a density enhancement resulting from a combination of both wave and non-wave processes. The CME drives the front through lateral expansion in the low corona, with this lateral expansion resulting in widespread magnetic reconnection as the coronal magnetic field reorganises. A similar result was obtained by \citet{Downs:2011p3523}, who used a similar model to study the evolution of an event from 2008~March~25. In this case, the signatures of the different EUV passbands available from \emph{STEREO}/EUVI were synthesised, allowing the simulation to be directly compared to observations. Two distinct signatures were observed, one identified as a fast-mode wave pulse, the other a separate compression front produced by the expansion of the CME into the surrounding magnetic structures. Much work remains to be done in this regard, and the high cadence data now available from \emph{SDO}/AIA will allow this issue to be resolved.	


\chapter{Instrumentation} 
\label{chap:inst}


\noindent 
\\ {\it 
These instruments have play'd me so many tricks that I have at last found them out in many of their humours.
\begin{flushright}
Sir William Herschel \\
\end{flushright}
 }


\vspace{15mm}
In this chapter, the different telescopes used to study ``EIT Waves'' are discussed, with special emphasis placed on the software required to process the raw data and prepare it for scientific analysis. The host spacecraft for each instrument have unique capabilities that are also discussed as these can influence how the data is approached. 

Although there have been some indications of large-scale disturbances in additional portions of the electromagnetic spectra (e.g.,\ radio and X-rays), the disturbances studied in this project were observed using the Extreme UltraViolet (EUV) part of the solar spectrum. The Earth's protective ozone layer makes this difficult as EUV radiation does not reach the ground, making it necessary to use space-based observatories to study the low corona. The two observatories primarily used in this work were the \emph{STEREO} and \emph{SDO} spacecraft.

\section{The \emph{STEREO}/Extreme UltraViolet Imager (EUVI)}
\label{sect:inst_euvi}

The \emph{STEREO} mission consists of two identical spacecraft, one following the Earth (\emph{STEREO}--Behind) and the other leading the Earth (\emph{STEREO}--Ahead) on its orbit around the Sun. \corr{\emph{STEREO}--A has an orbit slightly inside that of the Earth, with \emph{STEREO}--B following an orbit slightly outside that of the Earth. As a result, both} spacecraft are separating from the Earth and each other at rates of $\sim$22 and $\sim$44 degrees per year, respectively. \corr{This configuration was chosen to allow stereoscopic imaging of the Sun and the Sun--Earth line, thus allowing a determination of the feasibility of possible future missions.} They were launched on the same Delta~{\sc II} rocket in October~2006, with science operations beginning in December~2006 after a series of maneuvers allowed both spacecraft to escape Earth orbit. Quadrature was reached in February~2009, with both spacecraft at 90~degrees to each other, while opposition was achieved in February~2011 when the entire Sun was simultaneously visible for the first time. 

\begin{figure}[!t]
\begin{center}
\includegraphics[clip=,trim=0mm 90mm 0mm 90mm,width = 0.88\textwidth,angle=0]{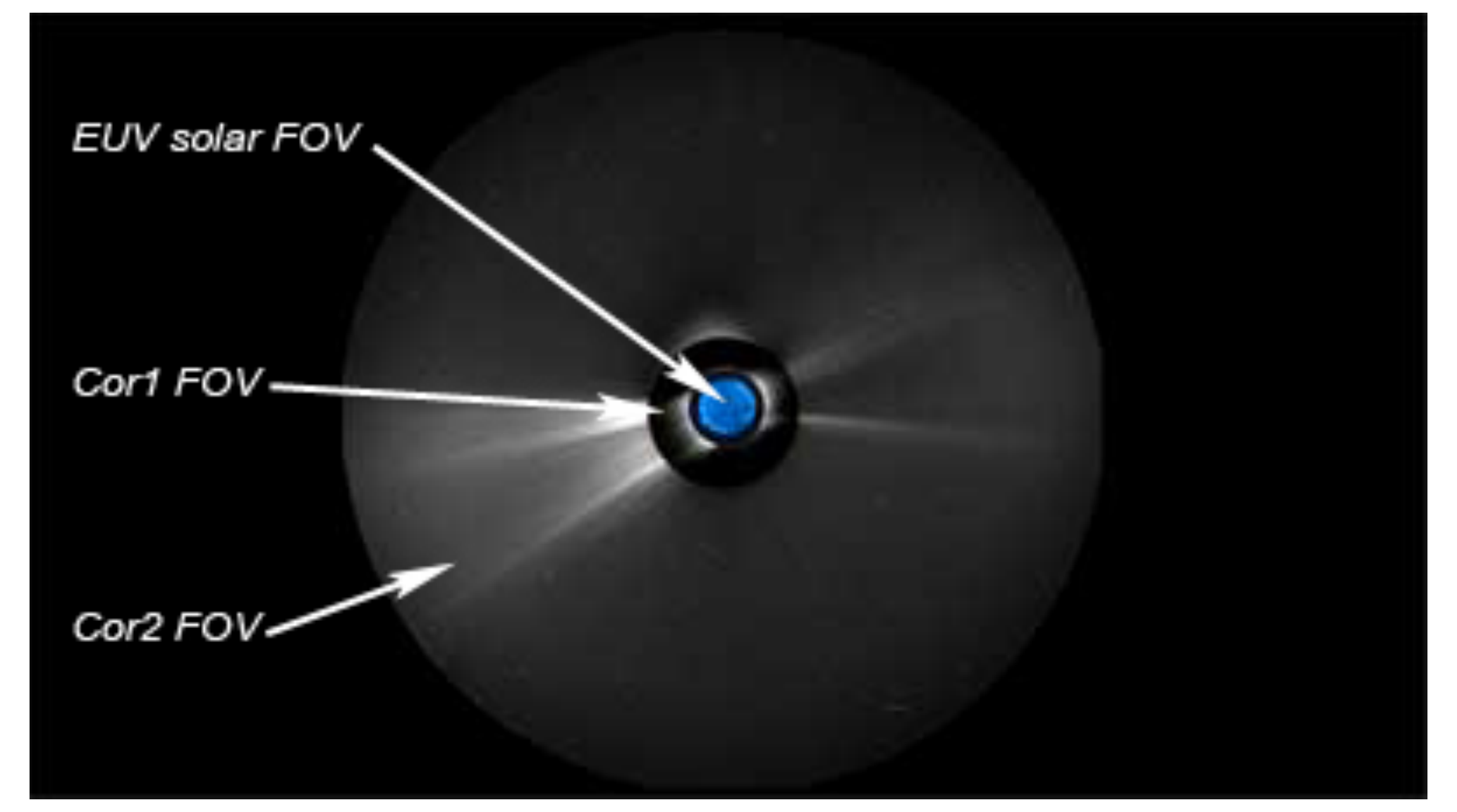}
\includegraphics[clip=,trim=20mm 0mm 20mm 0mm,width = 0.5\textwidth,angle=270]{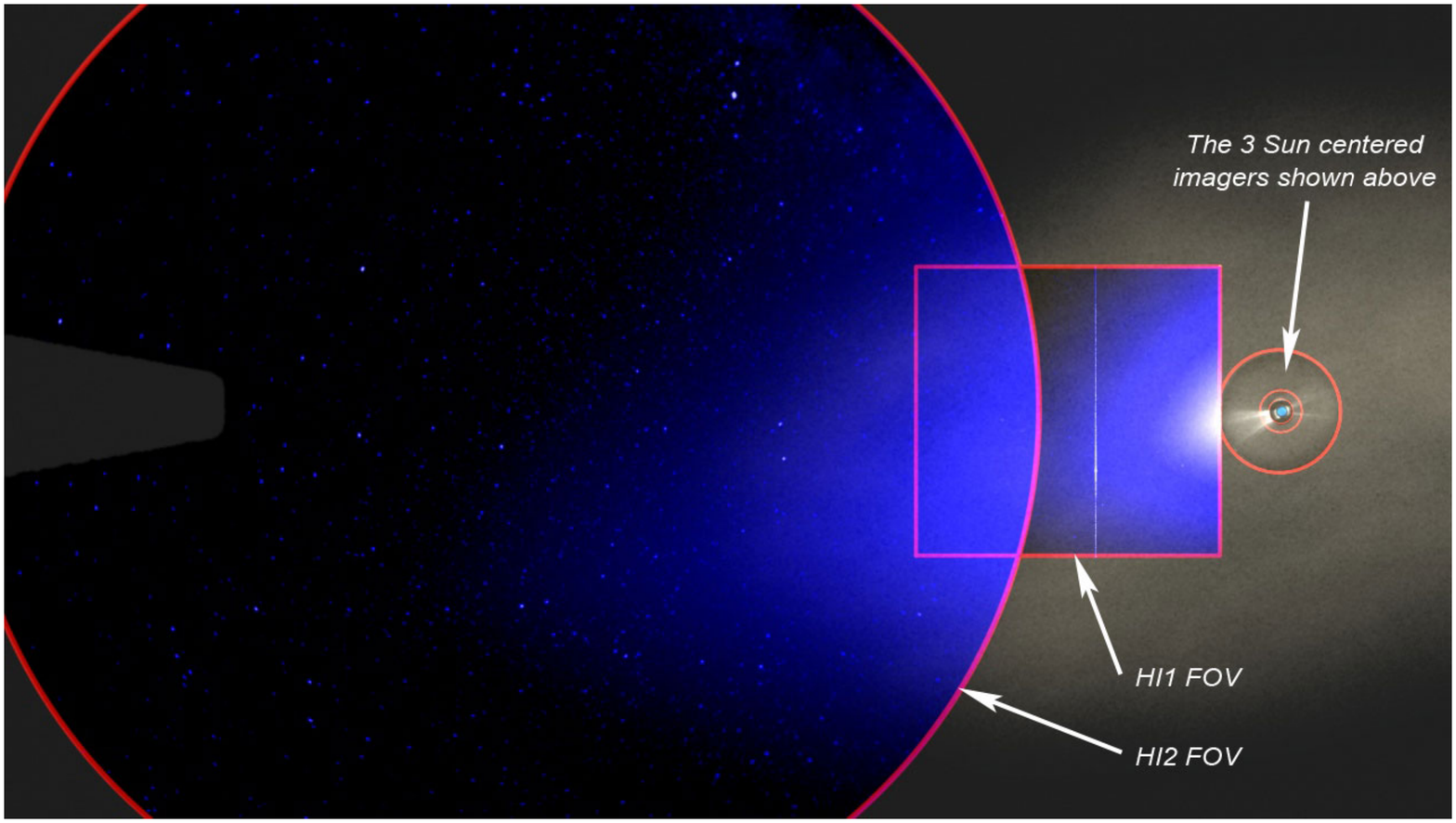}
\caption{The field-of-view of the SECCHI instrument suite as observed by \emph{STEREO}-A. Top panel shows EUVI and the coronagraphs Cor--1 and Cor--2. Bottom panel shows the field of view of the coronagraphs compared to the Heliospheric Imagers HI--1 and HI--2 \citep[cf.][]{Howard:2008p116}.}
\label{fig:stereo_secchi}
\end{center}
\end{figure}

The EUVI telescope is part of the Sun Earth Connection Coronal and Heliospheric Investigation \citep[SECCHI;][]{Howard:2008p116} instrument suite onboard \emph{STEREO}. SECCHI consists of an EUV imager (EUVI), two coronagraphs with different fields-of-view (1.4--4~\rsun\ for COR--1 and 2.5--15~\rsun\ for COR--2) and two heliospheric imagers observing the Sun--Earth line (at distances of 15--80~\rsun\ for HI--1 and 80--215~\rsun\ for HI--2) with the fields-of-view of each instrument shown in Figure~\ref{fig:stereo_secchi} for comparison. This allows solar eruptions to be tracked from initiation on the Sun through the corona and heliosphere to the Earth. EUVI has a similar design to the EIT instrument onboard the \emph{SOHO} spacecraft, but with a higher spatial and temporal resolution.

The increasing separation of the \emph{STEREO} spacecraft from each other and Earth has resulted in a gradual decrease in the amount of data being sent from the spacecraft to Earth and in the observing cadence of the different passbands. The scientific emphasis of \emph{STEREO} has also changed with the increased distance from Earth, with the initial focus being on the initiation of CMEs as observed using EUVI. The advent of \emph{SDO} and the present ability of \emph{STEREO} to observe the Sun--Earth line has resulted in an increased emphasis on CME observations using the coronagraphs and heliospheric imagers.

\subsection{EUVI Instrument Operation}
\label{subsect:inst_euvi_design}

\begin{figure}[!t]
\begin{center}
\includegraphics[clip=,trim=0mm 0mm 0mm 0mm,width = 1\textwidth]{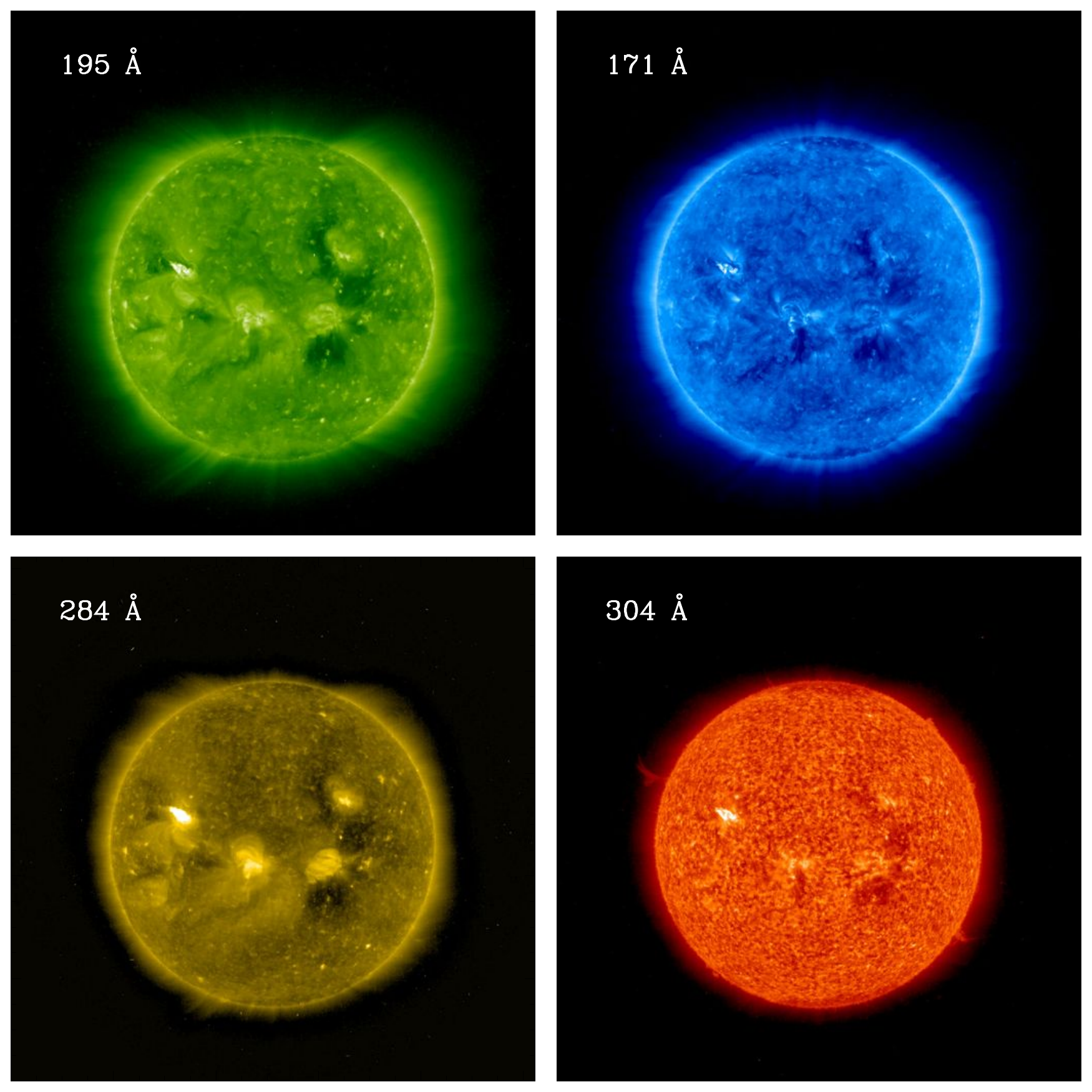}
\caption{EUVI images showing the 195~\AA\ (top left), 171~\AA\ (top right), 284~\AA\ (bottom left) and 304~\AA\ (bottom right) passbands. Each passband has been plotted with the appropriate colour table for clarity.}
\label{fig:euvi_images}
\end{center}
\end{figure}

EUVI is a normal incidence telescope with a Ritchey-Chr\'{e}tien construction, designed to allow a large field--of--view within a compact space by utilising hyperbolic primary and secondary mirrors. This layout allows an unvignetted, circular full-Sun field--of--view extending out to $1.7$~\rsun\ without any requirement for a focusing mechanism. The telescope images four separate EUV passbands \corr{peaking} at 304~\AA, 171~\AA, 195~\AA\ and 284~\AA, with an effective temperature range from $\sim$60~000~K (in the 304~\AA\ passband) to $\sim$2.5~MK (in the 284~\AA\ passband). Images from the four passbands observed by EUVI are shown in Figure~\ref{fig:euvi_images}. The different panels have been coloured using the appropriate colour tables for clarity.

\begin{figure}[!t]
\begin{center}
\includegraphics[width = 1\textwidth,clip=,trim= 32mm 125mm 32mm 125mm]{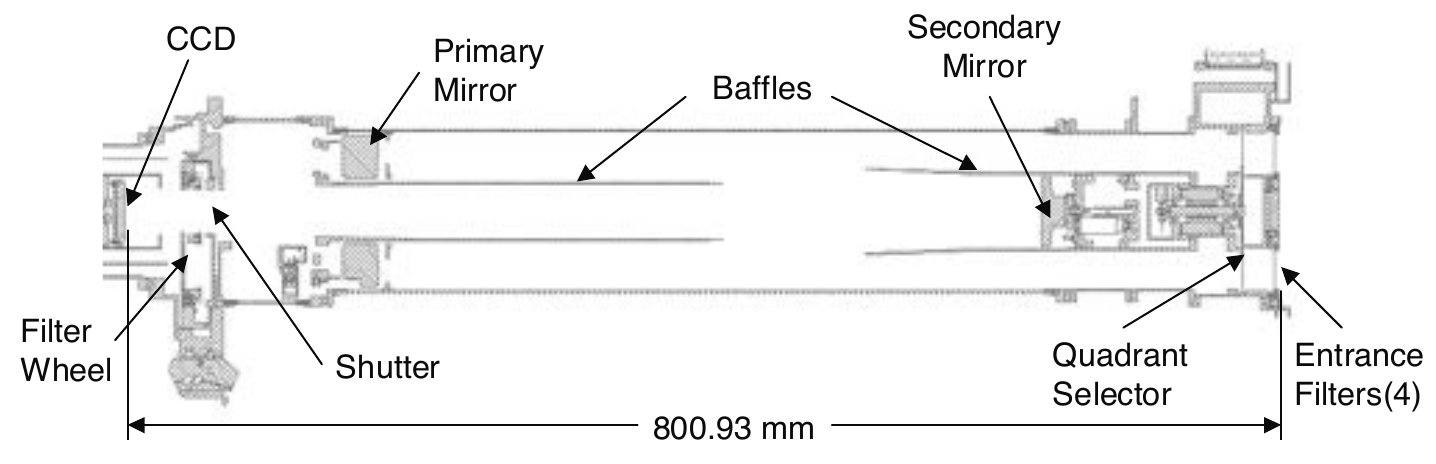}
\caption{Cross-section of the EUVI telescope onboard the \emph{STEREO} spacecraft \citep{Wuelser:2004bs}.}
\label{fig:euvi}
\end{center}
\end{figure}

The actual operation of the telescope is as follows; light enters the telescope at the front (the right of Figure~\ref{fig:euvi}), where thin metal film filters (aluminium--on--\corr{polyimide} on a coarse nickel grid for 171 and 195~\AA\ and aluminium on a fine nickel mesh for 284 and 304~\AA) remove undesired UV, IR and visible radiation, and keep heat out of the telescope. The required passband is selected through the use of a rotating quadrant selector allowing the radiation to pass through to one of the four quadrants of the optics. These are specially coated using narrow-band, multilayer MoSi coating of variable thickness to ensure optimal reflectivity for each passband, producing peak reflectivity values of $\sim$15--39~\%. While this may appear low, the very high intensities of the observed lines mean that this does not greatly affect performance. The telescope is also fully baffled to negate the effects of scattered light and charged particles entering the front aperture.

The CCD used by the EUVI telescopes is a backside-thinned, back-illuminated full-frame CCD. This arrangement allows improved low-light performance and short wavelength response. The CCD is also passively cooled to below $-60^{\circ}$C ($\sim$213~K) through the use of an aluminium ``cool finger'' to reduce dark current.

EUVI has a spatial resolution of 1.6~arcsec~per~pixel with an initial passband-dependent observing cadence of 75--300~s for 171~\AA, 300--600~s for 195~\AA\ and 304~\AA\ and 1200~s for 284~\AA. However, as the lifetime of the mission has continued and the spatial separation of the spacecraft has increased, the observing cadence has been greatly reduced to allow for decreased bandwidth. Currently, EUVI operates at a typical cadence of 300 and 600~s in the 195~\AA\ and 304~\AA\ passbands respectively, with both the 171~\AA\ and 284~\AA\ passbands operating synoptically (i.e.,\ one image every $\sim$4~hours).

\subsection{EUVI Temperature Response}
\label{subsect:euvi_response}

\begin{figure}[!t]
\begin{center}
\includegraphics[clip=,trim=5mm 15mm 10mm 45mm,width = 0.95\textwidth]{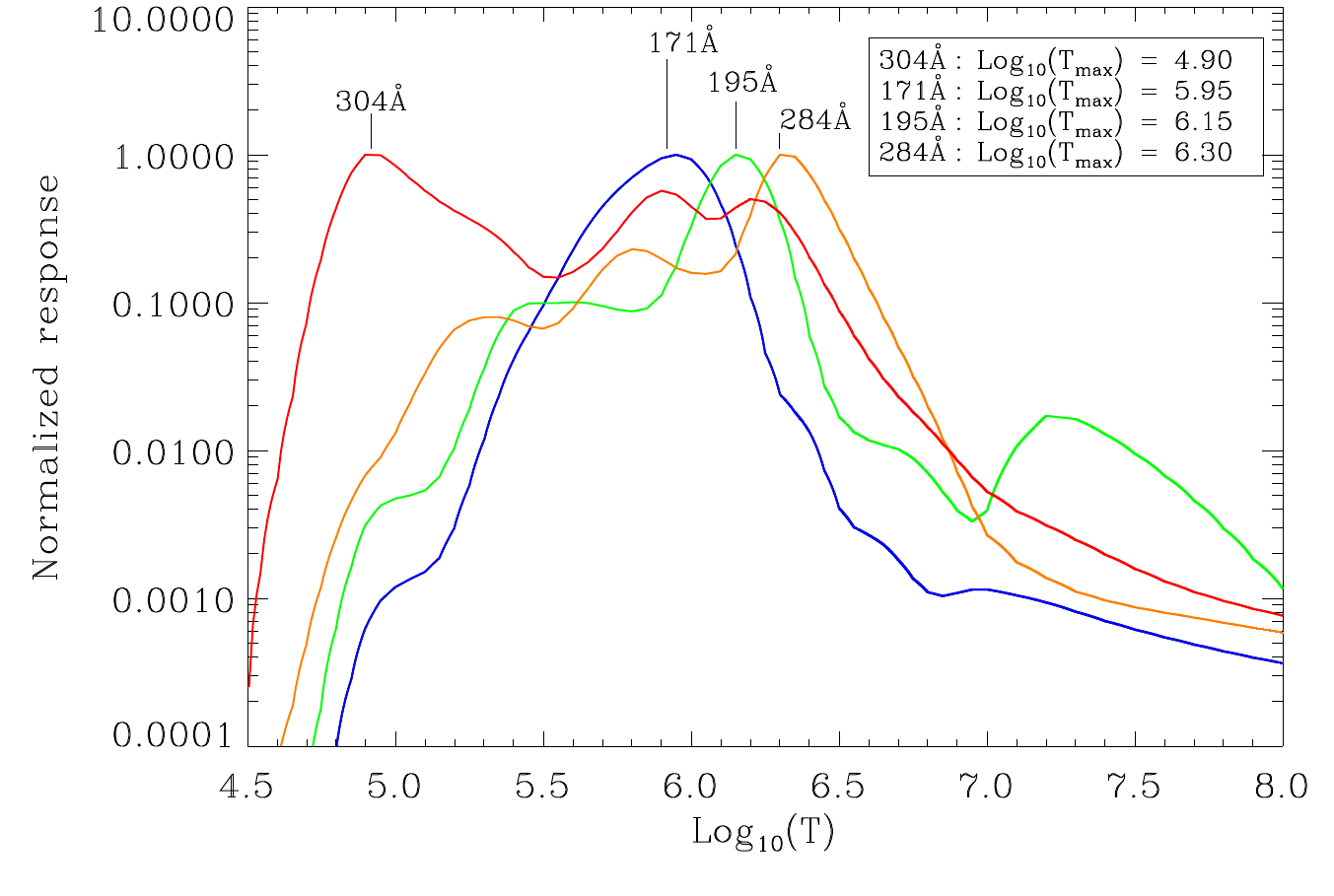}
\caption{Peak-normalised response curves versus temperature for each EUV passband observed by \emph{STEREO}/EUVI. Peak emission temperatures of the different passbands are given in the upper right for comparison \citep[cf.][]{Wuelser:2004bs}}
\label{fig:euvi_resp}
\end{center}
\end{figure}

The four passbands observed by EUVI cover a broad range of temperatures and hence different regions of the solar atmosphere. The peak-normalised temperature response curves for each of the four passbands are shown in Figure~\ref{fig:euvi_resp}, while the peak emission temperatures (T$_{peak}$) and dominant ions of the individual passbands are outlined in Table~\ref{tbl:euvi_emission}.

\begin{deluxetable}{ccc}
\tablecolumns{3}
\tablewidth{0pt}
\centering
\tablecaption{Temperature response characteristics of EUVI passbands\label{tbl:euvi_emission}}
\tablehead{
\colhead{Passband (\AA)} & \colhead{Dominant ion} & \colhead{T$_{peak}$ (MK)}
}
\startdata
304		& \ion{He}{2}, \ion{Si}{11}\tablenotemark{a}		& 0.08 \\
171		& \ion{Fe}{9}, \ion{Fe}{10} 					& 0.9 \\
195		& \ion{Fe}{12} 								& 1.4 \\
284		& \ion{Fe}{15} 								& 2.0
\enddata
\tablenotetext{a}{Although the hydrogen--like \ion{He}{2} is the dominant ion here, there is a small contribution from the coronal \ion{Si}{11} emission line.}
\end{deluxetable}

The range of values for T$_{peak}$ shown in Table~\ref{tbl:euvi_emission} show that EUVI is well-suited for studying the relatively cool quiet corona, while the 284~\AA\ passband allows an analysis of the hotter active regions. This is apparent in the different panels in Figure~\ref{fig:euvi_images}, where the cool 304~\AA\ passband shows the structure of the upper chromosphere, with cool structure visible in the 171~\AA\ passband and hotter material apparent in the hotter 195~\AA\ and 284~\AA\ passbands.

Although the individual passbands observed by EUVI have been tuned using multilayer coatings and filters, the temperature response curves remain quite broad with significant overlap between passbands as shown in Figure~\ref{fig:euvi_resp}. This means that EUVI tends to observe plasma at a typical temperature of 1--3~MK, consistent with the low solar corona and particularly the quiet Sun. This makes EUVI a valuable tool for studying the initiation of CMEs, in line with the original science objectives.

\subsection{EUVI Data Preparation Software}
\label{subsect:euvi_software}

Before the data from EUVI can be used for scientific analysis, the raw FITS\footnote{Flexible Image Transport System; a filetype typically used in astronomy that includes an informational header in addition to the observed data.} files must be calibrated and corrected for instrumental and other effects. The raw FITS file (classified as a Level 0.5 FITS file) is treated using the \emph{secchi\_prep.pro} calibration routine contained within the SolarSoftWare \citep[SSW;][]{Freeland:1998kx} software package.

\begin{landscape}
\begin{figure}[!p]
\begin{center}
\includegraphics[clip=,trim=10mm 10mm 10mm 75mm,width = 1.2\textwidth]{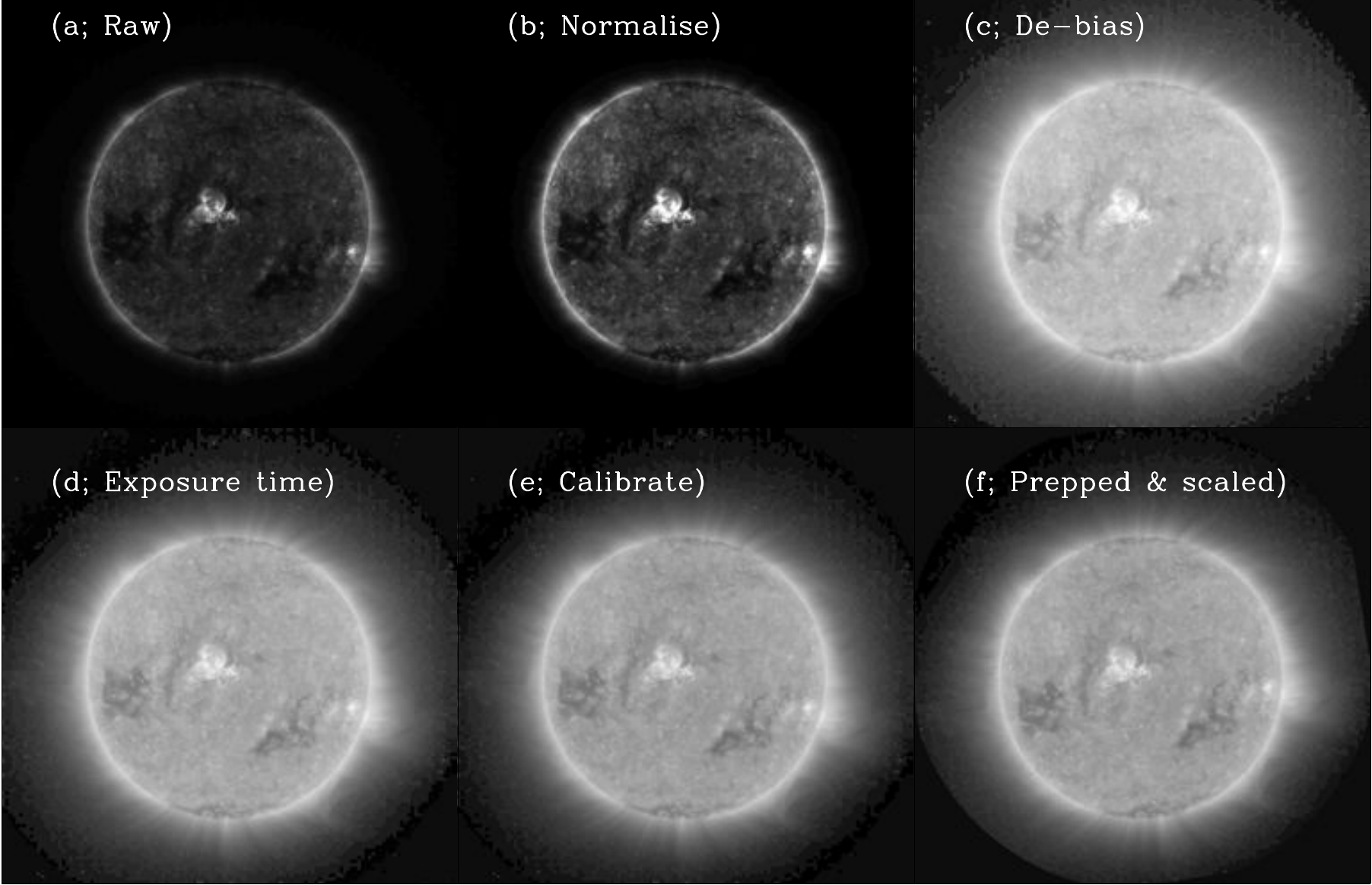}
\caption[The stages required to prepare EUVI data.]{The stages required to prepare EUVI data. The raw data (panel a) is first normalised (panel b), then de--biased (panel c), normalised with respect to exposure time (panel d), calibrated and a photometric correction applied (panel e). It is then corrected for image processing applied onboard the spacecraft and scaled appropriately to give a fully calibrated image (panel f). \corr{Note that all images have been byte--scaled to improve visual clarity.}}
\label{fig:euvi_calib}
\end{center}
\end{figure}
\end{landscape}

The \emph{secchi\_prep.pro} routine is designed to treat data from the SECCHI instrument suite onboard \emph{STEREO}, calibrating the data according to the instrument used. Once the EUVI image has been positively identified, the calibration involves the application of several corrections to the data to account for instrumental and downlink effects. Figure~\ref{fig:euvi_calib} shows the same image from 2007~May~19 at 12:42~UT in the 171~\AA\ passband for different stages of the calibration process, with panel (a) showing the original raw image as taken by \emph{STEREO}-A.

The first step in the calibration process involves identifying and removing cosmic rays hits and missing or corrupted pixels from the image which can be produced by hardware failure or transmission losses. Next, the image is normalised to account for the different passband-specific filters on the telescope, with the resulting image shown in panel (b). The next step involves identifying the image bias value defined in the file header and subtracting it from the image (panel c). Panel (d) shows the result of normalising the image with respect to the exposure time. This is a consequence of the different exposure times available for each passband and also for certain observing programs. Next, the effects of vignetting are removed using a calibration image and a photometric correction applied to the image values using a calibration factor; this is shown in panel (e). The final step involves correcting the image for any image processing that may have been carried out onboard the spacecraft and ensuring that solar north is aligned towards the top of the image. The image can then be scaled and displayed as shown in panel (f).

All of these calibration processes are applied to the data to produce a scientific quality Level 1.0 FITS file. The data may then be processed further as required, or plotted (as shown in Figure~\ref{fig:euvi_images}). Each of these processes is automatically applied, although it is possible to process the data without applying one or more of these steps using keywords. This is often required when analysing temperature or emission measures.

\section{The \emph{SDO}/Atmospheric Imaging Assembly (AIA)}
\label{sect:aia}

\begin{figure}[!t]
\begin{center}
\includegraphics[width = 0.52\textwidth,angle=90,clip=,trim=8mm 6mm 125mm 10mm]{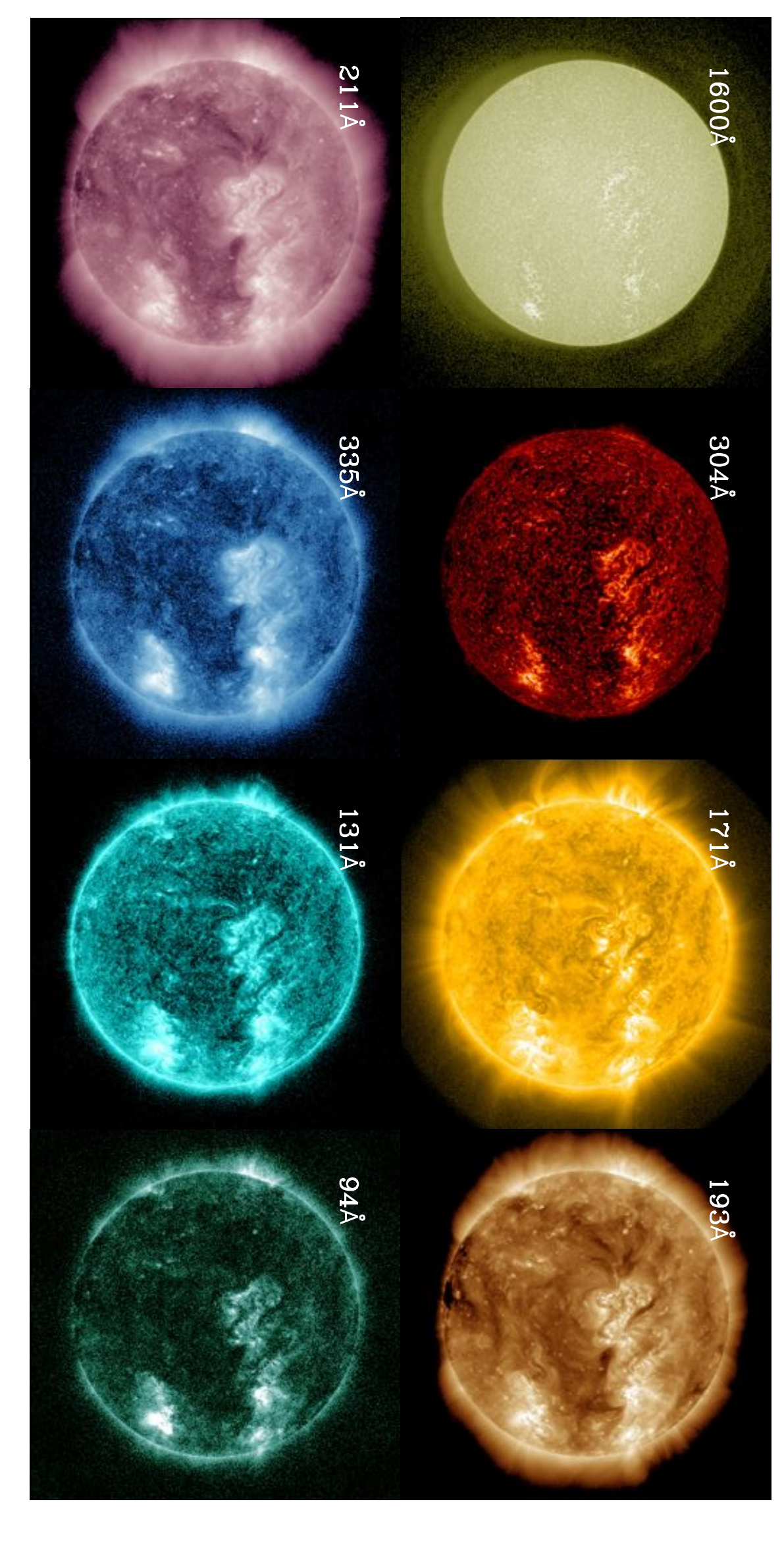}
\caption{Panels showing some of the different passbands available from \emph{SDO}/AIA for $\sim$06:15:00~UT on 2011~Jun~7. Each passband has been plotted with the appropriate colour table for clarity, with the passbands identified in the top left corner of each panel.}
\label{fig:aia_images}
\end{center}
\end{figure}

The \emph{SDO} spacecraft (launched in 2010~February, with science operations beginning in 2010~April) is the first part of NASA's ``Living with a Star'' program, designed to study the effects of the Sun on the near--Earth environment. \emph{SDO} is currently positioned in a geosynchronous orbit located above a dedicated ground station in White Sands, New Mexico. This is a necessary consequence of the very large data output of \emph{SDO}; $\sim2$~terabytes of data per day, every day. The largest contribution to this data flow comes from the Atmospheric Imaging Assembly (AIA) instrument, which uses four separate telescopes to take one 4096$\times$4096 pixel image in ten different passbands at a temporal cadence of $\sim12$~s per passband. Eight of the ten different passbands available from \emph{SDO}/AIA are shown in Figure~\ref{fig:aia_images}.

\subsection{AIA Instrument Operation}
\label{subsect:aia_design}

The Atmospheric Imaging Assembly instrument \corr{\citep{Lemen:2011uq}} consists of four Cassegrain telescopes designed to observe the low solar atmosphere in both UV and EUV emission. The telescopes have an F-number f/20\footnote{e.g.,\ a focal length that is 20$\times$ the pupil diameter} with a 20~cm primary mirror and a secondary mirror incorporating three piezoelectric transducers that offer tip-tilt stabilisation. Metal filters (aluminium for the 171~\AA\ and longer wavelength channels and zirconium for the shorter passbands) prevent stray EM radiation from entering the telescope, while each telescope is fully baffled to protect against scattered light. The metal filters are hosted on a 70 line-per-inch nickel mesh, which can result in diffraction patterns from high intensity point source emission (i.e.,\ flares).

\begin{figure}[!t]
\begin{center}
\includegraphics[clip=,trim=30mm 115mm 30mm 115mm,width = 1\textwidth]{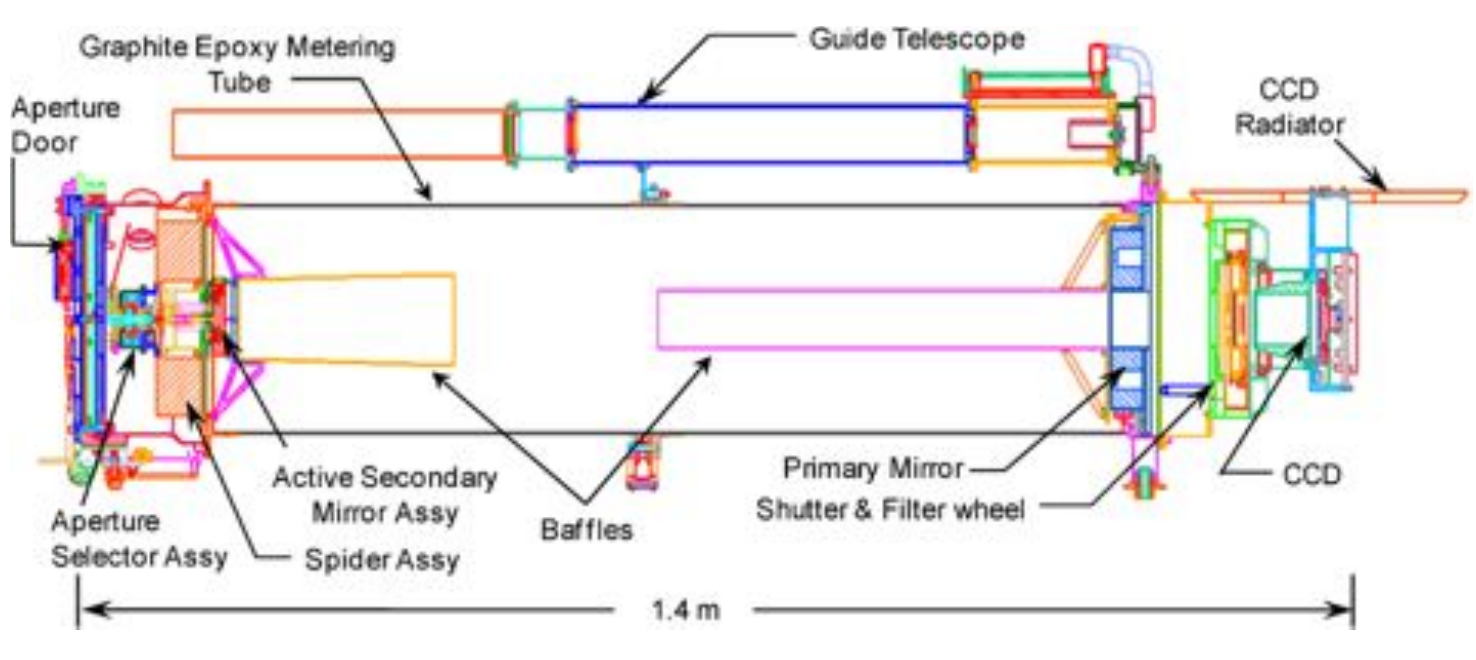}
\caption{Cross-section of one of the four AIA telescopes onboard the \emph{SDO} spacecraft \citep{Lemen:2011uq}.}
\label{fig:aia}
\end{center}
\end{figure}

A cross-section of one of the AIA telescopes is shown in Figure~\ref{fig:aia}. Radiation enters the telescope through the entrance aperture on the left of the figure and passes via the primary mirror and active secondary mirror to the shutter at the \corr{right} of the figure. The shutter is used to control the exposure time of the image; this varies according to passband but is of the order $\sim$2~s. AIA includes an Automatic Exposure Control (AEC) system, which automatically modifies the exposure time if the intensity of the image increases dramatically (as a result of e.g.,\ a flare). In this case, the exposure time can drop to $\sim$0.2~s. Each telescope in AIA is designed to study two distinct passbands, with the desired wavelength channel chosen using a filter wheel. 

The radiation then passes through to the CCD (which is identical for each telescope). Each CCD contains 4096$\times$4096 pixels, and is back-thinned and back-illuminated to improve signal. The telescope has a full-Sun spatial resolution out to $\sim$0.5~\rsun\ above the solar limb of 0.6~arcsec~per~pixel, superior to \emph{SOHO}/EIT, \emph{TRACE} and \emph{STEREO}/EUVI. 

AIA is designed to operate at a continuous cadence of 12~s in each passband, although it is possible to increase the cadence to 10~s if necessary. This combination of high temporal cadence and large image size produces a data volume that exceeds the telemetry allowance, so a lossless Rice compression algorithm is applied to the data onboard the spacecraft to ensure continuous data supply.

\subsection{AIA Temperature Response}
\label{subsect:aia_response}

\begin{figure}[!t]
\begin{center}
\includegraphics[clip=,trim=10mm 20mm 15mm 45mm,width = 0.95\textwidth]{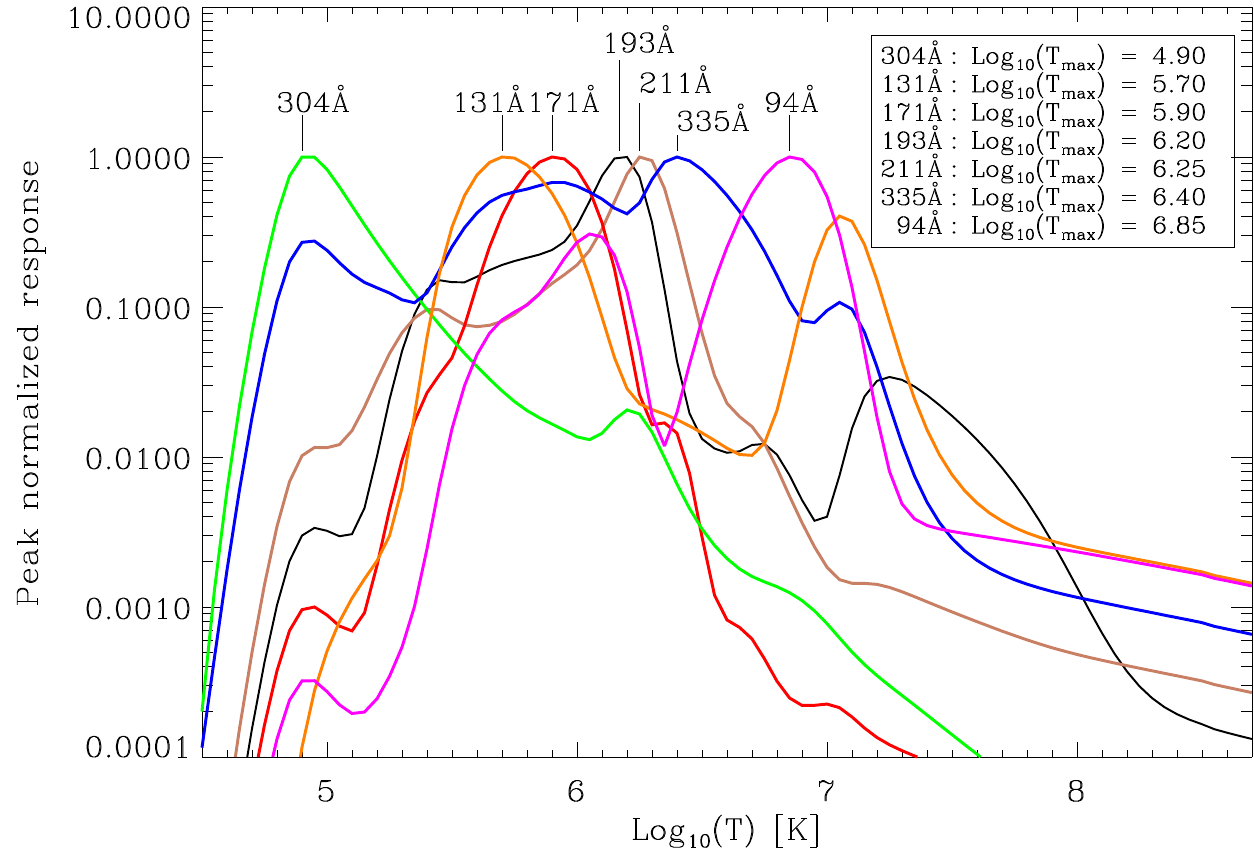}
\caption{Peak-normalised response curves versus temperature for each EUV passband observed by \emph{SDO}/AIA. The peak emission temperatures of the different passbands are given in the upper right for comparison. This figure is a peak-normalised version of that from \citet{Lemen:2011uq}.}
\label{fig:aia_resp}
\end{center}
\end{figure}

The large number of passbands observed by AIA allow it to continuously monitor a large part of the solar atmosphere and a wide range of phenomena. Table~\ref{tbl:aia_emission} outlines the different passbands studied along with the dominant ions and peak emission temperatures in each case.

\emph{STEREO}/EUVI and \emph{SDO}/AIA both observe the 304~\AA, 171~\AA, and 195/193~\AA\ passbands, making these passbands useful for confirming observations. However, the 171~\AA\ and 195/193~\AA\ passbands in particular have different temperature response curves, and as a result do not necessarily share observations of the same phenomenon. This is most apparent when studying ``EIT waves'', as the pulse appears in emission in EUVI 171~\AA, while simultaneously appearing in absorption in AIA 171~\AA. This observational discrepancy is discussed in more detail in Section~\ref{subsect:sdo_cbf_discrepancies}, along with an analysis of how this affects the interpretation of the pulse.

Although AIA observes multiple passbands using highly tuned filters, the broadband nature of the imager is apparent from the temperature response curves shown in Figure~\ref{fig:aia_resp}. As a result, determining temperature and differential emission measure estimates from AIA images is subject to misinterpretation; these issues are discussed in Section~\ref{subsect:sdo_cbf_discrepancies}.

\begin{deluxetable}{ccc}
\tablecolumns{3}
\tablewidth{0pt}
\centering
\tablecaption{Temperature response characteristics of AIA passbands.\label{tbl:aia_emission}}
\tablehead{
\colhead{Passband (\AA)} & \colhead{Dominant ion} & \colhead{\corr{log T$_{peak}$ (K)}}
}
\startdata
UV/Visible	& Continuum, \ion{C}{4}\tablenotemark{a}	& 3.7, 5.0 \\
304			& \ion{He}{2}								& 4.7 \\
171			& \ion{Fe}{9} 							& \corr{5.9} \\
193			& \ion{Fe}{12}, \ion{Fe}{24} 				& 6.2, 7.3 \\
211			& \ion{Fe}{14}							& 6.3 \\
335			& \ion{Fe}{16}							& 6.4 \\
94			& \ion{Fe}{18} 							& 6.8 \\
131			& \ion{Fe}{8}, \ion{Fe}{21} 				& 5.6, 7.0
\enddata
\tablenotetext{a}{Visible and UV light is observed in the 1600, 1700, and 4500~\AA\ passbands using one half of one telescope.}
\end{deluxetable}

\subsection{AIA Data Preparation Software}
\label{subsect:aia_software}

Once downloaded to the base station at White Sands \corr{in New Mexico}, all \emph{SDO} data is sent to the Joint \emph{SDO} Operations Centre (JSOC), located at Stanford University in California where the data is stored permanently. Here, a processing pipeline is used to convert the raw Level~0 data to fully--calibrated Level~1 data. This involves several distinct processes including,
\begin{itemize}
\item{The removal of over and under scan regions of the image}
\item{The removal of dark current effects}
\item{The application of a flat-field correction}
\item{The automated removal of bad pixels and cosmic ray hits}
\item{Flipping the image to place solar north at the top}
\end{itemize}
Once this pipeline processing has been completed, the Level~1 FITS files are sent to a number of data hosting stations worldwide (including the Harvard-Smithsonian Centre for Astrophysics in the U.S.A., the University of Central Lancashire in the U.K. and the Royal Observatory of Belgium) where it is made publicly available.

Once downloaded by the user, the Level~1 FITS files can be further processed using the \emph{aia\_prep.pro} processing routine. This applies additional corrections to the data that allow an intercomparison of the different passbands. This involves 
\begin{inparaenum}[(i)]
\item co-aligning the images to correct for residual offsets between the four telescopes;
\item adjusting the plate-scale to correct for fractional offsets in focal length; and
\item co-aligning the bore-sight pointing of the different telescopes.
\end{inparaenum}
With these adjustments applied to the data, the FITS header is adjusted to Level~1.5, and the images are now ready for scientific analysis.



\chapter{The Kinematics of a Globally-propagating Disturbance in the Low Corona}
\label{chap:first_obs}


\noindent 
\\ {\it 
In natural science the principles of truth ought to be confirmed by observation. 
\begin{flushright}
Carolus Linnaeus \\
\end{flushright}
 }

\vspace{15mm}
In this chapter, the first observations of a CBF made using data from the EUVI instrument onboard the \emph{STEREO} spacecraft are used to determine the kinematics of the pulse. It is shown for the first time that a CBF has similar kinematics in all four EUV passbands studied (304, 171, 195, and 284~\AA). In the 304~\AA\ passband the disturbance shows a velocity peak of 238$\pm$20~km~s$^{-1}$ within $\sim$28 minutes of its launch, varying in acceleration from 76~m~s$^{-2}$ to -102~m~s$^{-2}$. This passband contains a strong contribution from a \ion{Si}{11} line (303.32~\AA) with a peak formation temperature of $\sim$1.6~MK, suggesting that the 304~\AA\  emission may be coronal rather than chromospheric in origin. Comparable velocities and accelerations are found in the coronal 195~\AA\ passband, while lower values are found in the lower cadence 284~\AA\ passband. In the higher cadence 171~\AA\ passband the velocity varies significantly, peaking at 475$\pm$47~km~s$^{-1}$ within $\sim$20 minutes of launch, with a variation in acceleration from 816~m~s$^{-2}$ to -413~m~s$^{-2}$. The high image cadence of the 171~\AA\ passband (2.5 minutes compared to 10 minutes for the similar temperature response 195~\AA\ passband) is found to have a major effect on the measured velocity and acceleration of the pulse, which increase by factors of $\sim$2 and $\sim$10, respectively. This implies that previously measured values (e.g., using EIT) may have been underestimated. It is also noted that the disturbance shows strong reflection from a coronal hole in both the 171 and 195~\AA\ passbands. These observations are consistent with an impulsively generated fast-mode magnetoacoustic wave. The work presented in this chapter is based upon results published in \emph{The Astrophysical Journal Letters} by \citet*{Long:2008eu}.

\section{Introduction}
\label{sect:first_obs_intro}

The launch of the \emph{STEREO} spacecraft, with its EUVI instruments \corr{signalled} the beginning of a new age of solar physics. As noted in Section~\ref{sect:inst_euvi}, \emph{STEREO} consists of two separate spacecraft, designed to study the inner heliosphere from ahead and behind the Earth on its passage around the Sun. This represents the first time that the Sun can be observed stereoscopically, potentially allowing 3-D modeling of the inner Heliosphere. 

As well as offering two distinct locations for joint observations of the Sun and the inner Heliosphere, the suite of imaging instruments onboard \emph{STEREO} represent a major improvement compared to \emph{SOHO}. Whereas \emph{SOHO} carried an EUV imager (EIT) and several coronagraphs (LASCO C--1, --2 and --3), \emph{STEREO} carries the EUVI telescope, two coronagraphs (Cor--1 and --2) and two Heliospheric Imagers (HI--1 and --2) that allow direct imaging of the inner Heliosphere. Each of these instruments also operate at a higher cadence to those onboard \emph{SOHO}, while EUVI observes all four passbands at a high temporal cadence, unlike the synoptic operations of \emph{SOHO}/EIT. These instruments were designed specifically to allow a deeper understanding of CMEs and the on-disk phenomena associated with them. 

However, the launch of the \emph{STEREO} spacecraft in December~2006 coincided with the beginning of an unexpectedly long period of low solar activity. As a result, the first major eruption on the Sun did not occur until $\sim$6~months after launch\corr{,} in May~2007, by which time the \emph{STEREO} spacecraft were $\sim$8 degrees apart. This small separation meant that events occurring close to disk centre as observed from the Earth could be observed by both spacecraft with minimal projection effects.

The design of the \emph{STEREO} spacecraft and the EUVI telescopes make them ideal for studying the on--disk propagation of CBF pulses, particularly early in the mission when the imaging cadence is high and both spacecraft can observe the same event. Traditional analysis of CBFs has been hampered by the low imaging cadence available from \emph{SOHO}/EIT, which produced two or three observations per event. This makes it difficult to derive pulse kinematics, and forces a combination of the EUV observations with measurements from other passbands. While these combinations are vital for a full understanding of the pulse, the sparse nature of the data complicates interpretation. The true kinematics of a CBF pulse must therefore be determined using purely EUV observations, a situation heretofore impossible.

These difficulties in deriving the kinematics of CBF pulses have resulted in a number of alternative interpretations of this phenomenon. Although magnetohydrodynamic wave theory (see Section~\ref{sect:mhd_theory}) can explain many of the brightenings observed in EUV images as well as explaining some of the characteristics of EIT waves, including their slow expansion rates \citep{Wang:2000tg}, it is inconsistent with the low velocities derived from \emph{SOHO}/EIT observations (see Section~\ref{subsect:cbfs_waves} for more \corr{details}). However, there is some indication that CBFs exhibit wave properties, with \citet{Veronig:2006fy} noting signs of pulse refraction at the boundary of a coronal hole.

The inconsistency between the observed pulse behaviour and that predicted by MHD wave theory has resulted in some authors promoting a non--wave interpretation for CBFs. \citet{Delannee:2000ij}, \citet{Chen:2005xe} and \citet{Attrill:2007p99} have each proposed models where the observed CBF is not a true wave, but rather a byproduct of the magnetic field restructuring as the associated CME erupts away from the Sun. The specific details of each model differs slightly, with \citet{Delannee:2000ij} proposing heating in the separatrix between magnetic field lines as the CME erupts, an idea similar to that of \citet{Chen:2005xe} who proposed that the reconnecting magnetic field lines either side of the erupting CME produced the brightening observed as the CBF. In contrast, \citet{Attrill:2007p99} proposed that the reconnection of the erupting CME with quiet Sun loops either side of the CME resulted in small--scale brightenings seen as the CBF pulse.

Although the sets of both wave and pseudo--wave models are formed in different ways, they were proposed based on observation, with the result that it is difficult to distinguish them by observation alone. However, there are discrepancies in the predicted behaviour of the theoretical pulses. In particular, the kinematics and morphology of the pulses should provide an indication of the physical properties of the pulse. For the first time, the high cadence of \emph{STEREO}/EUVI allow the pulse kinematics to be determined to a high degree of accuracy using purely EUV observations, while the dual viewpoints available from \emph{STEREO} provide an independent confirmation of the kinematics.

In this chapter, an analysis of the CBF event from 2007~May~19 as observed by the EUVI instrument onboard both \emph{STEREO} spacecraft is presented. The observations made using the \emph{STEREO} spacecraft are outlined in Section~\ref{sect:first_obs_obs}, with the methods used to process the data discussed in Section~\ref{sect:first_obs_analysis}. The data are then analysed, with the results of this analysis presented in Section~\ref{sect:first_obs_results}. These results are discussed and some conclusions drawn in Section~\ref{sect:first_obs_conc}. Finally, Section~\ref{sect:first_obs_comments} discusses the results of this work in the context of follow--up work performed by other authors. 
     
\section{Observations}
\label{sect:first_obs_obs}

A GOES class B9 flare erupting from active region NOAA 10956 was observed on 2007~May~19, beginning at 12:34~UT, before peaking at 13:02~UT and then ending at 13:19~UT. A CME was initially observed in the LASCO C--2 field--of--view at 13:24~UT, with the \emph{SOHO}/LASCO CME catalogue classifying it as being associated with this eruption. A filament lift--off was also observed to begin at $\sim$12:31~UT, slightly before the first observation of the CBF pulse. The eruption was observed at high cadence in multiple EUV passbands by both \emph{STEREO} spacecraft, although \emph{SOHO}/EIT was undergoing CCD bake--out (to remove the effects of radiation damage) at the time of the eruption and was not available. 

As the first major solar eruption to be observed by \emph{STEREO} at high cadence, multiple aspects of this event have been studied in detail by other authors. In particular, the stereoscopic capabilities of \emph{STEREO} allowed a detailed examination of the filament eruption \citep{Bone:2009gf,Liewer:2009fr}, while the location of the erupting active region near disk centre made it possible to analyse the evolution of the magnetic field \citep{Li:2008p121,Conlon:2010mz}. The high cadence observations of the erupting CBF and the nature of its free propagation also meant that the CBF was further analysed by additional authors including \citet{Veronig:2008p216,Gopalswamy:2009p1527} and \citet{Attrill:2010ab}. 

\begin{figure*}[!t]
\centering
\includegraphics[clip=,trim=0mm 0mm 0mm 0mm, width=1\textwidth]{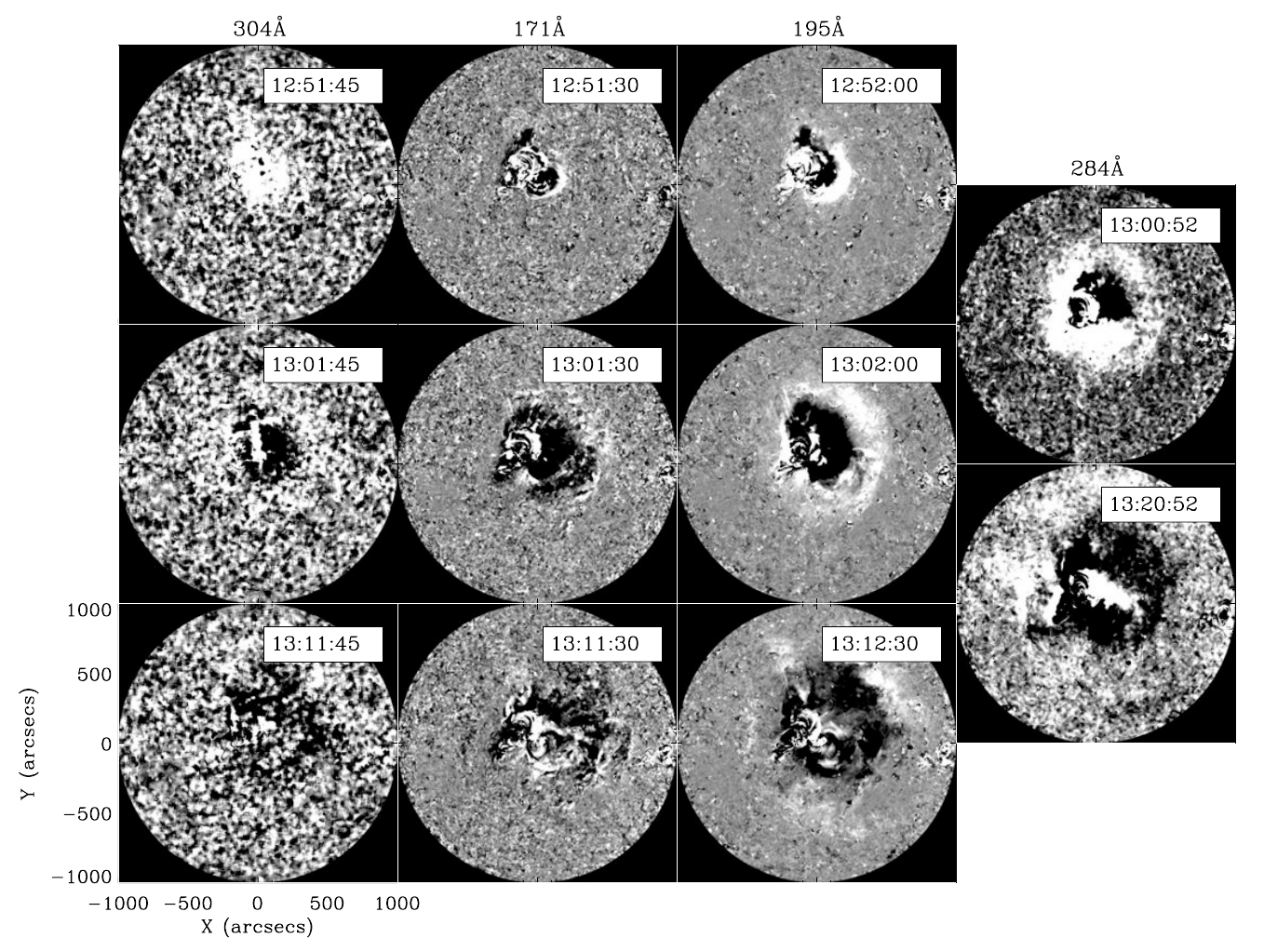}
\caption{Evolution of the disturbance with time in all four EUVI wavelengths from \textit{STEREO-A}. Wavelengths are arranged from left to right as 304~\AA, 171~\AA, 195~\AA\ and 284~\AA.}
\label{fig:images}
\end{figure*}

Although the imaging cadence of \emph{STEREO}/EUVI has shown some variation over the timescale of the mission, this event occurred soon after launch. As a result, the cadence was very high, at 150~s for the 171~\AA\ passband, 600~s for the 195 and 304~\AA\ passbands and 1200~s for the 284~\AA\ passband. The full-disk images were also analysed using the full spatial resolution of the instrument (1.6~arcsec per pixel) to ensure a thorough analysis.

Figure~\ref{fig:images} shows the CBF pulse observed using data from \emph{STEREO}-A and highlighted using the techniques outlined in Section~\ref{sect:first_obs_analysis}. The passbands have been arranged here from left to right as 304~\AA, 171~\AA, 195~\AA\ and 284~\AA, with the times of the leading image in each case shown in the top right of each image. These images were chosen to show the pulse just after formation, during its propagation and before it becomes too diffuse for clear identification.

\section{Methods}
\label{sect:first_obs_analysis}

Before proceeding with identification and analysis of the pulse, each image used was first processed using the SolarSoft calibration routine \emph{secchi\_prep.pro} (see Section~\ref{subsect:euvi_software} for details) to remove CCD bias, cosmic ray hits and missing or corrupted pixels that could affect the the image quality. The images were then flat-fielded to ensure uniform intensity variation and normalised to account for the different filters used for each passband. The effects of image processing carried out onboard the spacecraft were also accounted for. 

Once these initial processes had been applied to the data, each image was de-rotated to the same pre-event time to counter the effects of solar rotation. The propagating pulse was then highlighted using running difference images. This is a simple image processing technique \corr{traditionally used to highlight transient solar events \citep[cf.][]{Dere:1997p159,Thompson:1998sf,Zhukov:2004kh},} that allows the identification of motion by subtracting the previous image in time from each image (i.e.,\ $\Delta I = I_{t} - I_{t-1}$). Each difference image was then smoothed using a boxcar filter to highlight the diffuse pulse while retaining detail. The width of the boxcar filter applied was defined by the passband, with widths of 15, 9, 3 and 7 pixels for the 304, 171, 195 and 284~\AA\ passbands respectively.

\begin{figure}[!t]
\centering
\includegraphics[clip=,trim=0mm 5mm 10mm 77mm,width=1\textwidth]{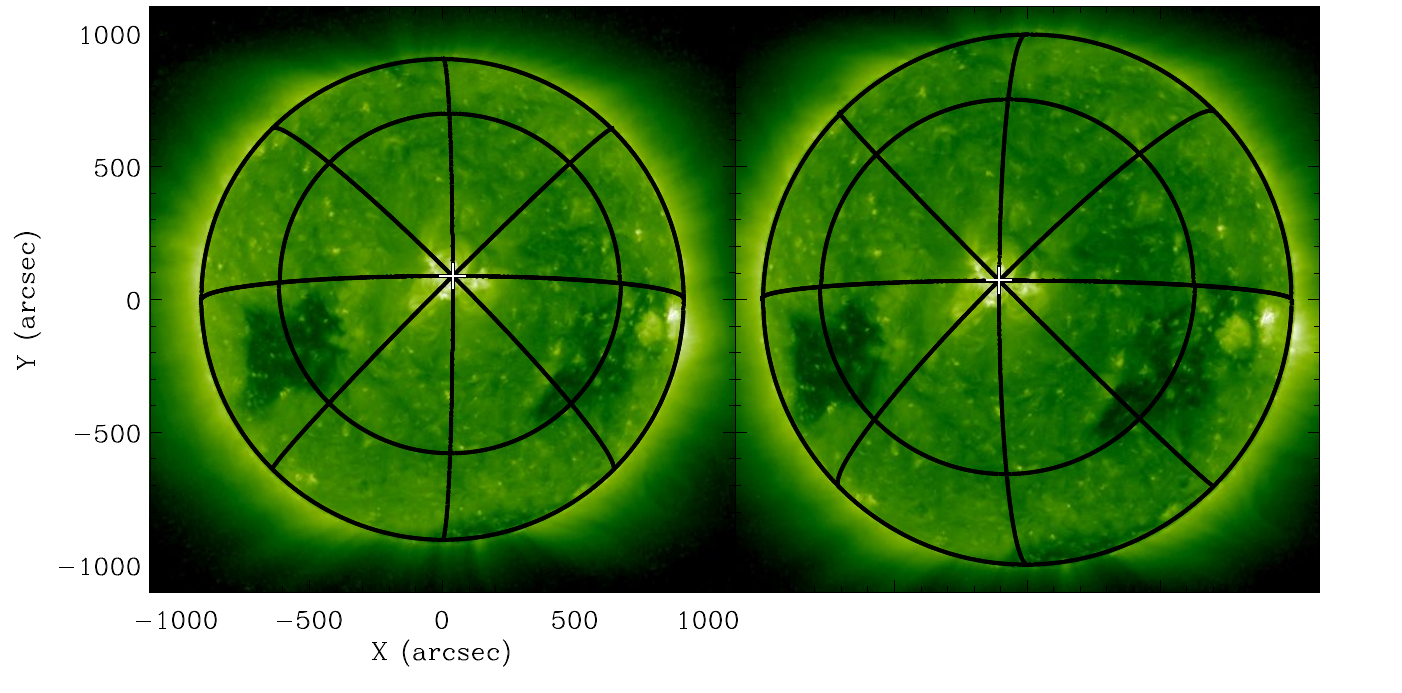}
\caption{\emph{STEREO}-B (left) and \emph{STEREO}-A (right) images from the 195~\AA\ passband with heliographic grid overlay. The north pole of the grid in each case is centered on the flare kernel.}
\label{fig:grid}
\end{figure}

Once the pulse had been visually identified in each individual image, a heliographic grid was applied to each image to allow the position of the pulse to be determined. This is shown in Figure~\ref{fig:grid} for both \emph{STEREO}-B (left) and \emph{STEREO}-A (right), with the grid oriented in each case so that the north pole was located at the flare kernel. The front of the pulse was then identified in subsequent images using a point-and-click technique with the heliographic grid lines used as a guide to ensure that the temporal variation of the pulse position was determined along the same great circle line. The distance travelled by the pulse could then be determined using the angular separation between the source and the identified front position for a given time. The same grid line towards solar west was used for each passband and for each spacecraft to ensure continuity. The variation in velocity and acceleration of the pulse were then derived using the three--point Lagrangian interpolation technique available from the deriv.pro routine in IDL \corr{(see Section~\ref{subsubsect:l_diff} for more details)}.

\section{Results}
\label{sect:first_obs_results}

The observed CBF pulse is apparent in Figure~\ref{fig:images}, particularly in the second row of difference images which clearly shows a bright front followed by a dimming region on the solar disk for all four passbands. Although this has been previously observed using the 195~\AA\ \citep{Thompson:1998sf}, 171~\AA\ \citep{Wills-Davey:1999ve} and 284~\AA\ \citep{Zhukov:2004kh} passbands, this is the first observation of the ``EIT wave'' phenomenon using the 304~\AA\ passband. This is also the first time that the pulse has been observed in all four EUV passbands simultaneously, and as a result is very important for a complete physical analysis of the disturbance. As each passband corresponds roughly to a different temperature and thus a different \corr{region} of the solar atmosphere, this observation should greatly assist with the 3--D modeling of CBFs.

\begin{figure}[!t]
\begin{center}
\includegraphics[width=0.99\textwidth,clip=,trim=2mm 1mm 0mm 0mm]{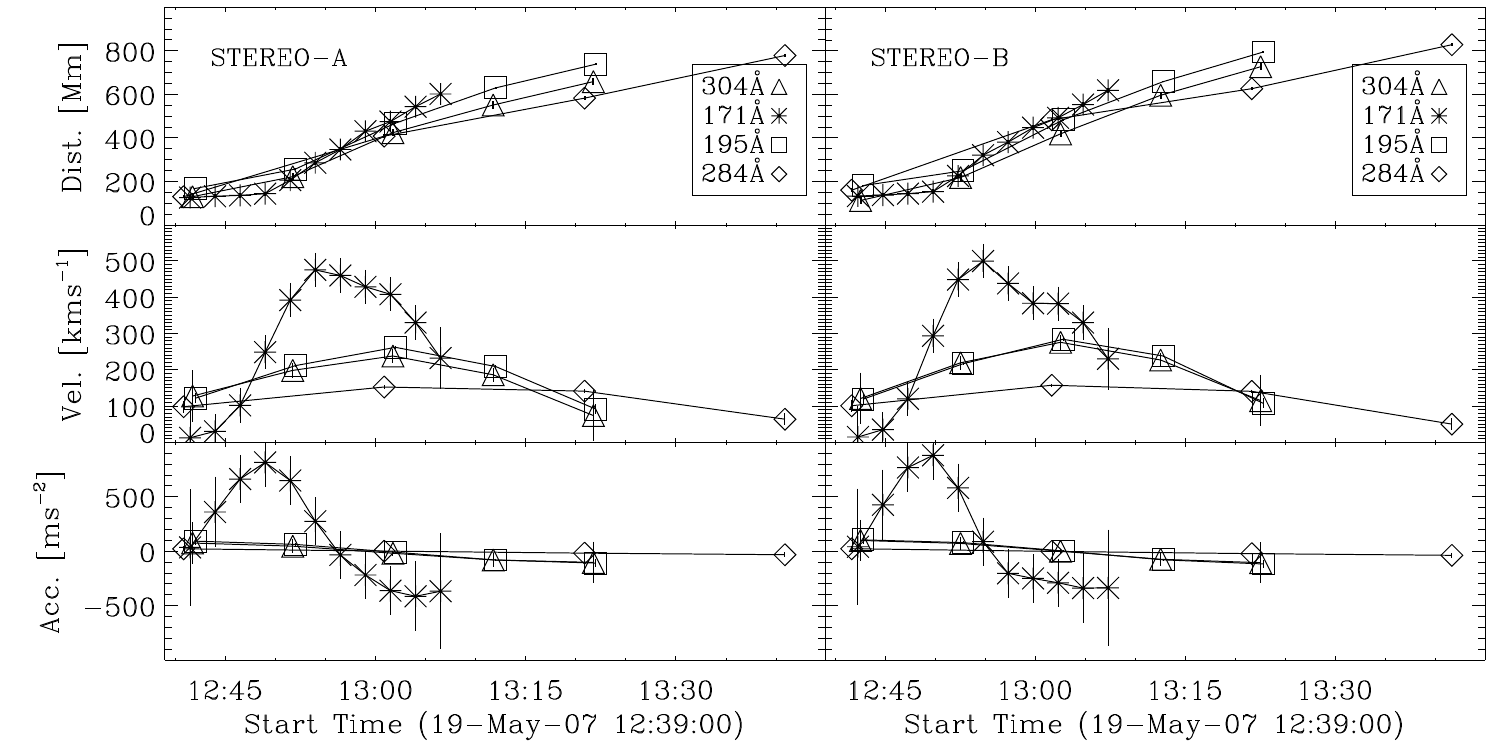}
\end{center}
\caption{Distance-time (top), velocity-time (middle), and acceleration-time (bottom) plots of the wavefront. Left-hand panels show data measured by \textit{STEREO-A}, while right-hand panels show \textit{STEREO-B}. All distances are measured to the leading edge of the running difference brightening along the great circle longitude line to solar west (see Fig.~\ref{fig:grid}). The first and last data point error bars in the 171~\AA\ data have been divided by two for display purposes.}
\label{fig:norm_graphs}
\end{figure}

The measured distance--time data is shown in the top row of Figure~\ref{fig:norm_graphs} for each passband at its original cadence. The high cadence 171~\AA\ measurements appear to show a more rapid increase in propagation distance compared to the lower cadence 195~\AA\ and 304~\AA\ measurements. Similarly, the lowest cadence 284~\AA\ measurements show a more gradual increase in propagation distance compared to the 195~\AA\ and 304~\AA\ passbands. The similarity in the observing cadence of the 195~\AA\ and 304~\AA\ passbands is also reflected in the similar temporal variation in propagation distance.

The difficulty in directly comparing measurements of unequal cadence is highlighted by the derived velocity--time graphs shown in the middle row of Figure~\ref{fig:norm_graphs}, where the variation in velocity appears to be strongly cadence--dependent. The curve corresponding to the 171~\AA\ images shows a significant peak in velocity (increasing from rest to 475$\pm$47~km~s$^{-1}$ within approximately 20 minutes), which is distinctly different from the more gradual peaks observed in the 195~\AA, 304~\AA\ and 284~\AA\ velocities. The 195~\AA\ and 304~\AA\ passbands show similar velocity variations; a peak velocity of 262$\pm$4~km~s$^{-1}$ within $\sim$28 \corr{minutes} for 195~\AA\ and a velocity peak of 238$\pm$20~km~s$^{-1}$ within $\sim$28 minutes in 304~\AA. However, the lowest cadence 284~\AA\ passband exhibits the smallest velocity variation, with a peak in velocity of 153$\pm$5~km~s$^{-1}$ within $\sim$27 minutes of launch.

\begin{deluxetable}{cccccc}
\tablecolumns{6}
\tabletypesize{\small}
\tablewidth{0pt}
\centering
\tablecaption{Derived pulse velocity and acceleration estimates.\label{tbl:first_cbf_kins}}
\tablehead{
\colhead{} & \colhead{Passband} & \colhead{Cadence} & \colhead{$v_{max}$} & \colhead{$a_{max}$} & \colhead{$a_{min}$} \\ 
\colhead{Spacecraft} & \colhead{(\AA)} & \colhead{(minutes)} & \colhead{(km~s$^{-1}$)} & \colhead{(m~s$^{-2}$)} & \colhead{(m~s$^{-2}$)}
}
\startdata
\emph{STEREO}-A & 304 & 10 & 238 & 76 & -102 \\
\nodata & 171 & 2.5 & 475 & 816 & -413 \\
\nodata & 195 & 10 & 262 & 93 & -109 \\
\nodata & 284 & 20 & 153 & 23 & -32 \\
\emph{STEREO}-B & 304 & 10 & 276 & 99 & -105 \\
\nodata & 171 & 2.5 & 500 & 881 & -339 \\
\nodata & 195 & 10 & 284 & 105 & -117 \\
\nodata & 284 & 20 & 158 & 23 & -38
\enddata
\tablecomments{\corr{Mean v}elocity errors are $\leq$50~km~s$^{-1}$, \corr{mean} acceleration errors $\leq$200~m~s$^{-2}$}
\end{deluxetable}

The highly variable nature of the kinematics is also apparent in the derived accelerations, with variations from 816~m~s$^{-2}$ to -413~m~s$^{-2}$, 93~m~s$^{-2}$ to -109~m~s$^{-2}$, 76~m~s$^{-2}$ to -102~m~s$^{-2}$ and 23~m~s$^{-2}$ to -32~m~s$^{-2}$ for the 171~\AA, 195~\AA, 304~\AA\ and 284~\AA\ passbands respectively. The derived kinematics are summarised in Table~\ref{tbl:first_cbf_kins} for the observations from \corr{both \emph{STEREO}--A and \emph{STEREO}--B}. The variations in velocity and acceleration can be seen in data from both \emph{STEREO} spacecraft, although the peak values differ slightly between the two spacecraft as a consequence of the different spacecraft positions. The errors associated with the derived velocity and acceleration values are the standard deviation of the derivative at each point. As a result, the error bars for the first and last value were larger than for the middle values, although the typical velocity errors were $\leq$50~km~s$^{-1}$ with the typical errors for the derived acceleration $\leq$200~m~s$^{-2}$.

\begin{figure}[!t]
\begin{center}
\includegraphics[width=0.99\textwidth,clip=,trim=2mm 1mm 0mm 0mm]{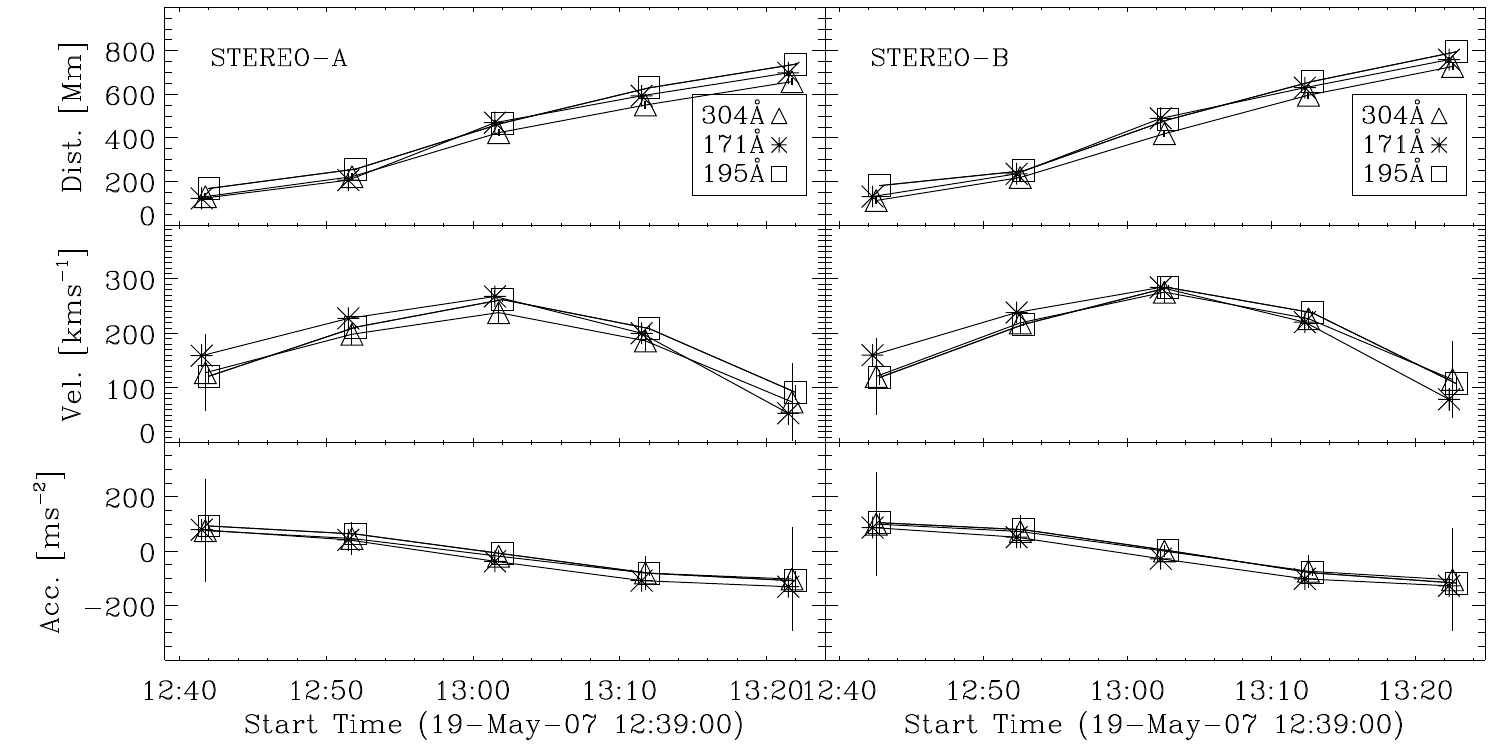}
\end{center}
\caption{Same as Fig.~\ref{fig:norm_graphs}, but with 171~\AA\ at 10 minute cadence and excluding 284~\AA. All data have the same cadence and running difference images have as close as possible effective times in each of the three passbands. The first and last data point error bars in the 171~\AA\  and 304~\AA\ data have been divided by two for display purposes.}
\label{fig:norm10_graphs}
\end{figure}

To overcome the difficulties in directly comparing measurements made using different image cadences, the cadence of the 171~\AA\ passband was degraded from 2.5~minutes to 10~minutes to match that of the 195~\AA\ and 304~\AA\ passbands. This was done by taking every fourth image from 171~\AA, with the effective times of the images chosen to be as close as possible to that of the 195~\AA\ images. \corr{The 284~\AA\ data was not examined here as its very low 20~minute cadence meant that it was not comparable to the other three passbands}. The same running difference analysis previously described was then applied to these images, yielding the results shown in Figure~\ref{fig:norm10_graphs}. \corr{Note that the reduced cadence had the effect of increasing the signal of the pulse for the 171~\AA\ passband, allowing the pulse to be identified after the time at which it could not longer be identified using the normal cadence images. This is a result of the longer time--step between images, allowing the pulse to travel beyond its previous extent and making it easier to track over a larger distance.}

It is apparent that the derived kinematics for the 171~\AA\ passband measurements made at lower cadence are comparable to the 195~\AA\ and 304~\AA\ measurements. The peak in velocity for the 10~minute cadence 171~\AA\ measurements is much more gradual than for the 2.5~minute cadence measurements and the acceleration--time plot, while still showing acceleration followed by deceleration, is not as pronounced as in Figure~\ref{fig:norm_graphs}. The discrepancy between the higher 2.5~minute and lower 10~minute cadence 171~\AA\ measurements suggests that the derived kinematics may be influenced by the observing cadence of the instrument. \corr{Although 2.5~minutes is the highest available cadence for this event, \emph{STEREO} is capable of observing at up to $\sim$75~s cadence, while the forthcoming \emph{Solar Dynamics Observatory} will have an observing cadence of $\sim$12~s. This will be vital for examining this effect in more detail.}

\section{Conclusions}
\label{sect:first_obs_conc}

The first major solar eruption with an associated coronal propagating front to be observed by the \emph{STEREO} spacecraft was studied using EUV images available from the EUVI instrument. The disturbance was observed in all four passbands covered by EUVI, namely the 171~\AA, 195~\AA, 284~\AA\ and for the first time, the 304~\AA\ passband. Despite the varying cadences, it was possible to derive the kinematics of the pulse in each passband, allowing the temporal variation in distance, velocity and acceleration to be examined. 

The kinematics derived from the higher cadence 171~\AA\ measurements indicate an average front velocity of 196~km~s$^{-1}$, peaking at 475$\pm$47~km~s$^{-1}$, with an initially positive acceleration that peaked at 816\corr{$\pm$220}~m~s$^{-2}$ before decreasing to -413\corr{$\pm$320}~m~s$^{-2}$; i.e., it initially accelerates away from the flare point before decelerating further away. This peaked velocity and positive to negative variation in acceleration was also noted for the other passbands studied, albeit with decreased strength. The observed rise and fall in velocity is in contrast to the model proposed by \citet{Delannee:2008uq} which indicates a constantly increasing expansion velocity. The peak in velocity derived for the 171~\AA\ passband is also at the higher end of the previously measured range of ``EIT wave'' velocities, while the peak velocities determined for the lower cadence 195~\AA\ and 304~\AA\ passbands (262$\pm$4~km~s$^{-1}$ and 238$\pm$20~km~s$^{-1}$ respectively) are comparable to the typically measured values. The results from \textit{STEREO}-A are qualitatively corroborated by those from \textit{STEREO}-B, with minor differences in the peaks probably due to geometric effects.

The cadence of the 171~\AA\ passband was artificially degraded from 2.5~minutes to 10~minutes to allow a direct comparison with the 195~\AA\ and 304~\AA\ passbands, with the resulting kinematics observed to decrease giving a peak velocity $\sim$300~km~s$^{-1}$ and a variation in acceleration from $\sim$100~m~s$^{-2}$ to $\sim-$100~m~s$^{-2}$. These values are comparable to those estimated for the 195~\AA\ and 304~\AA\ passbands, suggesting that the results obtained for these passbands are most likely due to temporal undersampling\corr{, analogous to the apparent smoothing with decreasing sampling rate exhibited in signal processing}. This indicates that the effects of temporal sampling must be accounted for when analysing observations of CBFs. These effects may explain the lower velocities previously measured in \emph{SOHO}/EIT observations of ``EIT waves''. The relatively low cadence available from EIT ($\sim$12~minutes) may have resulted in short time-scale variations of the propagation distance being missed.

Although the pulse is quite faint in the plain images, this event was visible in the 304~\AA\ passband, marking the first observation of a CBF pulse in this passband. This is a result of the consistently high cadence of the 304~\AA\ passband on \emph{STEREO}/EUVI and provides a new opportunity to understand the physical nature of the disturbance. The 304~\AA\ passband is used to observe the upper chromosphere using \ion{He}{2} emission so a pulse observed using this passband would appear to be chromospheric in nature. However, it should be noted that there can be a strong contribution from the coronal \ion{Si}{11} 303.32~\AA\ emission line in this passband. This \ion{Si}{11} line has a peak formation temperature of $\sim$1.6~MK \citep{Brosius:1996p54}, which is consistent with the observations made using the coronal 171~\AA, 195~\AA\ and 284~\AA\ passbands. 

The 304~\AA\ data required extensive \corr{processing} to make the disturbance visible above variable small-scale features present in the running difference images. These features are expected due to the dynamic region of the solar atmosphere sampled by 304~\AA, although the wavefront was discernible after filtering. This very small signal--to--noise is consistent with the small signal expected from the coronal \ion{Si}{11} emission line, and indicates that the observed pulse is coronal in nature.

The properties of this event should allow clarification as to the physical nature of the disturbance, in particular whether it was a fast--mode magnetoacoustic wave or the propagation of perturbation sources resulting from CME lift--off. The derived velocity of the pulse is higher than previous estimates, and may now be consistent with the fast--mode magnetoacoustic wave or propagating MHD shock interpretation. As the pulse moves through the quiet corona, it is propagating perpendicular to the predominantly radial magnetic field lines of the quiet Sun, indicating a fast--mode wave nature. This allows the characteristic wave velocity to be calculated using the equation,
\begin{equation}
v_{ms}=(c_{s}^2 + v_{A}^2)^{1/2},
\end{equation}
where the sound speed $c_{s}$ and the Alfv\'{e}n speed $v_{A}$ have been defined in Equations~\ref{eqn:sound_speed} and \ref{eqn:alfven_speed_1} respectively (see Section~\ref{sect:mhd_theory} for more details). It has been noted by \citet{Wills-Davey:2007oa} that the range of typical Alfv\'{e}n speeds in the quiet solar corona is of the order 200--1000~km~s$^{-1}$. The pulse velocities found in this work are higher than previously estimated values \corr{as a result of the higher image cadences available}, and are now more consistent with the fast magnetoacoustic wave interpretation.

It should also be noted that this event was associated with a partial halo CME and filament lift--off which began at $\sim$12:31~UT. The timescale over which the pulse accelerated to its peak velocity is comparable to that for the CME, possibly indicating a common launch mechanism. The pulse was also observed in the 171~\AA\ and 195~\AA\ passbands to exhibit some evidence of reflection and refraction when interacting with the boundary of an adjacent coronal hole. This is consistent with the interpretation of the pulse as a fast--mode magnetoacoustic wave pulse.

\section{Further Comments}
\label{sect:first_obs_comments}

The analysis of the 2007~May~19 event outlined above was the first discussion of this event to be published \citep{Long:2008eu} and as a result has been analysed in turn by other authors examining this event \citep[such as e.g.,][]{Veronig:2008p216,Gopalswamy:2009p1527,Attrill:2010ab}. Although the major conclusions of \citet{Long:2008eu} have been upheld in these re--examinations, some issues have been raised with regard to the techniques used.

The use of a running difference technique in particular, while common and historically used for the identification of ``EIT waves'', has been questioned by \citet{Attrill:2010ab}. The inherent nature of the running difference technique highlights relative motion, and may result in a misidentification of a moving pulse. 

The form of the kinematics derived by \citet{Long:2008eu} are also a source of concern. The highly variable nature of the acceleration is inconsistent with all previously proposed models, and invites further consideration. The kinematics derived by \citet{Veronig:2008p216} for this event, while comparable, do not show the same variability. However, both \citet{Veronig:2008p216} and \citet{Ma:2009ab} noted the effects of cadence on the temporal undersampling of pulse kinematics \citep[albeit for a different event in the case of][]{Ma:2009ab}. \corr{The form of the kinematics curve obtained for the 171~\AA\ passband in particular does not conform with results obtained by \citet{Veronig:2008p216}, who note that this apparent increasing acceleration is most likely due to a misinterpretation of expanding coronal loops as the initial propagation of the pulse. This highlights the difficulties in discriminating between rising loops and the initial propagation of the pulse, particularly when the pulse may be initiated by the rising loop. In addition, the initial velocity derived for the pulse in the 171~\AA\ passband is not consistent with a fast--mode magnetoacoustic wave, which should consistently have a velocity that exceeds the local Alfv\'{e}n speed.}

\corr{Once the results of these analyses had been taken} into consideration, it was decided to examine in detail the effects of running difference image processing and to identify an alternative technique for studying ``EIT waves''. Simultaneously, it was decided to test the capabilities of numerical differencing techniques when faced with small data--sets and determine a technique that could be relied upon to produce statistically accurate kinematics estimates regardless of sample size. This examination of image processing techniques and numerical differencing schemes is discussed in Chapter~\ref{chap:methods}.		


\chapter{Identification and Analysis Techniques}
\label{chap:methods}



\noindent 
\\ {\it 
You see, but you do not observe. The distinction is clear.
\begin{flushright}
Sir Arthur Conan Doyle (A Scandal in Bohemia) \\
\end{flushright}
 }

\vspace{15mm}
Coronal bright fronts are typically observed as broad diffuse features in coronal emission lines as previously discussed in Chapter~\ref{chap:cbfs}. This diffuse nature makes it difficult to identify CBF pulses in individual images, complicating analysis of the phenomenon. These issues have resulted in the development of several different image processing techniques for identifying CBFs, each with advantages and disadvantages. It is shown here that the techniques traditionally used for identifying CBFs are fundamentally flawed and prone to undefined user--dependent errors. A new semi--automated technique is proposed \corr{to allow} the rigorous identification of CBF pulses.

Once the pulse has been identified in subsequent images a thorough analysis requires a determination of the pulse kinematics. These can be derived using several different methods, either by direct numerical differentiation of the data or alternatively through the application of a model fit to the data. Again, each technique has drawbacks, particularly when dealing with the typical data-set sizes that are encountered when studying CBFs. 

In this chapter, the issues traditionally associated with identifying CBF pulses are discussed. The different image processing techniques used are presented in Section~\ref{subsect:diff_imgs} with a semi--automated technique for identifying CBFs outlined in Section~\ref{subsect:cor_pita}. The different numerical techniques that can be used to derive the pulse kinematics directly from distance--time measurements are then presented in Section~\ref{subsect:num_diff} with an examination of the effects of image cadence and pulse uncertainty outlined in Section~\ref{subsect:img_cad}. It is shown that the errors associated with the identification and analysis of CBFs can be minimised through the use of a percentage base difference image processing technique allied to a model fit using a bootstrapping approach.

\section{Pulse Identification}
\label{sect:identify_pulse}

The diffuse nature of CBFs has made them difficult to observe in single images, with identification improved in movies as a consequence of their motion. To overcome this issue, CBFs have traditionally been identified using difference images, an approach which involves subtracting one image from another in a bid to highlight the moving pulse. There are several different types of difference \corr{image} including the running difference (RD), base difference (BD) and percentage base difference (PBD) techniques.

\subsection{Difference Imaging}
\label{subsect:diff_imgs}

\begin{figure}[!t]
\begin{center}
\includegraphics[clip=,trim=0mm 10mm 0mm 60mm,width = 0.95\textwidth]{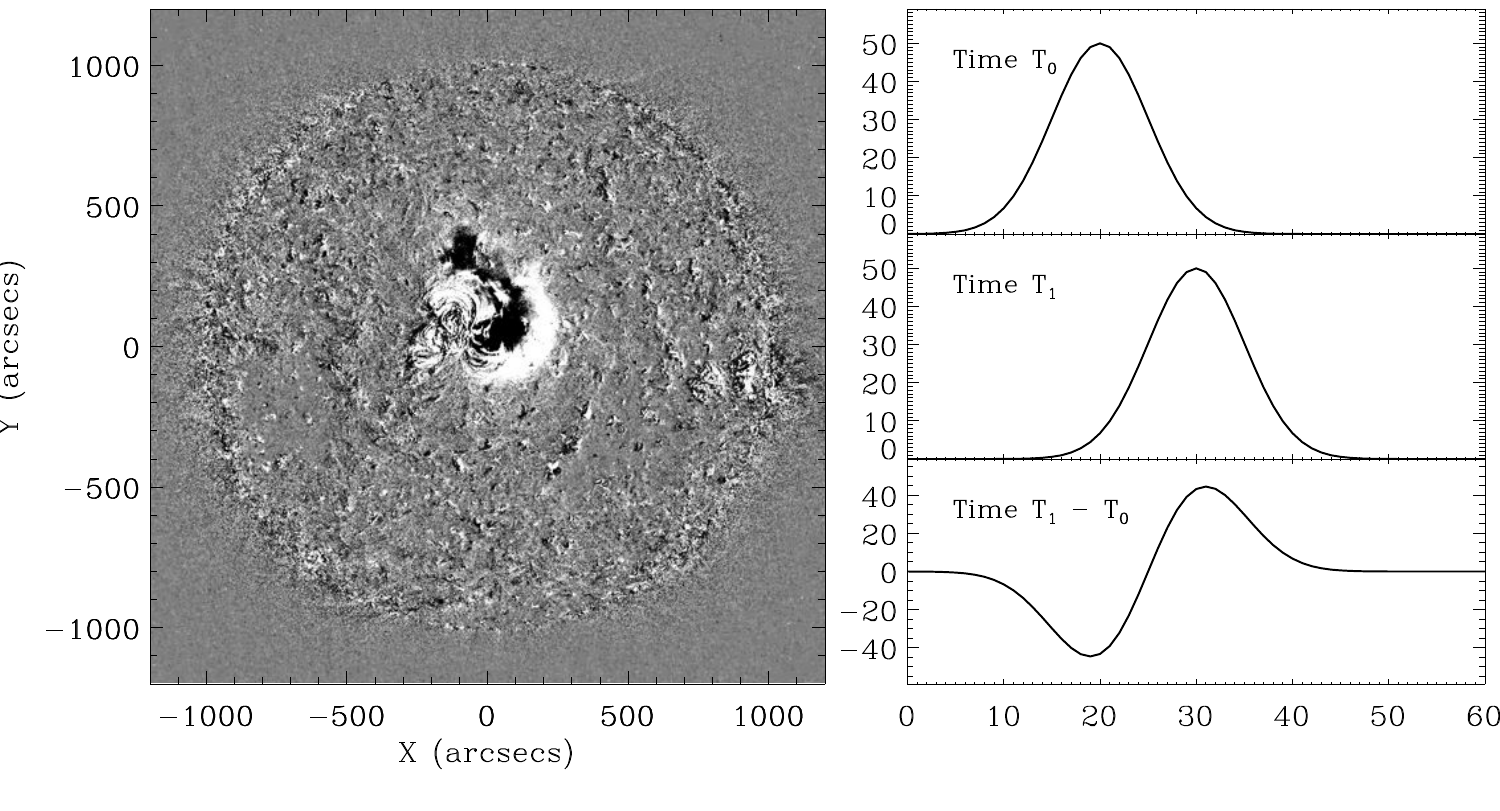}
\caption{\emph{Left}: Running difference image from 2007~May~19. White (black) is a positive (negative) intensity change relative to the previous image. \emph{Right}: Simulated intensity profiles showing a pulse at T$_{0}$ (top panel) T$_{1}$ (middle panel) and T$_{1}$ - T$_{0}$ (bottom panel).}
\label{fig:rd_images}
\end{center}
\end{figure}

Difference imaging is a simple image processing technique that is designed to highlight relative motion in images, making it particularly useful for identifying CBF pulses in subsequent images. Some examples of a differencing approach can be seen in Figures~\ref{fig:may1997} and \ref{fig:rd_images}, both of which show a CBF pulse highlighted using differencing.

\begin{figure}[!t]
\begin{center}
\includegraphics[clip=,trim=0mm 0mm 0mm 60mm,width = 0.99\textwidth]{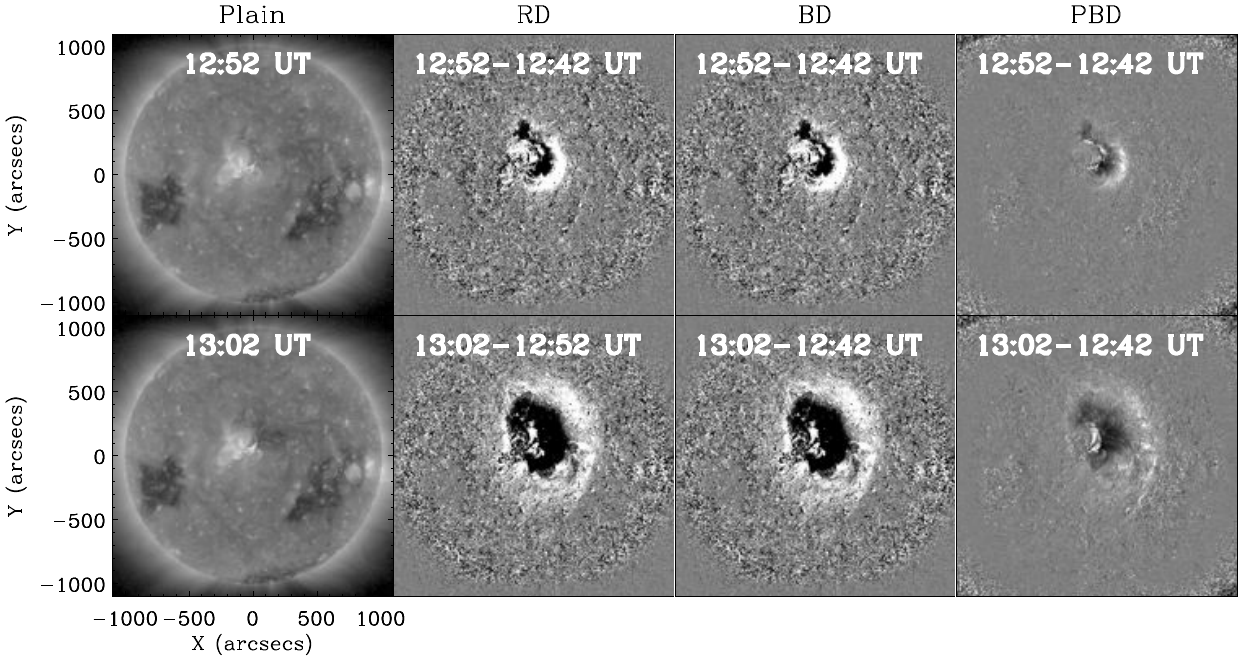}
\caption{Plain, running difference (RD), base difference (BD) and percentage base difference (PBD) images \corr{for the event from 2007~May~19 in the 195~\AA\ passband from \emph{STEREO}--A. The times used to produce the images are shown in the top of each panel.} The plain image is shown using logarithmic scaling, the RD and BD images using user-defined intensity scaling and the PBD image using its \corr{intrinsic} intensity \corr{range} (i.e., $-$100\% to 100\%).}
\label{fig:diff_imgs}
\end{center}
\end{figure}

The most common and simplest technique used to accentuate the pulse signature is the running difference image \corr{(see Section~\ref{sect:first_obs_analysis})}. This involves subtracting a following image from a leading image (i.e.,\ $\Delta I = I_{t} - I_{t-1}$), highlighting relative motion with an increase in intensity seen in white and a decrease in intensity seen in black (as shown in the left-hand panel of Figure~\ref{fig:rd_images} and in Figure~\ref{fig:diff_imgs}). While this technique \corr{clearly highlights} the moving pulse, there are \corr{several} issues that must be kept in mind when examining the resulting images \citep[see also][for some additional issues with the misinterpretation of RD images]{Attrill:2010ab}.

As shown in the right-hand panel of Figure~\ref{fig:rd_images}, the profile of the pulse seen in a RD image does not necessarily conform to that of the original pulse. The technique highlights the relative intensity change between images, which allows the general rather than specific motion of the pulse to be discerned. The nature of the data being analysed means that the resulting RD image will also have to be scaled to make the pulse apparent. This scaling is entirely user dependent and strongly influences the apparent size of the observed pulse. 

An alternative technique proposed to negate the influence of image--to--image variation is the base difference (BD) technique (see Figure~\ref{fig:diff_imgs}). Here, a pre-event image is subtracted from each subsequent image, highlighting the intensity changes relative to the pre-event image (i.e.,\ $\Delta I = I_{t} - I_{0}$). This is designed to remove only the coronal background, allowing an easier identification of the pulse itself. Although this \corr{is a better approach for enhancing the true pulse signature than running--difference}, the small-scale background structure of the corona is also highlighted using this technique. Consequently, this technique becomes ineffective over long timescales with the resulting image difficult to interpret.

The BD technique is also subject to user-defined scaling which can over-emphasise the pulse signature, while the diffuse nature of the pulse compared to the small-scale motion also highlighted by the technique means that image smoothing is often required. While this makes the pulse easier to identify, it does introduce an additional error that must be accounted for when identifying the pulse position.

A variation on this technique was originally proposed by \citet{Wills-Davey:1999ve} and involves dividing the BD image by the pre-event image (i.e.,\ $\Delta I = (I_{t} - I_{0})/I_{0} \times 100$ (see Figure~\ref{fig:diff_imgs}). This produces images with a relative intensity \corr{nominally} ranging from $-100$\% to 100\%, eliminating the possibility of arbitrary image scaling \corr{(although it should be noted that maximum intensities greater than 100~\% can be observed as a result of e.g., flares)}. The peak pulse intensity is also then given as a percentage of the pre-event intensity, allowing a better comparison from frame to frame. While this variation eliminates the issues with regard to the arbitrary image scaling, the assumption about a quiet pre-event coronal background \corr{remains}.

One additional approach to minimising the errors associated with the individual differencing \corr{techniques} is to account for the effects of the differential rotation of the Sun. This can be done by de-rotating all images to the same pre-event time before applying the differencing techniques. While the timescale of a CBF pulse is quite short, the de-rotation allows the errors associated with the small-scale coronal features to be minimised. This has the effect of reducing the additional error at the pulse edges due to small--scale coronal movements and solar rotation, allowing a better identification of the pulse.

\subsection{Automated Pulse Identification Techniques}
\label{subsect:cor_pita}

\begin{figure}[!t]
\begin{center}
\includegraphics[clip=,trim=0mm 0mm 0mm 47mm,width = 0.99\textwidth]{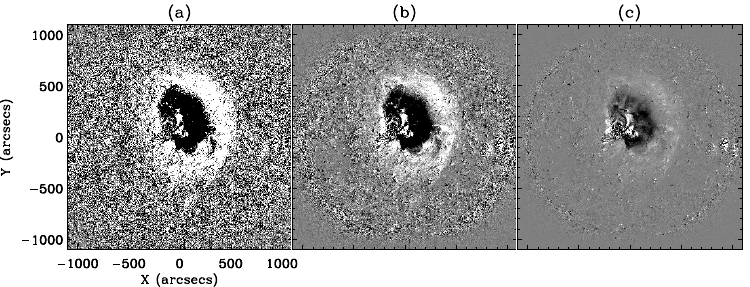}
\caption{Running difference images with variable photon flux intensity scaling \corr{(with units of detected photon difference)} of (a) $\pm$1, (b) $\pm$10 and (c) $\pm$50.}
\label{fig:pulse_scaling}
\end{center}
\end{figure}

Once identification of the pulse has been confirmed in a difference image, the next step in the pulse analysis is to determine the distance of the pulse from a defined source point at a given time. The traditional approach for this has been to use a point--and--click identification of the edge of the pulse using a RD or BD image with a user--defined scaling. However, as shown in Figure~\ref{fig:pulse_scaling}, the scaling used for the image defines the position and width of the pulse.

It is apparent from Figure~\ref{fig:pulse_scaling} that the pulse is most apparent when a harsh (i.e., $\pm1$) scaling is applied to the image. However, while this highlights the full extent of the pulse, it also results in an increase in the small--scale variation of the background corona. An additional problem with point--and--click pulse identification is defining the location on the pulse to be identified. This is generally the front edge of the pulse. However, it is clear from Figure~\ref{fig:pulse_scaling} that this varies with the scaling chosen. Consequently, point--and--click pulse identification is highly user--dependent, and not an efficient technique for a detailed analysis of CBFs. 

A more effective approach is to take an intensity profile across the pulse and identify the location of the peak intensity of the pulse, effectively allowing the centre of the pulse to be tracked rather than the edge. Whereas the edge of the pulse is loosely defined and highly variable, CBFs have been noted as having a roughly Gaussian shape \citep{Wills-Davey:1999ve}, meaning that the pulse centre is a more statistically significant indicator of the pulse position.

A technique was devised to allow semi--autonomous identification of the CBF pulse using derotated percentage base difference images. This approach was chosen as the \corr{nominal} inherent scaling of the technique used to produce them mitigated against any user--defined assumptions, while the derotation allowed the small--scale coronal variation to be minimised. To ensure that the signal--to--noise of the pulse was maximised, a great-circle sector (i.e., an area on the sphere bounded by two great circles) projected onto the Sun was defined with the source point of the pulse defined as the crossing point of the great circles. The intensity within this sector could then be averaged across the arc (in annuli of increasing radii with 1~degree width on the surface of the sphere) to produce an intensity profile as a function of \corr{great circle distance along the solar surface} away from the source location \citep[cf. similar techniques proposed by][]{Warmuth:2004ab,Podladchikova:2005ab,Wills-Davey:2006ab,Veronig:2010ab}.

To minimise the errors associated with identification of the source point, the \corr{first two observations of the CBF pulse in the 171 and 195~\AA\ passbands were visually located and their manually identified outline} fitted using an ellipse, with the mean centres of the four ellipses taken as the source point of the CBF pulse. The orientation of the arc sector was then chosen to maximise the number of pulse identifications for a given event, with an arc of 30~degrees chosen as this provided the greatest signal while minimising errors resulting from any change in the propagation direction of the pulse.

\begin{figure}[!t]
\centering{
              \includegraphics[width=0.99\textwidth,clip=,trim=0mm 5mm 0mm 70mm]{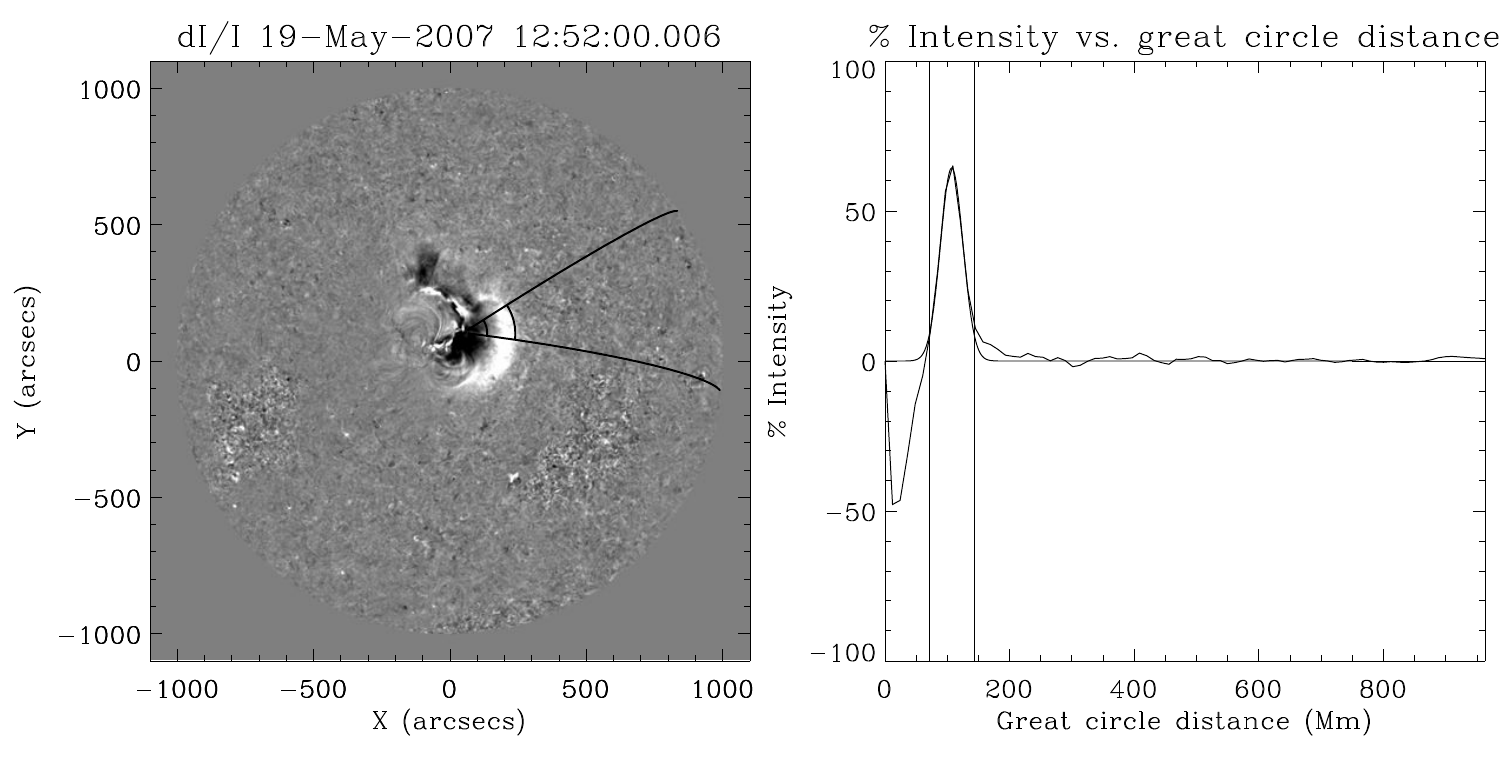}
              }
\caption{\emph{Left panel}: PBD image with arc sector over-plotted. \emph{Right panel}: Intensity profile across arc sector. The \corr{relative intensity change} at a given distance corresponds to the mean intensity \corr{enhancement} across the arc sector at that distance. This is then fitted using a Gaussian model and repeated for the next image. The vertical lines in the right-hand plot and arc lines in the left-hand plot correspond to the $2\sigma$ limits of the Gaussian fit.}
\label{fig:algorithm}
\end{figure}

Figure~\ref{fig:algorithm} shows how the algorithm works for the event from 2007~May~19 as observed by \emph{STEREO}-A. The left--hand panel shows the PBD image with the arc sector overplotted, while the right--hand panel shows the 1--dimensional intensity profile along the arc sector. The pulse is clearly visible in the arc sector, and has been fitted using a Gaussian model allowing the $2\sigma$ limits of the fit to be overplotted as arc--lines in the left--hand panel and vertical lines in the right--hand panel.

This process is repeated for each observation of the pulse, allowing the variation with time of the distance of the pulse from the source to be examined. This can then be used to determine the kinematics of the pulse, while the temporal variation in the pulse width can be studied for evidence of pulse broadening. The next step is to identify the optimal technique for deriving the pulse kinematics for a small data--set such as that typically produced by \emph{STEREO}/EUVI observations of a CBF.

\section{Statistical Analysis}
\label{sect:stats}

As well as minimising the errors associated with the actual identification of the CBF pulse, it was important to identify the optimum method for deriving the kinematics of the pulse. To do this, a number of different numerical differencing techniques and a residual resampling bootstrapping approach were tested and compared. This was required as a result of the sparse nature of the data obtained from \emph{STEREO}/EUVI in addition to concerns with regard to the approach previously used to derive the pulse kinematics.

\begin{figure}[!hp]
\begin{center}
\includegraphics[clip=,trim=0mm 5mm 5mm 0mm,width = 0.78\textwidth]{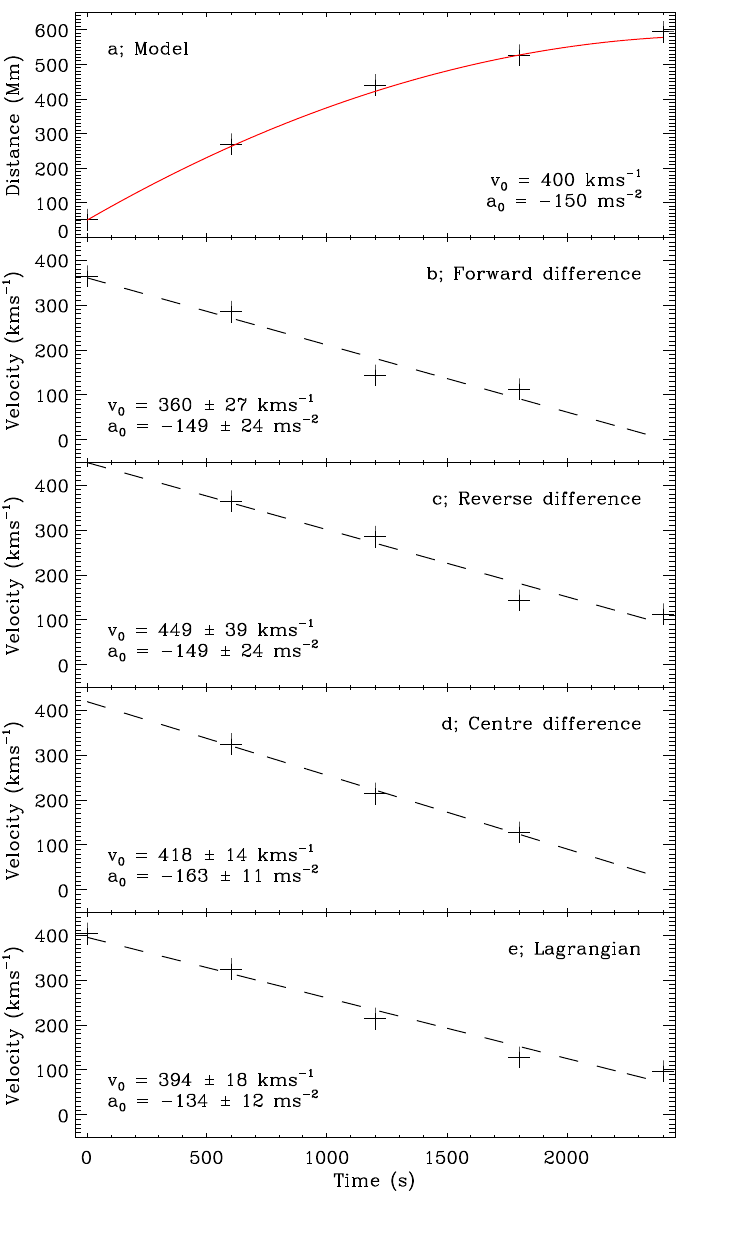}
\caption{\emph{Panel} (\emph{a}); Simulated data set (asterisks) with the model kinematics shown by the red line and given in the bottom right. \emph{Panels} (\emph{b})--(\emph{e}); Velocity derived using forward--difference (b), reverse--difference (c), centre--difference (d) and Lagrangian \corr{interpolation} (e) techniques. The kinematics determined by the fitted dashed line are shown in the bottom left in each case.}
\label{fig:num_diff}
\end{center}
\end{figure}

\subsection{Numerical Differencing}
\label{subsect:num_diff}

The simplest way to determine the pulse kinematics given distance--time measurements is to use a numerical differencing technique to directly derive the velocity and acceleration of the pulse. Here, four different numerical differencing techniques were identified; a forward, reverse and centre difference method, all of which can be derived using a Taylor expansion, and a three-point Lagrangian interpolation technique. To allow a direct comparison between the different techniques, a simulated data--set was created (shown in panel a of Figure~\ref{fig:num_diff}), with the numerical differencing techniques used to estimate the known kinematics.

This simulated data--set was constructed using an equation of the form,
\begin{equation}
r(t) = r_0 + v_0 t + \frac{1}{2}a t^2 + \delta r
\end{equation}
with $r_0 = 50$~Mm, $v_0 = 400$~kms$^{-1}$ and $a = -150$~ms$^{-2}$. The $\delta r$ term here corresponds to $\sim$5\% noise which was added to the simulated data to reproduce observational variations. \corr{This results in a relative error that increases with propagation distance and was chosen as it is representative of the increased difficulty in identifying an ``EIT Wave'' pulse at increasing distance from the source. For typical observations of an ``EIT Wave'', the generally high pulse intensities close to the source produce small positional uncertainty that increases with distance as the pulse intensity and spatial extent of the pulse decrease and increase respectively.} An observing cadence of 600~s was also used as this is the typical observing cadence of \emph{STEREO}. The resulting simulated data--set is shown as the \corr{crosses} in panel (a) of Figure~\ref{fig:num_diff}, with the known kinematics shown as the red line. This simulated data set was then used to compare the different numerical differencing techniques.

\subsubsection{Forward Differencing}
\label{subsubsect:f_diff}

The first numerical differencing technique to be tested was the forward-difference technique. This involves the computation of the derivative at the point $r[t + \Delta t]$ by extrapolating forward from the point $r[t]$. The derivative can be determined using the Taylor series,
\begin{equation}
r[t + \Delta t] = r[t] + r[t]'\Delta t +  \frac{r[t]''}{2!}(\Delta t)^{2} + \frac{r[t]'''}{3!}(\Delta t)^{3}  + ... \label{eqn:f_diff_0}
\end{equation}
By rearranging this equation, it is possible to find the derivative at time t, 
\begin{equation}
r[t]' = \frac{r[t + \Delta t] - r[t]}{\Delta t} -  \frac{r[t]''}{2!}(\Delta t) - \frac{r[t]'''}{3!}(\Delta t)^{2}  + ... \label{eqn:f_diff_1}
\end{equation}
Although this series continues to infinity, the correction values are dominated by the term of order $\Delta t$. As a result, this is usually written as,
\begin{equation}
r[t]' = v[t] = \frac{r[t + \Delta t] - r[t]}{\Delta t} + O(\Delta t) \label{eqn:f_diff_2}
\end{equation}
where $O(\Delta t)$ is the truncation error term. This can also be used to derive the acceleration in a similar fashion by replacing $r$ with $v$ to give,
\begin{equation}
v[t]' = \frac{v[t + \Delta t] - v[t]}{\Delta t} + O(\Delta t). \label{eqn:f_diff_3}
\end{equation}
However, $v[t]$ is already known from Equation~\ref{eqn:f_diff_2} above, so by substituting this into Equation~\ref{eqn:f_diff_3}, it is possible to derive the acceleration $a[t]$ as,
\begin{equation}
a[t] = \frac{r[t + 2\Delta t] - 2r[t+\Delta t] + r[t]}{\Delta t^{\text{\corr{2}}}} + O(\Delta t) . \label{eqn:f_diff_4}
\end{equation}
Equations~\ref{eqn:f_diff_2} and \ref{eqn:f_diff_4} can then be used to determine the instantaneous velocity and acceleration at the point $r[t]$ for each value of $t$. 

This approach does have some issues which make it unsuitable for analysing CBF observations. The technique calculates the derivative assuming a straight line gradient between points, while the final data point is \corr{removed} with each iteration as an inherent consequence of the method. This approach is also particularly susceptible to point--to--point variations, with the result that noisy data can produce strongly varying velocity and acceleration estimates.

The results of the forward--difference technique are shown in panel (b) of Figure~\ref{fig:num_diff}. The effects of the point--to--point variation are very apparent, while the inherent nature of the technique has removed the final data point. The kinematics obtained by fitting the data points also show a strong underestimation of $v_0$, while there are large errors associated with the estimates of both $v_0$ and $a$.

\subsubsection{Reverse Differencing}
\label{subsubsect:r_diff}

The reverse differencing approach uses a similar methodology, although in this case the derivative at the point \corr{$r[t]$ is calculated by extrapolating backwards to the point $r[t-\Delta t]$}. This can be determined using the Taylor series expansion,
\begin{equation}
r[t - \Delta t] = r[t] - r[t]'\Delta t +  \frac{r[t]''}{2!}(\Delta t)^{2} - \frac{r[t]'''}{3!}(\Delta t)^{3}  + ... \label{eqn:r_diff_0}
\end{equation}
Once again, this equation can be rearranged to find the derivative at time $t$,
\begin{equation}
r[t]' = \frac{r[t] - r[t - \Delta t]}{\Delta t} +  \frac{r[t]''}{2!}(\Delta t) - \frac{r[t]'''}{3!}(\Delta t)^{2}  + ... \label{eqn:r_diff_1}
\end{equation}
The correction values here are again dominated by the term of order $\Delta t$, allowing this equation to be rewritten as,
\begin{equation}
r[t]' = v[t] = \frac{r[t] - r[t - \Delta t]}{\Delta t} + O(\Delta t) \label{eqn:r_diff_2}
\end{equation}
where $O(\Delta t)$ is the truncation error term. As before, the acceleration term can also be derived by replacing $r$ with $v$ to give,
\begin{equation}
v[t]' = \frac{v[t] - v[t - \Delta t]}{\Delta t} + O(\Delta t). \label{eqn:r_diff_3}
\end{equation}
Again, $v[t]$ is already known from Equation~\ref{eqn:r_diff_2} above, so by substituting this into Equation~\ref{eqn:r_diff_3}, the acceleration $a[t]$ can be derived in terms of $r[t]$ as,
\begin{equation}
a[t] = \frac{r[t] - 2r[t-\Delta t] + r[t - 2\Delta t]}{\Delta t^{\text{\corr{2}}}} + O(\Delta t) . \label{eqn:r_diff_4}
\end{equation}
Both equation~\ref{eqn:r_diff_2} and \ref{eqn:r_diff_4} can then be used to determine the instantaneous velocity and acceleration at the point $r[t]$ for each value of $t$. 

Panel (b) in Figure~\ref{fig:num_diff} shows the result of this analysis for the simulated data--set. The reverse--difference approach is very similar to the forward--difference technique, and this is reflected in the derived data--points which have identical velocity values albeit at a different time. Once again, the point--to--point variation is significant, while the first data--point has also been removed. The fitted kinematics show a higher estimated initial velocity but identical acceleration, with the initial velocity not corresponding to that of the original model.

The same caveats that apply to the forward difference technique also apply to the reverse difference technique due to their inherent similarities in using two adjacent points ($r[t]$ \& $r[t+\Delta t]$ and $r[t]$ \& $r[t-\Delta t]$ for the forward and reverse difference respectively) to determine the derivative at a given point. Both techniques are quite simplistic and as a result are only applicable in simplistic cases. Both techniques also remove points from either edge with each iteration as an inherent consequence of the technique. As a result, neither technique is appropriate for determining the kinematics of a CBF pulse.

\subsubsection{Centre Differencing}
\label{subsubsect:c_diff}

A more accurate approach that utilises the Taylor series methodology is the centre difference technique. This uses the points either side of the point under examination $r[t]$ (i.e.,\ $r[t - \Delta t]$ and $r[t + \Delta t]$), smoothing the data and producing a more accurate estimate of the numerical derivative at that point. The Taylor series at the points $r[t - \Delta t]$ and $r[t + \Delta t]$ respectively are defined as,
\begin{align}
r[t - \Delta t] = r[t] - r[t]'\Delta t +  \frac{r[t]''}{2!}(\Delta t)^{2} - \frac{r[t]'''}{3!}(\Delta t)^{3}  + ... \label{eqn:c_diff_0_0} \\
r[t + \Delta t] = r[t] + r[t]'\Delta t +  \frac{r[t]''}{2!}(\Delta t)^{2} + \frac{r[t]'''}{3!}(\Delta t)^{3}  + ... \label{eqn:c_diff_0_1}
\end{align}
as shown previously in Equations~\ref{eqn:f_diff_0} and \ref{eqn:r_diff_0} above. The centre difference approach subtracts Equation~\ref{eqn:c_diff_0_0} from Equation~\ref{eqn:c_diff_0_1} to produce,
\begin{equation}
r[t + \Delta t] - r[t - \Delta t] = 2 r[t]'\Delta t +  2 \frac{r[t]'''}{3!}(\Delta t)^{3}  + ... ,
\end{equation}
which can then be rearranged in terms of $r[t]'$ to give,
\begin{equation}
r[t]' = \frac{r[t + \Delta t]- r[t - \Delta t]}{2\Delta t} +  \frac{r[t]'''}{3!}(\Delta t)^{2}  + ... 
\end{equation}
Following the notation from before, this can be rewritten as,
\begin{equation}
r[t]' = v[t] = \frac{r[t + \Delta t] - r[t - \Delta t]}{2\Delta t} + O(\Delta t^{2})
\end{equation}
where $O(\Delta t^{2})$ is the truncation error term. The instantaneous acceleration term $a[t]$ can then be determined as,
\begin{equation}
r[t]'' = a[t] = \frac{r[t + \Delta t] - 2r[t] + r[t - \Delta t]}{(\Delta t)^2} + O(\Delta t^{2})
\end{equation}
This allows the instantaneous velocity and acceleration at the point $r[t]$ to be determined. 

The results of applying this technique to the simulated data--set are shown in Panel (d) of Figure~\ref{fig:num_diff}. The centre--difference technique produces a smoother estimation of the kinematics than the forward-- and reverse--difference and the fitted kinematics are closer to the model values. However, the centre difference technique does remove both of the edge points as an inherent consequence of its operation. 

Despite producing a better estimate of the pulse kinematics, the inherent issues of the centre--difference technique are a cause for concern. The removal of edge points makes analysis of the kinematics difficult, particularly when considering that a given \emph{STEREO}/EUVI observation of a CBF event may have only 3 or 4 distance--time estimates. Consequently, the centre difference technique is usually unsuitable for this analysis.

\subsubsection{Lagrangian Differencing}
\label{subsubsect:l_diff}

A more advanced technique for determining the derivative at a point numerically is the three-point Lagrangian interpolation technique used by the built-in \emph{deriv.pro} routine in IDL \corr{\citep[see][for details]{Bevington:2003vn}}. This method uses three adjacent points ($r[t-\Delta t]$, $r[t]$ and $r[t+ \Delta t]$) to fit a Lagrangian polynomial function of the form,
\begin{align}
P(x) &= \frac{(x - t)(x - (t+\Delta t))}{((t-\Delta t) - t)((t-\Delta t) - (t+\Delta t))}r[t-\Delta t] \nonumber \\
&+ \frac{(x - (t-\Delta t))(x - (t+\Delta t))}{(t - (t-\Delta t))(t - (t+\Delta t))}r[t] \nonumber \\ 
&+ \frac{(x - (t-\Delta t))(x - t)}{((t+\Delta t) - (t-\Delta t))((t+\Delta t) - t))}r[t+\Delta t]
\end{align}
where $x$ is the abscissa of the point $r[t]$. The derivative of this is then given by,
\begin{align}
P'(x) &= \frac{2x - t - (t+\Delta t)}{((t-\Delta t) - t)((t-\Delta t) - (t+\Delta t))}r[t-\Delta t] \nonumber \\
&+ \frac{2x - (t-\Delta t) - (t+\Delta t))}{(t - (t-\Delta t))(t - (t+\Delta t))}r[t] \nonumber \\ 
&+ \frac{2x - (t-\Delta t) - t)}{((t+\Delta t) - (t-\Delta t))((t+\Delta t) - t))}r[t+\Delta t]
\end{align}
By substituting $t$ for $x$, the polynomial used to determine the derivative $P'(t)$ is given as,
\begin{equation}
P'(t) = \frac{1}{2\Delta t}(r[t+\Delta t] - r[t-\Delta t])
\end{equation}

This approach does not assume a straight line gradient between points (unlike the forward, reverse and centre difference techniques above), while also compensating for edge points by increasing the errors and retaining all of the original data points. The Lagrangian technique also smoothes the data, removing the spiky appearance produced by the Taylor series expansion techniques. This is apparent in panel (e) of Figure~\ref{fig:num_diff}, where the Lagrangian technique has retained all data--points.

Despite this, there are a number of issues associated with this approach, in particular as a result of the small data sets associated with these disturbances. The cadence of \emph{STEREO}/EUVI means that there are typically only 4 or 5 observations of a CBF pulse per event, of which three are required to determine the numerical derivative of one data point. The combination of this with the interpolation at the edges has the effect of skewing the edge points, giving the impression of kinematic variation that may not be real. This can be seen in the fit to the derived Lagrangian kinematics in panel (e) of Figure~\ref{fig:num_diff}, where the edges points have strongly influenced an underestimation of the kinematics. 

This skewing effect strongly suggests that the Lagrangian interpolation technique is inappropriate for the small data-sets typically obtained for CBF analysis. The problems associated with each numerical differencing approach studied here indicates that an alternative method must be used to determine the kinematics of a CBF pulse. A possible alternative is the fitting of a specified model to the distance-time measurements, although once again the small data-sets available can make this difficult. 

\subsection{Effects of Image Cadence and Positional Uncertainty}
\label{subsect:img_cad}

\begin{figure}[!t]
\begin{center}
\includegraphics[clip=,trim=0mm 5mm 0mm 0mm,width = 0.95\textwidth]{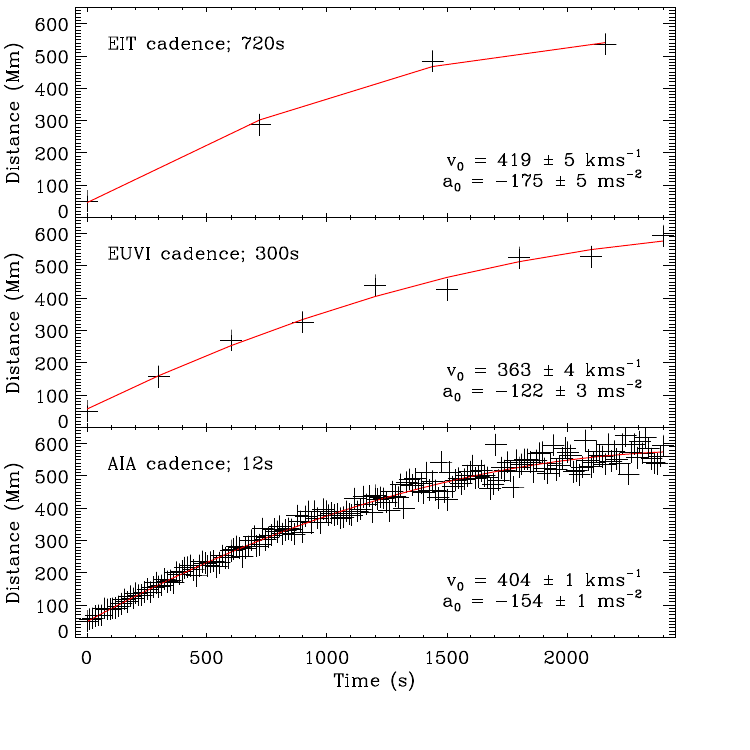}
\caption{Simulated data (crosses\corr{; derived using the model shown in the top panel of Figure~\ref{fig:num_diff}}) for cadences of 720~s (\emph{top}; comparable to EIT), 300~s  (\emph{middle}; comparable to EUVI) and 12~s  (\emph{bottom}; comparable to AIA). In each case, the fit to the data is shown in red with derived kinematics in the bottom right of each panel.}
\label{fig:num_diff_cad}
\end{center}
\end{figure}

\begin{figure}[!t]
\begin{center}
\includegraphics[clip=,trim=0mm 5mm 0mm 0mm,width = 0.95\textwidth]{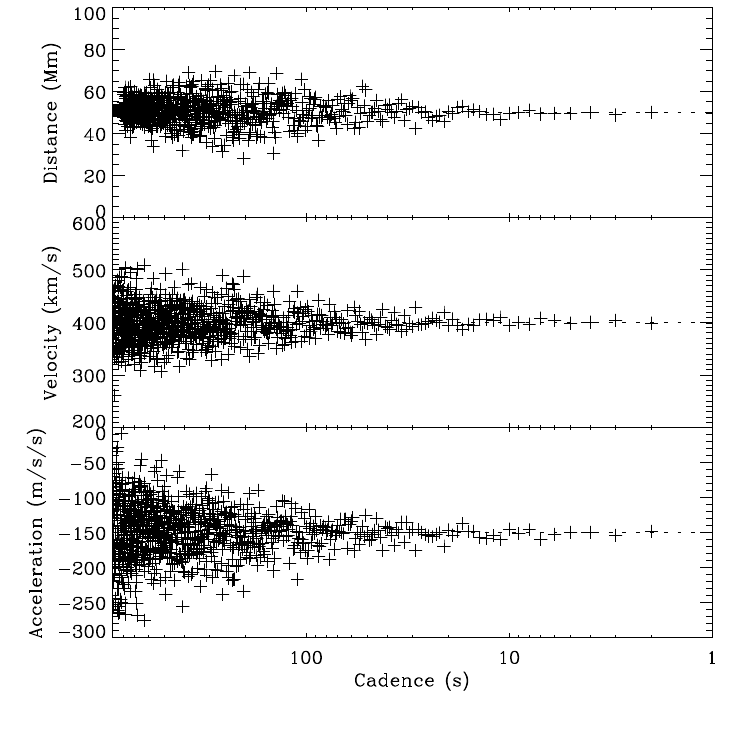}
\caption{Derived kinematics for varying image cadence with $\pm$5\% uncertainty. Distance, velocity and acceleration are shown in the top, middle and bottom panels respectively, with the x--axis shown using logarithmic scaling to highlight the effects of varying the image cadence.}
\label{fig:num_diff_vary_cad}
\end{center}
\end{figure}

In the absence of a numerical technique to determine the kinematics of a CBF pulse, the cadence of the observing instrument and the degree of uncertainty in the data must also be considered. Here, the influence of both cadence and uncertainty are examined using the simulated data discussed in Section~\ref{subsect:num_diff}. The effects of varying image cadence were the first to be examined; this is shown in Figures~\ref{fig:num_diff_cad} and \ref{fig:num_diff_vary_cad}. 

The same data--set was used for each of the panels shown in Figure~\ref{fig:num_diff_cad}, with the cadence varied to best reflect data from \emph{SOHO}/EIT (top panel at 720~s), \emph{STEREO}/EUVI (middle panel at 300~s) and \emph{SDO}/AIA (bottom panel at 12~s). The derived kinematics are given in the bottom right of each panel with the fit to the data shown by the red line. It is immediately apparent that the higher cadence data allows a better fit to the data despite the uncertainty in the data (which was kept at $\pm$5\% of the model for each data--set). This implies that higher cadence data is required to derive the true kinematics of a CBF pulse.

The variation in derived kinematics with cadence is shown in Figure~\ref{fig:num_diff_vary_cad} (again for $\pm$5\% uncertainty). As the cadence decreases, the derived velocity and acceleration approach the model values, with the scatter showing a dramatic reduction below $\sim$50~s cadence. These results are consistent with the observations made by both \citet{Long:2008eu} and \citet{Ma:2009ab} and show that the effects of image cadence must be accounted for when trying to derive the true kinematics of a CBF pulse.

\begin{figure}[!t]
\begin{center}
\includegraphics[clip=,trim=0mm 5mm 0mm 0mm,width = 0.95\textwidth]{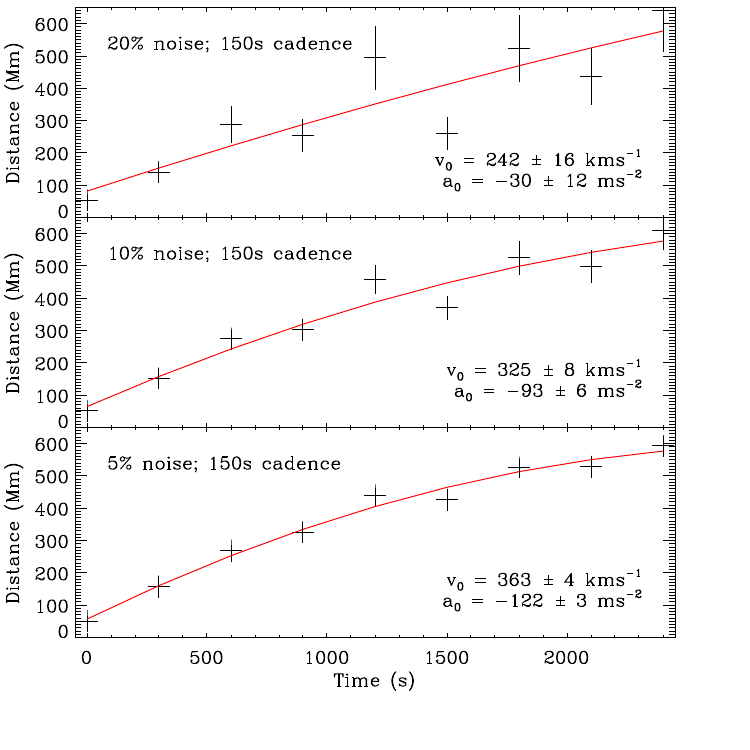}
\caption{Simulated data (crosses) for noise distributions with widths of $\pm$20\% (\emph{top}), $\pm$10\%  (\emph{middle}) and $\pm$5\%  (\emph{bottom}) of the data value. In each case, the fit to the data is shown in red with derived kinematics in the bottom right of each panel.}
\label{fig:num_diff_noise}
\end{center}
\end{figure}

\begin{figure}[!t]
\begin{center}
\includegraphics[clip=,trim=0mm 5mm 0mm 0mm,width = 0.95\textwidth]{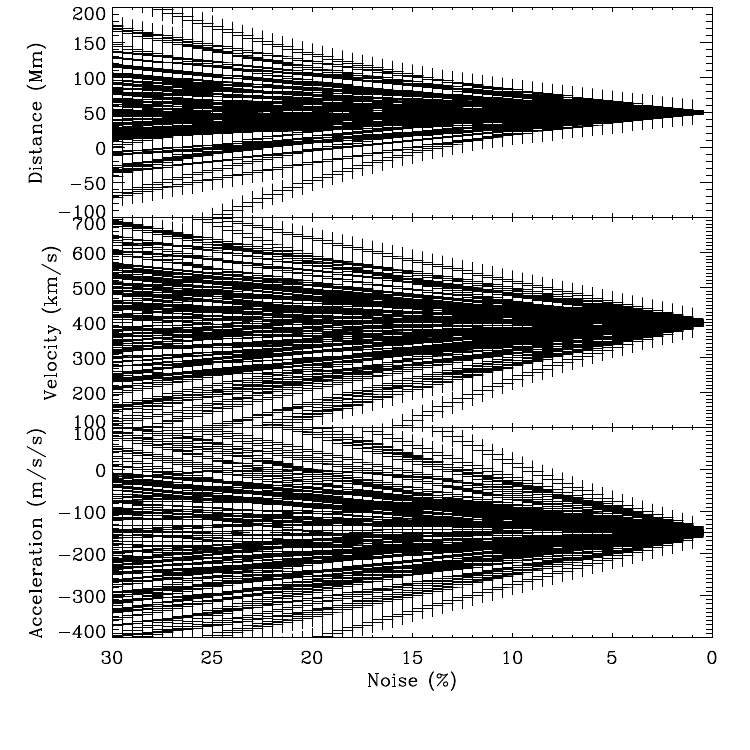}
\caption{Derived kinematics for varying uncertainty at 150~s image cadence. Distance, velocity and acceleration are shown in the top, middle and bottom panels respectively.}
\label{fig:num_diff_vary_noise}
\end{center}
\end{figure}

The effects of varying uncertainty in the data were also examined for constant image cadence with the results of this analysis shown in Figures~\ref{fig:num_diff_noise} and \ref{fig:num_diff_vary_noise}. Figure~\ref{fig:num_diff_noise} shows the derived kinematics for the simulated data--set with the top panel showing uncertainties of $\pm$20\%, the middle panel showing uncertainties of $\pm$10\% and the bottom panel showing uncertainties of $\pm$5\%. The variation in uncertainty has a noticeable effect on the derived kinematics in each case, with the acceleration in particular exhibiting significant variation with uncertainty. 

The variation in the derived kinematics with uncertainty is shown in Figure~\ref{fig:num_diff_vary_noise}, with the variation in offset distance (r$_{0}$), initial velocity (v$_{0}$) and acceleration (a$_{0}$) shown in the top, middle and bottom panels respectively. Some variation is apparent in each case, with the acceleration again showing the strongest variation. The derived kinematics are seen to approach the model kinematics as the noise reduces to 0\% as expected. 

It is clear that the variation in both noise and imaging cadence can strongly influence the derived kinematics of a CBF pulse, while the different numerical differencing techniques do not return accurate estimates of the pulse kinematics. However, it is possible to use a bootstrapping approach to overcome these issues as this is a statistically significant technique that has been optimised for small data-sets such as those typically obtained when studying CBF pulses.

\subsection{Bootstrapping}
\label{subsect:bootstrap}

\begin{figure}[!t]
\begin{center}
\includegraphics[clip=,trim=0mm 5mm 0mm 0mm,width = 0.95\textwidth]{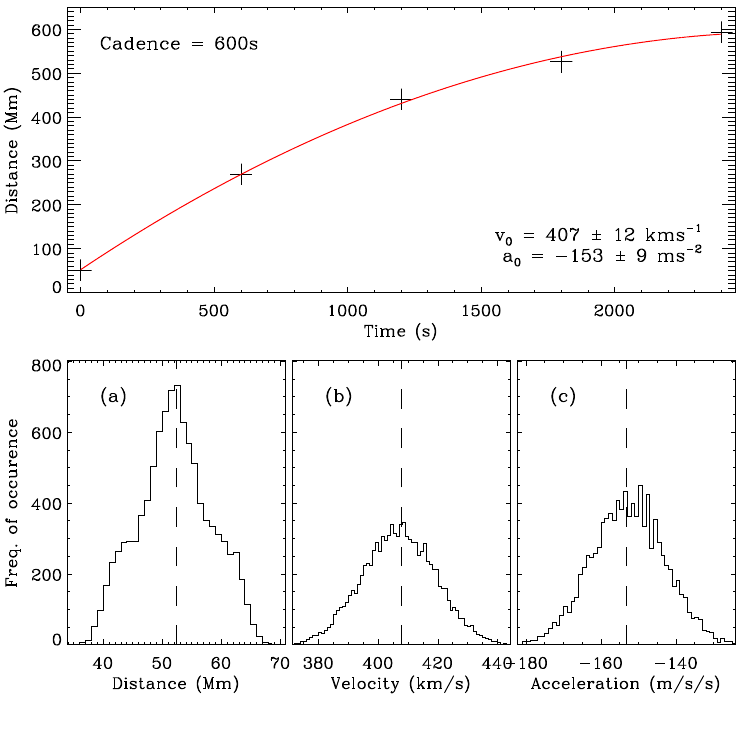}
\caption{\emph{Top}: Simulated data \corr{corresponding to $\sim$5\% noise added to the model} (asterisks) with bootstrapped fit shown in red and parameters in bottom right. \emph{Bottom}: Histograms showing the distance (a), velocity (b) and acceleration (c) derived using the bootstrapping technique.}
\label{fig:bootstrap}
\end{center}
\end{figure}

Bootstrapping was first proposed by \citet{Efron:1979p1831} as a general interpretation of the jackknife, allowing statistical quantities to be estimated from a limited sample to a high degree of accuracy. In principle this allows a statistical quantity (such as the mean, variance etc.) of a distribution to be estimated using a small sample of measurements from that distribution. The resulting estimates are statistically significant, with statistically defined associated error estimates. 

While there are \corr{several} different bootstrapping techniques, each of which is useful for a different purpose, in this case \corr{the residual resampling technique will be used}. Whereas the majority of bootstrapping techniques involve removing one point at random and recalculating the fit to the data, this is inappropriate in this case due to the small sample set available. The residual resampling technique however, does not require data-point removal, making it more appropriate for small data-sets such as those available here.

The residual resampling technique works as follows. The distance-time measurements ($y_i$; $i=1, 2, \ldots, n$) are first fitted using a quadratic equation of the form
\begin{equation}
r(t) = r_0 + v_0 t + \frac{1}{2}a t^2,
\end{equation}
where $r_0$ is the offset distance, $v_0$ is the initial velocity, $a$ is the constant acceleration of the pulse, and $t$ is the time elapsed since the first observation of the disturbance. This yields the fitted values ($\hat{y_{i}}$) and residuals ($\hat{\epsilon}=y_{i} - \hat{y_{i}}$) for each point. The residuals are then randomly ordered and randomly assigned a sign ($1$ or $-1$), before being added to the original fit values to produce a new data set. These new data are then fitted using the same model, with the resulting re-fitted parameters recorded. The process is then repeated a large number of times ($\sim$10\,000).

This has the effect of repeating the fit to slightly varying data a large number of times, allowing a distribution of fit values to be obtained. The mean and standard deviation of this distribution then provide the estimated value and associated error of the parameter. This technique is statistically rigorous and produces a more accurate result than a simple model fit to the given data. 

This technique was applied to the simulated data--set used in Section~\ref{subsect:num_diff} \corr{(corresponding to the model with $\sim$5\% noise added)} to allow a comparison of the effectiveness of the bootstrapping approach with the numerical differencing techniques. The results of this analysis are shown in Figure~\ref{fig:bootstrap}. The upper panel here shows the simulated data (indicated by the \corr{crosses}) with the bootstrapped fit to the data shown by the red line. The kinematics obtained by the bootstrapping approach are given in the bottom right. The bottom panels (a), (b) and (c) show the histograms for the distance, velocity and acceleration parameters respectively. These histograms returned by the bootstrapping approach for each parameter allow a statistical determination of the fit parameters, a fact reflected in the error term for the derived kinematics. 

The kinematics returned by the bootstrapping approach match the model kinematics within one standard deviation, while also producing an accurate estimate with well--defined errors, unlike those of the numerical differencing techniques. It should also be noted that none of the data--points have been removed, unlike the majority of the numerical differencing techniques tested. 

As a result of the simulations carried out to test the different approaches to determining the kinematics of a CBF pulse, the residual resampling bootstrapping technique was chosen as the best method. This fits a model directly to the distance--time measurements, allowing the kinematics of the CBF pulse to be determined to a high degree of accuracy. \corr{This is a statistically significant approach that removes much of the error inherent in the Lagrangian interpolation technique used in the previous analysis discussed in Chapter~\ref{chap:first_obs}. The kinematics for} each future event studied here therefore reflects the mean value of the bootstrapped distribution, with the associated error given by the standard deviation.

\corr{Note that the bootstrapping technique outlined here requires the application of a pre--defined model to the data, and as such does not solve the inherent problems associated with numerical differencing. However, bootstrapping does provide a statistically significant approach to determining the errors associated with the pulse kinematics without introducing any additional errors into the data. This makes it a very useful technique for determining the kinematics of CBF pulses, particularly considering the small data--sets available from \emph{STEREO}/EUVI.}

\section{Conclusions}
\label{sect:methods_conc}

As outlined in this chapter, there are many different issues beyond the simple identification and analysis of the pulse that must be examined and accounted for when studying CBFs. The choice of technique used to identify the pulse can have a strong effect on the estimated pulse parameters, while the resulting kinematics are influenced by the technique used to estimate them. The cadence of the observations and the uncertainty in the pulse position (which is influenced by the resolution of the instrument) can also affect the derived kinematics.

Of the different image processing \corr{techniques} considered here, the percentage base difference images offer the optimum approach for identifying the CBF pulse. This technique removes the background corona, highlighting the actual pulse rather than an intensity change relative to the previous observation. The inherent scaling afforded by the percentage intensity approach also removes any ambiguity with regard to the identification of the pulse. 

The combination of PBD images with the semi--automated algorithm presented here allows the CBF pulse to be identified in an unbiased way, negating the user--defined errors associated with traditional methods of pulse identification. The algorithm also uses a consistent approach to identify the pulse, with the use of a Gaussian model ensuring that the pulse position and width are comparable from both image to image and event to event. This will allow a direct comparison of multiple events and represents a opportunity to systematically examine CBFs.

Once the CBF pulse has been identified and tracked, the next step involves determining the pulse kinematics. This has \corr{previously} been done using numerical differencing techniques, but it has been shown here that this approach is incompatible with the typical cadence of CBF observations. The different numerical differencing techniques based on a Taylor expansion approach remove data points as a direct consequence of their operation, while the Lagrangian interpolation approach is strongly affected by the edge points. 

The derived kinematics have also been shown to be strongly affected by the cadence of the observations and the associated uncertainty of the pulse position. By comparing derived kinematics for a simulated data--set using cadences comparable to those available from \emph{SOHO}/EIT, \emph{STEREO}/EUVI and \emph{SDO}/AIA it was shown that the derived kinematics approached those of the model as the cadence decreased. \corr{This is analogous to the aliasing effect observed in signal processing and suggests that historical cadences were insufficient to derive the true kinematics of CBF pulses. The uncertainty in pulse position also influences the derived kinematics, with a decrease in the positional uncertainty producing derived kinematics that approach those of the model. This indicates that it will be possible to derive the true kinematics of a clearly identified CBF pulse using high cadence data from \emph{SDO}/AIA.}

The issues surrounding the pulse identification and derivation of the resulting kinematics can be negated by combining the semi--automated algorithm presented here with a residual--resampling bootstrapping technique. This allows any errors associated with the identification of the CBF pulse to be minimised while also producing statistically significant estimates of the pulse kinematics with defined associated errors.  This approach is used to examine a series of CBF events observed using the \emph{STEREO} spacecraft in the next chapter.		
			

\chapter{Deceleration and Dispersion of Large-scale Coronal Bright Fronts} 
\label{chap:wave_properties}


\noindent 
\\ {\it 
A man can't prove anything without statistics
\begin{flushright}
Mark Twain \\
\end{flushright}
 }

\vspace{15mm}
One of the most dramatic manifestations of solar activity are large-scale coronal bright fronts (CBFs) observed in extreme ultraviolet (EUV) images of the solar atmosphere. To date, the energetics and kinematics of CBFs remain poorly understood, due to the low image cadence and sensitivity of previous EUV imagers and the limited methods used to extract the features. In this chapter, work performed to determine the temporal variation of the properties of a sample of CBFs is outlined, with particular emphasis placed on the variation in their kinematics and pulse shape, dispersion and dissipation. This analysis followed the development of a semi-automatic intensity profiling technique to extract the morphology and accurate positions of CBFs in 2.5--10 min cadence images from \emph{STEREO}/EUVI. By applying this technique to sequences of 171~\AA\ and 195~\AA\ images from \emph{STEREO}/EUVI, it was possible to measure the wave properties of four separate CBF events from 2007~May~19 \corr{(see Chapter~\ref{chap:first_obs} for the original analysis of this event)}, 2007~December~07, 2009~February~12 and 2009~February~13. 

Following their launch at initial velocities of $\sim$240--450~km~s$^{-1}$ each of the four events studied showed significant negative acceleration ranging from $\sim -$290 to $-$60~m~s$^{-2}$. The spatial and temporal widths of the CBF pulses were found to increase from $\sim$50~Mm to $\sim$200~Mm and $\sim$100~s to $\sim$1500~s respectively, indicating that they are dispersive in nature. The variation in position-angle averaged pulse-integrated intensity with propagation shows no clear trend across the four events studied. These results are most consistent with CBFs being dispersive magnetoacoustic waves. This chapter contains results published in \emph{Astronomy \& Astrophysics} by \citet*{Long:2011ab}.

\section{Introduction}
\label{section:wave_properties_intro}

As the \emph{STEREO} mission continued, the number of observed CBF events increased with 10 wave events observed during the period 2006~Dec~12 to 2008~Nov~4 \citep{Aschwanden:2009p1108}. The increasing number of events combined with the questionable abilities of the typically used difference techniques to identify the true position of the propagating pulse \citep[cf.][]{Attrill:2010ab} has necessitated the development of a more automated approach. This was particularly important given the approaching launch of the \emph{SDO} spacecraft, with its $\sim$12~s cadence in multiple passbands implying $\sim$40--50~observations per event rather than the $\sim$\corr{5--10}~observations per event available from \emph{STEREO}/EUVI. 

While the launch of \emph{SDO} has focused attempts to automate detection of CBFs, this is not a new issue, with \citet{Podladchikova:2005ab} originally proposing a technique to identify and track ``EIT waves'' based on their associated dimming regions. This technique forms the basis of the online Novel EIT wave Machine Observing (NEMO) catalogue\footnote{\url{http://sidc.oma.be/nemo/}} which provided real time characterisation of coronal dimming regions until 2010~August~1. Here, the ``EIT wave'' is identified using the coronal dimming region formed by evacuation of plasma as the associated CME erupts. 

This work was followed by \citet{Wills-Davey:2006ab}, who proposed a technique to track a CBF pulse in data from the \emph{TRACE} spacecraft. While this technique does provide a way to track a pulse using Huygens tracking, the technique does not appear to have been fully automated at this stage. This approach was also never applied to data from \emph{SOHO}/EIT, with only one event described from \emph{TRACE}. 

Despite these initial attempts to automatically track CBF pulses, not much work has been done with regard to applying these techniques to the higher cadence data available from \emph{STEREO}/EUVI. The previously proposed techniques also suffer from some shortcomings, with the algorithm proposed by \citet{Podladchikova:2005ab} relying on detecting the dimming region rather than the actual pulse. While this is preferable for space weather forecasting and detection, it is also prone to false--positive detections and does not allow the properties of the actual pulse to be examined.

\corr{To eliminate} the issues with a point--and--click approach to pulse identification and to overcome the errors associated with running difference images, a semi--automated algorithm was devised to analyse CBF pulses. This algorithm (described in more detail in Section~\ref{sect:identify_pulse}) uses de--rotated percentage base difference images for pulse identification, neutralising the errors associated with running difference images. It should be noted that this algorithm, while effective at identifying the pulse, is not yet fully automated and does require user input in identifying optimal events for analysis. Once these events have been identified however, the actual detection and tracking of the pulse is fully automated.

Once the issues regarding the identification of the CBF pulse and the derivation of the kinematics from distance--time measurements had been resolved (as previously discussed in Chapter~\ref{chap:methods}), a sample of four CBF events was chosen for analysis. These events were well observed by the \emph{STEREO} spacecraft, with two of them (from 2007~May~19 and 2007~December~07) observed \corr{on--disk} by both \emph{STEREO} spacecraft\corr{. However, the $\sim$90~degree separation of the \emph{STEREO} spacecraft meant that the 2009~February events were observed in profile by \emph{STEREO}--A and on--disk by \emph{STEREO}--B. As a result, only the data from \emph{STEREO}--B were used here for both of these events.}

The algorithm described here is designed to return the estimated position of the centroid of the pulse, the pulse width and the \corr{integrated} intensity of the pulse for each image studied. This allows the kinematics of the pulse to be determined using the bootstrapping approach outlined in Section~\ref{subsect:bootstrap}, while the temporal variation in pulse width can be examined for evidence of pulse broadening. The variation in the \corr{integrated} intensity of the pulse can be used to determine if the pulse is dissipative or non--dissipative. This is important for identifying the true physical nature of the pulse.

The observations used for this work are outlined in Section~\ref{sect:wave_properties_obs}, with the algorithm used to identify and track the pulse summarised in Section~\ref{sect:wave_properties_cor_pita} (although a more detailed description can be found in Section~\ref{subsect:cor_pita}). The results of this analysis are then presented in Section~\ref{sect:wave_properties_analysis}, with the 2007~May~19 event discussed in detail, allowing a comparison to be made with the analysis previously presented in Chapter~\ref{chap:first_obs}. The application of the algorithm to the additional events is outlined in Section~\ref{sect:wave_properties_results}, before some conclusions are drawn about the resulting implications for CBFs in Section~\ref{section:wave_properties_concs}.

\section{Observations}
\label{sect:wave_properties_obs}

The different data discussed here were obtained using the EUVI telescope onboard the \emph{STEREO} spacecraft. Although EUVI observes the Sun in four EUV passbands (304~\AA, 171~\AA, 195~\AA, and 284~\AA), the nature of the passbands meant that only the 171~\AA\ and 195~\AA\ passbands were used here. This is primarily due to the high temporal cadence of both passbands (1.5--2.5 minutes for 171~\AA\ and 5--10 minutes for 195~\AA) and also as CBFs are more readily observed (i.e.,\ of higher contrast) in these two passbands. While CBFs have been observed in the 284~\AA\ passband \citep{Zhukov:2004kh} and the 304~\AA\ passband \citep{Long:2008eu}, the nature of the data and the lower cadence make it difficult to use these passbands for more rigorous analysis. 

\begin{figure}[!t]
\centering{
               \includegraphics[width=0.8\textwidth,clip=,trim=0mm 0mm 0mm 5mm]{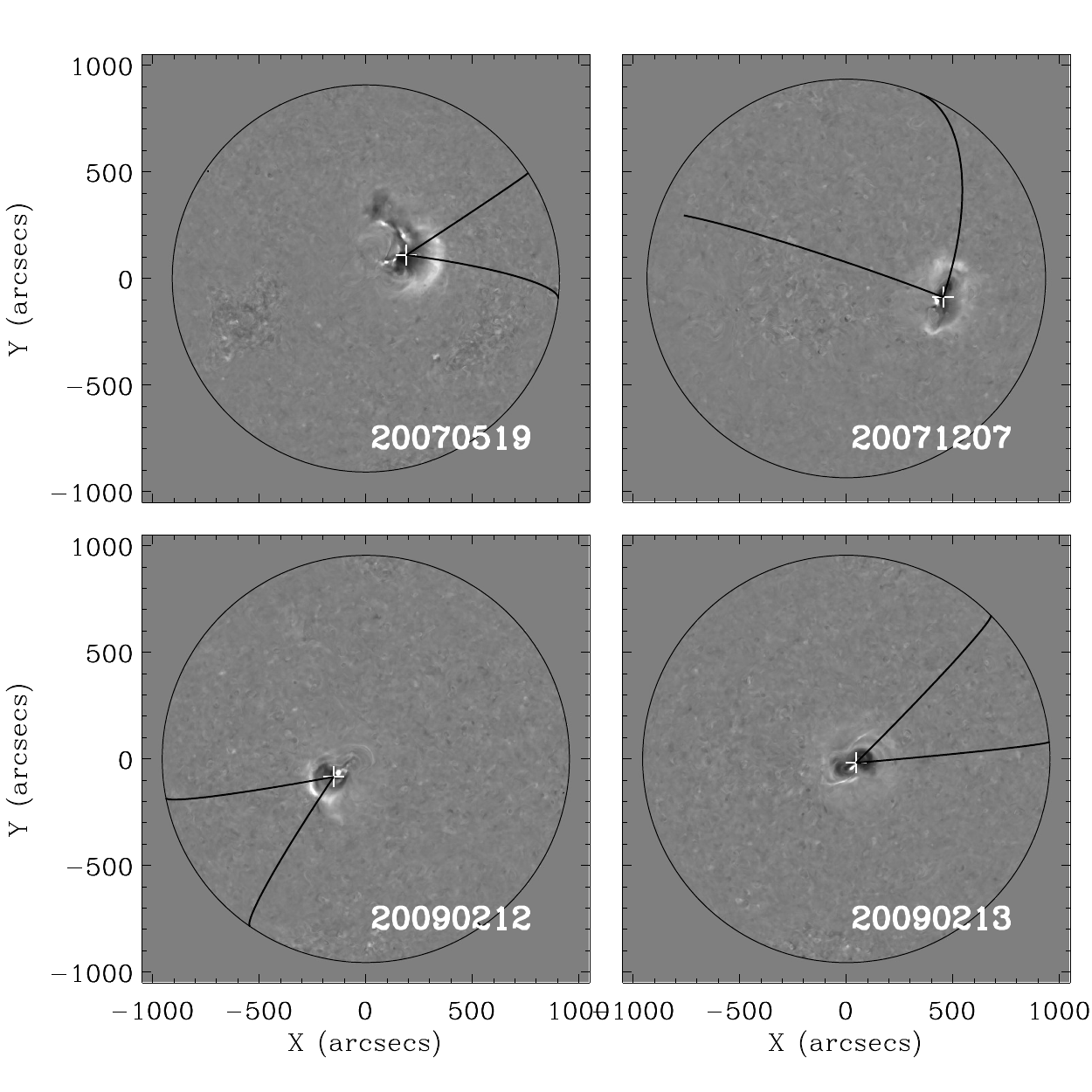}
               }
\caption{Percentage base difference images of the 2007~May~19 (top left), 2007~Dec~07 (top right), 2009~Feb~12 (bottom left) and 2009~Feb~13 (bottom right) CBFs, each in the 195~\AA\ passband as seen by \emph{STEREO}-B. The solid lines indicate the region in which the disturbance was identified using the intensity profile technique, while the cross marks the estimated origin of the event. White (black) is an increase (decrease) in intensity from the base image.}
\label{fig:wave_properties_image}
\end{figure}

Figure~\ref{fig:wave_properties_image} shows percentage base difference images of the four events used here, each taken using the 195~\AA\ passband onboard \emph{STEREO}--B. These events were chosen as the erupting active region is close to disk centre for each event, reducing the errors associated with measuring distances along the curvature of the Sun. The events from 2007~May~19 and 2007~December~07 occurred during the early phase of the \emph{STEREO} mission, with the result that the events could be observed close to disk centre by both \emph{STEREO} spacecraft. However \corr{as previously noted,} the events from 2009~February occurred later in the mission when the \emph{STEREO} spacecraft were \corr{near} quadrature. For these events, the erupting active region is at disk centre for the \emph{STEREO}--B spacecraft, but on the limb as observed by \emph{STEREO}--A\corr{, with the result that only data from \emph{STEREO}--B could be considered here for these events.}

Each of the events studied was associated with a CME eruption and a GOES X-ray flare, with each event also studied independently by other authors. The 2007~May~19 event (as noted in Chapter~\ref{chap:first_obs}) has been studied in detail by authors including \citet{Veronig:2008p216}, \citet{Gopalswamy:2009p1527} and \citet{Attrill:2010ab}, with the 2007~December~07 event analysed by \citet{Ma:2009ab}. The unique nature of the February~2009 quadrature events meant that they were also popular choices for further analysis, with \citet{Kienreich:2009ab} and \citet{Patsourakos:2009ab} in particular using the events to study the three--dimensional nature of the pulses. 

The clarity of the chosen events made them ideal candidates for examination using the detection algorithm, while the similar imaging cadence for each event minimised the effects of undersampling noted in Chapter~\ref{chap:first_obs}. The observing cadence of the 195~\AA\ passband was 600~s for each event, while the 171~\AA\ passband operated at a cadence of 150~s for each event with the exception of the 2009~February~13 event where it had a cadence of 300~s.

In each case, the CBFs were studied using de--rotated base-difference (BD) and percentage base-difference (PBD) images. The PBD images were used to identify the pulse and determine the variation in pulse position and width with time, while the BD images were used to determine the variation in integrated intensity with time. The peak intensity of the BD rather than the PBD intensity profile was used as this \corr{shows the actual intensity increase of the pulse above a constant background}, while the PBD intensity shows the \corr{ratio of the pulse intensity increase} with respect to the background\corr{, and consequently does not necessarily produce consistent results for different background values}. Both of these techniques also indicate the entirety of the CBF, rather than the portion of the pulse which has moved beyond its previous extent as is the case with running difference images.

\section{CBF Pulse Detection}
\label{sect:wave_properties_cor_pita}

The algorithm previously described in Section~\ref{subsect:cor_pita} was used to identify, track and analyse the observed CBF pulse for each event studied. In each case, the first two observations of the pulse in both the 171~\AA\ and 195~\AA\ passbands were fitted using ellipses, with the means of the four ellipse centres chosen as the projected source of the CBF disturbance. This point served as the crossing point of two great circles projected onto the Sun, with the sector between the great circles defined as the arc into which the CBF pulse propagated (see Figure~\ref{fig:algorithm}). The intensity of the PBD image was then averaged across the position angle (PA) within this arc sector for annuli of increasing radii with 1~degree width on the surface of the sphere. This was used to produce a 1--dimensional intensity profile as a function of distance away from the source location. This process was repeated using BD images to obtain the BD intensity profile.

The intensity profile produced by this technique was then fitted using a Gaussian profile \citep[as CBFs were noted by][to have a Gaussian form]{Wills-Davey:2006ab}. This allows the properties of the pulse to be identified, with the Gaussian fit providing the position of the pulse centroid, the width of the pulse and also the peak pulse intensity at a given time. The fit also identifies an associated error for each of these parameters, allowing the quality of each parameter to be examined.

\begin{figure}[!p]
\centering{
               \includegraphics[width=1\textwidth,clip=,trim=5mm 5mm 0mm 0mm]{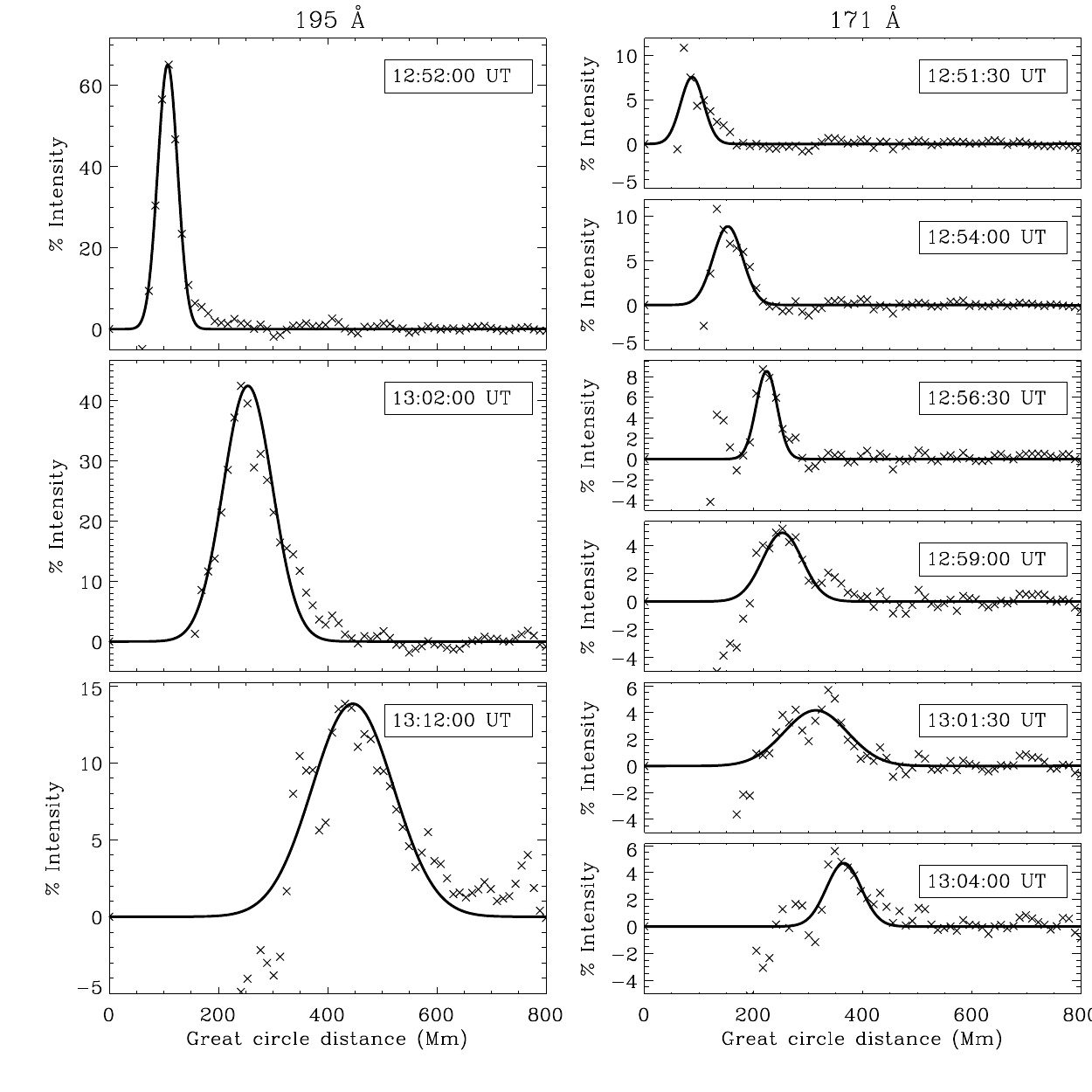}
               }
\caption{PBD image intensity profiles (crosses) for the 2007~May~19 event as obtained from \emph{STEREO}-A in the 195~\AA\ (left) and 171~\AA\ (right) passbands. The Gaussian fit (solid curve) to the positive section of each profile has been overplotted on each panel. The time of the leading image ($I_t$) in each case is on the upper right of the panel.}
\label{fig:profile_20070519_A}
\end{figure}

\begin{figure}[!p]
\centering{
              \includegraphics[width=1\textwidth,clip=,trim=0mm 5mm 0mm 0mm]{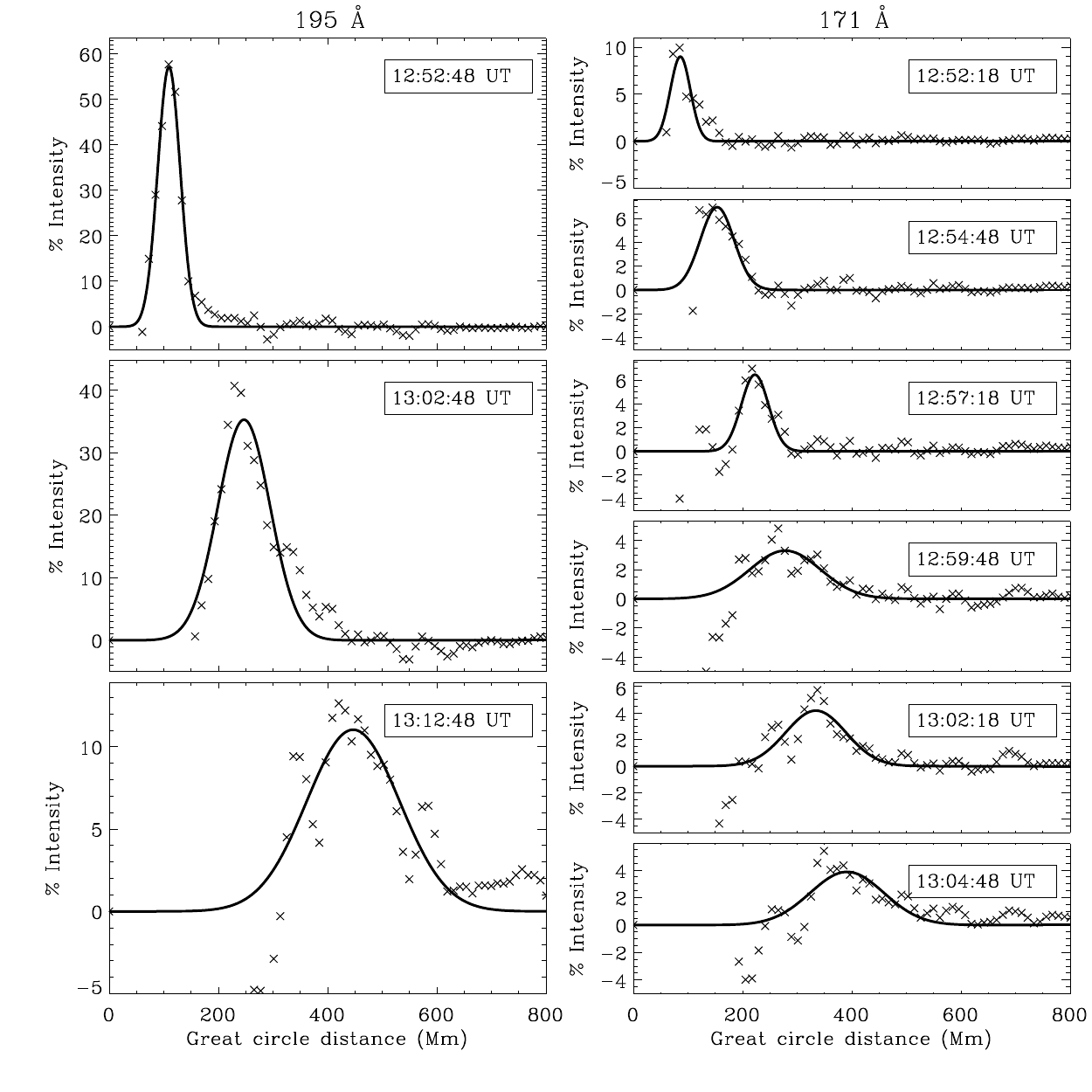}
		}
\caption{Same as Figure~\ref{fig:profile_20070519_A} but for the 2007~May~19 event observed by \emph{STEREO}-B.}
\label{fig:profile_20070519_B}
\end{figure}

The resulting intensity profiles from both passbands studied for the event of 2007~May~19 are shown in Figures~\ref{fig:profile_20070519_A} (for \emph{STEREO}-A) and \ref{fig:profile_20070519_B} (for \emph{STEREO}-B). In each case, the crosses show the PBD intensity values, with the solid line showing the Gaussian fit to the data. Only the first three observations in the 195~\AA\ passband and the first six observations in the 171~\AA\ passband from both spacecraft were used for this event. Beyond this, the $\chi^2$ values of the Gaussian fits were too large for accurate analysis. 

The pulse is clearly identifiable in the intensity profiles shown in both Figures~\ref{fig:profile_20070519_A} and \ref{fig:profile_20070519_B}, with the model providing a good fit to the data. As previously noted in Section~\ref{subsect:cor_pita}, the intensity profile technique is more accurate than a normal point-and-click technique as it is semi-automated and reproducible, with any associated errors quantifiable as they result from the fitting of a Gaussian to an intensity profile across the pulse. In contrast, the point-and-click methodology commonly used to identify CBFs is highly user-dependent, with large unquantifiable errors in the identification of the pulse position. Although a pulse may be tracked over larger distances using a visual method, increased image processing is often required to improve the visibility of the pulse, with the variable scaling making the comparison of measurements between images difficult. However, the intensity profile technique used here applies the same processing for all images, with the result that measurements of the pulse are directly comparable. The intensity profile technique can also be used to process large amounts of data (such as those from \emph{SDO}) rapidly as it does not require the same degree of user input as the visual method.

\section{Data Analysis}
\label{sect:wave_properties_analysis}

The CBF identification algorithm discussed in Section~\ref{sect:wave_properties_cor_pita} was used to study data from the 2007~May~19, 2007~December~07, 2009~February~12 and 2009~February~13 events, allowing the temporal variation of the pulse properties to be examined. The residual resampling bootstrapping technique outlined in Section~\ref{subsect:bootstrap} was then used to provide statistically accurate estimates of the pulse kinematics for each event, while the dispersion and dissipation of the pulse were studied using the temporal variation in pulse width and integrated intensity respectively.

The results of this analysis for the event from 2007~May~19 are discussed here in detail, with particular emphasis placed on the methods used to identify each parameter. The kinematics, dispersion and dissipation of the pulse are outlined in Sections~\ref{subsect:wave_properties_kinematics}, \ref{subsect:wave_properties_broadening} and \ref{subsect:wave_properties_intensity} respectively. This event is described in detail as it was the event originally discussed by \citet{Long:2008eu} and outlined in Chapter~\ref{chap:first_obs}. The additional events studied using this technique are presented in Section~\ref{sect:wave_properties_results}, with the event from 2007~May~19 used to provide additional context.

\subsection{Pulse Kinematics}
\label{subsect:wave_properties_kinematics}

\begin{figure}[!t]
\centerline{
    \includegraphics[width=0.8\textwidth,clip=,trim=0mm 5mm 60mm 73mm]{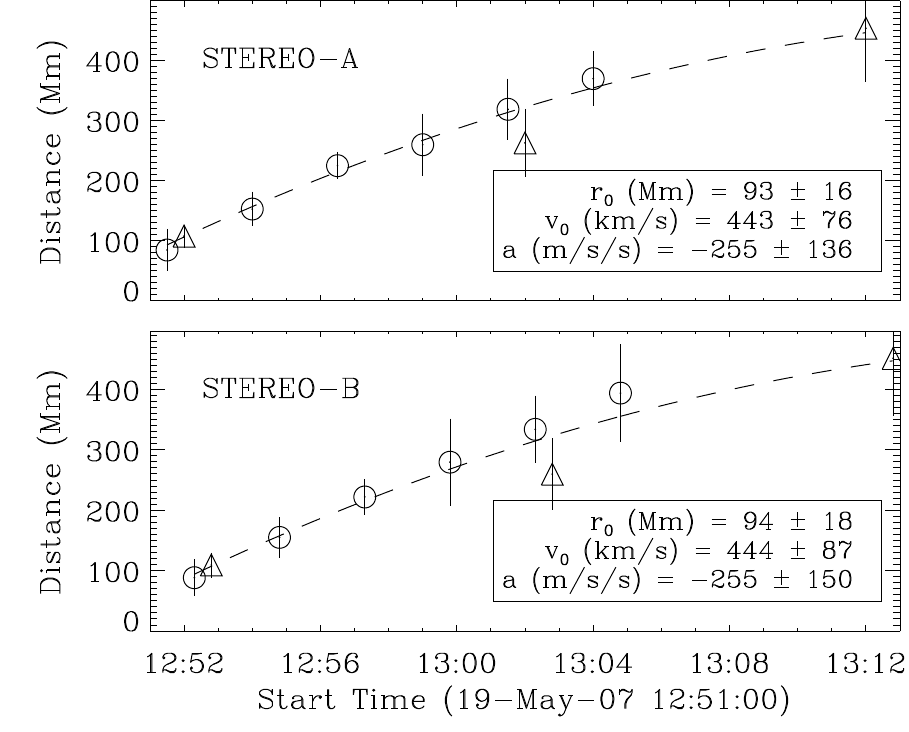}
}
\caption[Distance--time plots for the 2007~May~19 event obtained using \emph{STEREO}--A (top) and \emph{STEREO}--B (bottom)]{Distance-time plots for \emph{STEREO}-A (top) and \emph{STEREO}-B (bottom). The 171~\AA\ (circles) and 195~\AA\ (triangles) data have been combined as they follow similar kinematical curves. The mean offset distance, initial velocity, and acceleration terms resulting from the bootstrapping analysis (fit indicated by dashed line) are also stated in the bottom right of each panel, with errors represented by the standard deviation. The errors on each point are given by the error on the mean of the Gaussian fit applied to the intensity profile.}
\label{fig:cbf_kinematics_20070519}
\end{figure}

The Gaussian model applied to the positive section of the intensity profiles outlined in Section~\ref{sect:wave_properties_cor_pita} was used to plot the temporal variation of the pulse centroid distance from the source point. The resulting plot is shown in Figure~\ref{fig:cbf_kinematics_20070519} for \emph{STEREO}--A (top) and \emph{STEREO}--B (bottom), with the errors associated with each data-point given by the error associated with the mean of the Gaussian fit to the intensity profile in each case. Data from the 171~\AA\ and 195~\AA\ passbands \corr{were combined as} they have been observed to follow similar kinematical curves \citep{Warmuth:2004ab,Veronig:2010ab}. The relatively low observing cadence of the 195~\AA\ passband in particular also meant that the combination of data from both passbands was necessary to derive accurate pulse kinematics.

The distance--time data shown in Figure~\ref{fig:cbf_kinematics_20070519} were then fitted using a quadratic model of the form,
\begin{equation}
r(t) = r_0 + v_0 t + \frac{1}{2}a t^2,
\end{equation}
where $r_0$ is the offset distance, $v_0$ is the initial velocity, $a$ is the constant acceleration of the pulse, and $t$ is the time elapsed since the first observation of the disturbance. The data were analysed using the residual resampling bootstrapping technique shown in Section~\ref{subsect:bootstrap} to provide a statistically accurate estimate of the fit parameters given the small data--set.

The values given in the inset of Figure~\ref{fig:cbf_kinematics_20070519} show the mean and standard deviation of the bootstrapped parameter distributions (cf.\ the bottom row of Figure~\ref{fig:bootstrap}). The offset distance ($r_0$), initial velocity ($v_0$) and acceleration ($a$) are given using native SI units of Mm, km~s$^{-1}$ and m~s$^{-2}$ respectively for clarity. The estimated kinematics clearly show that the pulse has an initial velocity of $\sim$444~km~s$^{-1}$ and a deceleration of $\sim-$255~m~s$^{-2}$ as observed by both spacecraft. 

Although the kinematics estimated using both spacecraft are similar, the associated errors are slightly different with velocity errors of $\pm$76 \& $\pm$87~km~s$^{-1}$ and acceleration errors of $\pm$136 \& $\pm$150~m~s$^{-2}$ for \emph{STEREO}-A and \emph{STEREO}-B respectively. While the errors associated with the deceleration are quite large, in each case the acceleration is negative within the 1-$\sigma$ error range. The kinematics are also consistent between both spacecraft, indicating that these are the true kinematics of the pulse, with the slightly different errors most likely due to geometrical effects.

The initial velocity values given here are similar to previous estimates obtained by \citet{Long:2008eu} (see Chapter~\ref{chap:first_obs}) and \citet{Veronig:2008p216} who both studied this event, although in \corr{the case of \citet{Long:2008eu}} a three-point Lagrangian interpolation technique was used to \corr{examine the CBF kinematics}. The limitations of this technique have been discussed in detail in Section~\ref{subsubsect:l_diff}, with the artificial trends introduced to the data sometimes leading to a misinterpretation of the derived velocity and acceleration plots. In contrast, \corr{the kinematics} derived here are statistically significant and supported by quantifiable associated errors. 

\corr{The reexamination of the kinematics of the 2007~May~19 event presented here show a sharp contrast with those presented by \citet{Long:2008eu} and in Chapter~\ref{chap:first_obs}. A comparison with the top row of Figure~\ref{fig:norm_graphs} shows strong similarities with the positions of the pulse in both the 171 and 195~\AA\ passbands, while features which were originally identified as the pulse but have since been classified as oscillating coronal loops have been removed through the use of the semi--automated identification technique. The pulse kinematics derived here using the bootstrapping technique exhibit none of the variation observed in Figure~\ref{fig:norm_graphs} as a result of the model applied to the distance--time measurements. While there may be some point--to--point variation in pulse kinematics, it is not possible to test for this using the small data--sets currently available from \emph{STEREO}/EUVI.}

\subsection{Pulse Width}
\label{subsect:wave_properties_broadening}

\begin{figure}[!t]
\centerline{
   \includegraphics[width=0.7\textwidth,clip=,trim=0mm 6mm 60mm 98mm]{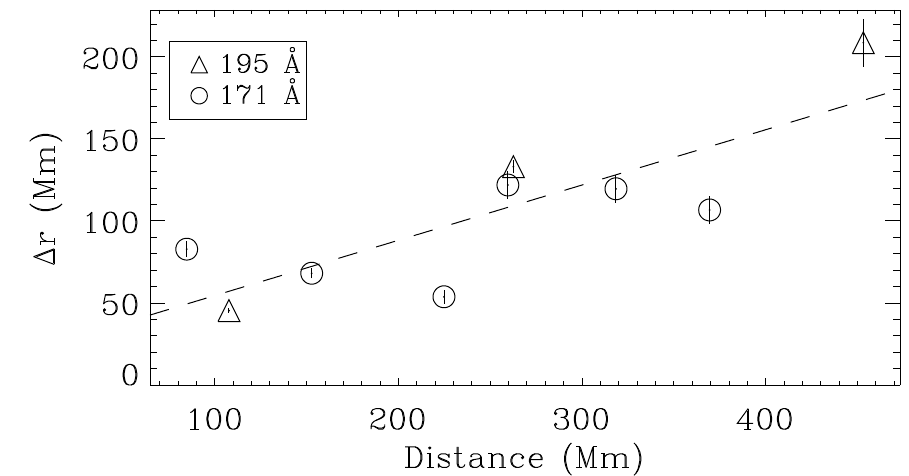}
   }
\centerline{
   \includegraphics[width=0.7\textwidth,clip=,trim=0mm 6mm 60mm 98mm]{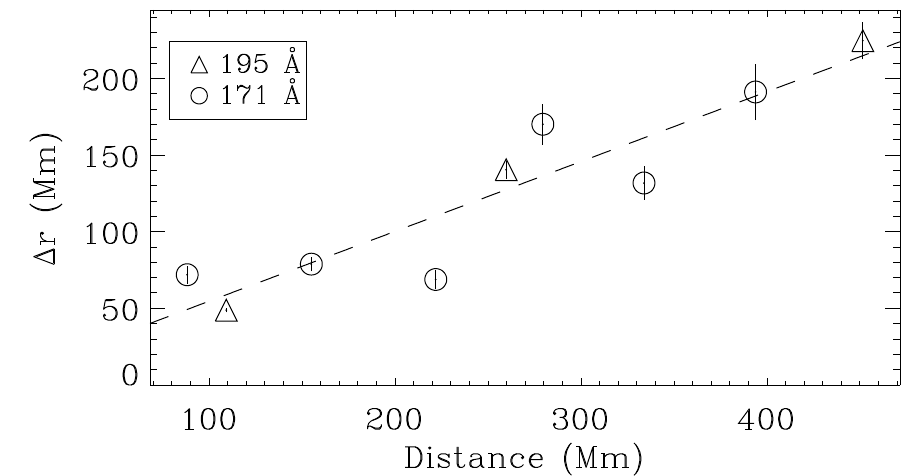}
               }
\caption[Variation in spatial pulse width with distance for the 2007~May~19 event observed by \emph{STEREO}--A (top) and \emph{STEREO}--B (bottom).]{Variation in pulse spatial width ($\Delta r$) with distance for the 2007~May~19 event observed by \emph{STEREO}-A (top) and \emph{STEREO}-B (bottom). Pulse spatial width here refers to the FWHM of the fitted Gaussian pulse (i.e., $\Delta r = 2\sigma\sqrt{2\mathrm{ln}2}$). The dashed line indicates the best linear fit to the data.}
\label{fig:space_broadening_20070519}
\end{figure}

The parameters returned by the Gaussian model fitted to the intensity profile also allowed the variation in pulse width to be examined in addition to the pulse kinematics. To do this, the full width at half maximum (FWHM; $\Delta r = 2\sigma\sqrt{2\mathrm{ln}2}$) of the Gaussian fit was determined for each intensity profile studied. The variation in pulse width with distance is shown in Figure~\ref{fig:space_broadening_20070519} for \emph{STEREO}-A (top) and \emph{STEREO}-B (bottom), with measurements from the differing passbands indicated by the different symbols. It is clear from these plots that an increase in pulse width with distance is present for both passbands as observed by both spacecraft. This is indicative of pulse broadening and shows that the CBF spreads out spatially as it propagates.

\begin{figure}[!t]
\centerline{
   \includegraphics[width=0.7\textwidth,clip=,trim=0mm 6mm 58mm 98mm]{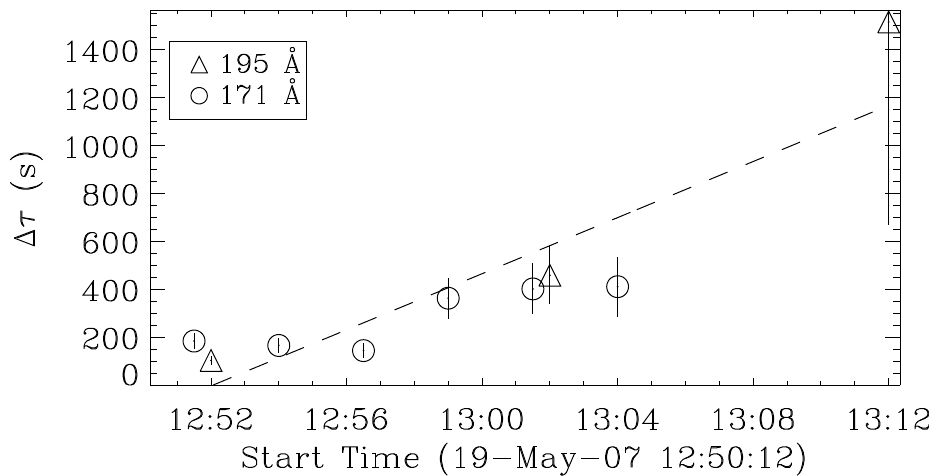}
   }
\centerline{
   \includegraphics[width=0.7\textwidth,clip=,trim=0mm 6mm 58mm 98mm]{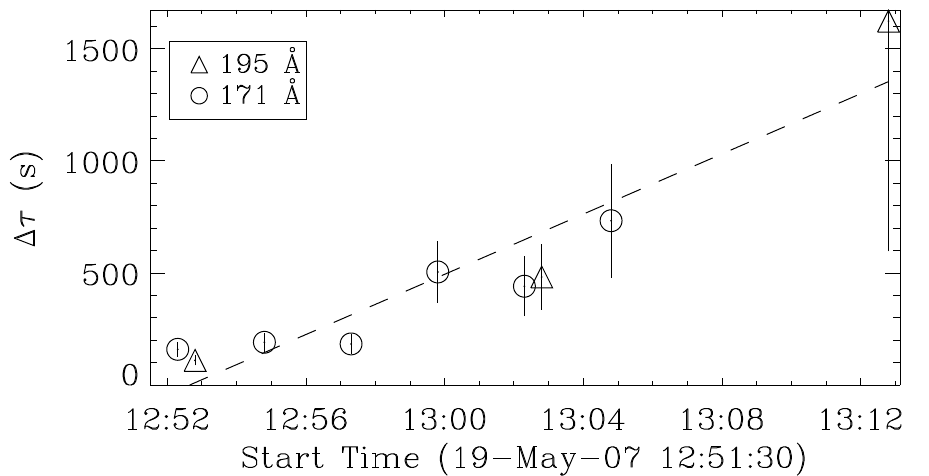}
               }
\caption[Variation in temporal pulse width with time for the 2007~May~19 event observed by \emph{STEREO}--A (top) and \emph{STEREO}--B (bottom).]{Variation in pulse temporal width ($\Delta\tau$) with time as observed by \emph{STEREO}-A (top) and \emph{STEREO}-B (bottom). The dashed line in all panels indicates the best linear fit to the data.}
\label{fig:time_broadening_20070519}
\end{figure}

The variation in the temporal width of the pulse with time was also considered in addition to the variation in the spatial width of the pulse with distance. The temporal width of the pulse, $\Delta\tau$, is defined as,
\begin{equation}\label{eqn:t_width}
\Delta\tau = \frac{\Delta r}{v_{\mathrm{pulse}}},
\end{equation}
where $\Delta r$ is the spatial width of the pulse and $v_{\mathrm{pulse}}$ is the pulse velocity. While an increase in the spatial width of the pulse may indicate pulse dispersion, this could be negated by increasing pulse velocity, producing a pulse with constant temporal width. By examining the variation of the temporal width of the pulse with time it is possible to determine if the pulse is indeed dispersive. The resulting plots showing the variation in pulse temporal width with time are given in Figure~\ref{fig:time_broadening_20070519} for \emph{STEREO}--A (top) and \emph{STEREO}--B (bottom).

The temporal width of the pulse is shown in Figure~\ref{fig:time_broadening_20070519} to increase with time for observations from both \emph{STEREO} spacecraft, indicating that the pulse broadens in both space and time with propagation. The clear increase in spatial and temporal pulse width in data from both \emph{STEREO} spacecraft strongly suggests that it is a true feature of the disturbance and not an observational artifact. The rate of change of the spatial and temporal width of the pulse with distance and time respectively ($d(\Delta r)/dr$ and $d(\Delta\tau)/dt$) are given in Table~\ref{tbl:events_characteristics}.

\begin{deluxetable}{cccccccc}
\tablecolumns{8}
\rotate
\tabletypesize{\footnotesize}
\tablewidth{0pt}
\centering
\tablecaption{Wave properties of studied CBFs.\label{tbl:events_characteristics}}
\tablehead{
\colhead{} & \colhead{} & \multicolumn{2}{c}{Kinematics} & \multicolumn{2}{c}{Dispersion} & \multicolumn{2}{c}{Integrated Intensity} \\ 
\colhead{Event} & \colhead{Spacecraft} & \colhead{$v_{0}$} & \colhead{$a$} & \colhead{$d(\Delta r)/dr$} & \colhead{$d(\Delta \tau)/dt$} & \multicolumn{2}{c}{$d(I_{tot})/dr$} \\
\colhead{} & \colhead{} & \colhead{km~s$^{-1}$} & \colhead{m~s$^{-2}$} & \colhead{} & \colhead{} & \colhead{171~\AA} & \colhead{195~\AA}
}
\startdata

2007~May~19	& Ahead 	& \corr{$443 \pm 76$}	& \corr{$-255 \pm 136$} 	& $0.3 \pm 0.1$ 	& $1.0 \pm 0.2$ 	& $1.08 \pm 0.53$ 	& $0.13 \pm 1.07$ \\
 				& Behind 	& \corr{$444 \pm 87$}	& \corr{$-255 \pm 150$} 	& $0.5 \pm 0.1$ 	& $1.1 \pm 0.2$ 	& $1.83 \pm 0.41$ 	& $0.24 \pm 0.61$ \\
2007~Dec~07 	& Ahead 	& $270 \pm 37$ 		& \corr{$-153 \pm 73$} 	& $0.4 \pm 0.1$ 	& $1.0 \pm 0.3$ 	& $0.71 \pm 0.37$ 	& $-0.049 \pm 0.002$ \\
				& Behind	& \corr{$249 \pm 61$}& \corr{$-119 \pm 109$} 	& $0.2 \pm 0.2$ 	& $0.8 \pm 0.3$ 	& $2.29 \pm 0.41$ 	& $-1.07 \pm 0.12$ \\
2009~Feb~12 	& Behind 	& \corr{$406 \pm 94$}& \corr{$-293 \pm 167$} 	& $0.4 \pm 0.2$ 	& $2.0 \pm 0.5$ 	& $-3.27 \pm 2.39$ 	& $-0.59 \pm 0.11$ \\
2009~Feb~13 	& Behind 	& \corr{$273 \pm 53$}& $-49 \pm 34$ 		& $0.4 \pm 0.1$ 	& $0.5 \pm 0.1$ 	& $4.33 \pm 0.66$ 	& $-0.74 \pm 0.36$ \\
\enddata
\tablecomments{Kinematic values refer to the mean and standard deviation error of the bootstrapping parameter distributions. Dispersion and integrated intensity values refer to the rate of change of the relevant parameters, resulting from the linear fits shown in Figures~\ref{fig:space_broadening_20070519}, \ref{fig:time_broadening_20070519} \& \ref{fig:app_space_broadening_20071207} to \ref{fig:app_time_broadening_200902} and \ref{fig:int_20070519_A}, \ref{fig:int_20070519_B} \& \ref{fig:app_intensity_20071207_A} to \ref{fig:app_intensity_20090213}.}
\end{deluxetable}

The observed broadening of the CBF pulse in both space and time confirms the previous observations of both \citet{Warmuth:2001p77} and \citet{Veronig:2010ab}. These observations, when taken together with the derived kinematics, are consistent with the interpretation of a CBF as a dispersive, decelerating pulse.

\subsection{Integrated Intensity}
\label{subsect:wave_properties_intensity}

\begin{figure}[!p]
\centerline{
    \includegraphics[width=0.7\textwidth,clip=,trim=10mm 0mm 45mm 0mm]{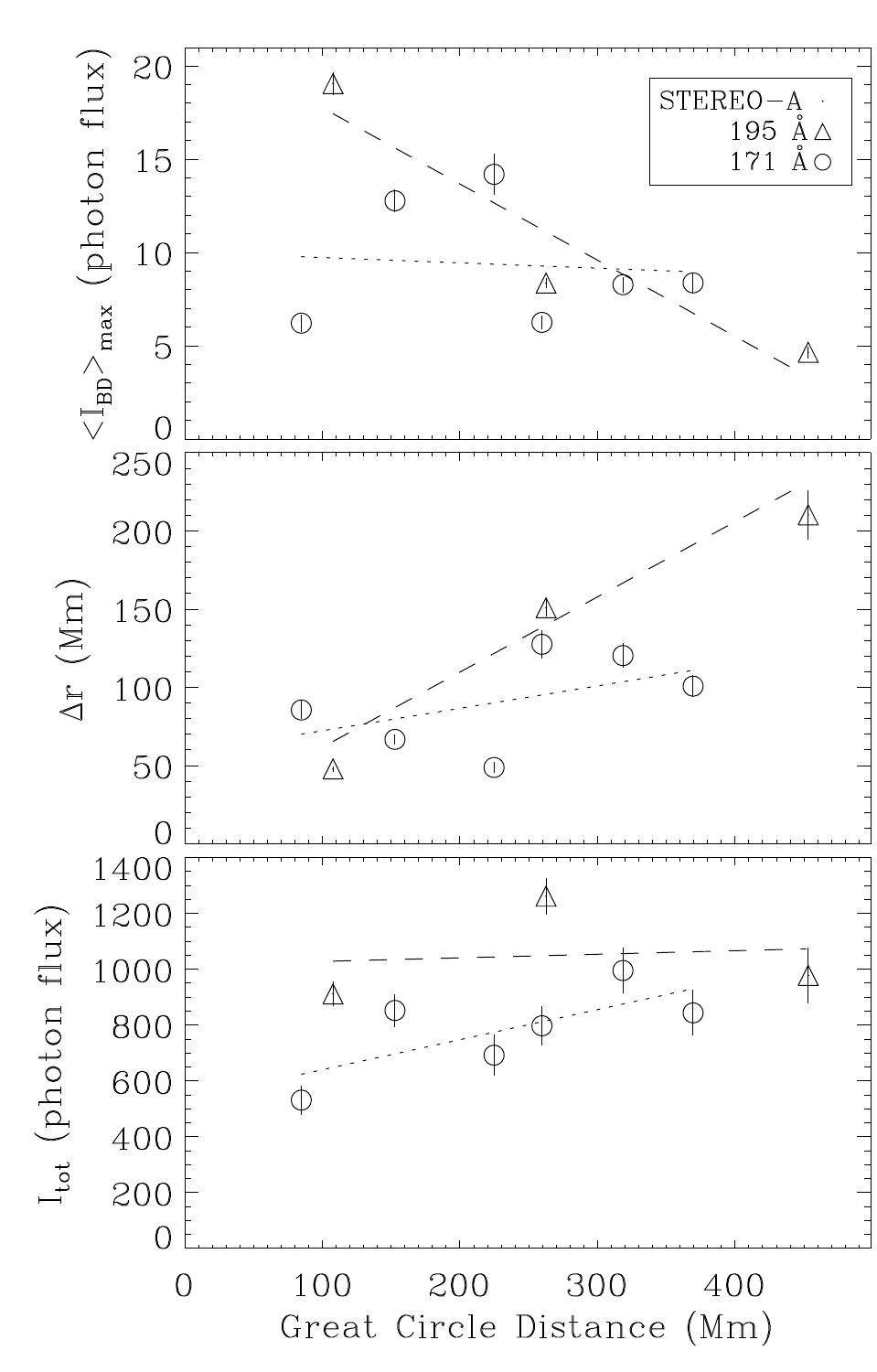}
               }
\caption{\emph{Top}: Variation in peak base difference intensity ($\langle I_{BD}\rangle_{\mathrm{max}}$) with distance. \emph{Middle}: Variation in FWHM of the Gaussian pulse with distance. \emph{Bottom}: Variation in integrated intensity ($I_{tot}$) with distance. All plots correspond to measurements from \emph{STEREO}-A, with the dashed (dotted) line in all panels indicating the best linear fit to the 195~\AA\ (171~\AA) data.}
\label{fig:int_20070519_A}
\end{figure}

\begin{figure}[!p]
\centerline{
    \includegraphics[width=0.7\textwidth,clip=,trim=10mm 0mm 45mm 0mm]{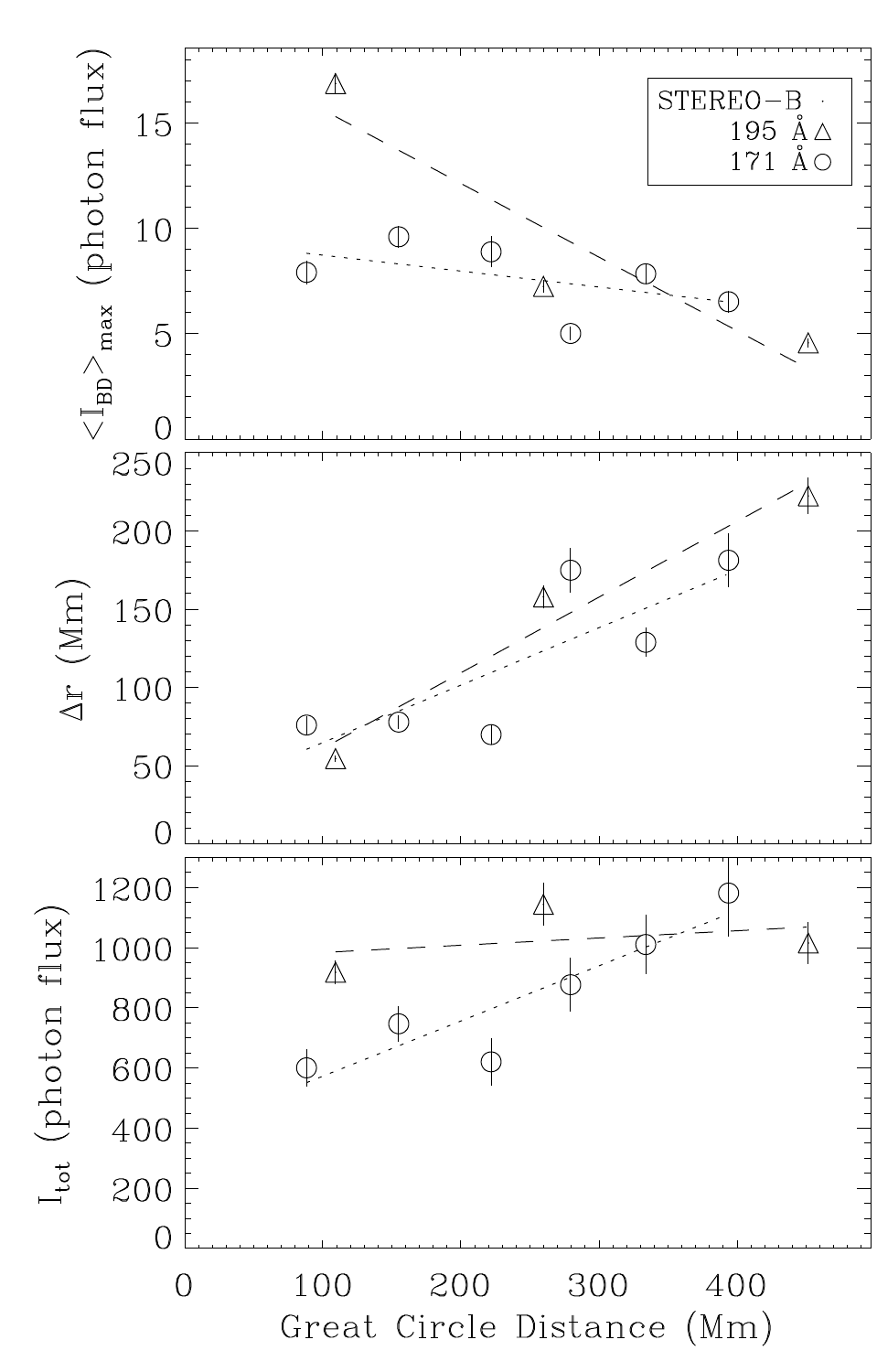}
               }
\caption{Same as Figure~\ref{fig:int_20070519_A} but showing the \emph{STEREO}-A data for the event from 2007~May~19.}
\label{fig:int_20070519_B}
\end{figure}

The temporal variation in the PA-averaged integrated intensity of the CBF pulse was also examined in addition to \corr{the} pulse kinematics and the temporal and spatial dispersion of the pulse. This provides a verification of the presence of pulse dissipation, allowing the physical nature of the disturbance to be identified. The PA-averaged integrated intensity of the pulse can be determined using the variation in peak intensity with distance ($\langle I_{BD}\rangle_{\mathrm{max}}$ versus r) and the variation in the FWHM of the pulse with distance ($\Delta r$ versus r). This allows the variation in PA-averaged integrated intensity to be estimated using,
\begin{equation}
I_{\mathrm{tot}} = \Delta r \langle I_{\mathrm{BD}}\rangle_{\mathrm{max}},
\end{equation}
and plotted as a function of distance. Figures~\ref{fig:int_20070519_A} and \ref{fig:int_20070519_B} show the variation with distance in peak BD intensity (top), FWHM of the fitted pulse (middle) and resulting integrated intensity (bottom) for \emph{STEREO}-A and \emph{STEREO}-B respectively for the 2007~May~19 event.

The variation in the integrated intensity rather than the peak intensity of the pulse was examined as the pulse has exhibited broadening (i.e., dispersion). This dispersion may influence the peak intensity of the pulse by causing it to decrease as the pulse broadens. In this case, the pulse would appear to show a drop in intensity but would contain the same amount of energy, giving the impression of dissipation without actually dissipating any energy. Similarly, the peak intensity of the BD intensity profile was used as this shows the actual emission of the pulse, while the PBD intensity shows the ratio of the pulse emission with respect to the background. As the coronal background is not uniform across the extent of the pulse, similar PBD intensity variations may not correspond to similar increases in emission.

The top panel in both Figures~\ref{fig:int_20070519_A} and \ref{fig:int_20070519_B} shows a large discrepancy between the peak BD intensity of the 171~\AA\ and 195~\AA\ observations. In both cases, the 195~\AA\ data shows a large decrease with time while the 171~\AA\ data shows no strong variation. As previously shown in Section~\ref{subsect:wave_properties_broadening}, the FWHM of the pulse follows a linearly increasing trend with distance for both passbands indicating pulse dispersion. However, the resulting variation in PA-averaged BD integrated intensity is inconclusive in each case shown in Figures~\ref{fig:int_20070519_A} and \ref{fig:int_20070519_B}. There does not appear to be any significant variation in 195~\AA\ measurements with distance, although some strong point-to-point variation is noticeable. The 171~\AA\ measurements however show a generally increasing trend in each case. The rate of change of $I_{tot}$ with distance (i.e.,\ d($I_{tot}$)/dr) for each event studied are given in Table~\ref{tbl:events_characteristics}.

The measurements shown here indicate that more analysis is required to understand the morphology of the pulse. The higher cadence observations available from \emph{SDO} should allow the variation in peak intensity and PA-averaged integrated intensity with distance to be determined with a much higher degree of accuracy than is possible using \emph{STEREO}. The variation between the different passbands observed by \emph{SDO} should also allow a better understanding of the temperature structure of the pulse. 

\section{Results}
\label{sect:wave_properties_results}

The analysis described in the previous Section~\ref{sect:wave_properties_analysis} indicates that the pulse identification algorithm described in Sections~\ref{subsect:cor_pita} and \ref{sect:wave_properties_cor_pita} is extremely effective at identifying the CBF pulse in observations. This is particularly apparent in the intensity profiles shown in Figures~\ref{fig:profile_20070519_A} and \ref{fig:profile_20070519_B}, where the model provides a good fit to the data in each case. The errors associated with the identification of the pulse are also quantified using this algorithm, unlike the unknown user--defined errors associated with the point--and--click approach. 

Once the source point and arc sector into which the pulse propagates have been identified, the algorithm provides accurate and reproducible estimates of the pulse characteristics, making these results more robust. The re--analysis of the CBF event of 2007~May~19 outlined in the previous section indicate that the pulse exhibited significant deceleration and dispersion. With the capabilities of the algorithm confirmed, it was then applied to three more events, from 2007~December~07, 2009~February~12 and 2009~February~13. The results of these analyses are discussed in this section, with all of the results summarised in Table~\ref{tbl:events_characteristics}.

\subsection{Pulse Kinematics}
\label{subsect:all_events_kins}

\begin{figure}[!p]
              \includegraphics[width=1\textwidth,clip=,trim=0mm 5mm 0mm 0mm]{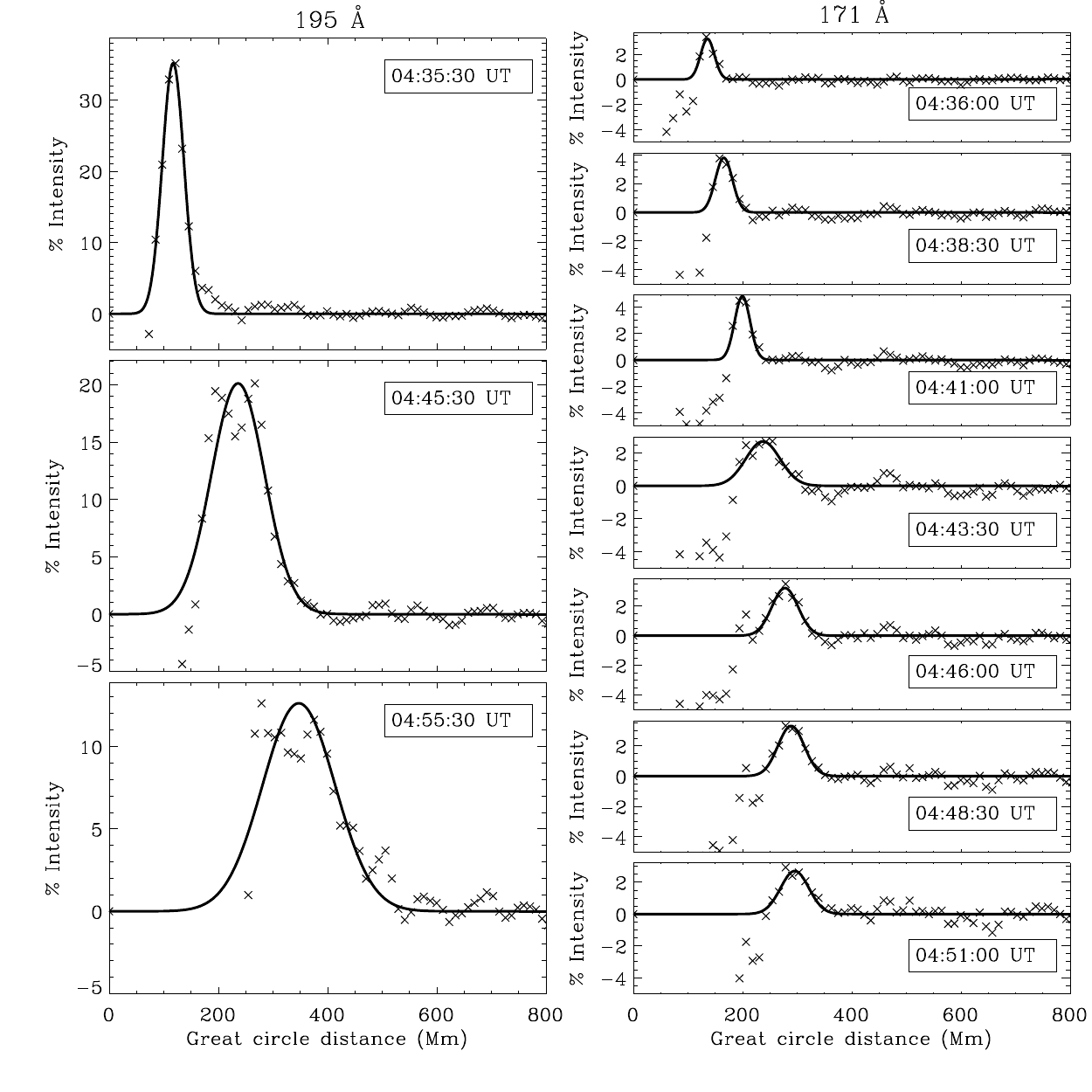}
\caption{Same as Figure~\ref{fig:profile_20070519_A} but for the event from 2007~Dec~07 observed by \emph{STEREO}-A.}
\label{fig:profile_20071207_A}
\end{figure}

\begin{figure}[!p]
              \includegraphics[width=1\textwidth,clip=,trim=0mm 5mm 0mm 0mm]{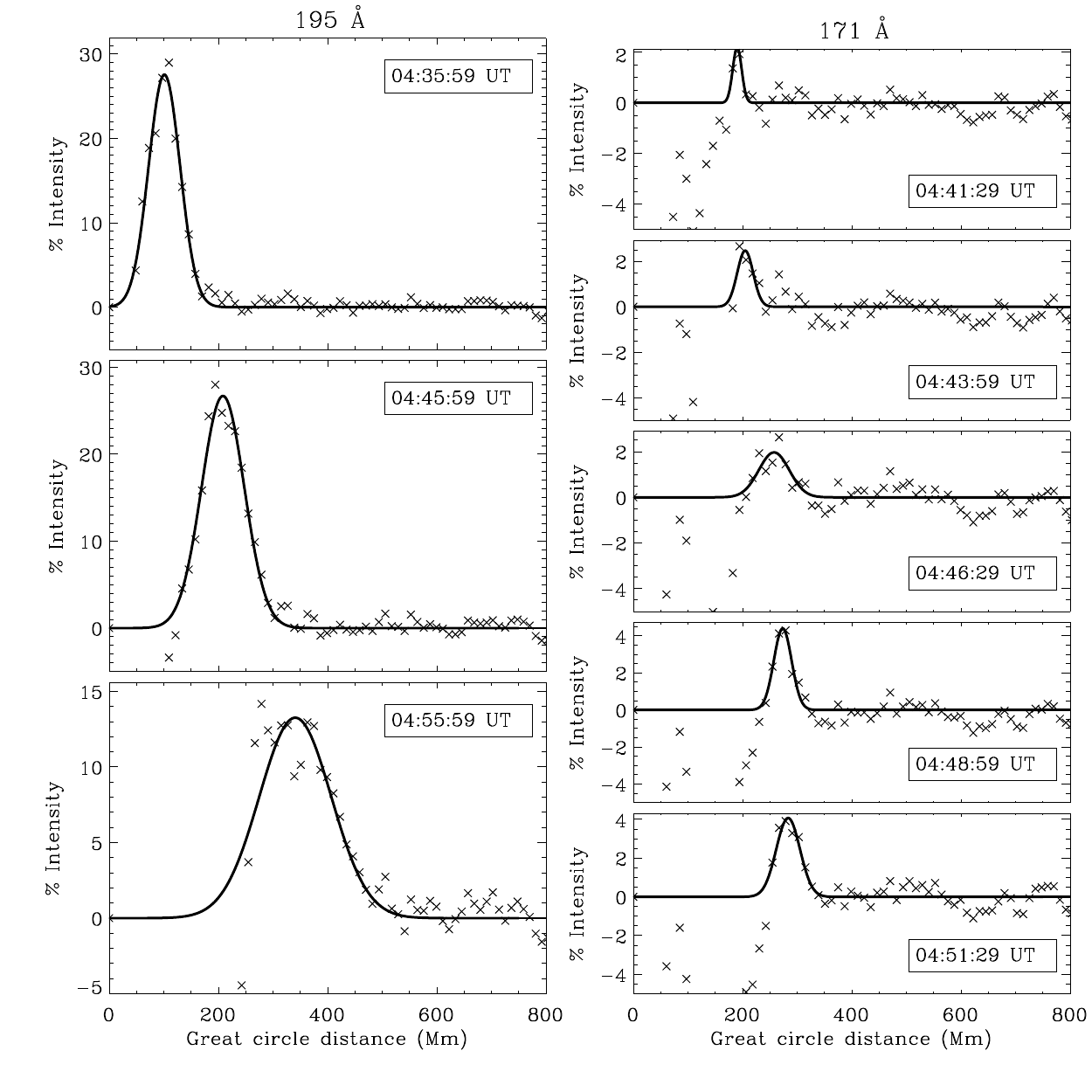}
\caption{Same as Figure~\ref{fig:profile_20070519_A} but for the event from 2007~Dec~07 observed by \emph{STEREO}-B.}
\label{fig:profile_20071207_B}
\end{figure}

\begin{figure}[!p]
              \includegraphics[width=1\textwidth,clip=,trim=0mm 5mm 0mm 0mm]{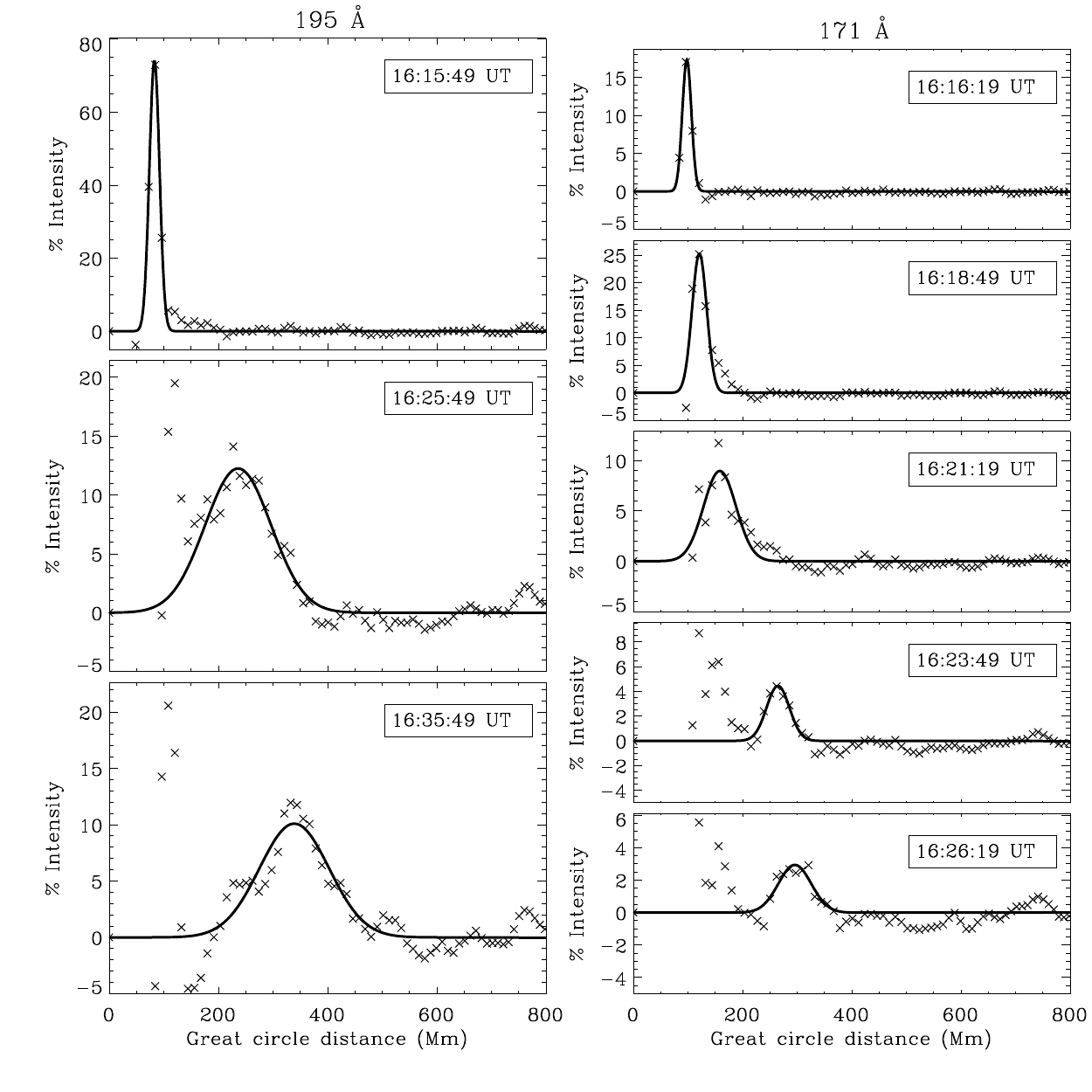}
\caption{Same as Figure~\ref{fig:profile_20070519_A} but for the event from 2009~Feb~12 observed by \emph{STEREO}-B.}
\label{fig:profile_20090212_B}
\end{figure}

\begin{figure}[!p]
              \includegraphics[width=1\textwidth,clip=,trim=0mm 5mm 0mm 0mm]{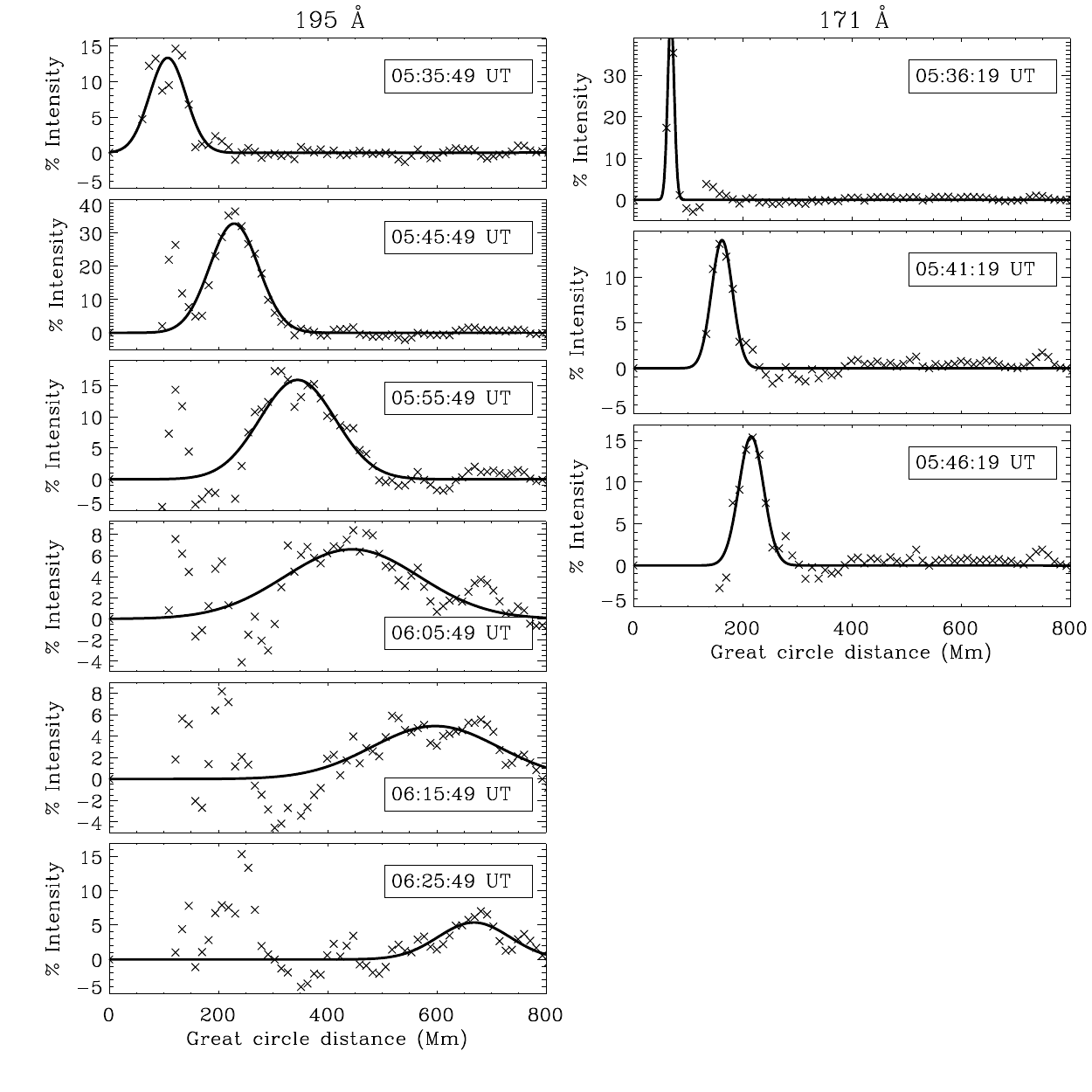}
\caption{Same as Figure~\ref{fig:profile_20070519_A} but for the event from 2009~Feb~13 observed by \emph{STEREO}-B.}
\label{fig:profile_20090213_B}
\end{figure}

The intensity profiles of the events from 2007~December~07, 2009~February~12 and 2009~February~13 were determined using the arc sectors shown in Figure~\ref{fig:wave_properties_image}, with the same orientation for the arc chosen for both \emph{STEREO} spacecraft to ensure that the different results were comparable. The resulting intensity profiles for the different events studied are shown in Figures~\ref{fig:profile_20071207_A}, \ref{fig:profile_20071207_B}, \ref{fig:profile_20090212_B} and \ref{fig:profile_20090213_B}. \corr{A difference in the peak intensity is apparent between the 195 and 171~\AA\ passbands, particularly for the 2007~December~07 event. This is most likely due to the different peak emission temperatures observed by the two passbands, and could be explained by plasma heating or cooling out of the 171~\AA\ passband \citep[cf.][]{Wills-Davey:1999ve}. It should also be noted that for the event of 2009~February~13, it was only possible to accurately detect the pulse in three consecutive images in the 171~\AA\ passband. After $\sim$05:46:19~UT, the errors associated with the pulse identification became too large, with the result that only the first three detections were used in this analysis.}

\begin{figure}
\centerline{
    \includegraphics[width=0.7\textwidth,clip=,trim=4mm 7mm 57mm 72mm]{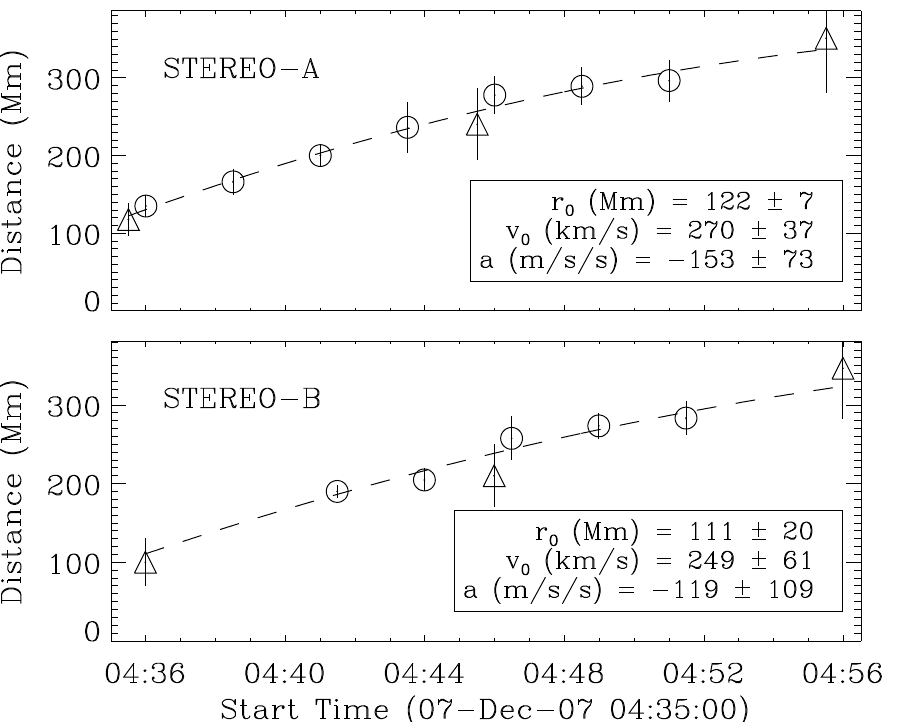}
}
\caption{Same as Figure~\ref{fig:cbf_kinematics_20070519} but for the event on 2007~Dec~07.}
\label{fig:app_kinematics_20071207}
\end{figure}

\begin{figure}
\centerline{
    \includegraphics[width=0.7\textwidth,clip=,trim=4mm 7mm 57mm 106mm]{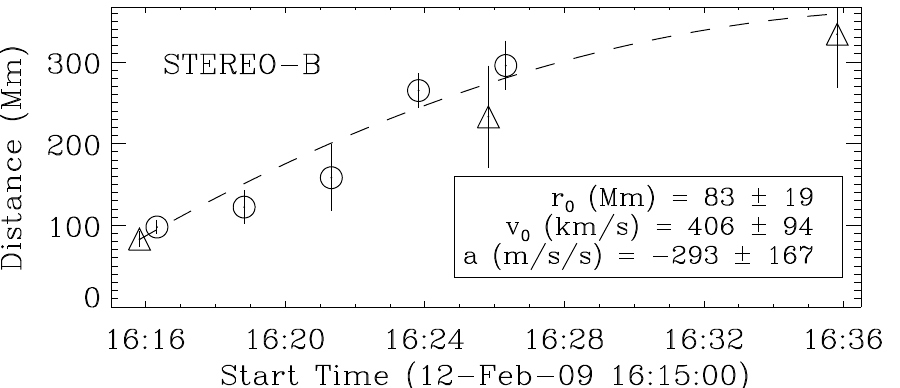}
}
\centerline{
    \includegraphics[width=0.7\textwidth,clip=,trim=4mm 7mm 57mm 106mm]{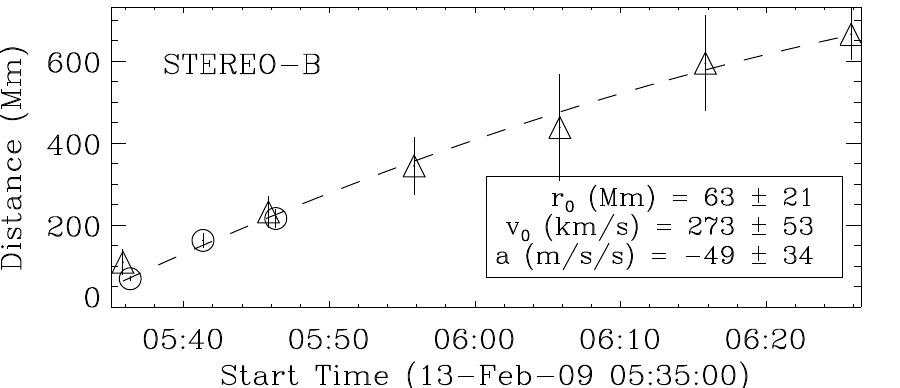}
}
\caption{Same as Figure~\ref{fig:cbf_kinematics_20070519} but for the events on 2009~Feb~12 (top) and 2009~Feb~13 (bottom), both showing \emph{STEREO}-B data.}
\label{fig:app_kinematics_200902}
\end{figure}

As with the 2007~May~19 event discussed in Section~\ref{sect:wave_properties_analysis}, the kinematics were determined for each pulse studied using the temporal variation of the centroid of the Gaussian fit to the intensity profile. The pulses exhibited comparable initial propagation velocities ranging from $\sim$249~km~s$^{-1}$ for the 2007~December~07 event as observed by \emph{STEREO}--B to $\sim$406~km~s$^{-1}$ for the 2009~February~12 event as observed by \emph{STEREO}--B. The distance--time plots measured for each event are shown in Figures~\ref{fig:app_kinematics_20071207} and \ref{fig:app_kinematics_200902}, with the kinematics values for each event shown in Table~\ref{tbl:events_characteristics}.

While the kinematics of the 2007~May~19 event as observed by both \emph{STEREO} spacecraft are very similar, the increased separation between the spacecraft is apparent in the kinematics estimated for the 2007~December~07 event, with the differences in initial velocity most likely due to geometrical effects. The quadrature events of 2009~February~12 and 13 were different from each other despite originating from the same active region. The 2009~February~12 event exhibited a higher initial velocity of $\sim$406~km~s$^{-1}$, while the 2009~February~13 event had a lower initial velocity of $\sim$273~km~s$^{-1}$. The 2007~May~19 event appears to have been a relatively fast event, while the range of \corr{velocities} ($\sim$240--450~km~s$^{-1}$) is consistent with previously estimated values. 

All of the events studied exhibited statistically significant non-zero deceleration, although the strength of the deceleration varied between events. The deceleration estimated for the 2007~December~07 event was also affected by the position of the \emph{STEREO} spacecraft, varying from $-153\pm73$~m~s$^{-2}$ in \emph{STEREO}-A data to $-119\pm109$~m~s$^{-2}$ in \emph{STEREO}-B data. The 2009~February~12 was also observed to exhibit stronger deceleration than the 2009~February~13 event ($-293\pm167$~m~s$^{-2}$ as opposed to $-49\pm34$~m~s$^{-2}$). The large errors associated with the acceleration terms given here indicate the difficulties associated with accurately determining the kinematics of CBFs from low cadence observations, despite the minimization of errors through the use of both the intensity profile technique and bootstrapping analysis.

\subsection{Pulse Broadening}
\label{subsect:all_events_broad}

\begin{figure}
\centerline{
   \includegraphics[width=0.6\textwidth,clip=,trim=0mm 6mm 60mm 98mm]{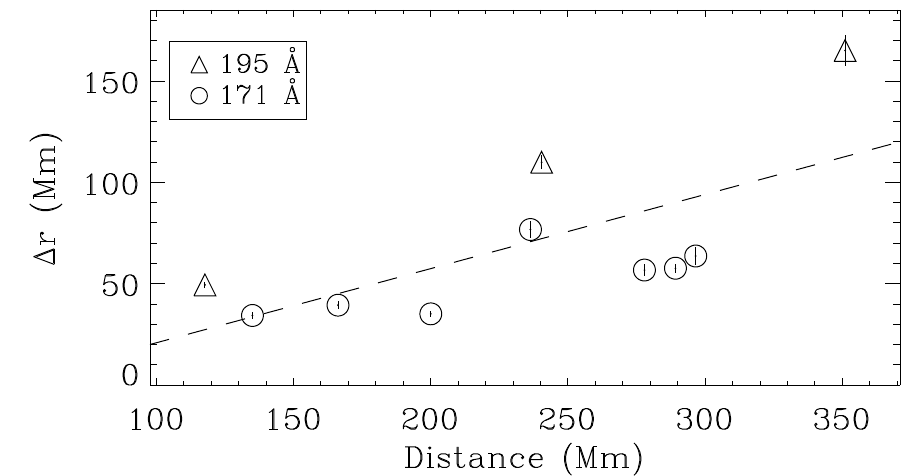}
               }
\centerline{
   \includegraphics[width=0.6\textwidth,clip=,trim=0mm 6mm 60mm 98mm]{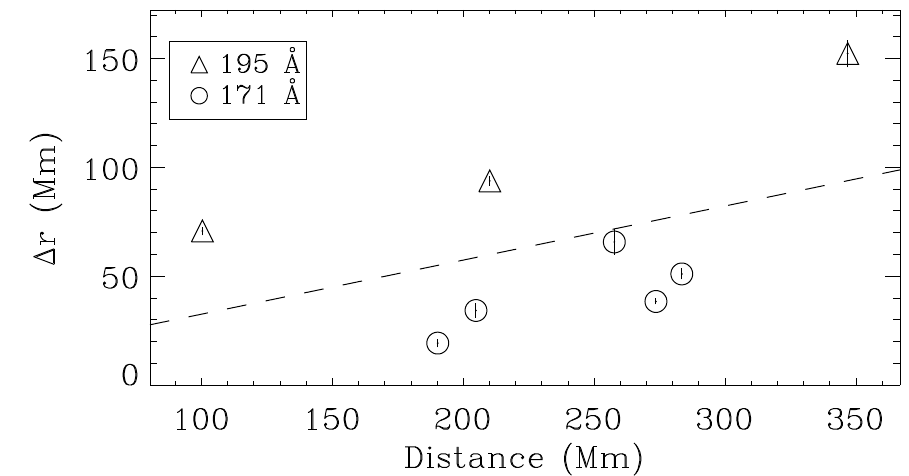}
               }
\caption{Same as Figure~\ref{fig:space_broadening_20070519} but showing the event on 2007~Dec~07.}
\label{fig:app_space_broadening_20071207}
\end{figure}

\begin{figure}
\centerline{
   \includegraphics[width=0.6\textwidth,clip=,trim=0mm 6mm 60mm 98mm]{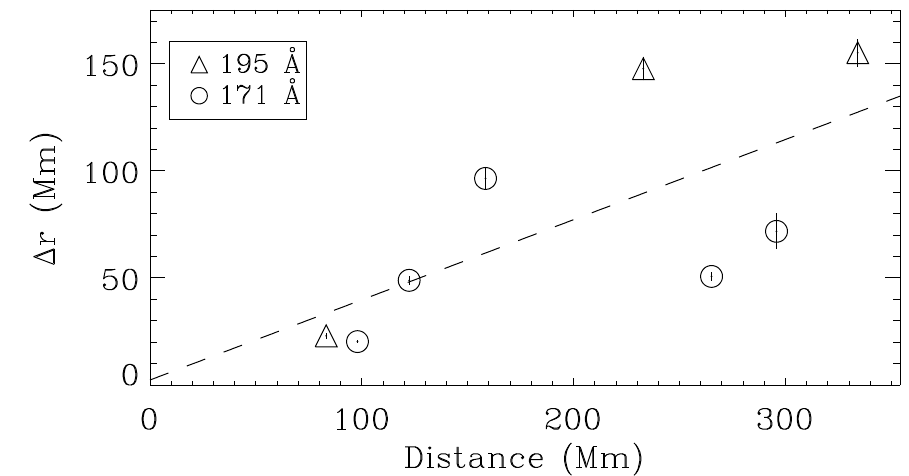}
}
\centerline{
   \includegraphics[width=0.6\textwidth,clip=,trim=0mm 6mm 60mm 98mm]{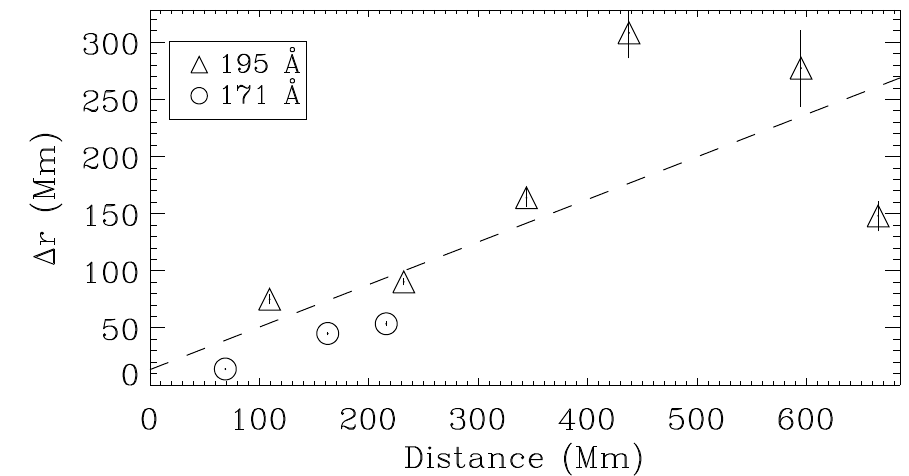}
               }
\caption{Similar to Figure~\ref{fig:space_broadening_20070519}, now only showing \emph{STEREO}-B data for the events on 2009~Feb~12 (top) and 2009~Feb~13 (bottom).}
\label{fig:app_space_broadening_200902}
\end{figure}

\begin{figure}
\centerline{
    \includegraphics[width=0.6\textwidth,clip=,trim=0mm 6mm 56mm 98mm]{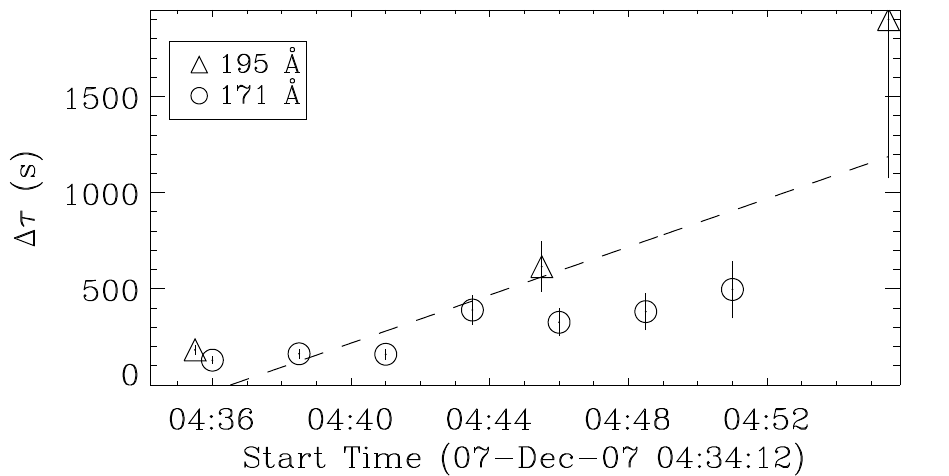}
               }
\centerline{
    \includegraphics[width=0.6\textwidth,clip=,trim=0mm 6mm 56mm 98mm]{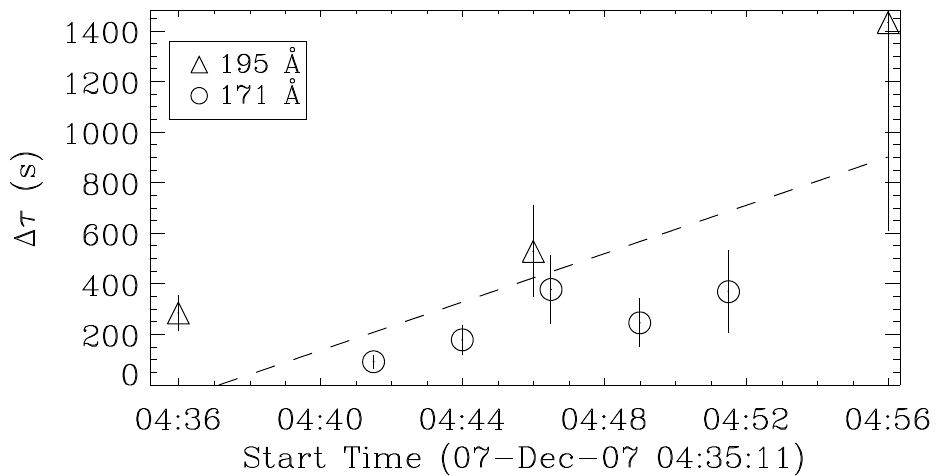}
               }
\caption{Same as Figure~\ref{fig:time_broadening_20070519} but showing the event on 2007~Dec~07.}
\label{fig:app_time_broadening_20071207}
\end{figure}

\begin{figure}
\centerline{
   \includegraphics[width=0.6\textwidth,clip=,trim=0mm 6mm 56mm 98mm]{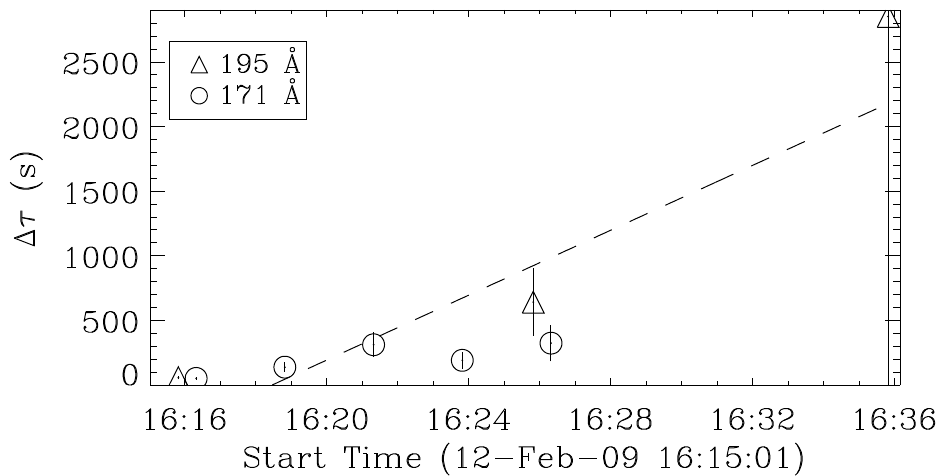}
}
\centerline{
   \includegraphics[width=0.6\textwidth,clip=,trim=0mm 6mm 56mm 98mm]{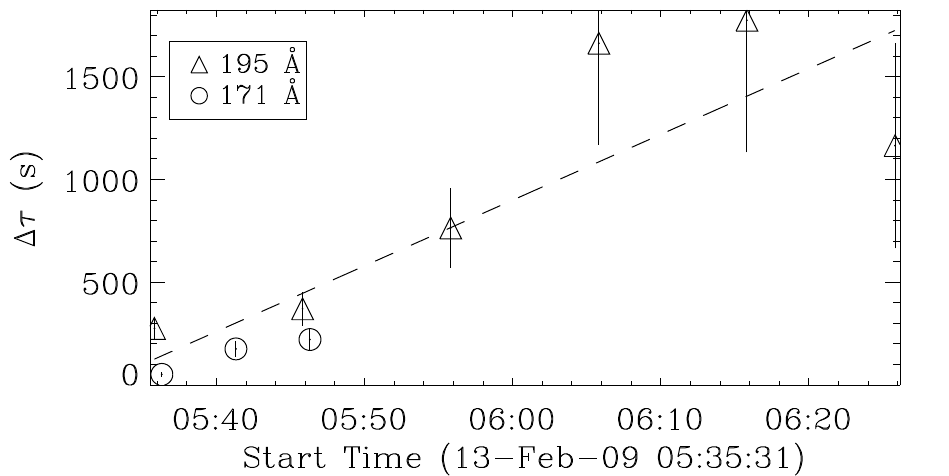}
               }
\caption{Similar to Figure~\ref{fig:time_broadening_20070519}, now only showing \emph{STEREO}-B data for the events on 2009~Feb~12 (top) and 2009~Feb~13 (bottom).}
\label{fig:app_time_broadening_200902}
\end{figure}

Next, the variation in both spatial and temporal pulse width with distance and time respectively was determined for each event studied. The resulting variation in spatial pulse width with distance is shown in Figures~\ref{fig:app_space_broadening_20071207} and \ref{fig:app_space_broadening_200902}, while Figures~\ref{fig:app_time_broadening_20071207} and \ref{fig:app_time_broadening_200902} show the variation in temporal pulse width with time.

It is clear from Figures~\ref{fig:app_space_broadening_20071207} to \ref{fig:app_time_broadening_200902} that each of the pulses studied exhibited evidence of pulse broadening in both spatial and temporal regimes with distance and time respectively. This suggests that pulse broadening is a general characteristic of CBFs and must be accounted for by all theories that seek to explain this phenomenon. Table~\ref{tbl:events_characteristics} shows the rate of change of both spatial and temporal pulse width $\Delta r$ and $\Delta \tau$ with distance and time respectively, with the $d(\Delta r)/dr$ and $d(\Delta \tau)/dt$ parameters positive for both the 171~\AA\ and 195~\AA\ passbands in all events studied. 

These results strongly imply that CBF pulses are dispersive pulses, something that may have been previously underestimated as a consequence of the techniques used to identify them. The presence of both pulse deceleration and dispersion is also suggestive of a wave interpretation for this phenomenon and indicates that the pulses are freely propagating. 

\subsection{Pulse Dissipation}
\label{subsect:all_events_dissip}

\begin{figure}[!p]
\centerline{
    \includegraphics[width=0.7\textwidth,clip=,trim=10mm 0mm 45mm 0mm]{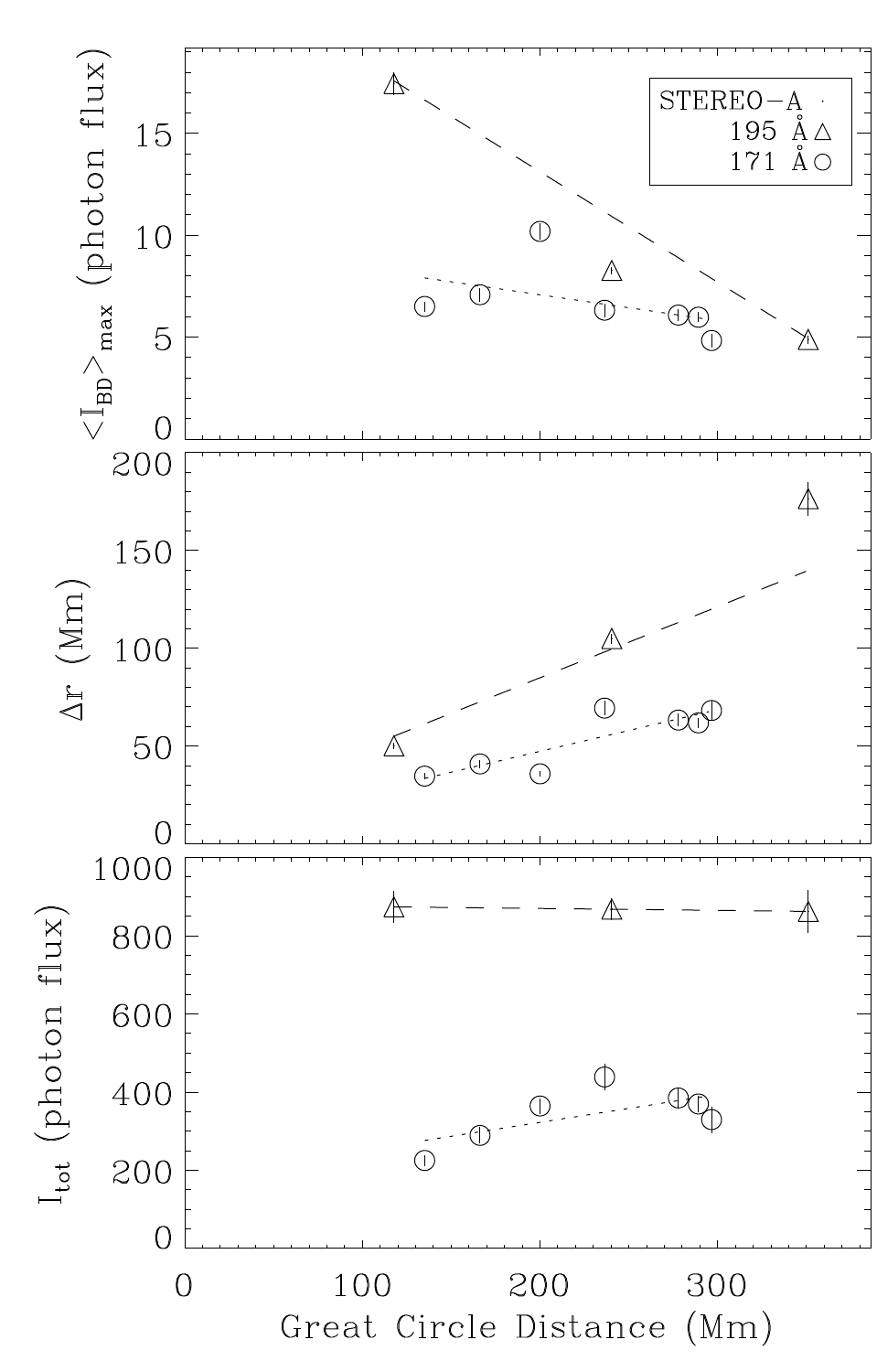}
               }
\caption{Same as Figure~\ref{fig:int_20070519_A} but showing the \emph{STEREO}-A data for the event of 2007~December~07.}
\label{fig:app_intensity_20071207_A}
\end{figure}

\begin{figure}[!p]
\centerline{
    \includegraphics[width=0.7\textwidth,clip=,trim=10mm 0mm 45mm 0mm]{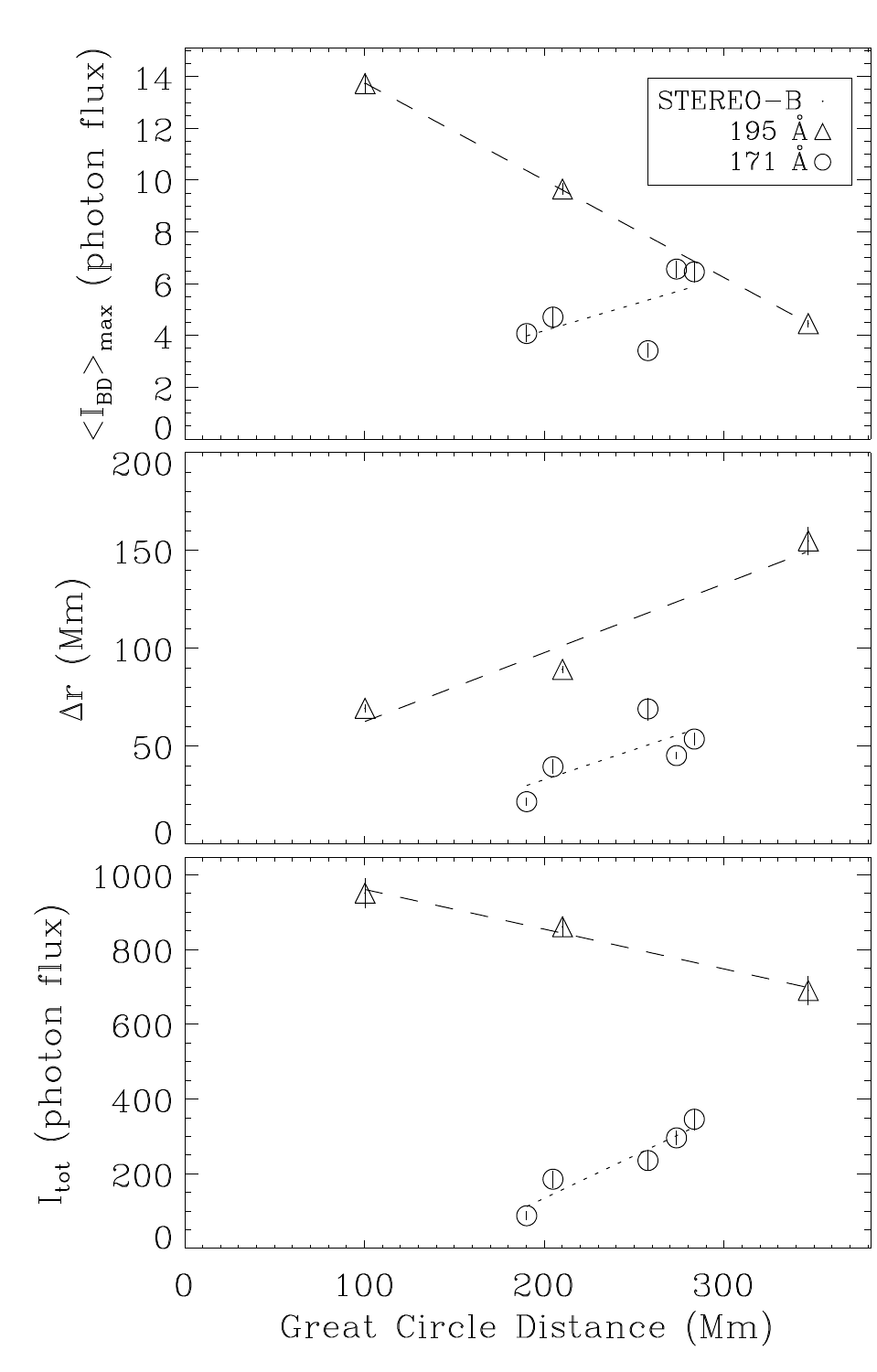}
               }
\caption{Same as Figure~\ref{fig:int_20070519_A} but showing the \emph{STEREO}-B data for the event of 2007~December~07.}
\label{fig:app_intensity_20071207_B}
\end{figure}

\begin{figure}[!p]
\centerline{
    \includegraphics[width=0.7\textwidth,clip=,trim=10mm 0mm 45mm 0mm]{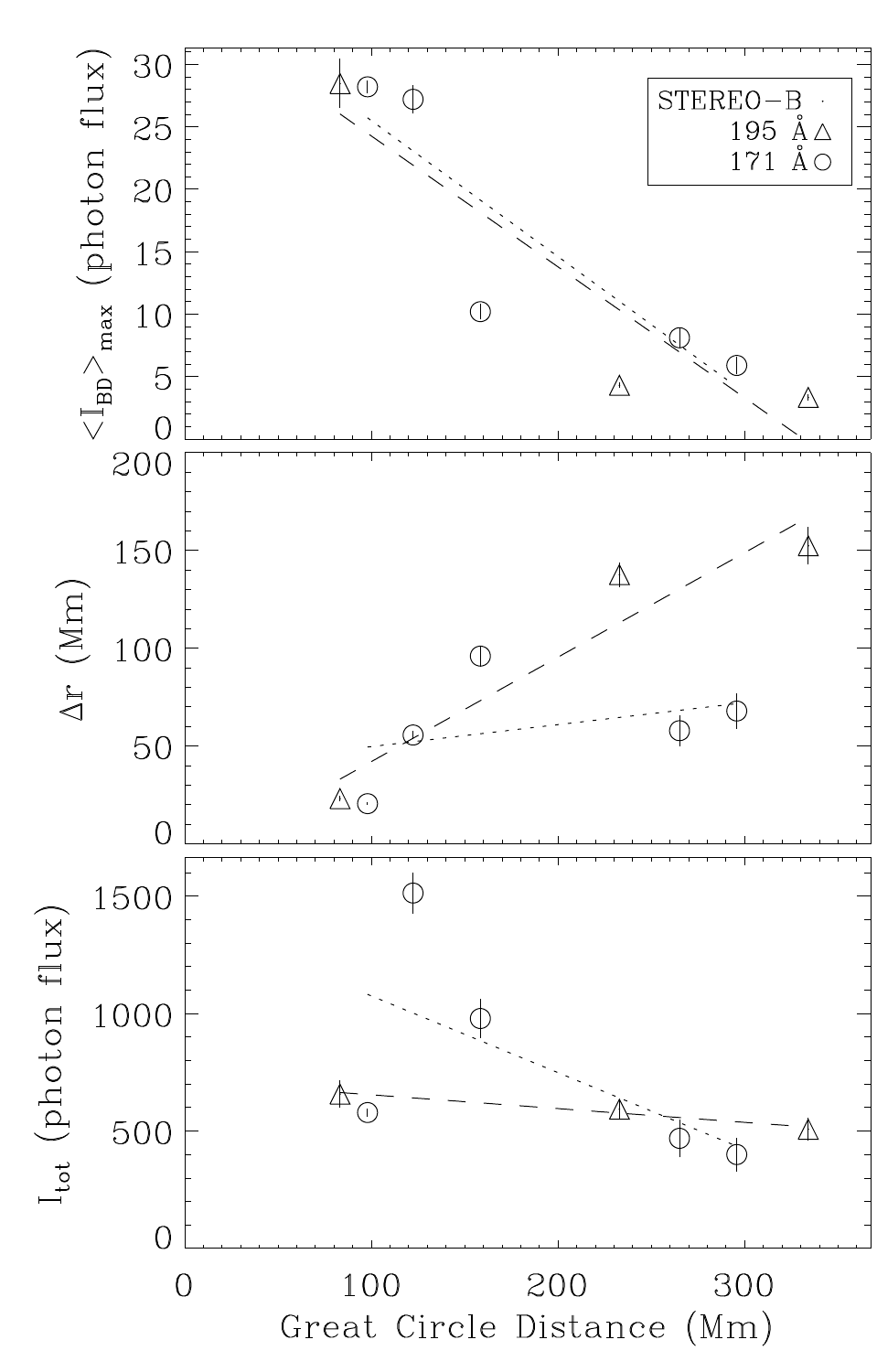}
               }
\caption{Similar to Figure~\ref{fig:int_20070519_A} but showing the \emph{STEREO}-B data for the event on 2009~February~12\corr{.}}
\label{fig:app_intensity_20090212}
\end{figure}

\begin{figure}[!p]
\centerline{
    \includegraphics[width=0.7\textwidth,clip=,trim=10mm 0mm 45mm 0mm]{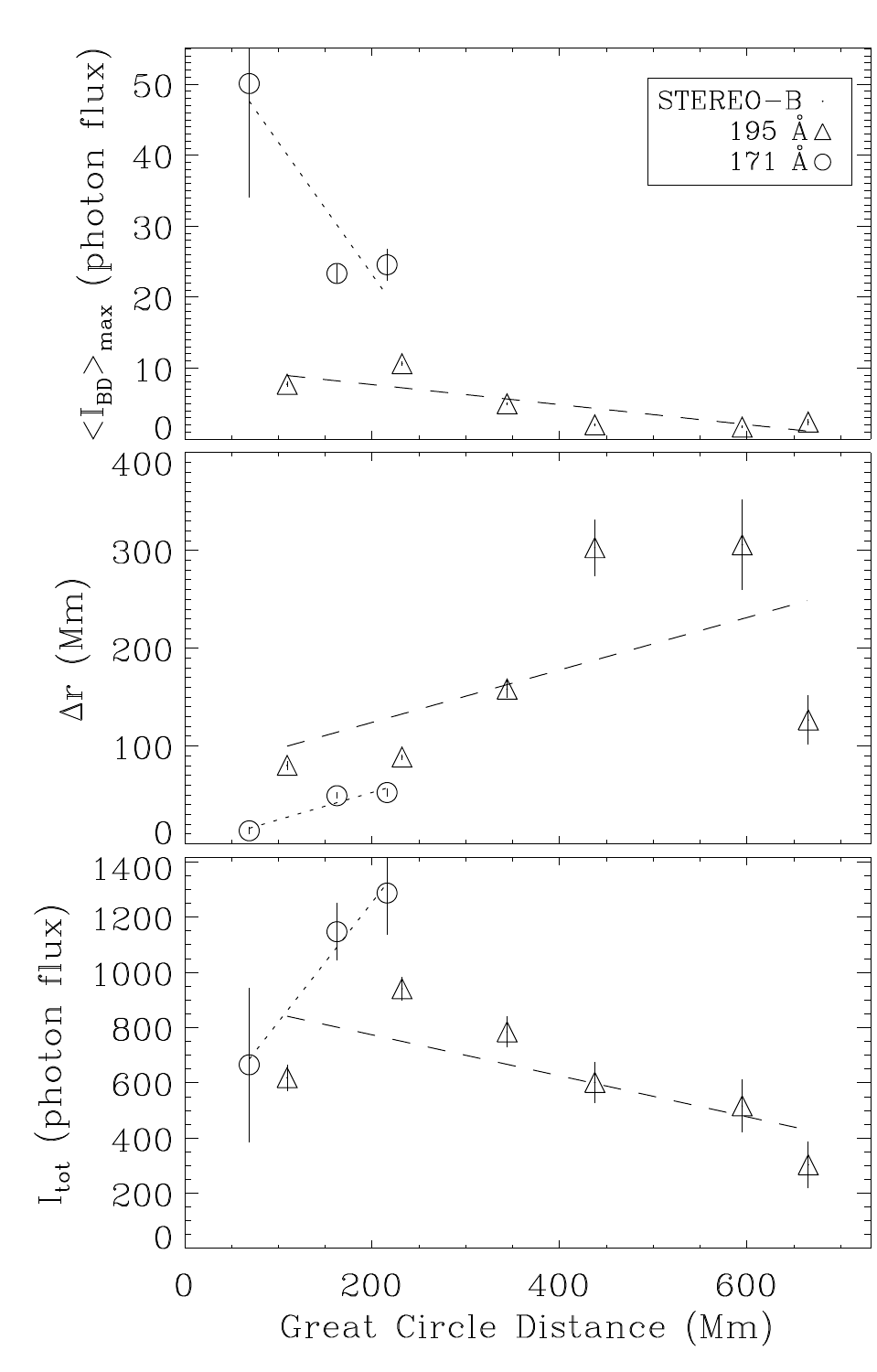}
               }
\caption{Similar to Figure~\ref{fig:int_20070519_A} but showing the \emph{STEREO}-B data for the event on 2009~February~13.}
\label{fig:app_intensity_20090213}
\end{figure}

The derived kinematics and dispersion of the pulse could then be used to calculate and examine the position--averaged integrated intensity of the pulse as discussed in Section~\ref{subsect:wave_properties_intensity}. The resulting plots showing the variation in peak intensity (top), FWHM (middle) and PA-averaged integrated pulse intensity (bottom) with distance for each event studied are shown in Figures~\ref{fig:app_intensity_20071207_A} (2007~December~07 \emph{STEREO}-A), \ref{fig:app_intensity_20071207_B} (2007~December~07 \emph{STEREO}-B), \ref{fig:app_intensity_20090212} (2009~February~12 \emph{STEREO}-B) and \ref{fig:app_intensity_20090213} (2009~February~13 \emph{STEREO}-B).

The variation in PA-averaged integrated pulse intensity with distance is difficult to interpret\corr{,} as apparent from Figures~\ref{fig:app_intensity_20071207_A} to \ref{fig:app_intensity_20090213}. The PA-averaged integrated intensity is flat or decreasing in the 195~\AA\ passband for each event, but the 171~\AA\ measurements are inconclusive. 

For the 2007~December~07 event, the 195~\AA\ measurements as observed by \emph{STEREO}-A appear flat, with a slight increase in the corresponding measurements for \emph{STEREO}-B. On the other hand \corr{in the 171~\AA\ passband,} an apparent increase can be observed in measurements from both \emph{STEREO} spacecraft. A large separation is also \corr{evident} between the 171~\AA\ and 195~\AA\ measurements for this event.

The events from 2009~February~12 and 2009~February~13 are also inconclusive. While both the 171~\AA\ and 195~\AA\ passbands exhibit a decrease for the 2009~February~12 event, the 2009~February~13 event shows an increase and decrease with distance for the 171~\AA\ and 195~\AA\ data respectively. The rate of change of the PA-averaged integrated intensity with distance for each passband is given in Table~\ref{tbl:events_characteristics} for each event studied along with the associated error in each case.

The inconsistency of these measurements between different events and indeed between the different passbands from the same spacecraft for the same event indicate that this property of CBFs is not currently well-constrained. This may be a consequence of the low observing cadence of the \emph{STEREO}/EUVI instrument\corr{, but could also be affected by the inhomogeneous propagation of the pulse and possible temperature variations in the front during its evolution. T}he multi-passband, high-cadence observations afforded by the \emph{Solar Dynamic Observatory} (\emph{SDO}) should allow the true variation (if any) to be determined to a high degree of accuracy. 

\section{Discussions and Conclusions}
\label{section:wave_properties_concs}

A semi-automated technique designed to identify, track and analyse ``EIT waves'' in the solar corona was applied to observations of four CBF events using data from the EUVI telescopes onboard both \emph{STEREO} spacecraft. The technique applies Gaussian fits to intensity profiles across the propagating CBFs, returning the position, width, and PA-averaged integrated intensity of the pulse for each observation. This approach is designed to minimise the errors typically associated with the identification of CBF pulses, allowing a better understanding of the physical properties of the pulse such as the kinematics and rate of pulse dispersion. While this method is similar to CBF identification techniques previously proposed by \citet{Warmuth:2004ab}, \citet{Podladchikova:2005ab}, \citet{Wills-Davey:2006ab} and \citet{Veronig:2010ab}, the combination of this approach with a statistically rigorous bootstrapping method and the high cadence observations afforded by \emph{STEREO}/EUVI, allow the errors typically encountered with the analysis of CBFs to be minimised.

The four events used for this analysis (from 2007~May~19, 2007~December~07, 2009~February~12 and 2009~February~13) were chosen as the erupting active region in each case was close to disk centre, minimising the projection effects of tracking a pulse across a spherical surface. The events were also well observed by the \emph{STEREO} spacecraft, with the close positioning of the \emph{STEREO} spacecraft for the 2007~May~19 and 2007~December~07 events in particular allowing the pulse kinematics to be derived using both spacecraft.

All of the pulses studied were found to exhibit statistically significant deceleration ranging from $\sim -300$ to $-50$~m~s$^{-2}$, with initial velocities ranging from $\sim$250--450~km~s$^{-1}$. This is comparable to previous estimates of CBF velocities, and is interesting given that most previous work has used point-and-click techniques applied to running-difference images. This approach identifies the forward edge of the CBF at a given time, using this to determine the kinematics of the disturbance as a whole. The analyses performed using such techniques have mainly returned kinematics that suggest a zero acceleration (i.e., constant velocity) interpretation of the CBF phenomenon. In contrast, our semi-automated technique uses the pulse centroid derived from percentage base difference intensity profiles to return a constant, non-zero acceleration (i.e., variable velocity) interpretation. 

This discrepancy can be understood by considering the physical processes at work in the propagating pulse. If the pulse is broadening with propagation, as found here, the forward edge of a decelerating pulse will appear to move faster than the pulse centroid. If the rate of pulse broadening is comparable to the deceleration of the pulse, the deceleration may be masked by this broadening. This suggests that the true kinematics of CBFs may have previously been disguised through use of the wrong position within the pulse profile to characterise its location. It should be noted that a variable acceleration may be present, but this can only be studied with an increased number of data points.

A positive increase in the variation of the width of the Gaussian fit to the CBF pulse in both the spatial and temporal domains was observed for all events studied. Although the increase in spatial pulse width is immediately suggestive of a dispersive pulse, this property is confirmed by the observed increase in the temporal width of the pulse. As shown in Equation~\ref{eqn:t_width}, an increase in the spatial pulse width may be negated by an increase in pulse velocity, producing a pulse with a constant temporal width. It is apparent from Figures~\ref{fig:space_broadening_20070519}, \ref{fig:time_broadening_20070519}, and \ref{fig:app_space_broadening_20071207} to \ref{fig:app_time_broadening_200902} that this is not the case and the observed CBF pulses are dispersive in nature. Although some previous studies have suggested a dispersive nature for ``EIT waves'' \citep[e.g.,][]{Warmuth:2004rm}, the true extent of this dispersion may have been disguised by the use of point-and-click techniques to analyse running-difference images. 

The confirmation of pulse broadening resulted in the variation in PA-averaged integrated pulse intensity with distance being studied rather than the variation in peak pulse intensity. Although the variation in the peak intensity of the pulse with distance is generally considered an indicator of the pulse energy, the dispersion can cause the peak of the pulse to drop without reducing the energy contained within it. The PA-averaged integrated pulse intensity was found by multiplying the FWHM by the peak pulse intensity at a given time. This was found to produce inconclusive results between different events and also between different passbands for the same event. While the peak intensity of the pulse was generally observed to decrease with distance and the FWHM was observed to increase with distance in both passbands, the PA-averaged integrated intensity typically showed no \corr{systematically} strong variation with distance in either passband, although a strong offset between the 171~\AA\ and 195~\AA\ data was typically observed. Contrasting results were found for the 2009~February~12 and 2009~February~13 events respectively, suggesting that further investigation using simultaneous analysis of multiple passbands at very high cadence is required, something that will be routinely available from the \emph{SDO} spacecraft.

The analysis outlined in this chapter indicates that the studied CBFs may be viewed as dispersive pulses exhibiting negative acceleration. This interpretation is consistent with the fast-mode magnetoacoustic wave interpretation for a freely-propagating ``EIT wave''. However, the inconclusive variation in the integrated intensity of the pulse indicates that more analysis is required to definitively determine the physical nature of the disturbance.

Although only four CBF events were considered here, the consistency of the results between the events studied suggests that the conclusions drawn here may be applicable to a larger sample of CBFs. The initial velocities of the CBFs are comparable to the lower range of estimated Alfv\'{e}n speeds proposed by \citet{Wills-Davey:2007oa}, suggesting that the randomly structured nature of the coronal magnetic field may have an important effect on the propagation of CBFs; an effect shown in simulations performed by \citet{Murawski:2001ab}. Although the kinematics and dispersive nature of the pulses suggest that a magnetoacoustic wave interpretation may be correct, further study is required. The $\sim$10~s temporal cadence across multiple EUV passbands available with the launch of the \emph{SDO} spacecraft will allow these results to be studied in much greater detail.		


\chapter{The Wave Properties of Coronal Bright Fronts Observed Using \emph{SDO}/AIA} 
\label{chap:sdo_cbf}


\noindent 
\\ {\it 
In a dispersive medium, the components of a wave group move with different speeds, and the phase relations between the components are altered
\begin{flushright}
R. W. Ditchburn\\
\end{flushright}
 }

\vspace{15mm}
Coronal bright fronts are large scale wavefronts that propagate through the solar corona at hundreds of kilometers per second. While their kinematics have been studied in detail, many questions remain regarding the temporal evolution of their amplitude and pulse width. In this chapter, contemporaneous high cadence, multi-thermal observations of the solar corona from the \emph{SDO} and \emph{STEREO} spacecraft are used to determine the kinematics and expansion rate of a CBF pulse observed on 2010~August~14. 

The CBF was found to have a lower initial velocity with weaker deceleration in \emph{STEREO} observations compared to \emph{SDO} ($\sim$340~km~s$^{-1}$ and $-72$~m~s$^{-2}$ as opposed to $\sim$410~km~s$^{-1}$ and $-279$~m~s$^{-2}$\corr{; cf. Chapter~\ref{chap:first_obs}}). The CBF kinematics from \emph{SDO} were found to be highly passband-dependent, with an initial velocity ranging from $379\pm12$~km~s$^{-1}$ to $460\pm28$~km~s$^{-1}$ and acceleration ranging from $-128\pm28$~m~s$^{-2}$ to $-431\pm86$~m~s$^{-2}$ in the 335~\AA\ and 304~\AA\ passbands respectively. These kinematics were used to estimate a quiet coronal magnetic field strength range of $\sim$1--2~G. Significant pulse broadening was also observed, with expansion rates of $\sim$130~km~s$^{-1}$ (\emph{STEREO}) and $\sim$140~km~s$^{-1}$ (\emph{SDO}). By treating the CBF as a linear superposition of sinusoidal waves within a Gaussian envelope, the resulting dispersion rate of the pulse was found to be $\sim$10--15~Mm$^2$~s$^{-1}$. These results are indicative of a fast-mode magnetoacoustic wave pulse propagating through an inhomogeneous medium. This chapter contains work by \citet*{Long:2011ys} \corr{published} in \emph{The Astrophysical Journal Letters}.

\section{Introduction}
\label{sect:sdo_cbf_intro}

Although CBFs were first noted by \citet{Moses:1997vn} and characterised by \citet{Thompson:1998sf}, they remain a source of much debate with many different theories proposed to explain the pheomenon. This is partly a consequence of their rare nature but is also due to the instruments used to identify and study them, with \emph{SOHO} \citep{Thompson:1999cd}, \emph{TRACE} \citep{Wills-Davey:1999ve} and \emph{STEREO} \citep{Long:2008eu} offering a choice between low cadence observations with a large field of view (\emph{SOHO} and \emph{STEREO}) or high cadence observations with a small field of view (\emph{TRACE}). 

Despite these restrictions, CBFs have been relatively well characterised. They are usually observed as diffuse bright fronts that propagate isotropically when unimpeded at typical velocities of 200--400~km~s$^{-1}$ across the solar disk \citep{Thompson:2009yq}. Optimum observations occur using the 195~\AA\ (\ion{Fe}{12}) passband, corresponding to a temperature of $\sim$1--2~MK and a height of $\sim$70--90~Mm above the photosphere \citep[i.e.,\ the low corona;][]{Patsourakos:2009ab,Kienreich:2009ab}. However, they have been observed using other coronal EUV passbands including the 171~\AA\ \citep[\ion{Fe}{9};][]{Wills-Davey:1999ve}, 284~\AA\ \citep[\ion{Fe}{15};][]{Zhukov:2004kh} and 304~\AA\ \citep[\ion{He}{2};][]{Long:2008eu} passbands.

Although the observations of CBFs have been mostly consistent, they have given rise to a multitude of theoretical interpretations. They have been alternatively interpreted as magnetohydrodynamic waves \citep{Wang:2000tg,Warmuth:2004ab,Wang:2009ab,Schmidt:2010ab}, solitons \citep{Wills-Davey:2007oa} and in terms of magnetic field restructuring during the eruption of an associated CME \citep{Chen:2002rw, Attrill:2007vn, Delannee:2008uq}. This proliferation of theories for interpreting CBFs is a direct consequence of the traditional techniques used to study the phenomenon, which produce strongly user--defined conclusions. This has resulted in different authors deriving different results for the same event \citep[e.g.,][]{Thompson:1998sf,Attrill:2007vn,Delannee:2008uq}. 

This combination of limited identification techniques with the low cadence of previous observations has produced results that imply a wave interpretation \citep[such as pulse reflection and refraction;][]{Veronig:2006fy,Gopalswamy:2009p1527}, but with \corr{velocities} that are lower than those predicted by this interpretation. However, observations of decelerating CBFs combined with the effects of low observing cadence \citep{Long:2008eu,Ma:2009ab,Long:2011ab} suggest that the low observing cadence and techniques used to identify the pulses may influence the derived kinematics. There have also been indications of CBF dispersion with propagation \citep{Warmuth:2004ab,Long:2011ab}, although this has been difficult to quantify. While these properties are inconsistent with ideal MHD wave theory, they have been shown in simulations by \citet{Murawski:2001ab} and \citet{Nakariakov:2005ab} to be a natural result of propagation through an inhomogeneous medium. \corr{These observations of deceleration, dispersion and dissipation are also consistent with the interpretation of CBFs as large--amplitude waves as a direct consequence of their inherently non--linear nature \citep{Vrsnak:2008p622,Vrsnak:2000p15,Warmuth:2010p2822}. In this case, the rapid expansion of an erupting CME or rising flare loops induce a large--amplitude disturbance that propagates as a simple wave before steepening and forming a discontinuity. As the shock propagates, its shocked edge moves faster than the trailing edge, while the amplitude drops as a consequence of geometrical expansion before the shock ultimately decays to an ordinary small--amplitude pulse.} A more complete description of the different observations of CBFs and their resulting theoretical interpretations can be found in Chapter~\ref{chap:cbfs}, as well as in the many recent reviews of the field by \citet{Gallagher:2011fk}, \citet{Zhukov:2011rt}, \citet{Warmuth:2010p2822} and \citet{Wills-Davey:2010ab}.

The launch of the \corr{\emph{Solar Dynamics Observatory}} (\emph{SDO}) spacecraft in 2011 marked the beginning of a new age of solar physics, with the Sun now visible in more detail than ever before. \emph{SDO}/AIA observes the Sun continuously at a cadence of $\sim$12~s in seven EUV passbands, an improvement on both \emph{SOHO}/EIT ($\sim$900~s in one of four passbands) and \emph{STEREO}/EUVI ($\sim$75--600~s in four passbands). As well as the 171~\AA, 193~\AA\footnote{The \ion{Fe}{12} passband peaks at 195~\AA\ in \emph{STEREO}/EUVI, but at 193~\AA\ in \emph{SDO}/AIA. As a result, all references to the 193~\AA\ passband will refer to \emph{SDO} data, with the 195~\AA\ passband referring to \emph{STEREO} data.} and 304~\AA\ passbands observed by \emph{SOHO} and \emph{STEREO}, \emph{SDO} also utilises the 94~\AA\ (\ion{Fe}{18}), 131~\AA\ (\ion{Fe}{8}), 211~\AA\ (\ion{Fe}{14}) and 335~\AA\ (\ion{Fe}{16}) passbands, allowing the Sun to be studied across a wide range of temperatures and heights.

The first observations of a CBF using \emph{SDO} were made by \citet{Liu:2010ab}, who noted the presence of the pulse in all EUV passbands monitored by the spacecraft\corr{, although the pulse was only analysed in detail using the 171~\AA\ and 193~\AA\ passbands. This was} partly due to the sheer volume of data available from \emph{SDO} ($\sim$1.5~TB per day) which is not conducive to manual data analysis. This data volume has necessitated the development of both automated and semi-automated CBF detection and tracking algorithms \citep{Podladchikova:2005ab,Wills-Davey:2006ab,Long:2011ab} to allow clarification of the true nature of CBFs. 

In this chapter, the semi-automated CBF algorithm discussed in Sections~\ref{subsect:cor_pita} and \ref{sect:wave_properties_cor_pita}, and published by \citet{Long:2011ab} is applied to \emph{SDO} and \emph{STEREO} observations of the 2010~August~14 CBF event. The observations and data used for this analysis are outlined in Section~\ref{sect:sdo_cbf_obs}. This allowed the physical nature of the pulse to be studied in detail, with the kinematics of the pulse and variation in pulse width discussed in Sections~\ref{subsect:sdo_cbf_kinematics} and \ref{subsect:sdo_cbf_broadening}. The multi-thermal nature of \emph{SDO} allowed the temperature dependence of the pulse to be examined; this discussion is presented in Section~\ref{subsect:sdo_cbf_discrepancies}. The variation in pulse kinematics, pulse width and the multi--thermal nature of the pulse were then combined to study the local plasma through which the pulse propagated, allowing the coronal magnetic field strength to be estimated (Section~\ref{subsect:sdo_cbf_seismology}). Finally, some conclusions regarding the implications of this work are discussed in Section~\ref{sect:sdo_cbf_disc_and_conc}.

\section{Observations and Data Analysis}
\label{sect:sdo_cbf_obs}

The CBF event observed on 2010~Aug~14\footnote{Solar Object Locator: SOL2010-08-14T09:38:00L353C79} erupted from NOAA active region (AR) 11093, with an associated coronal mass ejection (CME) and GOES C4.4 flare which started at 09:38~UT. The position of the active region (N11W65) meant that the on-disk CBF evolution was visible from \emph{STEREO}-A and \emph{SDO} but not \emph{STEREO}-B due to the relative positions of the \emph{STEREO} spacecraft. At the time of the eruption, \emph{STEREO}/EUVI-A was operating with an observing cadence of 300~s and 600~s in the 195~\AA\ and 304~\AA\ passbands respectively, although the 171~\AA\ and 284~\AA\ passbands were both taking synoptic data (i.e.,\ one image every two hours). In contrast, \emph{SDO}/AIA was taking observations with 12~s cadence in all seven EUV passbands (94, 131, 171, 193, 211, 304, and 335~\AA) over the same time period. 

As previously discussed in Sections~\ref{subsect:cor_pita} and \ref{sect:wave_properties_cor_pita}, the semi-automated detection algorithm used to identify and track the CBF has several steps. The algorithm was applied to data from both the \emph{STEREO} and \emph{SDO} spacecraft \corr{using an arc--sector of 30$^{\circ}$ width in both cases}, with the routines used to read in and prepare the data the only modifications made to the algorithm. This was necessary as the different data require different processing to achieve scientific quality (see Sections~\ref{subsect:euvi_software} and \ref{subsect:aia_software} for more details). 

\begin{landscape}
\begin{figure}[!p]
\centering
   \includegraphics[width=1.3\textwidth,clip=,trim=0mm 0mm 0mm 85mm]{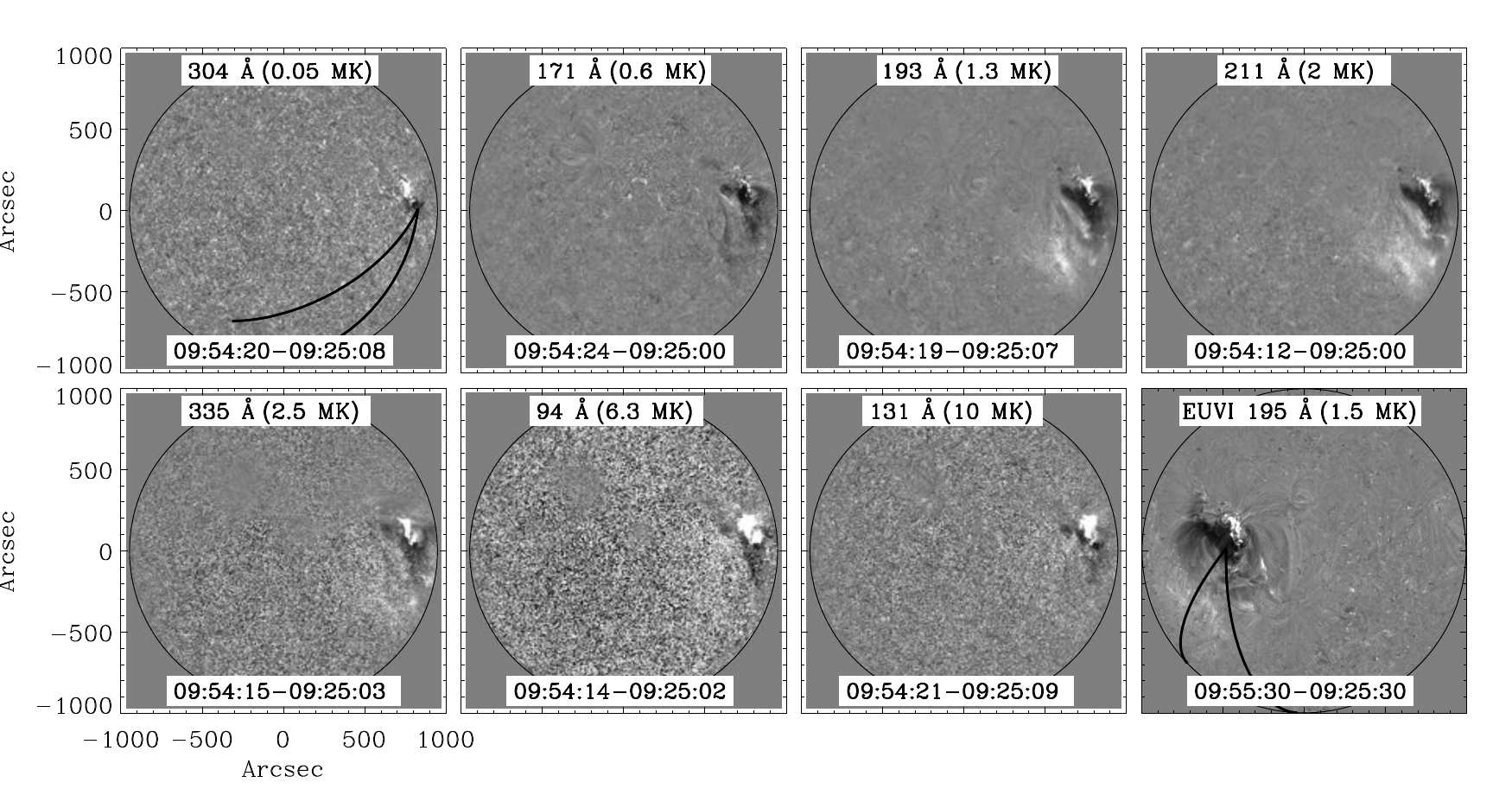}
\caption{PBD images for all \emph{SDO}/AIA passbands and the \emph{STEREO}/EUVI 195~\AA\ passband. Image times used are given on the bottom of each panel. The arc sectors used to identify the pulse are marked in the AIA 304~\AA\ and EUVI 195~\AA\ panels respectively.}
\label{fig:image_panel}
\end{figure}
\end{landscape}

The source point for the CBF was defined using the mean centre of ellipses fitted to the first three observations of the CBF in both 193~\AA\ (AIA) and 195~\AA\ (EUVI) data, giving a source unique to each spacecraft (although both sources are comparable when transformed between spacecraft). Percentage base difference images were used for this analysis, with the images de--rotated in each case to the same pre--event time (09:20:30~UT) to compensate for solar rotation. The base image for each passband was defined using a pre-event time of $\sim$09:25:00~UT, with a sample of the resulting images shown in Figure~\ref{fig:image_panel}. The arc sector employed by the algorithm to identify the pulse was then positioned to allow direct comparison of both AIA and EUVI observations and to ensure the maximum number of observations using both instruments.

The operation of the algorithm is similar to that previously described, with the PBD intensity of a given image averaged across the position angle of the arc sector in annuli of increasing radii, with the standard deviation giving the associated error. The width of the annuli used varied between the spacecraft as a consequence of the different spatial resolution available with 1~degree annuli used for \emph{STEREO} and 0.5~degree annuli used for \emph{SDO}. This had the effect of increasing the detail of the resulting intensity profiles from \emph{SDO}, but did not affect the results of the analysis.

The resulting intensity profile was then fitted using a Gaussian function, with the centroid and full width at half maximum (FWHM) giving the pulse position and width respectively. Each parameter was returned with an associated error allowing the ability of the algorithm to detect the pulse to be quantified. Although the source point position and orientation of the arc sector is determined by the user, all subsequent detection, tracking and analysis of the pulse is automated, allowing unbiased identification of the CBF. Once the intensity profiles for each image have been produced and fitted, the CBF is identified as a moving pulse, with any stationary bright features ignored.

\section{Results}
\label{sect:sdo_cbf_results}

\begin{figure}[!t]
\centering
   \includegraphics[width=1\textwidth,clip=,trim=0mm 0mm 0mm 0mm]{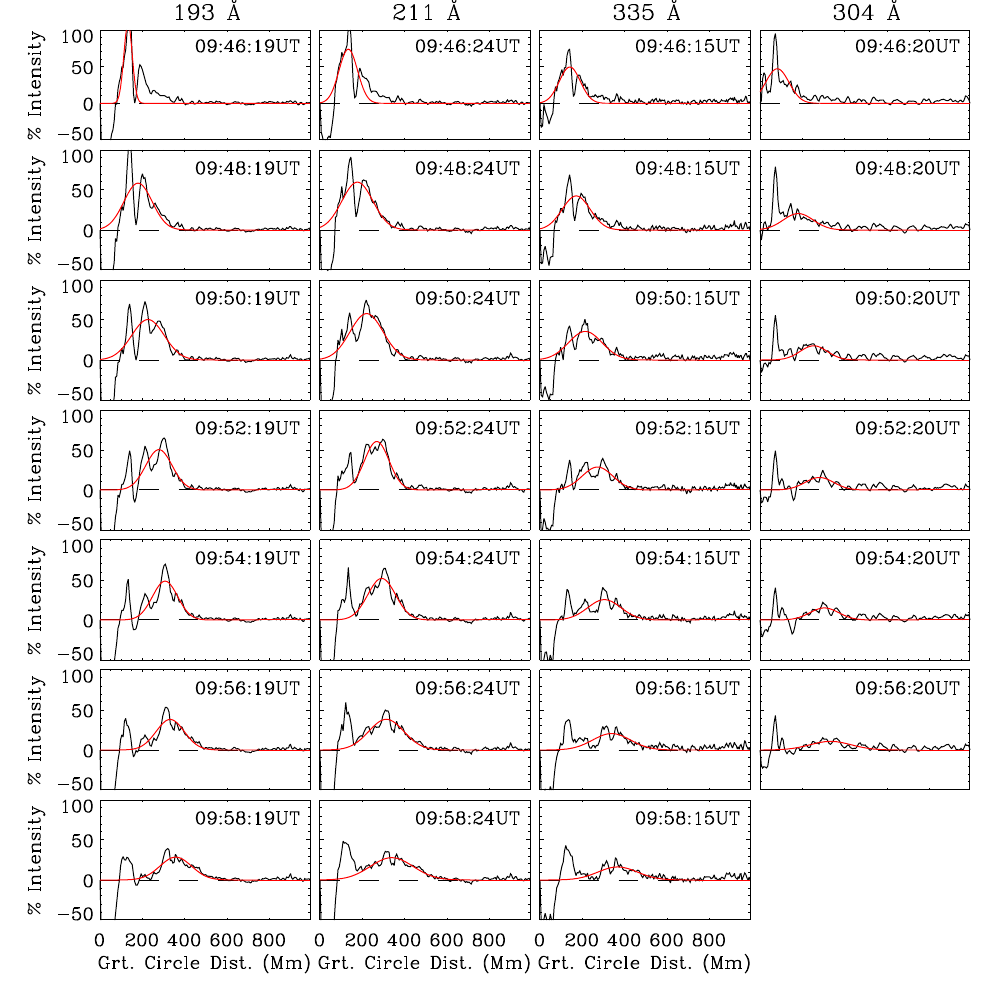}
\caption{\corr{Variation in \emph{SDO} percentage intensity profiles with Great Circle distance (Mm) in the (from left to right) 193, 211, 335 and 304~\AA\ passbands, with the Gaussian fit applied by the algorithm to the data in each case highlighted in red. Time is increasing from top to bottom, with the time of each intensity profile given in the upper right of each panel.}}
\label{fig:sdo_profs}
\end{figure}

The actual identification of the pulse was found to be strongly influenced by the passband used rather than instrument, with some evidence of a disturbance visible in all available passbands. The pulse was observed in the 195~\AA\ and 304~\AA\ passbands from \emph{STEREO}/EUVI, although only the 195~\AA\ data was used here due to the low cadence of the 304~\AA\ passband. In \emph{SDO}/AIA, it was possible to track the pulse in four of the seven passbands (193, 211, 304, and 335~\AA). The nature of the 94~\AA\ and 131~\AA\ passbands complicated identification of the pulse, while a slight intensity decrease was visually identified in the 171~\AA\ passband. However, it was not possible to track this using the detection algorithm. \corr{A sample of the resulting intensity profiles from the four passbands used in this work is shown in Figure~\ref{fig:sdo_profs}.}

\subsection{Pulse Kinematics}
\label{subsect:sdo_cbf_kinematics}

\begin{figure}[!t]
\centering
   \includegraphics[width=1\textwidth,clip=,trim=2mm 2mm 5mm 0mm]{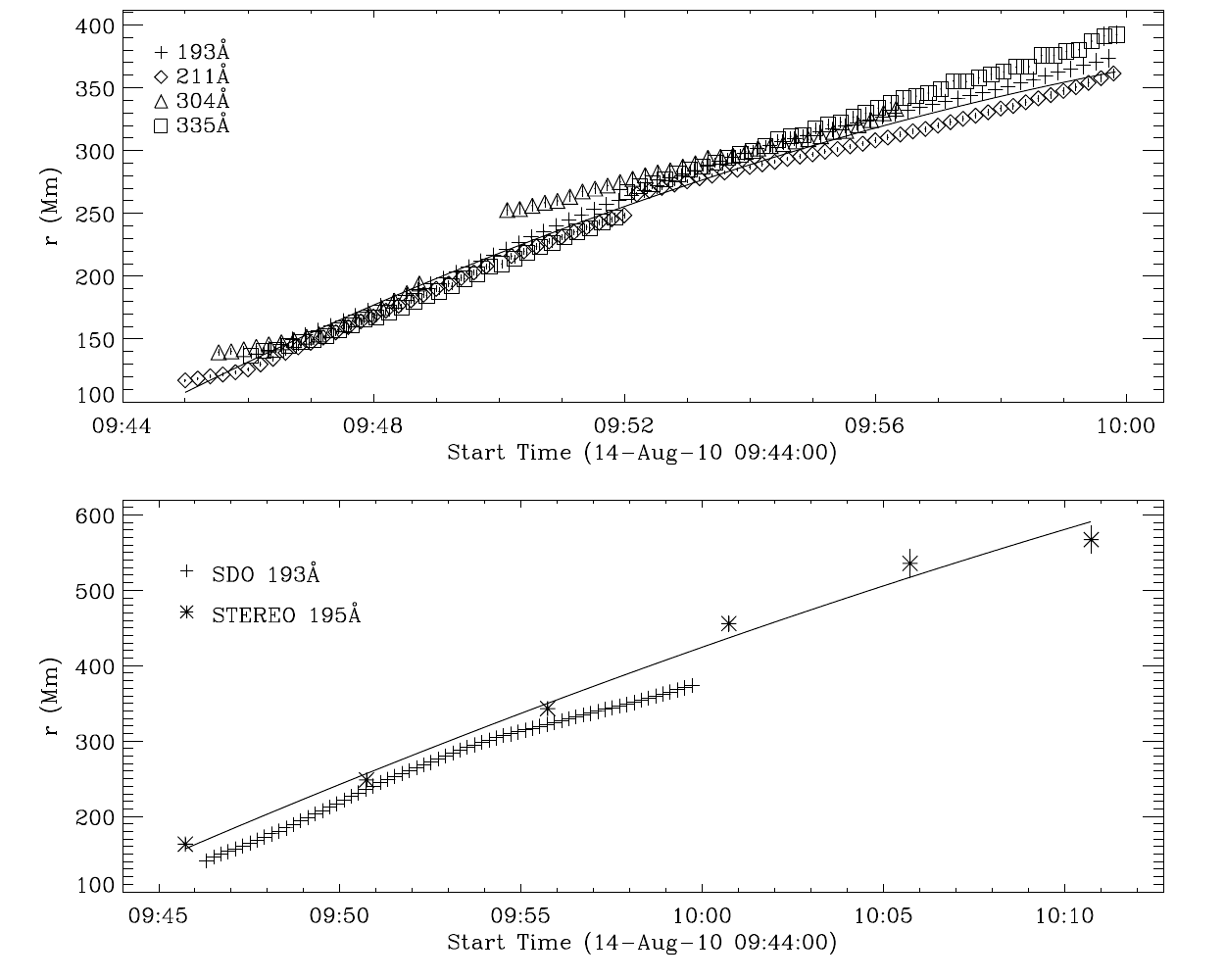}
\caption{\emph{Top}: Distance-time measurements from AIA 304, 193, 211, and 335~\AA\ passbands for the 2010~August~14 CBF event. \emph{Bottom}: AIA 193~\AA\ and EUVI 195~\AA\ distance-time measurements for the same event (line shows the best fit to EUVI measurements).}
\label{fig:cbf_kins}
\end{figure}

The temporal variation in the distance of the fitted pulse centroid from the defined source point was used to determine the kinematics of the pulse using measurements from all passbands studied. This was done by using the bootstrapping technique described in Section~\ref{subsect:bootstrap} to fit a model of the form,
\begin{equation}
r(t) = r_{0} + v_{0}t + \frac{1}{2}at^2
\end{equation}
to the distance--time measurements. Here $r_{0}$ describes the initial distance of the pulse from the source point, $v_{0}$ is the initial velocity and $a$ is the constant acceleration. 

The temporal variation in the distance measurements for each of the four \emph{SDO} passbands studied is shown in the top panel of Figure~\ref{fig:cbf_kins}, while the \emph{STEREO} 195~\AA\ and \emph{SDO} 193~\AA\ measurements are compared in the bottom panel of the same figure. The kinematics of the individual \emph{SDO} passbands are seen to be comparable, although there does appear to be some separation with propagation. \corr{The apparent discontinuities in individual \emph{SDO} passbands seen in Figure~\ref{fig:cbf_kins} are the result of a stationary bright feature close behind the detected pulse which influenced the detection of the pulse by the algorithm, while there was a data--gap in the 304~\AA\ passband which also complicated detection of the pulse. These effects were minimal for the distance--time measurements but exhibited a much stronger influence on the variation in pulse width with time as discussed in Section~\ref{subsect:sdo_cbf_broadening}.} The measurements from the similar 193~\AA\ and 195~\AA\ passbands \corr{shown in the bottom panel of Figure~\ref{fig:cbf_kins}} also appear homologous, with a slight positional offset due to the different spacecraft positions. 

\begin{deluxetable}{cccccccc}
\tablecolumns{7}
\rotate
\tabletypesize{\footnotesize}
\tablewidth{0pt}
\centering
\tablecaption{2010~August~14 CBF properties\label{tbl:cbf_kins}}
\tablehead{
\colhead{Spacecraft} & \colhead{Passband} & \colhead{$T_{peak}$\tablenotemark{a}} & \colhead{$v_{0}$} & \colhead{$a_{0}$} & \colhead{Expansion Rate} & \colhead{$d^2 \omega(k_{0})/dk^{2}$} & \colhead{$v_{\mathrm{final}}$} \\ 
\colhead{} & \colhead{\AA} & \colhead{MK} & \colhead{km~s$^{-1}$} & \colhead{m~s$^{-2}$} & \colhead{km~s$^{-1}$} & \colhead{Mm$^{2}$~s$^{-1}$} & \colhead{km~s$^{-1}$}
}
\startdata
{\it STEREO}-A 	& 195 					& 1.5		& $343\pm52$	& $-71\pm69$ 	& $130.6\pm12.3$ 	& $20.80\pm2.08$	& \nodata \\ 
{\it SDO} 		& All\tablenotemark{b} 	& \nodata	& $411\pm17$ 	& $-279\pm36$ 	& $222.0\pm1.8$ 	& $10.38\pm0.20$	& \nodata \\
 				& 335 					& 2.5		& $379\pm12$ 	& $-128\pm28$ 	& $211.5\pm4.9$ 	& $13.32\pm0.53$	& $273\pm35$ \\ 
 				& 211 					& 1.8		& $409\pm11$ 	& $-298\pm24$ 	& $238.1\pm2.3$ 	& $8.37\pm0.27$		& $144\pm32$ \\ 
 				& 193 					& 1.6		& $419\pm5$ 	& $-318\pm13$ 	& $190.4\pm5.2$	 	& $11.05\pm0.55$	& $163\pm15$ \\ 
 				& 304 					& 0.05		& $460\pm28$ 	& $-431\pm86$ 	& $214.4\pm7.1$ 	& $13.17\pm0.68$	& $181\pm84$ \\ 
\enddata
\tablenotetext{a}{$T_{peak}$ here refers to the peak emission temperature of each passband.}
\tablenotetext{b}{Distance-time measurements for all passbands observed by AIA were combined for comparison.}
\end{deluxetable}

The estimated kinematics derived using the bootstrapping technique are presented in Table~\ref{tbl:cbf_kins}. Each of the passbands studied exhibits statistically significant \corr{negative acceleration}, with the higher cadence of the \emph{SDO} data contributing to the lower associated errors for each parameter. A comparison of the similar \emph{STEREO} 195~\AA\ and \emph{SDO} 193~\AA\ passbands shows that the 195~\AA\ passband exhibits a lower initial velocity and much weaker acceleration relative to the comparable 193~\AA\ passband. The kinematic estimates from \emph{STEREO} 195~\AA\ are consistent with previous results derived using \emph{SOHO}/EIT and \emph{STEREO}/EUVI, while the higher initial velocity and acceleration from \emph{SDO} 193~\AA\ for the same event suggests a strong influence from the cadence of the observing instrument, confirming the results of \citet{Long:2008eu} and \citet{Ma:2009ab}.

The passbands studied by \emph{SDO} exhibit consistently higher initial velocities and stronger \corr{negative accelerations} than previously estimated for CBF pulses, suggesting that the pulses may be consistent with MHD waves. The larger uncertainties associated with the kinematics analysis of the 304~\AA\ passband may be explained by the nature of the passband and also by a data gap, which complicated detection of the pulse. The discrepancy between the kinematics derived for each \emph{SDO} passband is interesting and will be discussed further in Section~\ref{subsect:sdo_cbf_discrepancies}.

\subsection{Pulse Broadening}
\label{subsect:sdo_cbf_broadening}

\begin{figure}[!t]
\centering
   \includegraphics[width=1\textwidth,clip=,trim=2mm 2mm 4mm 0mm]{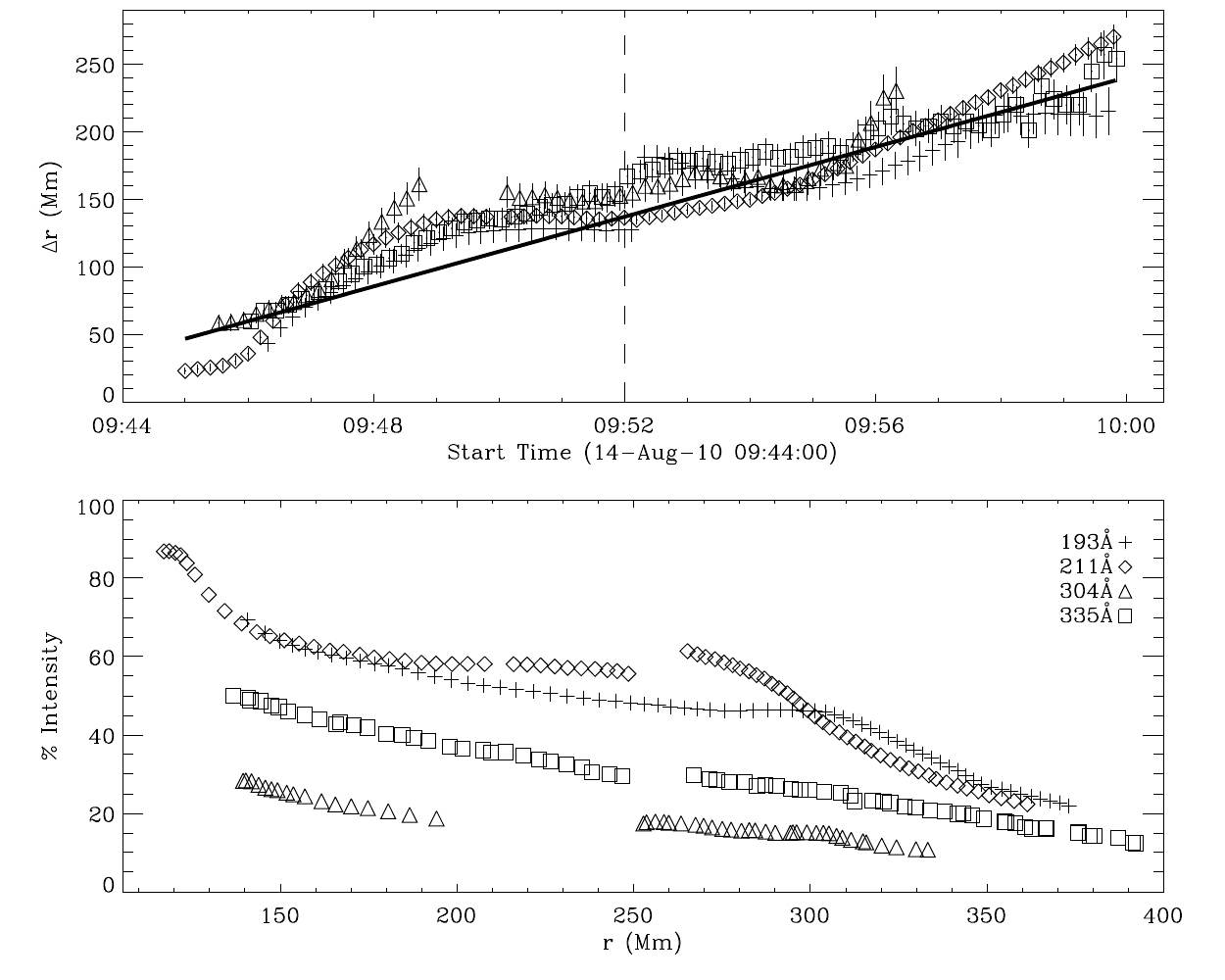}
\caption{\emph{Top}: Variation in FWHM with time for AIA passbands; EUVI 195~\AA\ measurements show a similarly increasing trend over a much longer time range. The black line shows the fit to the combined \emph{SDO} data. \emph{Bottom}: Peak PBD pulse intensity variation with distance.}
\label{fig:cbf_disp}
\end{figure}

The temporal variation in the FWHM of the fit to the intensity profile was next examined for evidence of pulse broadening. This variation is shown in the top panel of Figure~\ref{fig:cbf_disp}, which indicates that the pulse width changes from $\sim$70~Mm to $\sim$270~Mm over a time period of $\sim$900~s. This increase was observed for all passbands studied, with the \emph{STEREO} 195~\AA\ measurements also showing a similar increase, albeit over a longer timescale.

The data prior to $\sim$09:52~UT (to the left of the dashed vertical line) has been corrected to remove the effects of a stationary bright feature close behind the CBF \corr{\citep[cf.][]{Muhr:2011fk}}. This feature initially exerts a strong influence on the Gaussian fit, but was negated by subtracting a constant offset value for each passband from the FWHM measurements. From $\sim$09:52~UT onward, the CBF is sufficiently far from this feature that the fit to the data is no longer affected. The effects of this bright feature can also be seen in the bottom panel in Figure~\ref{fig:cbf_disp}, which shows the peak \% intensity variation with distance. While the bright feature does initially influence the pulse width and peak intensity variation, a general increase and decrease is apparent for the pulse width and peak intensity respectively.

To try \corr{to} understand the physical nature of the observed pulse dispersion, the CBF was treated as a linear superposition of sinusoidal waves within a Gaussian envelope. This can be represented using the equation,
\begin{equation}
\psi(r,t) \simeq \textrm{exp}\left(-\frac{(r - v_{g}t)^{2}}{2\sigma^{2}_{r}}\right)\textrm{cos}(k_{0}r - \omega_{0} t)
\end{equation}
where $k_{0}$ is the wavenumber, $\omega_{0}$ is the angular frequency, $\sigma_{r}$ is the characteristic width and $v_{g}$ is the group velocity of the pulse ($v_{g} = d\omega/dk$). The pulse extends in Fourier space from $k_{0} - \Delta k/2$ to $k_{0} + \Delta k/2$ ($\Delta k \sim \sigma_k$ where $\sigma_k = 1/\sigma_r$), so that the velocity varies from $v_{g}(k_{0} - \Delta k/2)$ to $v_{g}(k_{0} + \Delta k/2)$ across the pulse. As the pulse propagates, this velocity gradient produces an increase in the spatial extent of the pulse, where the spatial extent (FWHM) is defined as,
\begin{equation}
\Delta r(t) = \Delta r_{0} + \left[v_{g}\left(k_{0} + \frac{\Delta k}{2}\right) - v_{g}\left(k_{0} - \frac{\Delta k}{2}\right)\right]t
\end{equation}
and $\Delta r_{0}$ is the initial pulse width. This can be rewritten in terms of the change in group velocity $v_{g}$ as,
\begin{equation}
\Delta r(t) \sim \Delta r_{0} + \frac{dv_{g}(k_{0})}{dk}\Delta kt.
\end{equation}
As the group velocity $v_{g} = d\omega/dk$, this can be redefined to give the width of a dispersive pulse at any time $t$ as,
\begin{equation}
\Delta r(t) = \Delta r_{0} + \frac{d^2 \omega(k_{0})}{dk^{2}}\frac{t}{\Delta r_{0}} \label{eqn:delta_r}
\end{equation}
where $d^2 \omega(k_{0})/dk^{2}$ is the rate of change of the group velocity of the pulse with respect to $k$. Equation~\ref{eqn:delta_r} was fitted to the FWHM measurements given in the top panel of Figure~\ref{fig:cbf_disp}, allowing $d^2 \omega(k_{0})/dk^{2}$ to be determined for each passband.

The rate of expansion of the pulse $\Delta r(t)$ and the resulting value of $d^2 \omega(k_{0})/dk^{2}$ determined using the top panel of Figure~\ref{fig:cbf_disp} are given in Table~\ref{tbl:cbf_kins} for each passband studied. It is clear that the general expansion rate in each case is positive within error, indicating statistically significant pulse broadening. This implies that CBFs are dispersive pulses, confirming the results of \citet{Warmuth:2004ab,Veronig:2010ab} and \citet{Long:2011ab}.

\subsection{Temperature Dependence}
\label{subsect:sdo_cbf_discrepancies}

As a result of the very high cadence of \emph{SDO} across multiple passbands at varying peak emission temperatures, it was possible to derive the accurate kinematics of a CBF for each passband studied. These values are given in Table~\ref{tbl:cbf_kins}, with a spread in the kinematics estimated for the \emph{SDO} passbands, from $\sim$380 to $\sim$460~km~s$^{-1}$ and $\sim -128$ to $\sim -430$~m~s$^{-2}$ for the initial velocity and acceleration respectively. Although this variation could be ascribed to measurement errors, there is a statistically significant difference between the derived values, indicating that this discrepancy is real. This variation was examined by making a comparison with the peak emission temperatures ($T_{peak}$) of the different AIA passbands \citep[as given in Table~\ref{tbl:cbf_kins}, Figure~\ref{fig:aia_resp} and discussed by][]{Odwyer:2010ab}. 

An inverse relationship between the kinematics of the CBF and the peak emission temperature $T_{peak}$ for each passband was noted. \corr{As shown in Figure~\ref{fig:solar_model}, the general trend in the quiet corona is for temperature to increase with height while density and magnetic field strength decrease. This} implies that in cooler, more dense plasma the CBF has a higher velocity\corr{, although it should be noted that this is a general trend as plasma elements of different temperatures may co--exist at the same height in the corona}. While this result appears unusual at first glance, it is characteristic of a compressive pulse and combined with the observed pulse dispersion and deceleration indicates that the CBF is best described as a magnetohydrodynamic wave pulse. The randomly structured nature of the quiet corona suggests that any globally-propagating pulse must traverse magnetic field lines, indicating a fast-mode rather than slow-mode interpretation for the CBF.

However, the morphological signatures of the CBF across different passbands show some discrepancies that invite further investigation, particularly the simultaneous intensity decrease at 171~\AA\ and increase in the cooler 304~\AA\ passband. While an intensity decrease in the 171~\AA\ passband has previously been noted for \emph{TRACE} \citep{Wills-Davey:1999ve} and \emph{SDO} \citep{Liu:2010ab}, this was not apparent in \emph{STEREO} data due to the broad nature of the temperature response curve for this passband \citep[see e.g.,][]{Raftery:2009zr}. This drop in 171~\AA\ emission  (visually identified here but not tracked) has previously been characterised as evidence of plasma heating from 171~\AA\ into the 193, 211, and 335~\AA\ passbands \citep{Wills-Davey:1999ve,Liu:2010ab}. This heating implies that the CBF pulse is coronal, an observation consistent with the height measurements made by \citet{Kienreich:2009ab} and \citet{Patsourakos:2009ab}.

However, this implication is complicated by the observed increase in 304~\AA\ emission, as this passband is dominated by two chromospheric \ion{He}{2} lines at 303.781 and 303.786~\AA. Although there is also a coronal \ion{Si}{11} emission line at 303.33~\AA\ (cf. Section~\ref{sect:first_obs_conc}), \citet{Odwyer:2010ab} have noted that this line does not make a significant contribution to AIA quiet Sun observations. This suggests that the observed intensity increase is most likely due to \ion{He}{2} emission. 

The formation mechanism of \ion{He}{2} emission has been the subject of detailed investigation over many years \citep[see e.g.,][]{MacPherson:1999ab,Andretta:2003ab,Jordan:2007ab} as a result of its complex nature. Analysis has suggested that it is formed by collisional excitation resulting from thermal electrons in the quiet corona. This thermal excitation effect could be enhanced by the increased temperature gradient caused by passage of a compressive pulse in the solar corona, producing the observed 304~\AA\ intensity increase. Consequently, the CBF would be a coronal pulse, consistent with EUV observations, height measurements and the observed drop in 171~\AA\ intensity.

\subsection{Coronal Seismology}
\label{subsect:sdo_cbf_seismology}

The passband--dependent kinematics and the observed temporal variation in the width of the CBF pulse indicate that the pulse morphology is significantly influenced by the plasma through which it propagates. While the variation in pulse morphology could be used to examine the physical nature of the pulse itself, an examination of how the plasma affects the kinematics for each passband can be used to directly quantify the characteristics of the quiet coronal plasma. For example, as noted in Section~\ref{subsect:mag_acous_waves}, the fast-mode wave speed is defined as,
\begin{equation}\label{eqn:v_fast_mode}
v_{fm} = \sqrt{v_A^2 + c_s^2},
\end{equation}
where the Alfv\'{e}n speed and sound speed are $v_A = B/(4 \pi n m)^{1/2}$ and $c_s = (\gamma k T/m)^{1/2}$ respectively. Here $B$ is the magnetic field strength, $n$ is the particle density, $m$ is the proton mass, $\gamma$ is the adiabatic index (typically $5/3$), $k$ is Boltzmann's constant and $T$ is the peak emission temperature ($T_{peak}$; the values used are given in Table~\ref{tbl:cbf_kins}).

The pulse kinematics and dispersion are consistent with the properties of a fast--mode wave, allowing the CBF to be treated as such. In this case, the final pulse velocity (i.e.:\ the velocity of the pulse when it can no longer be detected by the algorithm) can be taken as the fast-mode velocity of the given passband, since the pulse can not propagate below this velocity. These values are given in Table~\ref{tbl:cbf_kins} for each \emph{SDO} passband studied.

By taking the peak emission temperature of each passband as the temperature, only the magnetic field strength and density are unknown in the above equations for the sound speed and Alfv\'{e}n speed. Coronal magnetic field strength estimates typically involve extrapolating photospheric magnetic field measurements into the corona and are not very well constrained (particularly in the quiet Sun). However, coronal densities can be estimated using density sensitive line ratios \citep{Gallagher:1999ab} and are well-constrained.

The above equations can be rearranged in terms of the magnetic field strength $B$ as,
\begin{equation}\label{eqn:b_n_variation}
B = \sqrt{4 \pi n (m v^2_{fm} - \gamma k T_{peak})},
\end{equation}
allowing the quiet coronal magnetic field strength to be estimated using the derived CBF kinematics. This was done by combining the final velocity values given in Table~\ref{tbl:cbf_kins} with a range of typical quiet coronal densities \citep[$\sim$2--6$\times10^{8}$~cm$^{-3}$; see][for more details]{Wills-Davey:2007oa}. This produced an estimated magnetic field strength range of $\sim$1--2~G for the quiet solar corona. 

This estimated range of values is comparable to the value previously derived by \citet{West:2011ab}. In this case, \citet{West:2011ab} used detailed \emph{STEREO}/EUVI estimates of the kinematics of a CBF pulse and combined them with accurate measurements of the density of the corona through which the pulse propagated made using the Extreme ultraviolet Imaging Spectrometer (EIS) instrument onboard the \emph{Hinode} spacecraft. \corr{The value of 1--2~G is also comparable to the value of $\sim$1.5--3.5~G observed by \citet{Kienreich:2011ab}, who used four consecutive CBF eruptions originating from the same active region over a period of eight hours and observed by \emph{STEREO}--B. By determining the velocity of each pulse, \citet{Kienreich:2011ab} were able to derive the magnetosonic Mach number for each pulse and estimate the coronal magnetic field strength using a Saito coronal density model. The good agreement of the range estimated here with the work of \citet{West:2011ab} and \citet{Kienreich:2011ab} indicates that the assumptions made here are correct and CBFs can be used to probe the physical characteristics of the plasma through which they propagate. }

\section{Discussion and Conclusions}
\label{sect:sdo_cbf_disc_and_conc}

The CBF detection algorithm outlined by \citet{Long:2011ab} was applied to contemporaneous EUVI and high cadence AIA observations of the 2010~August~14 CBF event. This allowed the accuracy of previous CBF kinematics estimates to be examined and was used to probe the physical properties of the pulse and the plasma through which it propagated. A comparison of pulse kinematics estimated using the same passband but at different cadences indicated that the estimates of the pulse kinematics are influenced by the observing cadence, supporting the conclusion of \citet{Long:2008eu}. Significant pulse deceleration was measured for this event in both EUVI and AIA data (despite the different cadence and spacecraft positions), suggesting that it is characteristic of the phenomenon.

Analysis of the high cadence observations available for this event also suggested that the techniques previously used to estimate pulse kinematics may have been flawed. These techniques typically involved combining distance-time measurements from different passbands to calculate the general pulse kinematics for an event due to a paucity of data \citep[e.g., ][]{Long:2011ab,Patsourakos:2009ab,Kienreich:2009ab,Veronig:2010ab}. While this was necessary to derive kinematics from the small data-sets available, the results presented here indicate that this approach underestimated the kinematics of the CBF. It may have also masked the amount of acceleration experienced by the pulse, and did not detail the effect of the plasma on the pulse. 

In addition to the kinematics derived for the pulse, the algorithm used also allowed the temporal variation in pulse width to be examined using both AIA and EUVI data. Clear dispersion was observed in both sets of data, confirming the observations of \citet{Warmuth:2004ab,Veronig:2010ab} and \citet{Long:2011ab}. These repeated measurements of significant pulse broadening strongly indicate that CBFs have a dispersive nature which, allied to the traditional point-and-click techniques for identifying them, may have contributed to the uncertainty surrounding their acceleration. 

When both the dispersion and deceleration are considered, CBFs may be best described using a wave interpretation. Although ideal \corr{linear} MHD wave theory predicts a constant velocity pulse of constant width, deceleration and dispersion are consistent with the propagation of a pulse through a randomly structured medium. This behaviour has been observed in simulations performed by \citet{Murawski:2001ab} and \citet{Nakariakov:2005ab}, and is congruous with the randomly structured nature of the quiet corona. \corr{The observed deceleration and dispersion are also consistent with the interpretation of the CBF pulse as a large amplitude simple wave \citep[cf.][]{Vrsnak:2000p15,Vrsnak:2008p622}. In this case, the pulse is initially driven by the rapid expansion of the erupting CME or by rising flare loops, producing a non--linear pulse which steepens to form a shock front. Through geometric expansion, this shock front exhibits deceleration, dispersion and dissipation with propagation \citep[see e.g.,][]{Muhr:2011fk}.}

The observed dispersion also allowed an insight into the physical nature of the pulse. This was possible by treating the pulse as a linear superposition of sinusoidal waves within a Gaussian envelope. The velocity gradient across the pulse could then be quantified and combined with the measured variation in pulse width to determine the dispersion relation of the pulse. This is the first time that the dispersion relation of a CBF pulse has been determined and provides a unique tool for the analysis of CBFs and the solar corona.

The high cadence observations available from \emph{SDO} allowed the kinematics of the pulse to be determined in each of the four passbands studied. Statistically significant differences were noted for both the initial velocity and acceleration estimated from each passband; a unique result that supports the wave interpretation of CBFs. In particular, the pulse exhibited a compressive nature, appearing to propagate at a higher velocity with stronger deceleration in cooler, denser plasma. This is the first time that this property of CBFs has been observed, and is a direct consequence of the very high observing cadence capabilities of \emph{SDO}/AIA. 

The kinematic variation indicated that the propagation of the pulse was strongly affected by the plasma through which it propagated, and provided a simple diagnostic of the emitting plasma in each passband. This variation was used to determine the physical properties of the plasma through the application of coronal seismology \citep[e.g.,][]{Ballai:2008p1116}. Although this approach has previously been proposed as a way of directly probing the structure of the solar corona, it has been complicated by the uncertain physical nature of CBFs. The results presented here, in addition to recent work by \citet{Patsourakos:2010ab,Kienreich:2011ab} and \citet{Long:2011ab} strongly indicate that CBFs are fast-mode MHD waves, allowing them to be used to examine the environment through which they propagate. 

This analysis relies on the ability to determine typical coronal densities and the temperature of the different passbands using alternative techniques. With these values known or well--constrained, the magnetic field strength can be estimated using the CBF. This was done here for a range of typical coronal density estimates, producing an estimated magnetic field strength range of $\sim$1--2~G for the quiet Sun. This range of values is comparable to the value estimated by \citet{West:2011ab} \citep[and typically assumed for the quiet corona, e.g.,][]{Wills-Davey:2007oa}, indicating that CBFs can be used to directly probe the plasma through which they propagate. 

The results of this analysis are most compatible with the wave interpretation of a CBF pulse. The observed dispersion implies that CBFs are not accurately described by the soliton model proposed by \citet{Wills-Davey:2007oa}, while the CBF height range (on-disk near the limb over an extended time period in both \emph{SDO} and \emph{STEREO} observations) is inconsistent with the progressively higher emission predicted by \citet{Delannee:2008uq}. The multi-temperature emission does not match the low foot-point signature predicted by \citet{Attrill:2007vn} and there was no indication of the additional coronal Moreton wave predicted by \citet{Chen:2002rw}. 

Despite this, the initial driver producing the CBF remains uncertain and the subject of some debate. However, the CBF could be the result of the rapid over-expansion of the erupting CME bubble \citep[cf.][]{Patsourakos:2010bc} low down in the solar corona before decoupling and propagating freely. The high cadence observations available from \emph{SDO} will allow this issue to be resolved.		


\chapter{Conclusions and Future work} 
\label{chap:concs}


\ifpdf
    \graphicspath{{8/figures/PNG/}{8/figures/PDF/}{8/figures/}}
\else
    \graphicspath{{8/figures/EPS/}{8/figures/}}
\fi


\noindent 
\\ {\it 
What's next?
\begin{flushright}
Aaron Sorkin (The West Wing)\\
\end{flushright}
 }

\vspace{15mm}
Although ``EIT waves'' have been studied for almost 15~years, the relatively low observing cadence available from \emph{SOHO}/EIT has traditionally made analysis difficult, and has necessitated combination with observations from other instruments and passbands. While this was and is necessary for a thorough analysis, it has resulted in a widely varying description and interpretation of this phenomenon. Here, data from the \emph{STEREO} and more recently \emph{SDO} spacecraft \corr{have} been used to study coronal bright fronts in a more statistically rigorous manner and across a wider range of EUV passbands than previously possible.

The quality of data such as that available from \emph{STEREO} and \emph{SDO} has resulted in a deeper understanding of the inherent physics of the ``EIT wave''. It has been shown that CBF pulses exhibit statistically significant deceleration with propagation, while pulse broadening has also been displayed. This pulse broadening combined with the techniques traditionally used to study CBFs may have contributed to the uncertainty regarding the true kinematics of these disturbances.

The higher time cadence available from more recent spacecraft has also been shown to influence the estimated kinematics of the CBF pulses. Although this was suggested by the initial observations of \citet{Long:2008eu}, the contemporaneous CBF observations from \emph{STEREO} and \emph{SDO} (with its higher cadence) of the 2010~Aug~14 event have shown that this effect must be accounted for when analysing this phenomenon. The availability of higher cadence observations from \emph{SDO} is also necessitating the development of more rigorous, automated approaches to the identification of individual disturbances. The technique outlined here has been shown to provide accurate determinations of pulse kinematics, allowing the physical nature of individual pulses to be examined.

\section{Principal Results}
\label{sect:res_principle}

The original aim of this work was to study the ``EIT wave'' phenomenon, with a view to determining its true nature using purely observable basic physics. Although the traditional theoretical interpretation of this phenomenon is complex and divided between alternative points--of--view, this work was driven by observation, with the launch of the \emph{STEREO} spacecraft at the start of this project allowing a new perspective on these disturbances. It should be possible to distinguish between different interpretations by determining and examining the physical properties of the disturbances as different theories propose different observable properties. The \emph{STEREO} spacecraft also offered the chance to identify previously unknown characteristics and use these as an additional discrimination between different interpretations.

To this end, the project has been successful. Whereas original observations of CBF pulses returned kinematics estimates that were inconsistent with a fast-mode wave, it has been shown here that these underestimations may have been a consequence of the low observing cadence of \emph{SOHO}/EIT. It has also been shown that CBF pulses display clear deceleration and dispersion with propagation, both of which are wave properties consistent with the propagation of a pulse through a randomly structured medium. 

The observations of dispersion also allowed the physical nature of the pulse to be examined. Dispersion in a single Gaussian pulse can be studied by treating the pulse as a linear superposition of sinusoidal waves, allowing the dispersion relation of the pulse to be determined. This has important implications for the analysis of both CBFs and also seismology of the solar corona as it allows the variation in the morphology and kinematics of individual CBFs to be used to directly probe the structure of the solar corona. It also allows the basic physics of the pulse to be examined theoretically, potentially providing a way to improve theoretical modeling of CBFs.

CBFs have also been shown to be present in the \ion{He}{2} 304~\AA\ emission line for the first time. While this was originally considered to be due to the coronal \ion{Si}{11} emission line at $\sim$1.6~MK, it has been \corr{suggested} that this emission line is too weak in the quiet Sun to be a viable candidate. The pulse signature \corr{implies} that the CBF may not be a purely coronal feature, although more recent results indicate that the temperature gradient resulting from the propagation of a coronal pulse would be expected to produce an increase in \ion{He}{2} emission similar to that observed. The CBF signature observed in the 304~\AA\ passband \corr{can therefore be taken as} an indication that the CBF is a coronal pulse which produces a heating effect. This is consistent with the observations of passband--dependent kinematics made using \emph{SDO}.

\subsection{Wave Kinematics and Morphology}
\label{subsect:res_kins_morph}

This work has shown that the kinematics of CBFs are more complex than originally perceived. While they were originally modeled as constant velocity single pulses, it has been shown here that CBFs do exhibit deceleration with propagation\corr{, a result consistent with the previous observations by \citet{Warmuth:2004rm,Warmuth:2004ab,Warmuth:2005p757,Vrsnak:2006fk} and \citet{Veronig:2008p216}}. This deceleration is also passband dependent, indicating that the CBF is strongly influenced by the plasma through which it passes. This may help to explain the wide variation of CBF kinematics noted by multiple authors including \citet{Thompson:2009yq} and \citet{Wills-Davey:2007oa}.

As well as the clear deceleration shown by the pulses studied here, it was also shown that the observing cadence used to study the different CBFs strongly influences the derived kinematics. The very low pulse \corr{velocities} determined using data from \emph{SOHO}/EIT may have been a direct consequence of its low observing cadence. \corr{Although these low kinematics were a contributing factor in the number of theories proposed to interpret CBFs, the results presented here indicate that they are not inconsistent with the fast--mode wave model originally proposed to explain the Moreton and ``EIT wave'' phenomenon.} The effects of the low observing cadence are particularly evident for the event studied in Chapter~\ref{chap:sdo_cbf}, where the same event exhibits different kinematics when observed using \emph{STEREO}/EUVI and \emph{SDO}/AIA. 

The estimation of the CBF kinematics was also greatly assisted by the use of a statistically rigorous bootstrapping technique. This represents a huge improvement on the numerical differentiation techniques typically used to determine the kinematics of CBF pulses. It has been shown that the inherent nature of these techniques make them inappropriate for use with the small data sets typically derived from CBF observations.

The physical properties of CBFs have also been determined to a high degree of accuracy. It has been shown \corr{here} for multiple events that CBF pulses show broadening with propagation. While this \corr{had} been previously \corr{shown in observations made by other authors \citep[e.g.,][]{Warmuth:2004ab,Warmuth:2005p757,Muhr:2011fk,Veronig:2010ab,Warmuth:2001p77}}, the rigorous technique used here to identify the CBF pulse has shown conclusively that this is an inherent property of the disturbances. This has major implications for the study of CBFs as it implies that the theoretical soliton model proposed by \citet{Wills-Davey:2007oa} in particular is not appropriate for describing CBFs.

The deceleration and dispersion of the CBF pulses also raises issues for the pseudo-wave interpretations proposed by \citet{Chen:2002rw}, \citet{Attrill:2007vn} and \citet{Delannee:2008uq}. None of these proposed models are consistent with the observations of pulse reflection observed in the 2007~May~19 event \citep{Gopalswamy:2009p1527}, while neither deceleration or dispersion is predicted by the authors of these theories. The passband dependence of the pulse kinematics is at odds with the low coronal nature of the pulse predicted by \citet{Attrill:2007vn}, while only single pulses were observed in each case studied, with no evidence of the coronal Moreton wave predicted by \citet{Chen:2002rw}.

Thus, basic physical principles have been used to show that the best representation of the CBF pulse is as a linear superposition of sinusoidal waves within a Gaussian wavepacket propagating in a randomly structured medium. This interpretation is driven by the various observations presented here, and is consistent with the observed deceleration and dispersion. This interpretation has also been used to quantify the dispersion relation of the pulse and the effect of the pulse on the local medium (and vice versa).

\subsection{CBF Identification}
\label{subsect:res_ident}

Although the kinematics and morphology of CBF pulses have important implications for our understanding of the solar corona, these results are predicated on an unbiased approach to the identification and tracking of ``EIT waves''. This phenomenon is inherently difficult to identify and track in individual images as a direct consequence of the diffuse nature of the disturbance. The result of this has been to complicate analysis of these disturbances and produce the wide variation in theoretical interpretations discussed in Section~\ref{chap:cbfs}. 

Whereas previous work has relied on point-and-click identification of a CBF in running difference images, it has been \corr{suggested} that this produces an inaccurate estimate of the pulse location \citep{Attrill:2010ab}. This approach is also entirely user-dependent, with the scaling required to highlight the pulse and the actual identification of the pulse itself highly variable and unquantifiable. \corr{However, it has been suggested by \citet{Muhr:2011fk} that the kinematics obtained using both an intensity profile approach and a visual tracking technique are comparable. In this case, the front of the pulse was used in both cases rather than the pulse centroid, with the pulse positions obtained from both approaches matching within error. However, this approach can mask the presence of deceleration in a dispersive pulse and give rise to inaccurate pulse properties (cf. Section~\ref{subsect:wave_properties_broadening}). The use of running--difference images can also complicate pulse identification if the pulse position overlaps between images.} This issue was overcome \corr{in this work} through the use of percentage base difference images, which remove the coronal background to highlight the pulse using \corr{a nominally} inherent scaling.

These images were combined with an unbiased, semi-automated technique which was proposed for the identification and tracking of CBFs. This allowed the pulses to be identified in a consistent manner, and enabled their deceleration and dispersion to be quantified. The technique was subsequently applied to data from both the \emph{STEREO} and \emph{SDO} spacecraft, exhibiting an ability to return accurate pulse kinematics and morphological variation regardless of spacecraft or event. \corr{However, it should be noted that there are issues with the intensity profile technique as currently implemented. The orientation of the arc used to determine the intensity profile along with the source point used to define the crossing point of the arc boundary are defined by the user. The detection of the pulse can also be influenced by stationary bright features as noted in Section~\ref{subsect:sdo_cbf_broadening}. As discussed in Section~\ref{subsect:future_algorithm}, it is intended to improve the detection algorithm to negate these issues.}

\section{Outstanding Questions and Future Work}
\label{sect:future_work}

Although this work has provided many answers to the ``EIT wave'' question, there remain many unresolved issues with these disturbances. CBFs remain rare phenomena, even by comparison with solar flares and coronal mass ejections. Their exact relationship with these other solar phenomena also remains uncertain, although observations from \emph{SDO} may help to resolve this issue \citep[e.g.,][]{Patsourakos:2010ab}. From observation, CBFs are strongly associated with specific active regions, with AR~10956 producing $\sim$6 CBFs over the course of $\sim$5 days in May~2007 while similar active regions produce no eruptions. \corr{Similarly, \citet{Kienreich:2011ab} reported observations of four CBF pulses erupting from AR~11067 over the course of an 8~hour period from 28--29~April~2010.} The reasons for this selectivity remain unknown, but could potentially have a major role to play in predicting space weather. 

\subsection{What Causes ``EIT Waves''?}
\label{subsect:future_eit_waves}

Although the evidence continues to mount that CBFs are waves as originally suggested, the exact nature of their initiation remains enigmatic. This is partly a consequence of the previously low temporal cadence of the various observing instruments but also arises from the rare nature of CBFs. The strong relationship of CBFs to CMEs suggests that it is most likely the initial driver, although the low rate of occurrence suggests that this is not a simple relationship. As with the association of CBFs with active regions, there may be a number of criteria that must be met before a CME eruption from an active region produces a CBF.

These criteria can only be identified through the use of a large sample of CBF eruptions. While the analysis of a single eruption may suggest some criteria, these may be negated by the analysis of an alternative eruption event. This effect can be seen in the previous work done on this topic, where theories have been proposed based on single events. A thorough analysis of CBFs must be performed on a large sample of events using an automated approach to negate individual bias.

\subsection{How do ``EIT Waves'' Affect the Corona?}
\label{subsect:future_temp_analysis}

The effect of the passage of a CBF on the background corona is most apparent in the work outlined in Chapter~\ref{chap:sdo_cbf}, where the plasma variation produces different derived kinematics for each passband studied. This effect has never been previously noted and may provide a link between the coronal ``EIT wave'' and the chromospheric Moreton wave. This has important implications for the study of CBFs and also for the study of the solar corona through the use of coronal seismology. By quantifying the effect of the plasma on the pulse parameters it should be possible to provide a way to study the corona directly using CBFs.

The different passbands observed by the AIA instrument onboard \emph{SDO} also allow the temperature and density variation of the solar corona and a CBF to be studied. By examining these parameters before and after the passage of a CBF, it should be possible to identify the processes involved in the propagation of the CBF pulse, as well as providing a way to determine the validity of different theoretical interpretations. The launch of \emph{SDO} makes this an area of rich potential over the next few years.

\corr{Although \emph{SDO} provides seven EUV passbands with very high cadence and spatial resolution, the true variation in temperature and plasma properties with the passage of a CBF pulse can only be studied to a high degree of accuracy using spectroscopy due to the broadband nature of EUV imagers. As a result, instruments such as the Extreme ultraviolet Imaging Spectrometer (EIS) onboard the \emph{Hinode} spacecraft will play a vital role in the determination of the basic physics at play in the corona and in a CBF pulse.}

\subsection{How are ``EIT Waves'' Related to Other Propagating Disturbances?}
\label{subsect:future_moreton}

The historical questions regarding the relationship between the coronal ``EIT wave'' and the chromospheric Moreton wave still remain unanswered. Although Moreton wave kinematics are typically estimated to be much larger than those of an ``EIT wave'', the results in Chapter~\ref{chap:sdo_cbf} suggest that this could be influenced by plasma variations. However, observations of Moreton waves remain rare and as a result the relationship between the two phenomena remains elusive.

The \emph{SDO} spacecraft provides an opportunity to observe CBFs with a cadence comparable to that used for studying Moreton waves. As a result, a well--observed CBF/Moreton wave event should provide a unique insight into the processes involved in this phenomenon.

\subsection{Automated CBF Detection}
\label{subsect:future_algorithm}

The study of CBFs and the different open questions raised in the previous sections is entirely dependent on a rigorous technique for identifying and tracking CBFs. Quantifying the basic physical principles at play in a CBF pulse requires removing all associated error from the different measurements (where possible). Although there are inherent errors associated with the identification of features in images, the larger, unquantifiable human errors can be removed through the use of an automated approach similar to that proposed here.

Although rigorous, the algorithm proposed here is only semi-automated and does have room for improvement. In particular, the identification of the source point and the orientation of the arc sector are currently user-defined. This needs to be fully automated, with the source taken as the location of the associated flare or some appropriate alternative, and the direction of the arc modified to include analysis of the entire Sun from that point. The sheer size of the \emph{SDO} data also needs to be accounted for to streamline the algorithm and speed up processing.

In addition to the scientific level analysis, this algorithm could be modified to identify and track CBFs as part of a pipeline. This would ensure a consistent, bias-free CBF identification process and would allow the \emph{SDO} data archive (along with those from \emph{SOHO} and \emph{STEREO}) to be rigorously examined and processed. Answering the outstanding issues in this field requires a consistent technique that can be provided using this approach.		


\cleardoublepage
\phantomsection
\addcontentsline{toc}{chapter}{Bibliography}
\addtocounter{chapter}{1}

\chapter*{Bibliography}
\label{chap:bibliography}
\chaptermark{Bibliography}


\newenvironment{hanglist}[1][\parindent]{%
    \begin{list}{}{%
        \setlength{\leftmargin}{#1}
        \setlength{\labelwidth}{0pt}
        \setlength{\labelsep}{0pt}
        \setlength{\itemindent}{-#1}}
    }{%
        \end{list}
    }

\begin{hanglist}[0.3cm]
\item {\sc Golub, L. \& Pasachoff, J.~M.} (2009), \emph{The Solar Corona} (2nd ed.; Cambridge: Cambridge University Press)
\item {\sc Goosens, M.}, ed. (2003), \emph{An Introduction to Plasma Astrophysics and Magnetohydrodynamics} (Vol. 294 of {\it Astrophysics and Space Science Library}; Dordrecht: Kluwer Academic Publishers)
\item {\sc Korobeinikov, V.~P.} (1991), \emph{Problems of Point--Blast Theory} (1st ed.; New York, NY: American Institute of Physics)
\item {\sc Landau, L.~D. \& Lifshitz, E~M.} (1959), \emph{Fluid Mechanics} (Vol. 6 of {\it Course of Theoretical Physics} 1st ed.; Oxford: Pergamon Press)
\item \corr{{\sc Phillips, K.~J.~H., Feldman, U. \& Landi, E.} (2008), \emph{Ultraviolet and X-ray Spectroscopy of the Solar Atmosphere} (1st ed.; Cambridge: Cambridge University Press)}
\item {\sc Priest, E.~R.} (1987), \emph{Solar Magnetohydrodynamics} (1st ed.; Dordrecht: D.~Reidel)
\end{hanglist}




%


\bibliographystyle{jmb} 
\renewcommand{\bibname}{References} 
\renewcommand{\bibpreamble}{
\noindent
\corr{
Please note that the following references take the general form:\\
{\sc author(s)} (date). {\it Title}. Journal, {\bf Volume}, journal page. Thesis location page}
}
\bibliography{thesis_ref} 







\end{document}